\documentclass[en,12pt,fancy,hide,superscriptaddress]{eiccbook}
\usepackage{lineno}
\usepackage{float}
\usepackage{changes}
\usepackage{adjustbox}
\usepackage{rotating}
\usepackage{array}
\usepackage{multirow}
\usepackage{booktabs}
\usepackage{xcolor}
\usepackage{threeparttable}

\DeclareMathAlphabet{\mathpzc}{OT1}{pzc}{m}{it}

\usepackage{relsize}
\def\babar{\mbox{\slshape B\kern-0.05em{\smaller A}\kern-0.05em
    B\kern-0.05em{\smaller A\kern-0.05em R}}}
    
\usepackage{newtxtext}
\usepackage{newtxmath}    

\title{{\Huge\bf Electron-Ion Collider in China
}\\ \vfill{\large \today}\newpage
}

\author[1]{Daniele P. ANDERLE}
\author[2]{Valerio BERTONE}
\author[3,4]{Xu CAO}     
\author[5]{Lei CHANG}                
\author[6]{Ningbo CHANG}              
\author[7]{Gu CHEN}                
\author[3,4]{Xurong CHEN}
\author[8]{Zhuojun CHEN}               
\author[9]{Zhufang CUI}              
\author[8]{Lingyun DAI}               
% ...

\affil[1]{Guangdong Provincial Key Laboratory of Nuclear Science, Institute of Quantum Matter, South China Normal University, Guangzhou 510006, China} 
\affil[2]{IRFU, CEA, Universit\'e Paris-Saclay, F-91191 Gif-sur-Yvette, France}
\affil[3]{Institute of Modern Physics, Chinese Academy of Sciences, Lanzhou 730000, China}    
\affil[4]{University of Chinese Academy of Sciences, Beijing 100049, China}
\affil[5]{Nankai University, Tianjin 300071, China}
\affil[6]{Xinyang Normal University, Xinyang 464000, China}
\affil[7]{School of Physics and Materials Science, Guangzhou University, Guangzhou 510006, China}
\affil[8]{Hunan University, Changsha 410082, China}
\affil[9]{Nanjing University, Nanjing 210093, China}  
% ...

\begin{document}

\date{}
\pagestyle{empty}
\pagestyle{fancy}
\maketitle



\begin{center}
\vspace{2cm}
{\bf\large Abstract}
\end{center}

\indent
Lepton scattering is an established ideal tool for studying inner structure of small
particles such as nucleons as well as nuclei. As a future high energy nuclear physics
project, an Electron-ion collider in China (EicC) has been proposed. It will be
constructed based on an upgraded heavy-ion accelerator, High Intensity heavy-ion
Accelerator Facility (HIAF) which is currently under construction, together with a
new electron ring. The proposed collider will provide highly polarized
electrons (with a polarization of $\sim$80\%) and protons (with a
polarization of $\sim$70\%) with variable center of mass energies from 15 to 20 GeV and the luminosity of (2-3)
$\times$ 10$^{33}$ cm$^{-2}$ s$^{-1}$. Polarized deuterons and Helium-3, as well as unpolarized ion beams from Carbon to Uranium, will be also available at the EicC.      

The main foci of the EicC will be precision measurements of the structure of
the nucleon in the sea quark region, including 3D tomography of nucleon; 
the partonic structure of nuclei and the parton interaction with the
nuclear environment; the exotic states, especially those with heavy flavor
quark contents. In addition, issues fundamental to understanding the origin of mass
could be addressed by measurements of heavy quarkonia near-threshold
production at the EicC. In order to achieve the
above-mentioned physics goals, a hermetical detector system will be constructed with
cutting-edge technologies. 

This document is the result of collective contributions and valuable inputs from experts across the globe.
The EicC physics program complements the ongoing scientific
programs at the Jefferson Laboratory and the future EIC project in the United States.
The success of this project will also advance both nuclear and particle physics as
well as accelerator and detector technology in China.

\tableofcontents

\chapter{Executive summary}


\section{Physics highlights}

\def\xslash#1{{\rlap{$#1$}/}}
\def \p {\partial}
\def \dd {\psi_{u\bar dg}}
\def \ddp {\psi_{u\bar dgg}}
\def \pq {\psi_{u\bar d\bar uu}}
\def \jpsi {J/\psi}
\def \psip {\psi^\prime}
\def \to {\rightarrow}
\def\bfsig{\mbox{\boldmath$\sigma$}}
\def\DT{\mbox{\boldmath$\Delta_T $}}
\def\xit{\mbox{\boldmath$\xi_\perp $}}
\def \jpsi {J/\psi}
\def\bfej{\mbox{\boldmath$\varepsilon$}}
\def \t {\tilde}
\def\epn {\varepsilon}
\def \up {\uparrow}
\def \dn {\downarrow}
\def \da {\dagger}
\def \pn3 {\phi_{u\bar d g}}

\def \p4n {\phi_{u\bar d gg}}

\def \bx {\bar x}
\def \by {\bar y}

\begin{figure}[htbp]
\centerline{%
\includegraphics[clip, width=0.60\textwidth]{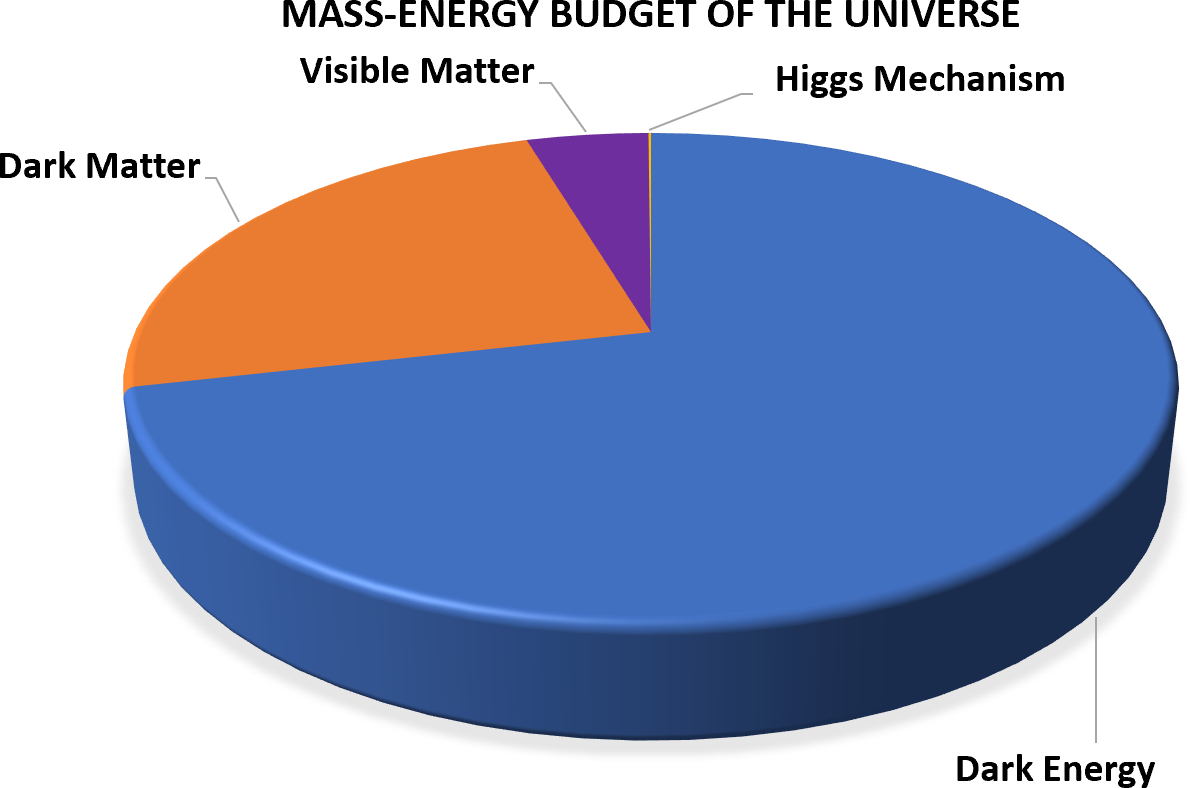}}
\caption{\label{FigMassEnergy}
The mass-energy budget of the Universe determined by Wilkinson 
Microwave Anisotropy Probe (WMAP) \cite{Seife2038}.
}
\end{figure}

The study on the inner structure of matter and fundamental laws of interactions
has always been one of the research forefronts of natural science. 
It not only allows mankind to understand the underlying laws of nature, 
but also promotes various advances in technologies. Considering the mass-energy budget of the Universe, illustrated in Fig.\,\ref{FigMassEnergy}: dark energy constitutes 71\%; dark matter is another 24\%; and the remaining 5\% is visible material.  Little is known about the first two: science can currently say almost nothing about 95\% of the mass-energy in the Universe. On the other hand, the remaining 5\% has forever been the source of everything tangible, which can be beautifully described within the Standard Model. 

\begin{figure}[htbp]
\begin{center}
\includegraphics[width=0.95\textwidth]{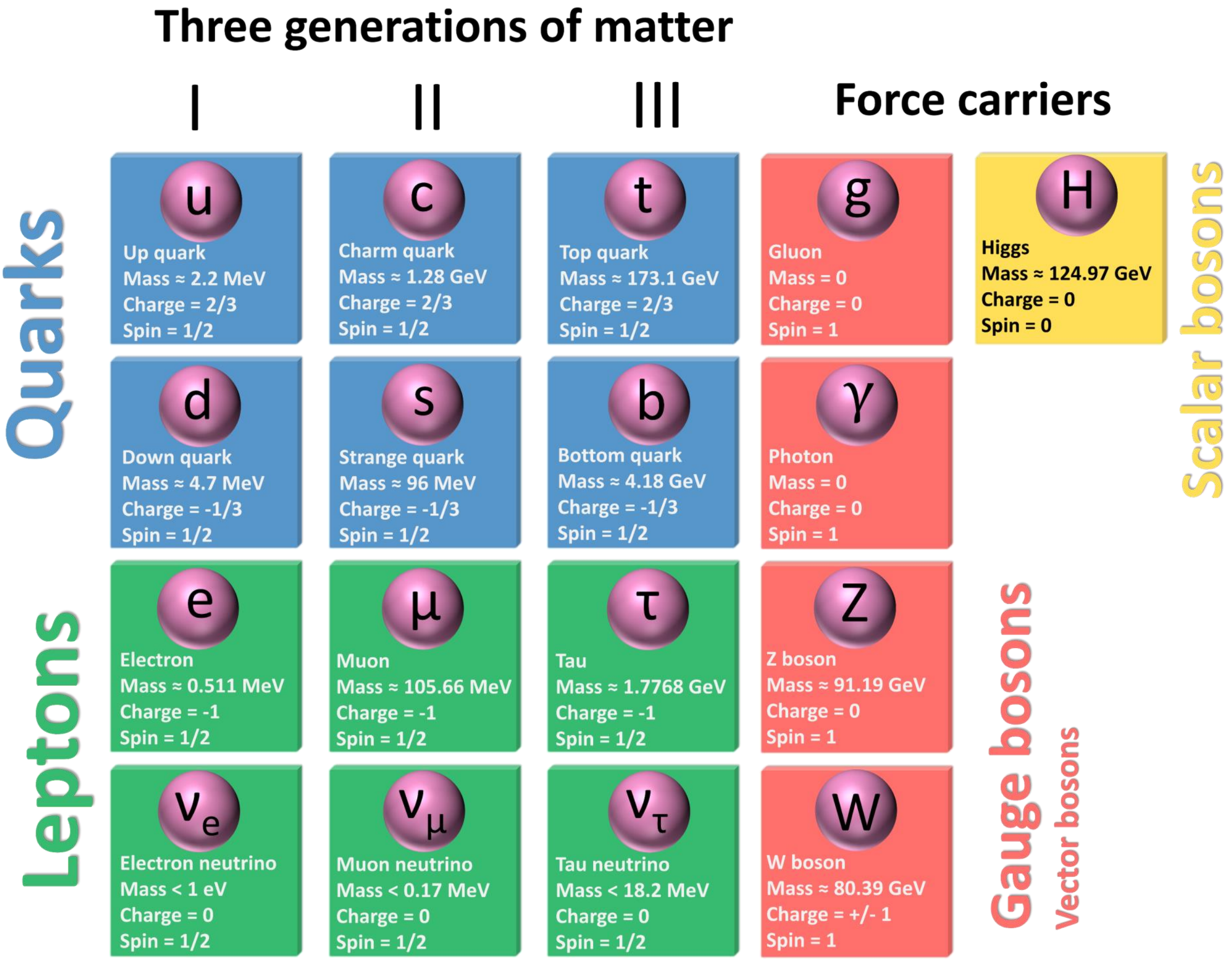}
\end{center}
\caption{The Standard Model of elementary particles. }
\label{figSM}
\end{figure}

One of the greatest achievements of physics in the 20th century is the invention of 
the Standard Model~\cite{Weinberg:1967tq,Salam:1959zz,Glashow:1959wxa,Gross:1973ju,Gross:1974cs,Weinberg:1979sa}.
It is the theory describing the strong, electromagnetic, and weak interactions 
among elementary particles that make up the visible Universe. As shown in Fig.~\ref{figSM}, now we know that there are three generations of quarks and leptons in nature. 
The forces in the Standard Model are carried by the so-called force mediating gauge bosons, which 
are $\gamma$, $W^{\pm}$ and $Z^0$ for electro-weak interaction, and gluons $g$ for the strong interaction. 
The Higgs boson $\textrm{H}$ was introduced in the famous Higgs mechanism~\cite{Englert:1964et,Higgs:1964pj}
to explain the mass origin 
of the $W^{\pm}$ and $Z^0$ bosons, and it also generates the masses of quarks and leptons. Yet, amongst the visible matter, less-than 0.1\% is tied directly to the Higgs boson; hence, even concerning visible matter, too much remains unknown.

In particular, it is still challenging to quantitatively explain the origins of nucleon \emph{mass and spin}, which are two fundamental properties of building blocks of the visible matter. First, about 99\% of the visible mass is contained within nuclei~\cite{10.1093/ptep/ptaa104}. Within Standard Model, the protons and neutrons in nuclei are composite particles, built from nearly massless quarks ($\sim 1\%$ of the nucleon mass) and massless gluons. An immediate question then arises: How does $99$\% of the nucleon mass emerge? 
Besides the mass issue, despite of many years of theoretical and experimental efforts, the quantitative decomposition of nucleon spin in terms of quark and gluon degrees of freedom is not yet fully understood. To address these fundamental issues, we have to understand the nature of the subatomic force between quarks and gluons, and the internal landscape of nucleons.

The underlying theory, which describes the strong interactions between quarks and gluons, is 
known as Quantum Chromodynamics (QCD)~\cite{Callan:1977gz}. As a non-Abelian gauge theory, QCD has the extraordinary properties of 
asymptotic freedom at short distance \cite{Gross:1973id,Politzer:1973fx} and color confinement at long distance. The strong force mediated by gluons is weak in hard scatterings with large momentum transfers. On the other hand, it has to be incredibly strong to bind quarks together within the tiny space of a nucleon. Confinement is crucial because it ensures stability of the proton. Without confinement, protons in isolation could decay; the hydrogen atom would be unstable; nucleosynthesis would be accidental, with no lasting consequences; and without nuclei, there would be no living Universe. All in all, the existence of our visible Universe depends on confinement. 

In QCD, the proton mass is usually decomposed into several elements in terms of quark and gluon degrees of freedom.
Specifically, it is believed that the nucleon mass can be almost entirely derived from the kinetic energy of quarks and gluons, interactions between them, as well as other novel dynamical effects of QCD. 
Similarly, despite being composite particles, nucleons have a constant spin of $1/2$ which is an intrinsic property like electric charge. It is extremely fascinating to note that proton spin can manifest itself from the many-body system of quarks and gluons. In addition to the spin contributions of quark and gluon, which has been measured in certain kinematic regions, the orbital angular momentum contributions due the orbital motions of quark and gluon have been shown to be indispensable for the proton spin.     

Hence, QCD should be the physical mechanism responsible for the majority of visible matter in the Universe.  To gain a more comprehensive understanding of the internal partonic structure of a nucleon, explore the nature of color confinement and ultimately explain the emergence of the nucleon mass and spin, we certainly need to expand the scope of our current experiments and enrich our knowledge on the dynamics of the strong interaction, especially the non-perturbative aspects of QCD. 

In the following, a few highlighted physics topics, highly relevant to above mentioned essential QCD physics, that EicC can significantly contribute to 
will be discussed briefly. For the detailed discussions regarding physics, accelerators, and detectors for the EicC project, please refer to the following chapters of this document \footnote{By default, the natural unit system is used in all the physics discussions and plots.}.

\par\vskip20pt 
\noindent
\subsection{Partonic structure and three-dimensional landscape of nucleon} 

\par
\begin{figure}[htbp]
\begin{center}
\includegraphics[width=0.9\textwidth]{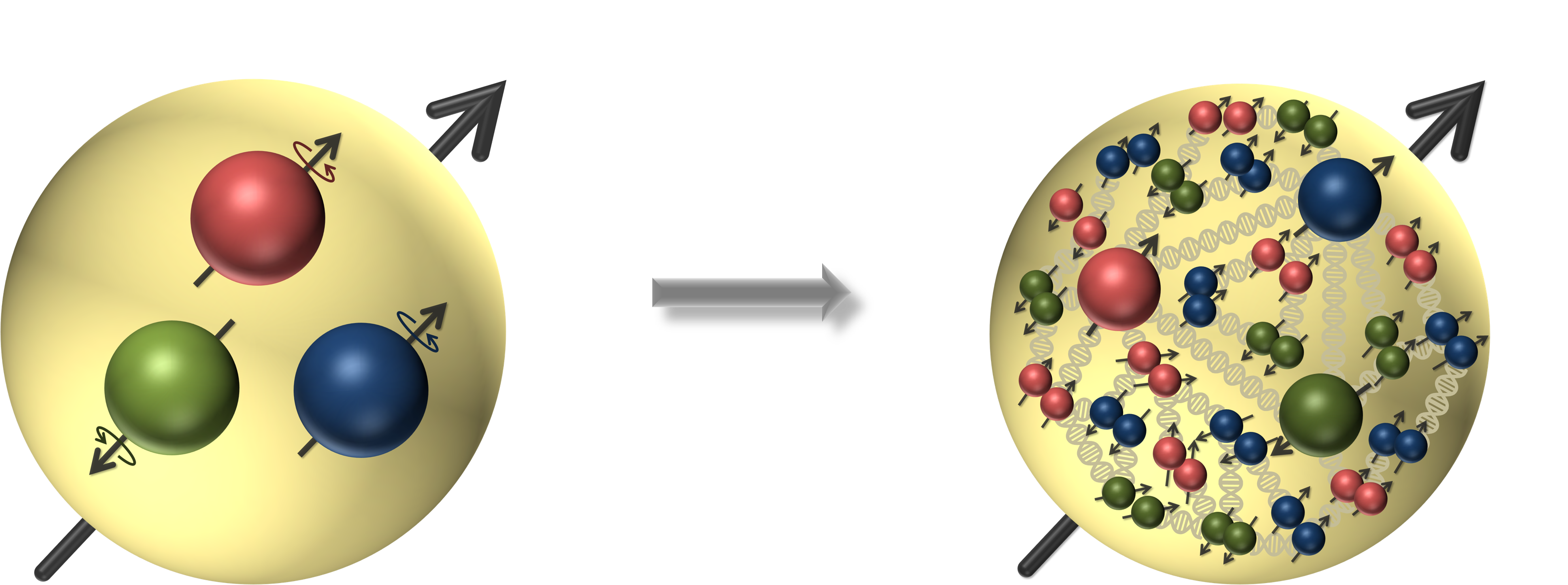}
\end{center}
\caption{Illustration of the quark and the partonic structure of the proton. }
\label{figparton}
\end{figure}

 
In the naiive constituent quark model~\cite{GellMann:1964nj,Zweig:1981pd}, nucleons are considered as the bound states of $u$- and $d$- quarks. 
The proton (neutron) corresponds to a $uud$-state ($udd$ state). These quarks are known as valence quarks.  
However, due to the quantum property of QCD, quarks
can radiate gluons, and these gluons, in turn, can fluctuate into quark-antiquark pairs. 
Therefore, a nucleon is a composite object containing quarks, antiquarks, and gluons.
Besides valence quarks (and possible intrinsic quarks), there are also sea quarks coming out of quantum fluctuations.
Especially, when the probing scale becomes smaller as the energy scale goes higher, one 
sees more sea quarks comparing to valence quarks, as illustrated in Fig.~\ref{figparton}. 
Moreover, compared to the simple 
picture of the constituent quark model, the underlying dynamics among quarks/gluons is a lot more interesting and intricate, and offers much more important information regarding the internal structure of nucleons as a composite many-body system.

In high-energy scatterings, the proton can be viewed as a cluster of high energy quarks and gluons, which are collectively referred to as partons. The probability distributions of partons within the proton are called the 
parton distribution functions (PDFs). In general, PDFs give the probabilities of finding partons (quarks and gluons) in a hadron as a function of the momentum fraction $x$ w.r.t. the parent hadron carried by the partons. Due to the QCD evolution, quarks and gluons can mix with each other, and their PDFs depend on the resolution scale. When the resolution scale increases, the numbers of partons and their momentum distributions will change according to the evolution equations. These evolution equations can be derived from the perturbation QCD, although PDFs themselves are essentially non-perturbative objects. Thanks to QCD factorization theorems, PDFs can be extracted from measurments of
cross-sections and spin-dependent asymmetries.

The partonic structure of the nucleon was firstly studied in experiments of electron-nucleon Deeply Inelastic Scattering(DIS). 
Since electrons are point-like particles and they do not participate in the strong interaction,
they are the perfect probe for studying the internal structure of hadrons in high energy scatterings. Therefore, the DIS experiment is also known as the 
``Modern Rutherford Scattering Experiment'', which opens up a new window to probe the subatomic world. In 1969, the pioneer DIS experiments at 
SLAC discovered the so-called Bjorken scaling~\cite{Bloom:1969kc}, which showed that the proton is composed of point-like partons with 
spin $1/2$ (which are known as quarks afterward). Starting from DIS with unpolarized fixed targets, DIS experiments are later extended to 
unpolarized collider experiments and fixed-target experiments with polarized beam and targets.
These DIS experiments have revolutionized our understanding of the subatomic structure of nucleons and nuclei. Later on, high energy DIS
experiments observed the violation of Bjorken scaling~\cite{Chang:1975sv}, which indicates the existence of gluon and QCD evolution mentioned above. All these results across a 
wide range of energy scales have verified that QCD is the correct theory for the strong interaction between quarks and gluons within hadrons. In addition, 
within the current experimental accuracy, lepton and quark are still point-like particles at the scale of $10^{-3}$ fm, which is 
one-thousandth of the size of the proton.

With better experimental precisions, our understanding of nucleon structure
continues to improve even in unpolarized PDFs. Furthermore, many interesting phenomena, such as the isospin asymmetry of $\bar{u}$ and $\bar{d}$ quark distributions and the asymmetry between strange and anti-strange quark distributions in the proton, were discovered. These phenomena are still compelling issues in medium and high energy physics research. 

In the wake of the development of polarized source in the 1970s, the study of the nucleon spin structure became possible by exploring the helicity distributions of quarks and gluons, also defined as the longitudinally polarized PDFs analog to their unpolarized counterparts discussed above, from high-energy scattering processes involving polarized leptons and/or polarized nucleons.  
A lot more interesting phenomena have been unraveled by polarized DIS experiments.  
One of them is the so-called ``proton spin crisis''.  
Experimental data showed that the sum of the spin from quarks and anti-quarks is only a small fraction of the total spin of a proton.   
It triggered a series of experimental and theoretical investigations on the origin of the proton spin.  
From the QCD perspective, we now know that the proton spin is built up from the spin and orbital 
angular momenta of quarks and gluons. 
Currently, except the quark spin contribution, other decomposed contributions in the spin sum rule, especially the ones from orbital angular momenta, are largely unexplored. 
Through semi-inclusive DIS and other interesting processes, recent experimental and theoretical developments have enabled us to extend our research on nucleon structure from one-dimensional PDFs to three-dimensional imaging. These have been providing us new insights into the proton spin puzzle. 

\par
Currently, there are two immediate and important issues in the research frontier of nucleon structure:
1) The precision measurement of the one-dimensional spin structure of the polarized nucleon;
2) The study on the three-dimensional imaging of the partonic structure of the nucleon. 

An interesting question when studying the one-dimensional spin structure of the nucleons is how to clearly decompose the individual contributions from different quark flavors. 
Despite the large uncertainty, the recent measurement at Relativistic Heavy Ion Collider (RHIC)
implies that the sea quark helicity distributions also have flavor asymmetries. 
Furthermore, the polarized quark distribution of different flavors, especially for sea quarks, still have large uncertainties. 
This directly imposes a challenge to our efforts to understand the proton spin structure. 
Therefore, the precise determination of various quark helicity distributions
is a fundamental issue which is needed to be addressed.

\par 

In the meantime, three-dimensional imaging of the parton structure has attracted a lot of attention as well. By additionally measuring the transverse momentum and angular distribution of the final hadron in DIS, 
one can extract important information about initial transverse momentum distributions of partons in the incoming nucleon, thus 
explore the internal three-dimensional structure of the nucleon in the momentum space. 
Meanwhile, through some exclusive processes, in which all the particles are measured, 
one can access the three-dimensional spatial distributions of partons.
In general, the internal three-dimensional structure of the nucleon in the momentum and coordinate space can be characterized by the Transverse-Momentum-Dependent parton distribution functions (TMDs) and Generalized Parton Distributions (GPDs), respectively. 
Compared to one-dimensional PDFs, 
these more sophisticated parton distribution functions
encode much more abundant information about the internal structure of the nucleon. For example, they can allow us to access the orbital angular momenta of partons and the quantum effect of multi-parton correlations. Future experimental efforts, especially the high precision measurements, can certainly have a profound impact on the theoretical development of TMD and GPD physics. 

EicC, together with existing experiments at Jefferson Lab, CERN COMPASS, BNL RHIC, Fermi-Lab, and
the proposed EIC in the US, can offer significant insights into the three-dimensional landscape of internal 
structure of the proton and other hadrons, and provide us important clues on how the mass and spin as well as other interesting properties
of proton emerge from the quark and gluon degrees of freedom. 
\par 

\par\vskip20pt 
\noindent
\subsection{Partonic structure of nuclei} 
\par
One of the biggest challenges in nuclear physics is how to study the nuclear structure in the partonic level using the QCD theory that has successfully described the partonic structure of a free nucleon. While the focus of studying partonic structure in free nucleons has been extended from the precisely known one-dimensional PDFs to the three-dimensional distributions such as TMD and GPD, the knowledge of the partonic structure in nuclei, however, remain largely unknown. 

The most outstanding reason for this gap is that the nucleons with their sizes much smaller than the size of a nucleus are interacting weakly with each other through the long-range interactions. Conventional models, such as the mean-field theory, can describe the nuclear structure in the nucleonic degree of freedom without introducing the partonic pictures. On the other hand, the partonic structure of a bounded nucleon in a heavy nucleus had been naively treated as the same as one in a free nucleon, until the discovery of the EMC effect.  In the 1980s, the European Muon Collaboration (EMC) at CERN used heavy nuclei as a high-density target to measure the PDFs. They discovered the measured cross-sections differed from ones using free nucleons, and meanwhile, observed these differences strongly depend on the nuclear numbers~\cite{Aubert:1983xm}. This lately-called EMC effect has been further studied at SLAC, HERMES, Fermi-Lab and lately JLab ~\cite{Benvenuti:1987az,  Gomez:1993ri, Geesaman:1995yd, Seely:2009gt} in the valance quark region ($0.3<x<0.7$) and the correlation with nuclear numbers were obtained. These experimental results also reveal much richer details at lower $x$ where anti-shadowing and shadowing effects are present. However, the physics origin of how the nuclear PDFs (nPDF) are modified in nuclei is still puzzling us and no single theoretical interpretation is satisfactory. A full understanding of the physics behind the EMC, anti-shadowing, and shadowing effects will open a door to describe the nuclear structure in QCD. An encouraging development in the last few years was the suggestion of possible connection between the EMC effect and the short-range correlations (SRC) which describe a case when nucleons are largely overlapping and strongly interacting with each other\cite{Weinstein:2010rt,Arrington:2012ax,Hen:2016kwk}. This new finding sheds a light to cover the gap between studying nuclear structure in the nucleonic level and the partonic level. 

The EMC effect implies that the distributions of valence quarks in the nucleus are modified. However, no existing experimental evidence suggests that the distributions of sea quarks and gluons in bounded nucleons are also modified in the nuclear medium. Joint research of theory and experiment is eagerly needed to obtain the precise global description of nPDFs of different quark and gluon flavors in the entire $x$ region for a wide range of nuclei, and finally unveil the physics origin of the EMC, anti-shadowing and shading effects. A power tool in the last many decades is to utilize the high-energy electrons in colliding with light to heavy nuclei and measure the inclusive DIS cross-sections by only detecting outgoing electrons. On top of that, one can also detect the additional outgoing hadrons which contain the information of the initial quark or gluon, and study their semi-inclusive DIS (SIDIS) cross-sections or other observable to decouple the nPDFs of different partonic flavors.

\par 
Another hot topic in high-energy eA physics is to understand the quark confinement. Quarks cannot exist alone but have to be combined with other quarks to form color-neutral hadrons such as mesons and nucleons. When a quark is struck by a high-energy particle, it will continuously interact with its surroundings via strong interaction, generate additional quarks and gluons, and eventually "fragment" into color-neutral hadrons or jets inside the nucleus or in the vacuum. This process is called hadronization. Studying the hadronization process has important implications for the formation of matter and even the evolution of the Universe. One can perform a detailed study of the hadronization physics by measuring the SIDIS processes in eA collision. With a wide variety of nuclei that serve as QCD laboratories, one can control the sizes of different nuclei so that the hadronization happens at varying depth inside the nuclei or the vacuum.

\par\vskip20pt
\noindent
\subsection{Exotic hadronic states} 

\begin{figure}[tb]
\begin{center}
\includegraphics[width=0.95\textwidth]{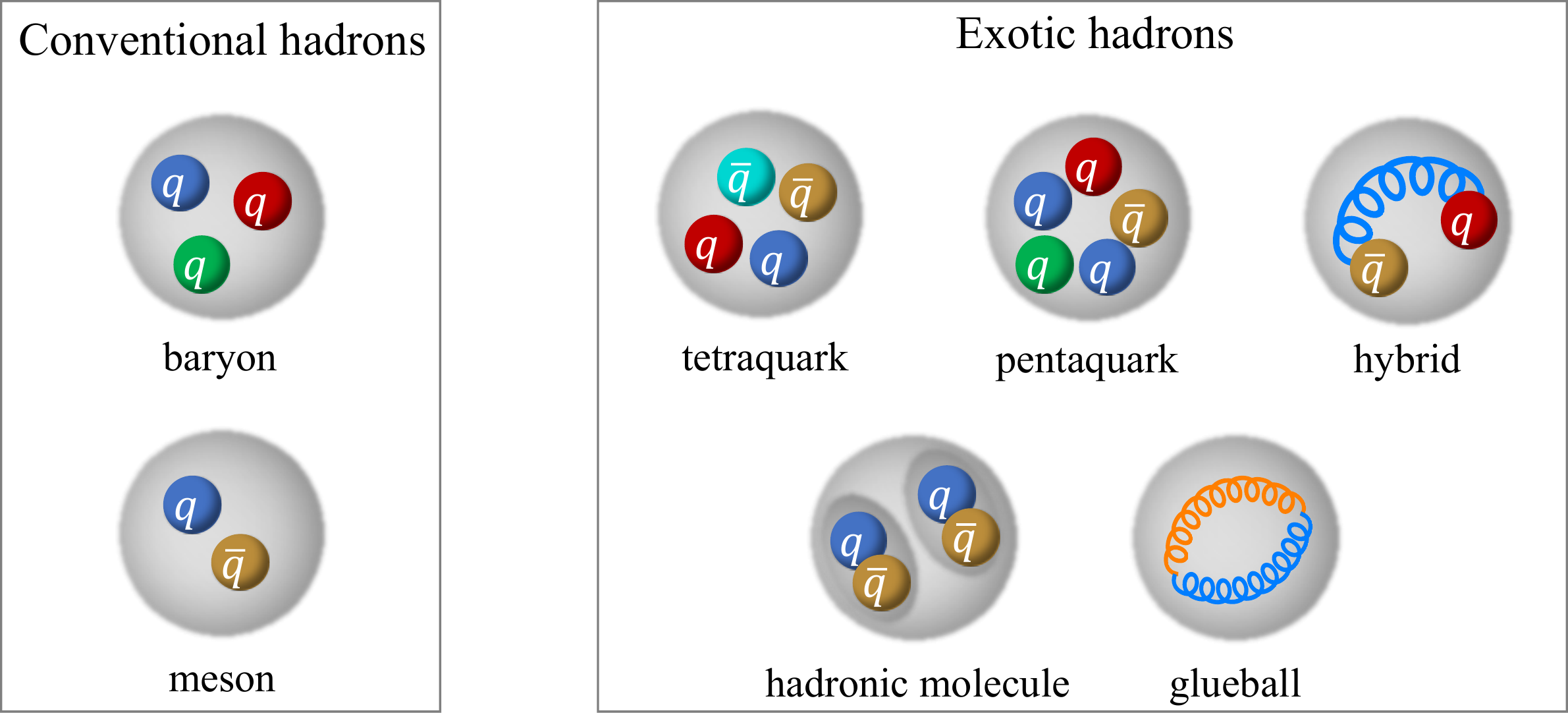}
\end{center}
\caption{Illustration of conventional and exotic hadrons. }
\label{figexotics}
\end{figure}

Quark model was invented before QCD to classify various hadrons composed of light (up, down and strange) quarks~\cite{GellMann:1964nj,Zweig:1981pd}. After incorporating the QCD dynamics, it was able to provide an excellent description of the mass spectrum of hadrons up to a few exceptions (see, e.g., Refs.~\cite{Godfrey:1985xj,Capstick:1986bm}).
In the traditional quark model, a meson is formed by a quark and an antiquark, and a baryon is formed by three quarks. Most of the hadrons discovered in the last century can be classified into flavor multiplets in the quark model. But quarks and gluons can constitute other types of hadronic objects: the so-called compact tetraquark  and pentaquark states contain more than three (anti-)quarks as a single colorless cluster; hadronic molecules are bound states of hadrons formed by the mediation of the strong force, just like that the deuteron is a proton-neutron bound state; there can be colorless states with both quark and gluonic excitations, {\it i.e.}, the hybrid states; glueballs composed of gluons. These different types of hadrons are shown in Fig.~\ref{figexotics}.
Such hadrons beyond the traditional quark model are collectively called exotic hadron states. 
Although such a classification is a quark model notation, the hadron spectrum as observed presents a grand challenge to understand from QCD, and the experimental search of exotic hadrons is one of the most important handles towards understanding how the massive hadrons emerge from the underlying nonperturbative strong interactions among quarks and gluons.

Since the beginning of the 21st century, experimental study on hadron states has made significant
progresses. Experiments such as BESIII (Beijing
Spectrometer III) at Beijing Electron-Positron Collider (BEPC) in China, Belle at KEK in Japan, \babar~at the SLAC National Accelerator Laboratory in US, LHCb at the Large Hadron Collider (LHC) in Europe and many others have reported  fascinating discoveries of candidates of exotic hadron states. These discoveries 
have opened up a new exciting window in the nonperturbative regime of QCD at the low-energy frontier of the Standard Model.
However, until now there is no unified picture for understanding the new experimental discoveries, and the internal structure of these states is still a mystery to be resolved.


EicC can contribute significantly in studying exotic hadron states, especially the charmonium-like states and hidden-charm pentaquarks, which can be produced abundantly. 
EicC has a unique place for studying their photoproduction, beyond the JLab 12 GeV programme.
In particular, given the existing measurements, the interpretation of some of the prominent candidates of hidden-charm tetraquarks and pentaquarks (either compact or of hadronic molecular type) is not unambiguous due to the 
the so-called triangle singularity contribution. Such singularities are due to the simultaneous on-shellness and collinearality of all intermediate particles in a triangle diagram and are able to produce resonance-like signals when the special kinematics required by the Coleman-Norton theorem~\cite{Coleman:1965xm} is fulfilled~\cite{Guo:2019twa}. 
However, for the photoproduction processes at EicC, the production mechanism is free of such kinematics singularities. 
Therefore, one can investigate the properties of pentaquark states and other hidden-charm hadrons in a more clear way. The energy coverage of EicC also allows for the seek of hidden-bottom exotic hadrons.
A clearer picture of the hadron spectrum is foreseen with the inputs from EicC.

\section{Polarized electron ion collider in China (EicC)}
The polarized electron ion collider in China (EicC) aims at achieving the highlighted physics goals presented above. It will be based on the existing High Intensity heavy-ion Accelerator Facility (HIAF). HIAF is the major national facility focusing on nuclear physics, atomic physics, heavy ion applications and interdisciplinary researches in China. It is designed to provide intense beams of primary and radioactive ions for a wide range of research fields. HIAF will be a scientific user facility open to researchers from all over the world that enables scientists with concerted effort to explore the hitherto unknown territories in the nuclear chart, to approach the experimental limits, to open new domains of physics researches in experiments, and to develop new ideas and heavy-ion applications beneficial to the society. HIAF is located in Huizhou City of Guangdong Province in south China. It is funded jointly by the National Development and Reform Commission of China, Guangdong Province, and Huizhou City. The total investment is about 2.5 billion in Chinese Yuan, including about 1.5 billion Yuan from the central government for facility construction and 1.0 billion Yuan from the local governments for infrastructure. The construction is scheduled for seven years, and the beam commissioning is planned for 2025. HIAF is a completely new facility that a series of upgrades for EicC in the future have been taken into consideration during the design stage, and its capability to run with EicC concurrently is also reserved.

\begin{figure}[htbp]
    \centering
\includegraphics[width=0.99\textwidth]{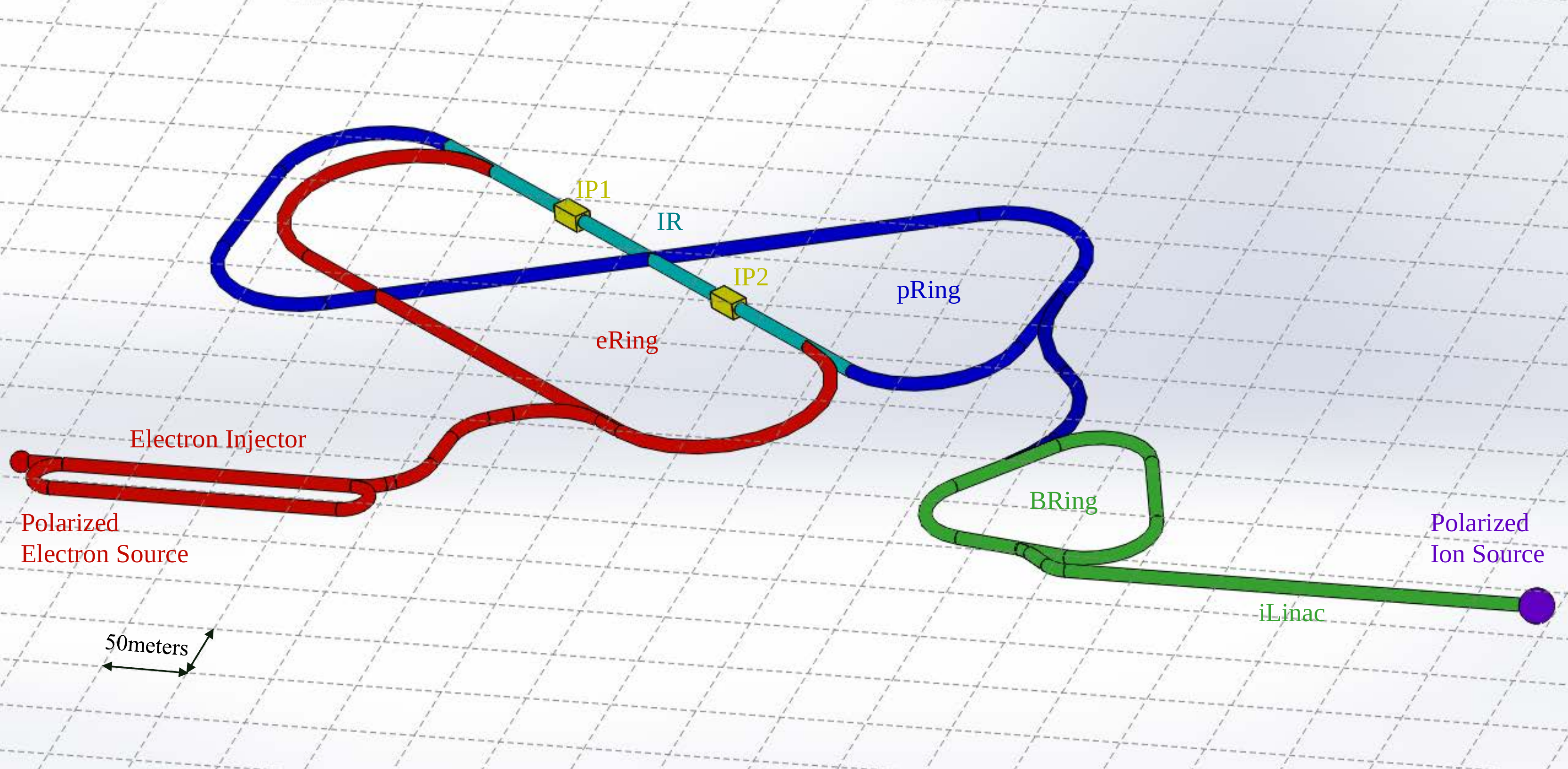}
    \caption{Accelerators in the EicC accelerator facility.}
    \label{fig:EicC_layout}
\end{figure}

EicC will adopt the scheme of circular colliders which includes a figure-8 shaped ion collider ring (pRing), an electron injector as well as a racetrack electron collider 
ring (eRing), as shown in Fig.\ref{fig:EicC_layout}. The center of mass energy of the EicC will range from 15 GeV to 20 GeV, with the luminosity higher than $2.0\times 10^{33}$  $\mathrm{cm^{-2}s^{-1}}$, and the average proton polarization about $70\%$, the average electron polarization about $80\%$ in the collisions of electrons with protons. The integrated luminosity is higher than $50$ $\mathrm{{fb}^{-1}}$ when the operating time accounts for $80\%$ of the entire year. All these parameters can satisfy the physics goals required. Available particles, including heavy ions, and their corresponding energy, polarization, luminosity, and integrated luminosity are listed in  Tab.\ref{tab:1.1}.

\begin{table}[htbp]
\centering
\caption{Available particles and their corresponding energy, polarization, luminosity and integrated luminosity.}
\resizebox{\textwidth}{!}{
\begin{tabular}
{p{0.12\textwidth}p{0.15\textwidth}p{0.15\textwidth}p{0.15\textwidth}p{0.2\textwidth}p{0.15\textwidth}}
\hline
Particle &
Momentum (GeV/c/u) &
CM energy (GeV/u) &
Average Polarization&
Luminosity at the nucleon level ($\rm{cm^{-2}s^{-1}}$) &
Integrated luminosity ($\mathrm{{fb}^{-1}}$)
\\ \midrule
$\rm{e}$ &
3.5&
&
80\%&

\\

$\rm{p}$&
20&
16.76&
70\%&
$\rm{2.00\times 10^{33}}$&
50.5
\\ 

$\rm{d}$&
12.90&
13.48&
Yes&
$\rm{8.48\times 10^{32}}$&
21.4
\\

$\rm{^{3}He^{++}}$&
17.21&
15.55&
Yes&
$\rm{6.29\times 10^{32}}$&
15.9
\\

$\rm{^{7}Li^{3+}}$ &
11.05&
12.48&
No&
$\rm{9.75\times 10^{32}}$&
24.6
\\

$\rm{^{12}C^{6+}}$ &
12.90&
13.48&
No&
$\rm{8.35\times 10^{32}}$&
21.1
\\

$\rm{^{40}Ca^{20+}}$ &
12.90&
13.48&
No&
$\rm{8.35\times 10^{32}}$&
21.1
\\

$\rm{^{197}Au^{79+}}$ &
10.35&
12.09&
No&
$\rm{9.37\times 10^{32}}$&
23.6
\\

$\rm{^{208}Pb^{82+}}$ &
10.17&
11.98&
No&
$\rm{9.22\times 10^{32}}$&
23.3
\\

$\rm{^{238}U^{92+}}$&
9.98&
11.87&
No&
$\rm{8.92\times 10^{32}}$&
22.5
\\ \bottomrule
\end{tabular}}%
\label{tab:1.1}%
\end{table}%

In general, the ion accelerator complex of the EicC accelerator facility mainly consists of a polarized ion source, iLinac, BRing, and pRing, while the electron accelerator complex is composed of an electron injector and eRing. Two interaction regions will be available. 
Several key accelerator designs are listed below.

\begin{itemize}
\item \textbf{Generate low emittance ion beams}. Reducing the beam emittance to the one required by the design specifications is of crucial importance for achieving the targeted luminosity in the EicC accelerator facility. To this end, the scheme of staged electron cooling will be adopted. In the first stage, the cooling of the proton beams with low energy will be performed by a DC electron cooler in the BRing. In the second stage, the proton beam with the energy will be cooled by the high energy bunched beam electron cooler based on an energy recovery linac (ERL). This scheme ensures optimum efficiency for the cooling system in the ion accelerator complex of the EicC accelerator facility.
\item \textbf{Maintain and control beam polarization}. The physics goals of the EicC project put high requirements on the average polarization and the polarization direction of the beams. Relevant beam polarization control schemes should be made for both the ion accelerator complex and the electron accelerator complex. Specifically, the Siberian snake, which is a control system of spin tune, will be installed to keep high polarization in the acceleration process in the BRing, where depolarization resonances exist. For the acceleration in the pRing, only weak solenoid magnetic fields are required to keep high polarization of the ion beams, 
thanks to the figure-8 shaped design of the pRing. The polarization direction control system will be set up along the beamlines and at both sides of the interaction regions. Such a design makes it possible to perform the rotation of beam polarization directions arbitrarily, as well as control the polarization directions of two beams accordingly.
\item \textbf{Optimize interaction regions (IR)}. A full-acceptance detector will be built to detect and identify almost $100\%$ of reaction products at one of two IPs while the second IP will be reserved for upgrading. The specifications of the detector put forwards many constraints on the design and optimization of the interaction region (IR) . The IR will be designed to be asymmetrical since there are lots of differences between electron beams and ion beams. Such a design will not only reduce the background of the detector but also ensure the features of the full acceptance of the detector. Furthermore, the beamlines related to the forward reaction products will be placed downstream of the interaction point (IP) in the pRing and eRing.
\end{itemize}

For the design specifications listed above, a pre-research will be carried out, including the polarized ion source, the photocathode polarized electron gun, the high energy bunched beam electron cooler based on the energy recovery linac (ERL), the Siberian snake, the spin rotator as well as the preservation of the polarization in the figure-8 shaped synchrotron. All of them will certainly provide the technical underpinnings for the construction of the EicC accelerator facility in the future.


\section{Complementarity of EicC and EIC-US}

\begin{figure}[htbp]
\begin{center}
\includegraphics[width=0.95\textwidth]{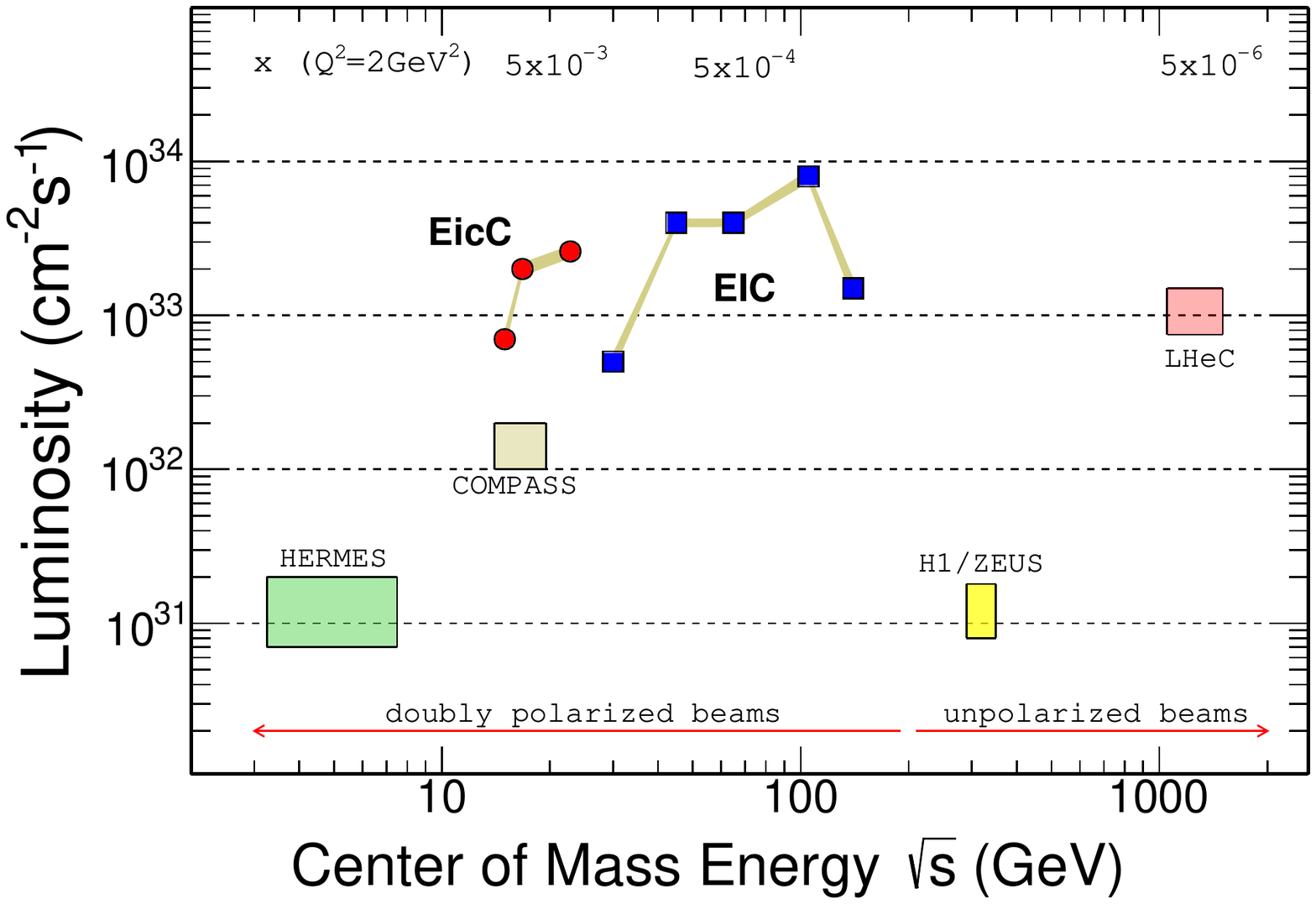}
\caption{\label{fig:lumi} Luminosity and center-of-mass energy of the proposed electron ion colliders\cite{Accardi:2012qut, Abelleira_Fernandez_2012,Gautheron:2010wva,vanderSteenhoven:2004xk,BRAUNSCHWEIG1993206}. For the EicC, the three data points are corresponding to electron-proton
collisions with energy 3.5~GeV (electron) $+$ 16~GeV (proton), 3.5~GeV $+$ 20~GeV, 
5~GeV $+$ 26~GeV, respectively.
}
\end{center}
\end{figure}

\begin{figure}[htbp]
\begin{center}
\includegraphics[width=0.8\textwidth]{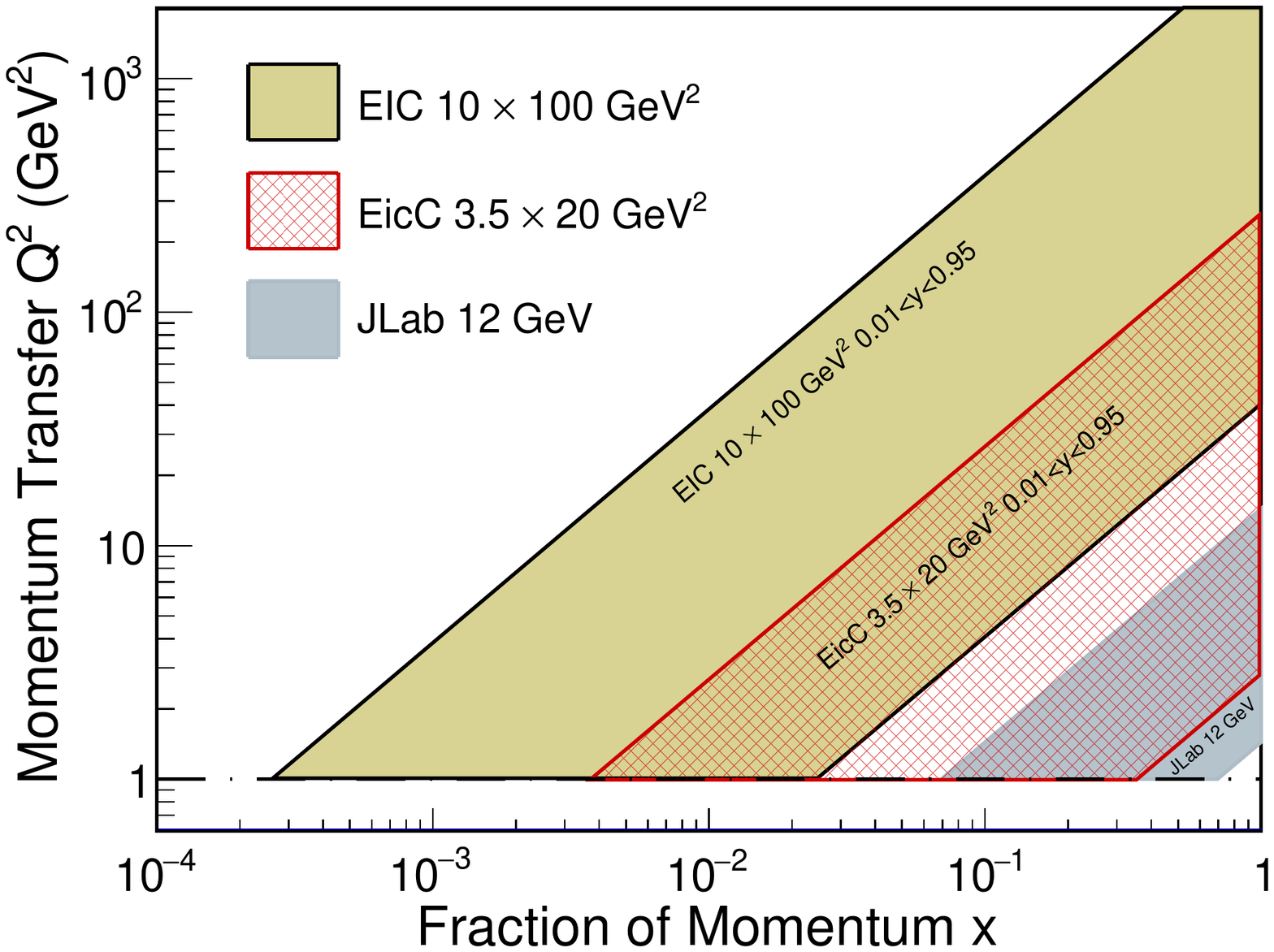}
\caption{\label{fig:Q2} Kinematic coverage of deep inelastic scattering process for different beam energy 
configurations at two proposed electron ion colliders as well as JLab.
Note that there are other energy configurations for both electron ion colliders, as shown in Fig.\ref{fig:lumi}.
}
\end{center}
\end{figure}

Both electron-ion colliders aim at the precision exploration of the partonic structure of nucleon/nucleus, but focus on different kinematics and perspectives. 
The design parameters and the luminosity versus center-of-mass energy of two colliders are shown in 
Table \ref{tab:EIC_comparison} and Figure \ref{fig:lumi}, respectively.
As shown in Figure \ref{fig:Q2}, the $x$-$Q^2$ coverage puts EicC at a sweet spot to systematically study the behavior of sea quarks. EIC-US~\cite{Accardi:2012qut} is a higher energy machine with an emphasis on low and moderate-$x$ region. Combining the measurements at both colliders will provide systematically controlled physics interpretation. Here are some examples.

\begin{itemize}

\item \textbf{Nucleon Spin}. With wide kinematic coverage and hermetic detector designs, EICs will provide a final answer to this decades-old question. One of the major goals of the EIC-US focuses on the gluon helicity contribution at small-$x$. EicC is optimized to systematically explore the nucleon spin including sea quark helicity contribution and orbital angular momentum contributions from quarks and gluons in the moderate $x$ regime. The unique $Q^2$ range will position EicC at a crucial place between JLab and EIC-US to unambiguously interpret and determine the orbital angular momentum contributions, hence providing a comprehensive 3-D imaging for the sea quarks inside a nucleon . 

\item \textbf{Proton Mass Decomposition}. Electroproduction and photoproduction of heavy quarkonia near threshold have been proposed to study the proton mass decomposition. EicC can contribute these important physics uniquely, through systematically investigating the $\Upsilon$ production with high luminosity near its threshold, where the optimal energy range of EicC is. Because of the 3 times larger mass of $\Upsilon$, the physics behind the measurement becomes much cleaner as compared to that of $J/\psi$ production at JLab 12GeV. Because of different kinematic coverage, EicC and EIC-US will be complementary to each other for $\Upsilon$ near-threshold production.

\item \textbf{Exotic Hadron States}. Both EicC and EIC-US can contribute to understanding the challenge posed by the unexpected $XYZ$ structures in the heavy-quarkonium mass region. The hidden-charm pentaquarks observed at LHCb need independent confirmation, and their hidden-bottom analogues are hard to be found at LHC but can be sought at EicC and EIC-US. The events of these states at EIC-US are expected to be more than those at EicC due to the larger energy and higher luminosity. For exclusive productions of exotic hadrons, the final state particles at EicC are within the middle rapidity range, facilitating the detection with relatively low background.
   
\item \textbf{Partonic Structure in Nuclear Environment}. Nuclear modification of the structure functions and hadron production in deep inelastic scattering $eA$ collisions are major focuses at both EIC-US and EicC. The kinematics at EicC provide a unique perspective to investigate the details of fast parton/hadron interactions with cold nuclear matter and shed light on energy loss and hadronization mechanisms. New information on the parton distribution in nuclei can be achieved at EicC at moderate $x$, whereas EIC-US concentrates in the small-$x$ region.

\end{itemize}

\begin{table}[htbp]
\footnotesize
\caption{\label{tab:EIC_comparison}The comparison between the parameters of the electron-ion
colliders proposed in China and in the US\cite{Accardi:2012qut}.}
\centering
\begin{tabular}{c c c c c}
\hline \hline
Facility&	CoM energy & lum./10$^{33}$cm$^{-2}$s$^{-1}$ & Ions & Polarization \\ \hline
EicC &  15 - 20  & 2 - 3 & $p \rightarrow$U& $e^-$, $p$, and light nuclei\\
EIC-US & 30 - 140 & 2 - 15& $p\rightarrow$U& $e^-$, $p$, $^3$He, Li\\
\hline \hline
\end{tabular}
\end{table}

\chapter{EicC physics highlights}

\section{One-dimensional spin structure of nucleons}
\label{OneDSpinStructure}

EicC will enable us to study the one-dimensional structure of the nucleons in various aspects, and to a great extent help search answers to many fundamental questions concerning the structure of nucleons with unprecedented precision. In particular, through a large amount of data, EicC can provide us the direct and precise information regarding the distributions of valence quarks, sea quarks, and gluons inside nucleons in the moderate and large $x$ regime. Furthermore, it can reveal the internal landscape of nucleons and deepen our understanding of their structure, and give us excellent opportunities for important discoveries in high energy nuclear physics. In addition to the physics significance by themselves, accurate parton distribution functions are extremely important for the precision study of particle physics and the exploration of new physics at the Large Hadron Collider.

How to understand the spin of protons in terms of the quark and gluon degrees of freedom has been an important cutting-edge research problem in high energy nuclear physics. In the 1980s, the EMC collaboration\cite{Ashman:1987hv} used a muon beam as a probe, and found that the sum of the spin contributions of all quarks inside the proton is very small comparing to the spin of the 
proton: $\Delta \Sigma = \Delta u + \Delta d + \Delta s = 12 \pm 9 (stat) \pm 14 (syst) \%$. This measurement has then precipitated the so-called ``proton spin crisis" in nuclear physics research. The current understanding of the structure of the proton spin is that the spin of proton consists of the spin contributions from quarks and gluons, and the orbital angular momenta of quarks and gluons. In addition to the spins of the valence quarks, many experimental results show that the sea quarks inside the proton also have non-zero spin contributions. Nowadays, the pressing issue is that the current measurement of the sea quark spin distribution is not particularly accurate. Through the double polarized collision processes, the spin distribution of different flavors of sea quarks can be precisely measured at the EicC, and elaborate experimental analysis on the spin distribution of sea quarks can be carried out, which will help to further study the spin structure of nucleons and enrich our understanding of non-perturbative properties of quantum chromodynamics.

With its designed high luminosity, EicC can generate an enormous amount of experimental data, which helps to clarify some intriguing problems and phenomena observed in experiments in the past few years. The first phenomenon is the asymmetry in the distribution of light sea quarks. In high-precision unpolarized scattering experiments, we have observed that the unpolarized $ \bar {u} $ and $ \bar {d} $ are asymmetrically distributed inside the proton, and the measured asymmetry is larger than what people have expected~\cite{Amaudruz:1991at,Arneodo:1994sh,Ackerstaff:1998sr,Baldit:1994jk,Towell:2001nh}. The theories and models which explain this asymmetry also predict the asymmetry for polarized light sea quarks~\cite{Bhalerao:2001rn,Peng:2003zm,Bourrely:2001du}. 
Moreover, the measurement of the longitudinal spin asymmetries for weak boson production in
proton–proton collisions at RHIC \cite{Adam:2018bam} suggests a difference between the $\Delta \bar u$ and $\Delta \bar d$ helicity distributions as well. Another interesting issue is the polarized distribution of the strange ($ s $) quark and its contribution to the proton spin. Assuming the $SU (3)$ flavor symmetry, one finds that the analysis of DIS data \cite{Leader:2006xc, Hirai:2006sr} indicates that the strange quark contribution to the proton spin is roughly $ -0.1 $. It is well-known that the strange quark contribution can be directly probed by semi-inclusive DIS (SIDIS). However, in SIDIS experiments, it is difficult to distinguish the current fragmentation process from the target fragmentation process. Also, the fragmentation function which describes the strange quark to a hadron transition process is not sufficiently precise. Due to the above-mentioned difficulties, it is still challenging to draw any firm conclusions on the polarization of strange quarks\cite{Airapetian:2004zf,Airapetian:2008qf}. In addition, whether the distribution function of the polarized $s$ quark as a function of $ x $ changes its sign or not is also an interesting research question\cite{deFlorian:2008mr}.

\begin{figure}[htbp]
\begin{center}
\includegraphics[width=0.8\textwidth]{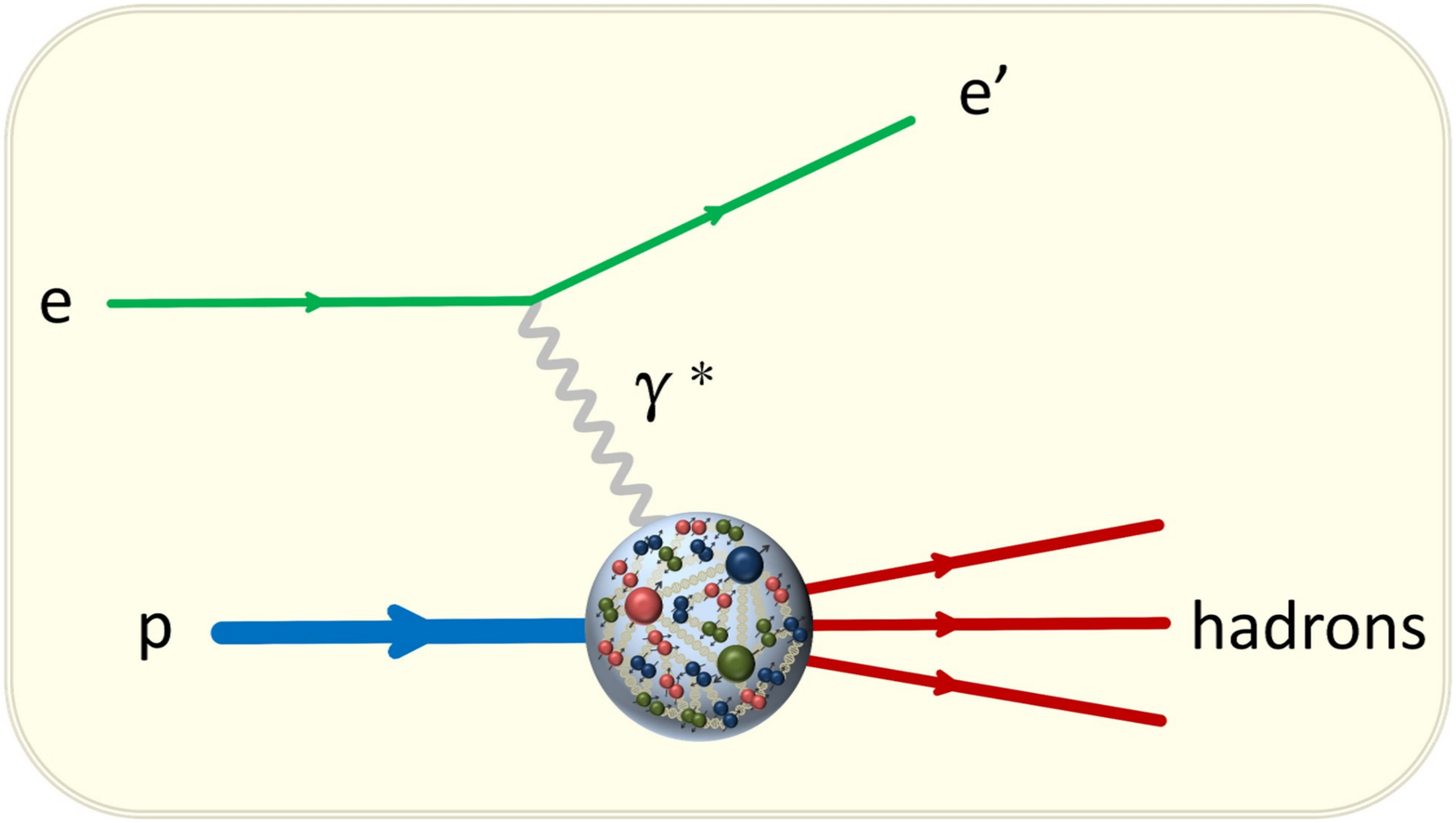}
\vspace*{-3mm}
\caption{\label{fig:DIS} Diagram of the Deep Inelastic Scattering (DIS) process. 
In this process, if the four-momentum of the incoming and outgoing electron are $k$ and $k'$, the
four-momentum of a nucleon is $p$, the relevant kinematic variables can be defined as: the squared
e+p collision center-of-mass energy $s = (p+k)^2$, the squared momentum transfer of the electron
$Q^2=-q^2=-(k-k')^2$, the Bjorken variable $x=\frac{Q^2}{2p \cdot q}$, the 
inelasticity $y=\frac{q \cdot p}{k \cdot p}$. In addition to these Lorentz invariants, there are two other
important kinematic variables: the invariant mass of the produced hadronic system $W=\sqrt{(q+p)^2}$, the
energy lost by the electron in the nucleon rest frame $\nu = \frac{q \cdot p}{M}$ where M is the nucleon mass.
}
\end{center}
\vspace{-0.5cm}
\end{figure}

\begin{figure}[htbp]
\begin{center}
\includegraphics[width=0.98\textwidth]{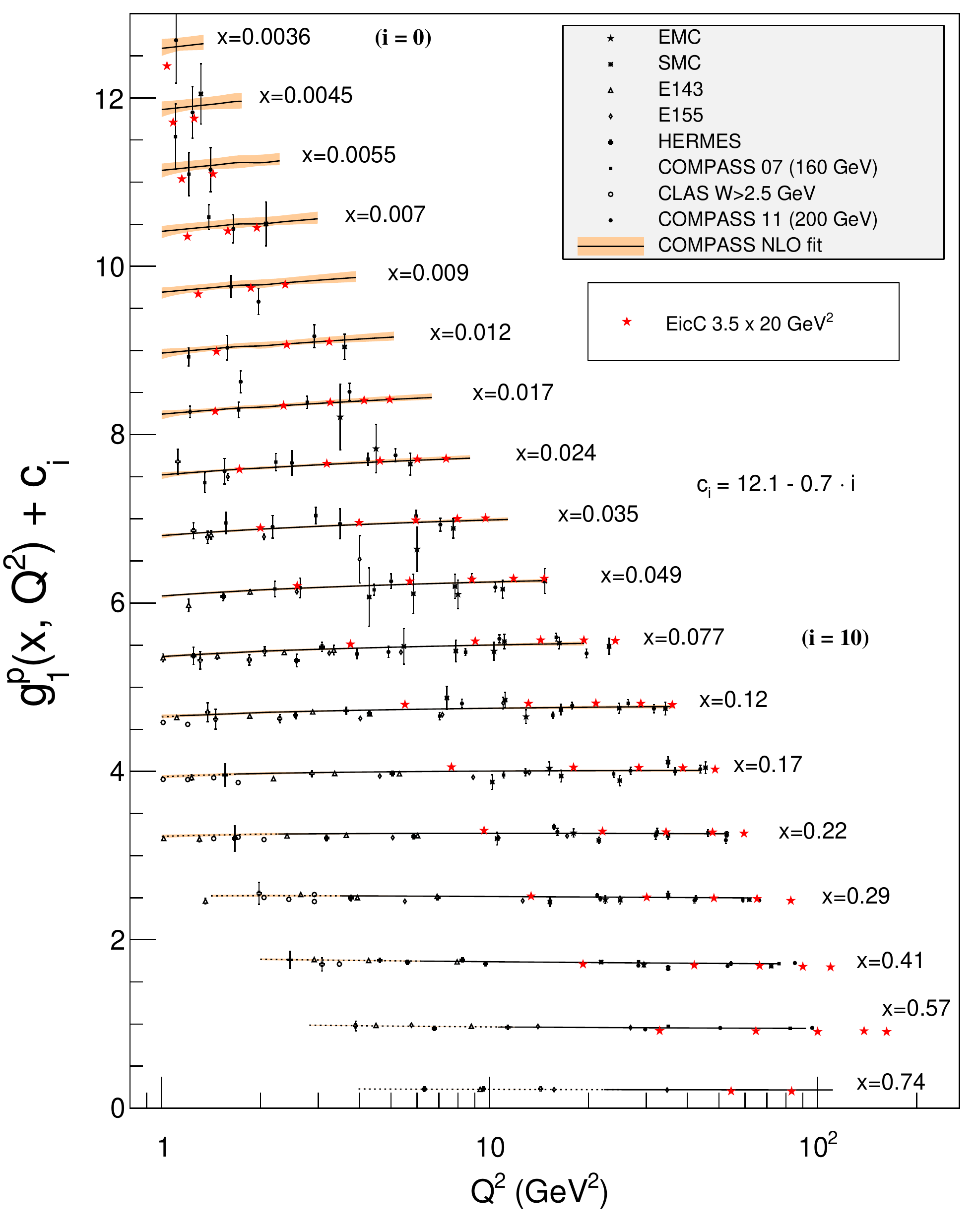}
\caption{\label{fig:g1p} Global data of the polarized proton structure function $g_1$ from inclusive DIS measurements compared with the projected EicC data based on the integrated luminosity of 50$fb^{-1}$ (about one year of running at the EicC). 
}
\end{center}
\vspace{-0.5cm}
\end{figure}

In the inclusive double polarized DIS process (Fig. \ref{fig:DIS}), where only the final state electrons are measured, the spin-dependent $g_1$ structure function can be extracted from the double spin asymmetry measurements. In the parton model, the $g_1$ structure function can be expressed as the sum of the contributions of various flavor of quarks
\begin{equation}
g_1(x,Q^2)=\frac{1}{2}\sum_{q=(u,d,s)} e_q^2\big[\Delta q(x,Q^2)+\Delta\overline{q}(x,Q^2)\big] ,
\end{equation}
 where the contribution from the light favor quarks is summed over. Fig.~\ref{fig:g1p} shows the kinematical coverage of the $g_1$ structure function at the EicC, as compared to the currently available experimental data. Generally speaking, $ g_1 $ is often measured through the inclusive DIS experiment, and it allows us to extract the polarized distribution functions of quarks of various flavors based on the assumption of the $SU (3)$ flavor symmetry. However, this method has a strong model dependence, and it mixes the contributions from quarks of different flavors. 
 
 Another method \cite{Airapetian:2003ct} is to use SIDIS processes to extract more quark and hadron flavor information from experimental data,
 where a leading hadron among the final state hadrons in Fig. \ref{fig:DIS} is detected in coincidence with the scattered electron. When a quark inside a proton absorbs a virtual photon emitted by an electron, the quark gets struck out of the proton and becomes a final state jet, which consists of many hadrons clustered inside a narrow cone. This final state hadronization process can be described by fragmentation functions. The final-state hadron contents in the jet carry the flavor information of the initial state quark, therefore this process offers a way to tag the flavor of the produced quark. If one measures a pion or a kaon in the SIDIS process in addition to the recoiled electron, one can separate spin contributions from quarks of different flavors. In this case, the polarized structure function in the parton model can be written as
\begin{equation}
g_1(x,Q^2,z)=\frac{1}{2}\sum_qe_q^2\big[\Delta q(x,Q^2) D^{q \to h}(Q^2,z)+\Delta\overline{q}(x,Q^2) D^{ \overline{q} \to h}(Q^2,z) \big] ,
\end{equation}
where $D^{q \to h}(Q^2,z)$ describes the fragmentation process from a quark $q$ to a hadron $h$. $z$ represents the momentum fraction of the final state hadron with respect to the momentum of the produced quark, experimentally, it is defined as 
$z=\frac{P_{hadron} \cdot p}{q \cdot p}$.

\begin{figure}[htbp]
\begin{center}
\includegraphics[width=0.49\textwidth]{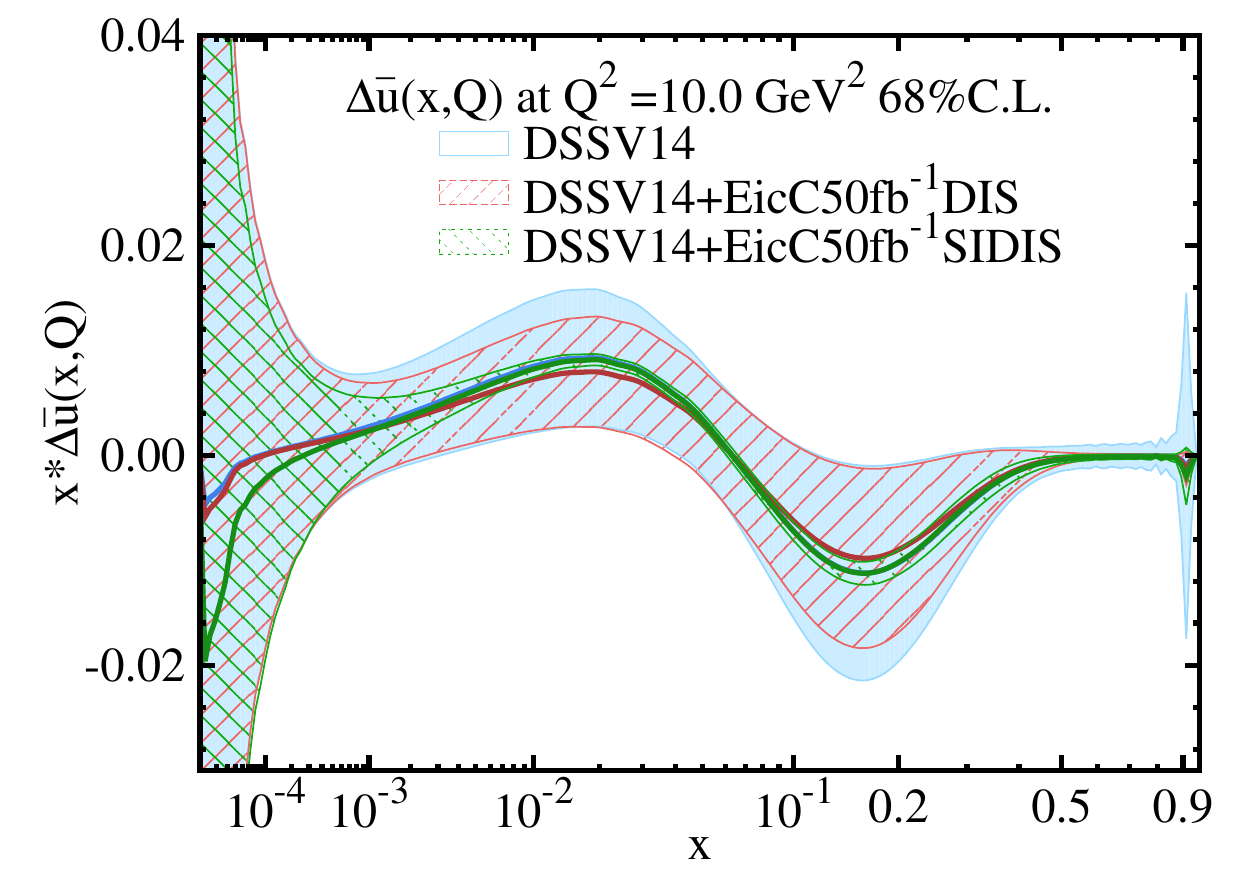}
\includegraphics[width=0.49\textwidth]{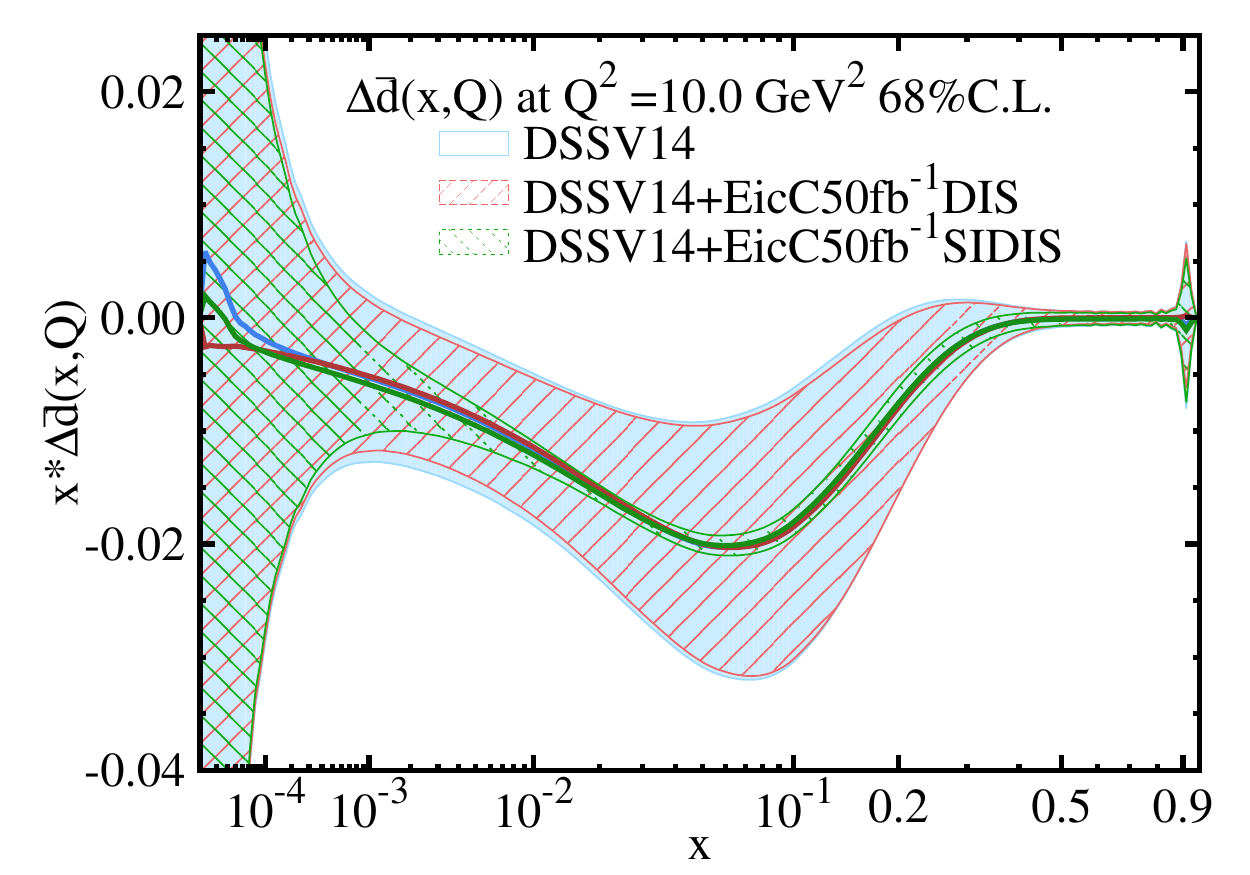}
\includegraphics[width=0.49\textwidth]{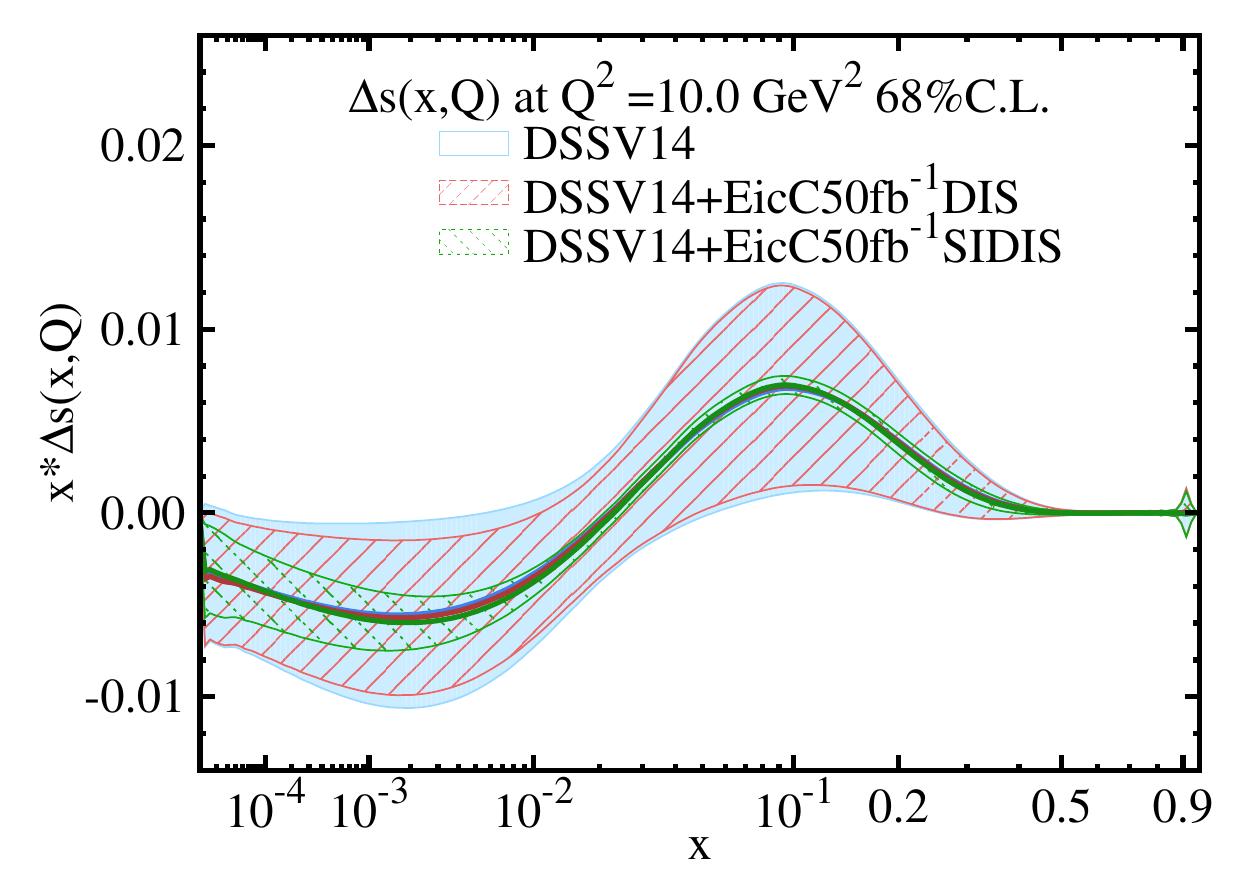}
\includegraphics[width=0.49\textwidth]{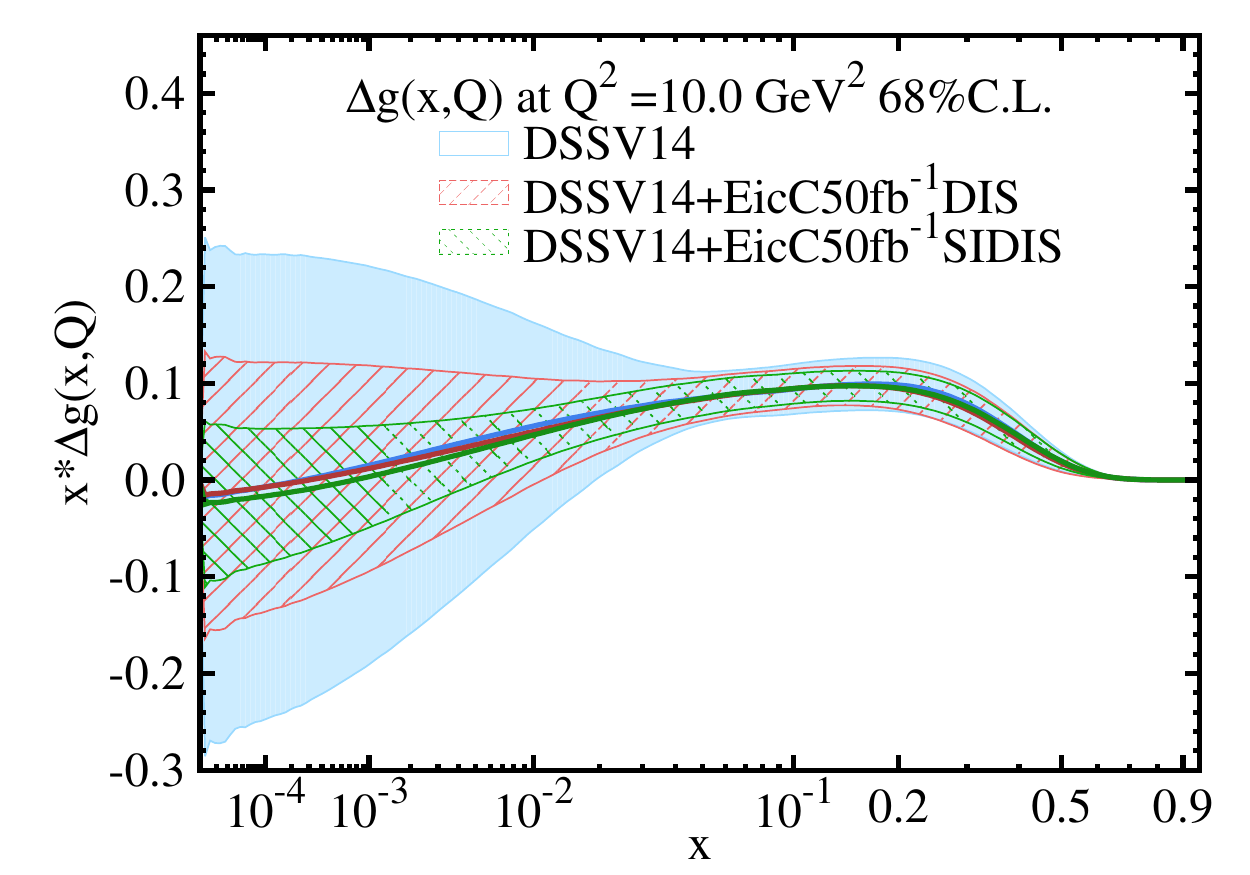}
\caption{\label{fig:seaquark}
Results on the uncertainty band of polarized sea quark and gluon distributions after a next-to-leading
order fit by including EicC pseudodata. The light blue band represents the original DSSV14 global fit.
The red (green) band shows the results by adding DSSV14 fit with EicC DIS (SIDIS) pseudodata with integrated luminosity of 50 fb$^{-1}$ (10 months of running at 2$\times$10$^{33}$/cm$^2$/s instantaneous luminosity) for both electron-proton (3.5 GeV + 20 GeV) and electron-$^3$He 
collisions (3.5 GeV + 40 GeV).
During the pseudodata analysis, the following cuts were applied:
$Q^2 > 2 GeV^2$, $W^2 > 12 GeV^2$, $0.05 < y < 0.8$, $0.05 < z < 0.8$.}
\end{center}
\end{figure}

Through measurements in $e^+e^-$ and $e-p$ scatterings, we have been studying and extracting various hadron fragmentation functions. Using these hadron fragmentation functions as inputs, we can further separate and extract the polarized quark distributions of certain flavor accurately from polarized SIDIS data measured at EicC. Figs. ~\ref{fig:seaquark} show the EicC projection of the polarized sea quark and gluon distributions, respectively, for various flavors of quarks obtain from longitudinally polarized double spin asymmetry measurements via DIS and SIDIS processes. 
In these figures, the light blue band represents the original uncertainty of the DSSV14 global data fit \cite{deFlorian:2019zkl}. The red (green) dashed band is the uncertainty from a next-to-leading 
order fit using ePump \cite{Schmidt:2018hvu,Hou:2019gfw} by adding DSSV14 fit with EicC DIS (SIDIS) pseudodata with integrated luminosity of 50 fb$^{-1}$ for both electron-proton (3.5 GeV + 20 GeV) and electron-$^3$He 
collisions (3.5 GeV + 40 GeV). One can tell that the SIDIS data, taking advantage of $\pi^\pm$ and $K^\pm$ final states from both proton and effective neutron targets, is 
more powerful comparing to DIS data in the flavor separations.
The plots clearly show that EicC can significantly improve the precision of helicity distributions of sea quarks and gluons in the $x>0.005$ region.
This can have an impact on the understanding of the proton spin puzzle, since the current sea quark contribution to the proton spin $\int \Delta q (x) dx $ ($ q = \bar {u}, \bar {d}, s $) has an uncertainty of $ 100-200 \% $. The measurement at EicC can help to improve precision by more than a factor of five.

As a short summary, thanks to the particular energy range of EicC, the high-luminosity and versatile capability of the accelerator machine design, and the $4\pi$ coverage layout of the detector, SIDIS at EicC allows us to measure the polarized sea quark and gluon distributions with remarkable precision. Moreover, using various polarized hadron beams (protons and helium-3) together with the detector with the capability of particle identification, EicC can help to significantly improve the flavor separation and thus extract polarized quark distributions of different flavor reliably.

After one year of running at the EicC, about 50$fb^{-1}$ integrated luminosity will be obtained, as
one can see from the above statistical analysis that the measurement can be significantly improved
comparing to the existing world data. Therefore, it is critical to control the systematic uncertainty.
According to the study in the ongoing experiments, the major sources of systematic uncertainty are 
from the precision of measurements on the beam polarization, luminosity fluctuation of beam bunches
in different spin states, contamination of photon-induced electrons in the scattered electron 
detection, and so on. These sources are also applied to the following physics topics and will 
be further investigated quantitatively while the detector design is refined in the following years.   


\section{Three-dimensional tomography of nucleons}
\label{ThreeDnucleons}

The conventional parton distribution functions (PDFs) first introduced by Feynman~\cite{Feynman:1973xc} and formalized by Bjorken and Paschos~\cite{Bjorken:1969ja} only contain the information on the longitudinal motion of partons inside a nucleon. To gain more comprehensive knowledge about partonic structures of the nucleon, one may introduce multi-dimensional distributions, including transverse momentum dependent parton distributions (TMDs)~\cite{Collins:1981uk,Collins:1981uw} and generalized parton distributions (GPDs)~\cite{Mueller:1998fv,Ji:1996nm,Ji:1996ek,Radyushkin:1997ki}. For a given longitudinal momentum fraction $x$ carried by a parton, TMDs represent the transverse momentum distribution of the partons and GPDs encode the transverse spatial distribution of the partons. Both TMDs and GPDs provide three-dimensional images of the nucleon, allowing us to access much richer partonic structures, especially when the spin degrees of freedom are taken into account. Therefore, the measurement of TMDs and GPDs will lead us to a more profound understanding of strong interaction. 

Experimental studies of TMDs and GPDs have been carried out in the existing facilities during the last two decades. Although valuable data have been collected for a first exploration, TMDs and GPDs are still far from well constrained, especially for sea quarks and gluons, due to the low luminosity and the limited kinematic coverage. Recently approved electron-ion collider to be built at BNL is designed to reach a high center-of-mass energy region, which makes the quantitative exploration of sea quark TMDs and gluon TMDs possible for the first time. On the other hand, EicC as a facility at the intensity frontier with relatively high center-of-mass energy, and versatile beam species, will be an ideal machine for exploring the internal landscape of the nucleon in the sea quark region. 

In this section, we describe the TMD and GPD programs at EicC via the semi-inclusive deep inelastic scattering (SIDIS), deeply virtual Compton scattering (DVCS), and deeply virtual meson production (DVMP) processes.

\subsection{Transverse momentum dependent parton distributions}

The extraction of partonic structures of the nucleon from high energy scattering processes relies on the QCD factorization, which provides the link between the observed hadrons and the partons that participate in the hard scattering. In inclusive DIS, where only the scattered lepton is identified, the large momentum scale $Q$ mediated by the virtual gauge boson, {\it i.e.} photon or $W^\pm/Z$, serves as a short-distance probe, allowing us to ``see'' the quarks and gluons indirectly. The cross section can be factorized into the lepton-parton scattering at short-distance convoluted with the PDFs in which the active parton's transverse momentum $k_T$ is integrated. Overall corrections are suppressed by inverse powers of $Q$. This is known as the collinear factorization.

\begin{figure}[htbp]
\centering{\includegraphics[width=0.8\textwidth]{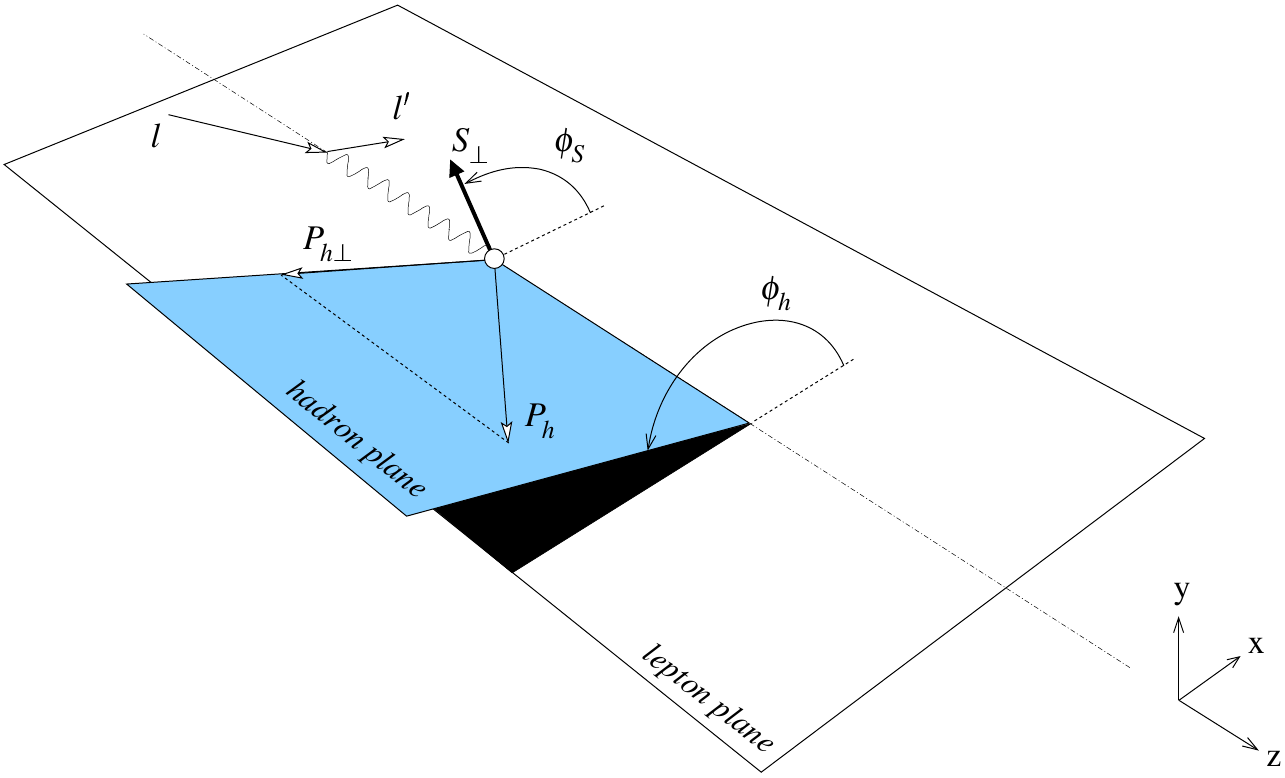}}
\caption{ The Trento convention of SIDIS kinematic variables~\cite{Bacchetta:2004jz}.
The $z$-direction is defined by the virtual photon (or $W^\pm/Z$), momentum, also referred to as the photon-target frame.
The incoming and outgoing leptons define the lepton plane, and the detected final-state hadron together with the virtual photon defines the hadron plane. The azimuthal angle $\phi_h$ is defined from the lepton plane to the hadron plane. For a transversely polarized nucleon beam/target, the azimuthal angle of the transverse spin $S_\perp$ is define from the lepton plane to the transverse spin direction.}
\label{fig:SIDIS}
\end{figure}

Apart from the scattered lepton, one final-state hadron with momentum $P_h$ is identified in SIDIS (Fig. \ref{fig:SIDIS}),
which not only allows us to detect quark and gluon distributions in the nucleon as in the inclusive DIS, 
but also provides the opportunity to explore the hadronization process, the emergence of color neutral hadrons from colored quarks and gluons. It also allows us to learn the flavor dependence by selecting different hadrons, {\it e.g.} pions and kaons, in the final state, as explained in the previous section.

In addition to the large momentum scale $Q$, an adjustable momentum scale is given by the transverse momentum of the observed hadron in the final state. In the Breit frame, where the virtual gauge boson and the nucleon are headed on, the SIDIS process is naturally dominated by the small transverse momentum region, $P_{h_\perp} \ll Q$. In this regime, the hard momentum scale $Q$ localizes the probe to see the particle feature of quarks and gluons in the nucleon and the soft scale $P_{h_\perp}$ is sensitive to the confined motion of partons perpendicular to the colliding direction. The SIDIS cross section can be factorized into the lepton-parton short-distance scattering convoluted with transverse momentum dependent parton distribution functions and fragmentation functions. This is known as the TMD factorization. Overall corrections are suppressed by powers of $P_{h_\perp}/Q$. For events with large transverse momentum comparable to $Q$,  the SIDIS process is effectively characterized by a single large momentum scale, no longer sensitive to the transverse motion of partons. The cross section is described by the collinear factorization as in the previous section, where double polarized SIDIS events are utilized to extract helicity distributions.

\begin{figure}[htbp]
\centering{\includegraphics[width=0.98\textwidth]{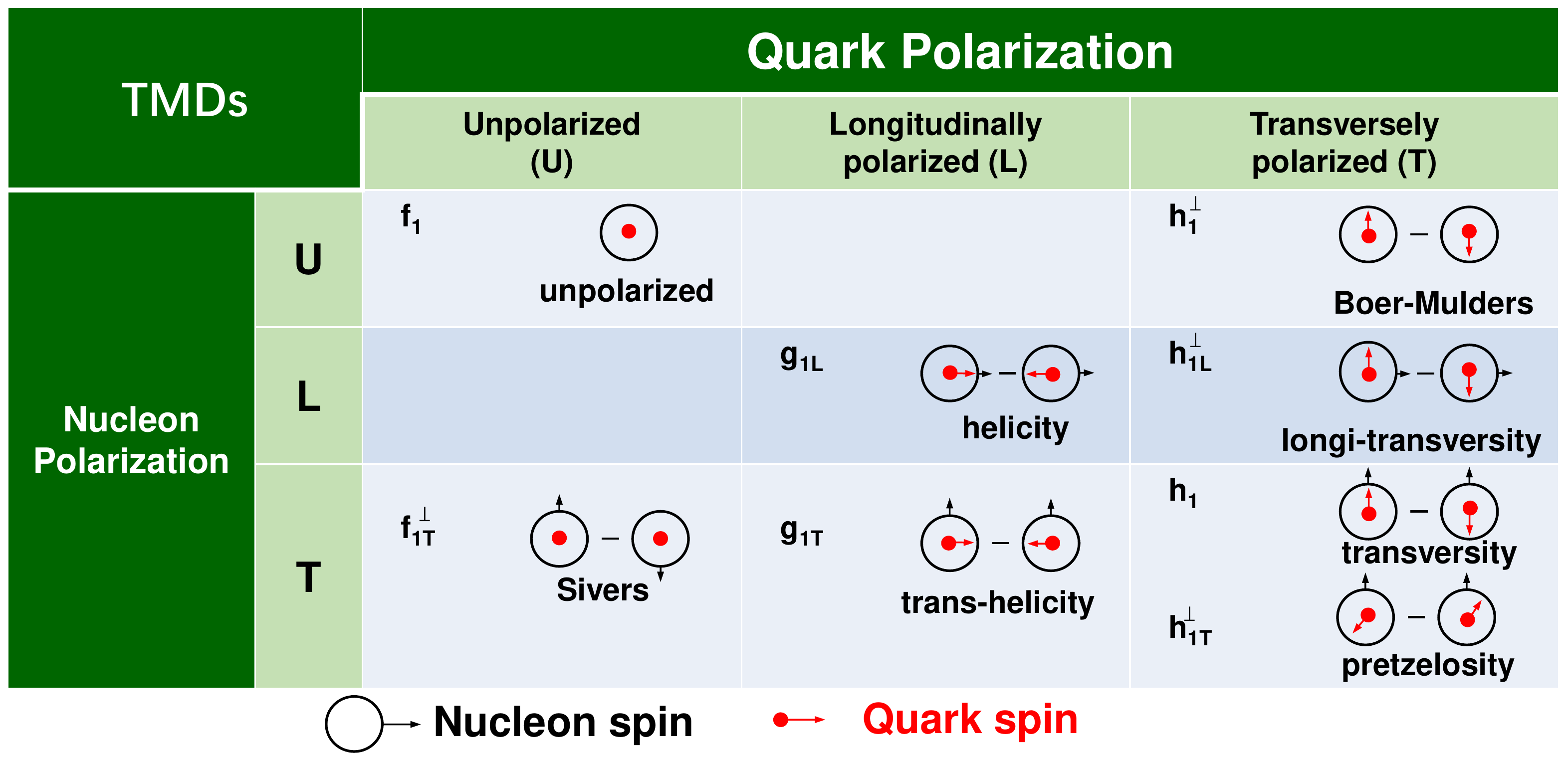}}
\caption{ The leading-twist quark TMD distributions. }
\label{fig:8tmds}
\end{figure}

Here we focus on the small transverse momentum regime where one can apply the TMD factorization~\cite{Ji:2004wu,Ji:2004xq}. By introducing the spin degrees of freedom, one can define eight independent leading-twist quark/gluon TMDs~\cite{Mulders:1995dh} as shown in Fig.~\ref{fig:8tmds}. The spin-dependent TMDs encode rich information of the nucleon structure, and in particular can shed light on our understanding of parton orbital motions and spin-orbit correlations. When integrating out the transverse momentum $k_T$ of the parton, three out of the eight leading-twist TMDs, the unpolarized distribution $f_1(x,k_T)$, the helicity distribution $g_{1L}(x,k_T)$ and the transversity distribution $h_{1}(x,k_T)$ reduce to their collinear counterparts, while the other five that describe the correlations between parton transverse momentum and the parton/nucleon's spin  vanish. 

Under the single-photon exchange approximation, the SIDIS cross section can be expressed in terms of 18 structure functions, corresponding to different polarization configurations and final-state azimuthal modulations. 
The spin-dependent TMDs are usually extracted by measuring particular azimuthal asymmetries, while
the unpolarized TMDs can be obtained by analyzing unpolarized SIDIS cross section or multiplicity.
Among the leading-twist spin-dependent TMDs, the Sivers function $f_{1T}^\perp(x,k_T)$~\cite{Sivers:1989cc}, as well as related phenomenologies, stimulates tremendous theoretical progress and experimental investigations in recent years. It describes the correlation between quark transverse momentum and the transverse spin of the nucleon.
To put it simply, the Sivers function reflects the left-right asymmetry of quark transverse momentum distribution in a transversely polarized nucleon. One of the most distinguished features of the Sivers function is its unique universality property exhibited in different processes. 
If final/initial-state interactions, which are formally summarized into the gauge link, were absent between the active quark and the remnants of the nucleon, the time reversal invariance requires the Sivers function to be zero~\cite{Collins:1992kk}, and thus it is commonly referred to as naive time-reversal odd (T-odd) distribution. Once turning on QCD interactions, the Sivers function can arise from the final-state interaction in the SIDIS process and from the initial-state interaction in the Drell-Yan process.
As the staple like Wilson line flips the direction between the final-state and initial-state interactions, the quark Sivers functions are predicted to have an exact sign change between SIDIS and Drell-Yan processes, $f_{1T}^\perp(x,k_T)|_{\rm SIDIS}=-f_{1T}^\perp(x,k_T)|_{\rm DY}$~\cite{Brodsky:2002cx,Collins:2002kn}. Although recent $W$ production data from STAR~\cite{Adamczyk:2015gyk} and the $\pi N$ Drell-Yan data from COMPASS~\cite{Aghasyan:2017jop} support this prediction,
 the current uncertainties are too large to confirm the sign change. 
Future precise measurements of the Sivers function in different processes are of great importance to test this prediction associated with the QCD factorization. Moreover, theoretical studies have suggested that the Sivers function is closely related to parton orbital angular momentum~\cite{Ji:2002xn}. Therefore, the experimental studies of the Sivers function are not only crucial for unveiling the spin structure of the nucleon, but also important for deepening our understanding of the strong interaction and/or QCD.

\begin{table}[htbp]
\caption{The existing measurements of the Sivers asymmetry in SIDIS.}
\begin{center} \small
\begin{tabular}{lcccc}\hline
collaboration & $\sqrt{s}$ ~ GeV& target  & final state hadron  & literature \\\hline
COMPASS (CERN) & 18 & Deuterium & $h^{\pm},\pi^{\pm},K^{\pm},K^0$ & ~\cite{Ageev:2006da,Alekseev:2008aa}\\
 &  & proton & $h^{\pm}$ & \cite{Alekseev:2010rw}\\
 &  & proton & $\pi^{\pm},K^{\pm}$ & \cite{Adolph:2014zba}\\\hline
HallA (JLab) & 3.5  & neutron & $\pi^\pm,K^{\pm}$ & \cite{Qian:2011py}  \\\hline
HERMES (DESY) & 7.4  & proton  & $\pi^{\pm}$ & \cite{Airapetian:2004tw} \\
 &  & proton & $\pi^{\pm},(\pi^+ - \pi^-),\pi^{0},K^{\pm}$ & \cite{Airapetian:2009ae} \\\hline
\end{tabular}
\label{tab:siv}
\end{center}
\end{table}

In SIDIS, one can access the Sivers function by measuring a transverse single-spin asymmetry, known as the Sivers asymmetry. Within the TMD factorization, the corresponding structure function can be expressed as the convolution of the Sivers function and the unpolarized fragmentation function.
During the past two decades, great efforts have been made to extract the Sivers function as well as other TMDs via the SIDIS process at many experimental facilities around the world, including HERMES, COMPASS, and JLab, as summarized in Table~\ref{tab:siv}.
However, TMDs, especially the spin-dependent ones, are still very poorly determined due to various difficulties. The JLab experiments were carried out at relatively low energies, where high-twist effects and target mass corrections are expected to be sizable. HERMES data were mostly collected in the so-called valence quark region, where the contribution to the cross section is dominated by valence quarks, and hence the data are not quite sensitive to sea quark distributions. 
The ongoing and upcoming SIDIS experiments at the 12-GeV upgraded JLab aim at unprecedented precise measurements of valence quark TMDs. Recently approved EIC-US at BNL is designed with a high center-of-mass energy and will have quantitative measurements of gluon and sea quark TMDs for the first time. EicC given the center-of-mass energy in between has unique advantages to study sea quark TMDs and fills the energy gap from JLab to EIC-US.
In addition, the separation of current fragmentation and target fragmentation remains as a challenging task at the existing fixed target facilities. The large experimental acceptance at EicC will provide a wide kinematic coverage, which is crucial to make the clean selection of events in the current fragmentation region, allowing us to apply more strict kinematic cuts, especially for $K$ meson productions that play an important role in flavor separation due to its sensitivity to strange quark distributions. Currently available $K$ meson production data from polarized SIDIS are rather limited. EicC SIDIS experiments will have high statistics measurements of both charged pion and charged kaon productions. Together with the combination of the proton beam and the $^3$He beam, EicC will allow a full separation of all light quark flavors, $u$, $d$, $s$, $\bar u$, $\bar d$, and $\bar s$. 

\begin{figure}[htp]
    \centering
    \includegraphics[width=0.48\textwidth]{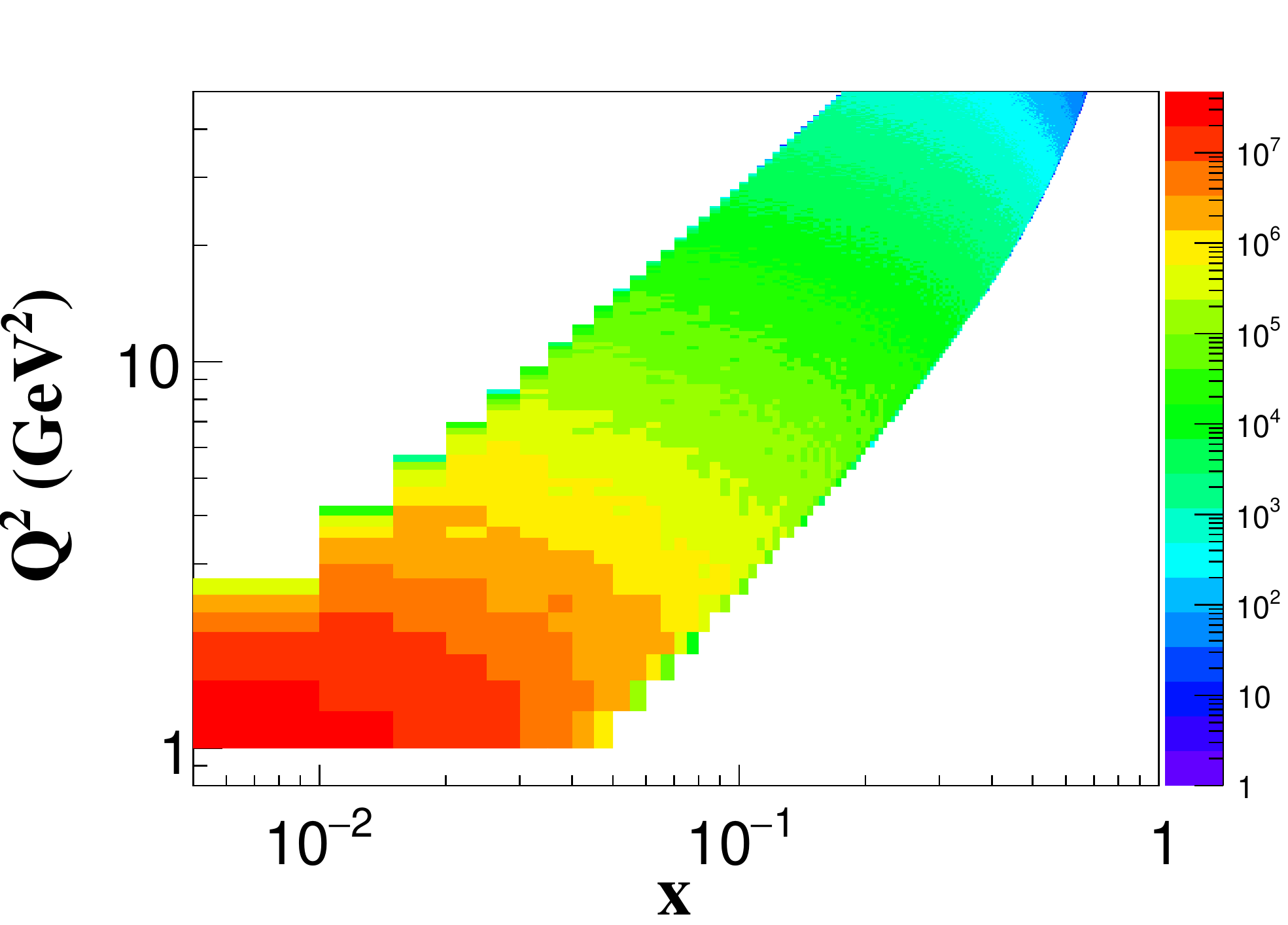}
    \includegraphics[width=0.48\textwidth]{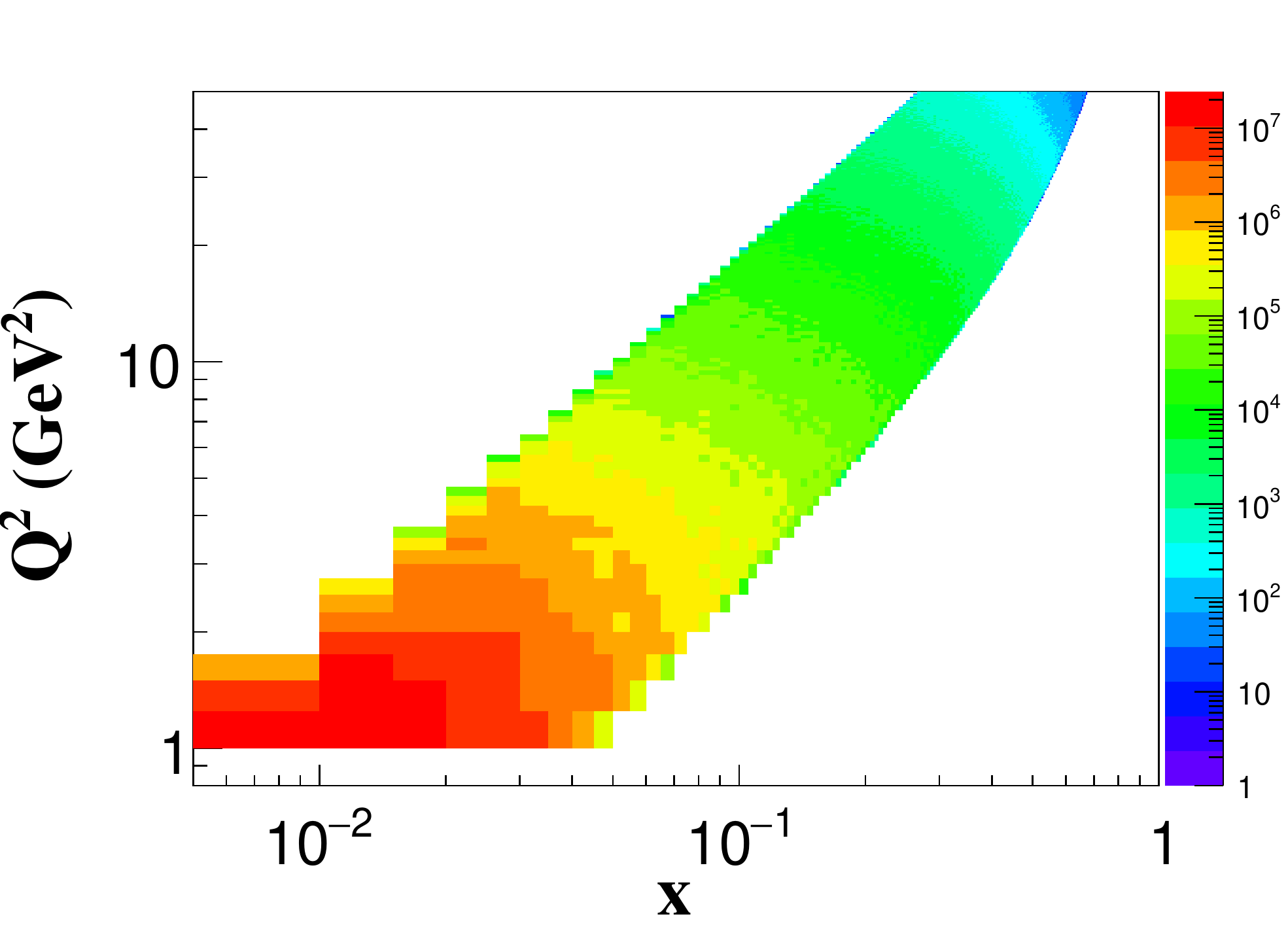}
    \caption{$x-Q^2$ coverage of EicC SIDIS events from simulation. Left: $\pi^+$ production from the proton beam. Right: $K^+$ production from the $^3$He beam. Kinematic cuts are described in the text.}
    \label{fig:sidiskincut}
\end{figure}

In Fig.~\ref{fig:sidiskincut}, we show the $x-Q^2$ distribution of EicC SIDIS events. Instead of presenting all cases, we only select two examples, the $\pi^+$ production from the proton beam and the $K^+$ production from the $^3$He beam, to cover both $ep$ and $e^3{\rm He}$ collisions and both pion and kaon productions. Kinematic cuts, including $Q^2>1\,\rm GeV^2$, $W>5\,\rm GeV$, $W'>2\,\rm GeV$, $0.3<z<0.7$, and the current fragmentation cut as described in~\cite{Boglione:2019nwk}, have been applied. 
 One can see that with the current EicC design, the relatively high $Q^2$ coverage in the typical sea quark region ($x \sim 0.05$) ensures a reliable extraction of sea quark TMDs. Moreover, compared to the fixed-target experiments, the collider mode of EicC allows a wide kinematic coverage, which provides the opportunity to quantitatively estimate power suppressed corrections.

Among the $k_T$-odd TMDs, quark Sivers function is the most extensively studied, but it is still poorly constrained, particularly in the sea quark region where the sign is even not yet determined without ambiguity~\cite{Boglione:2018dqd}. Here we take the Sivers function as an example to demonstrate the impact of EicC SIDIS experiments.

\begin{figure}[htbp]
\includegraphics[page=1,scale=0.2]{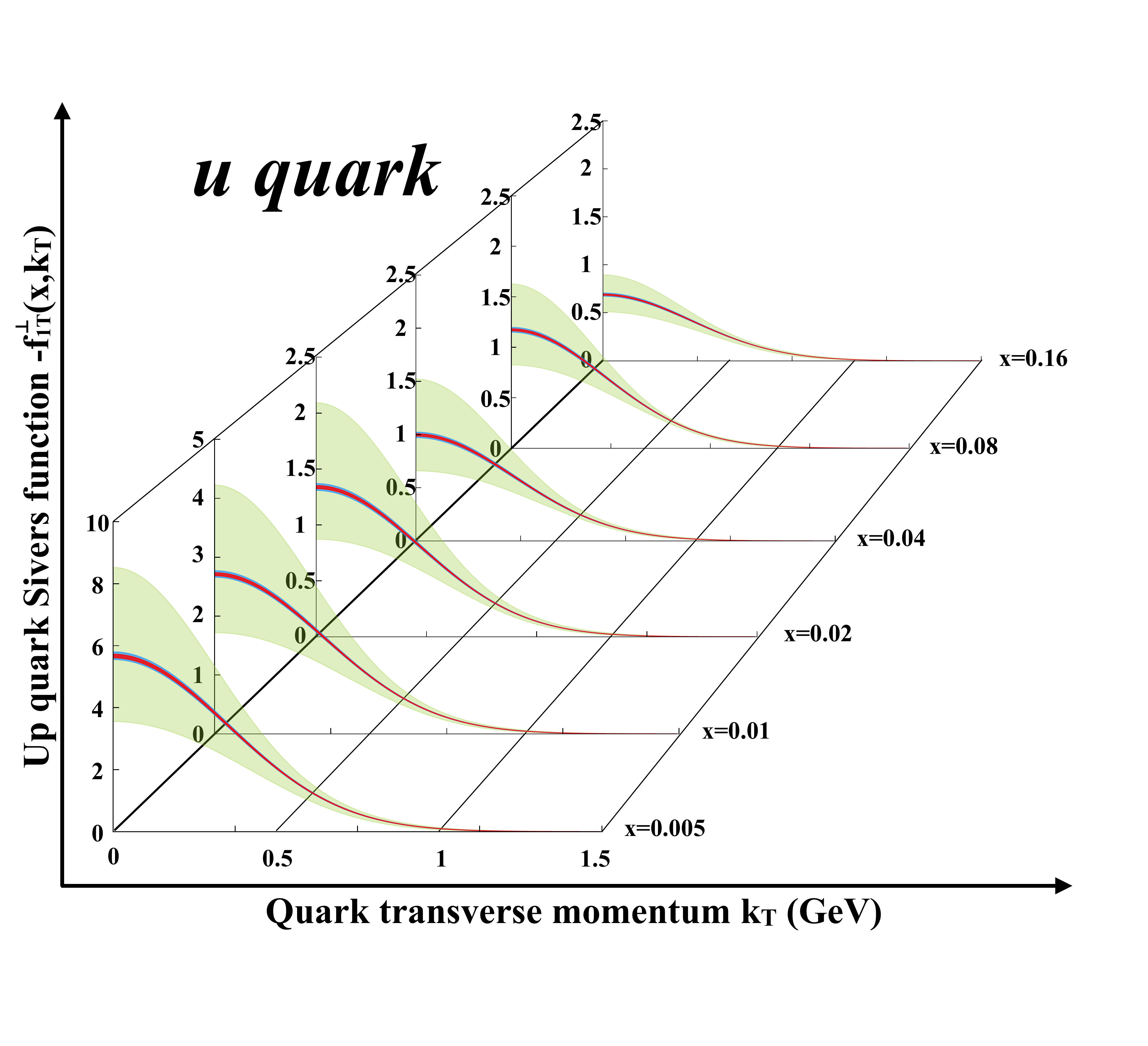}
\includegraphics[page=2,scale=0.2]{figures/chapter2/sivers.pdf}
\caption{The precision of extractions of up and down quark Sivers functions. The light green bands represent the accuracy from the currently available SIDIS data, the red bands represent the accuracy by including the projected EicC data with statistical uncertainty only, and the blue bands represent the accuracy by including the projected EicC data with part of systematic uncertainties as described in the text. Integrated luminosities of $50\,\rm fb^{-1}$ for $ep$ and $50\,\rm fb^{-1}$ for $e\,^3\rm He$ are adopted in this projection.}
\label{fig:sivers-updown}
\end{figure}

\begin{figure}[htbp]
\centering
\includegraphics[page=3,scale=0.25]{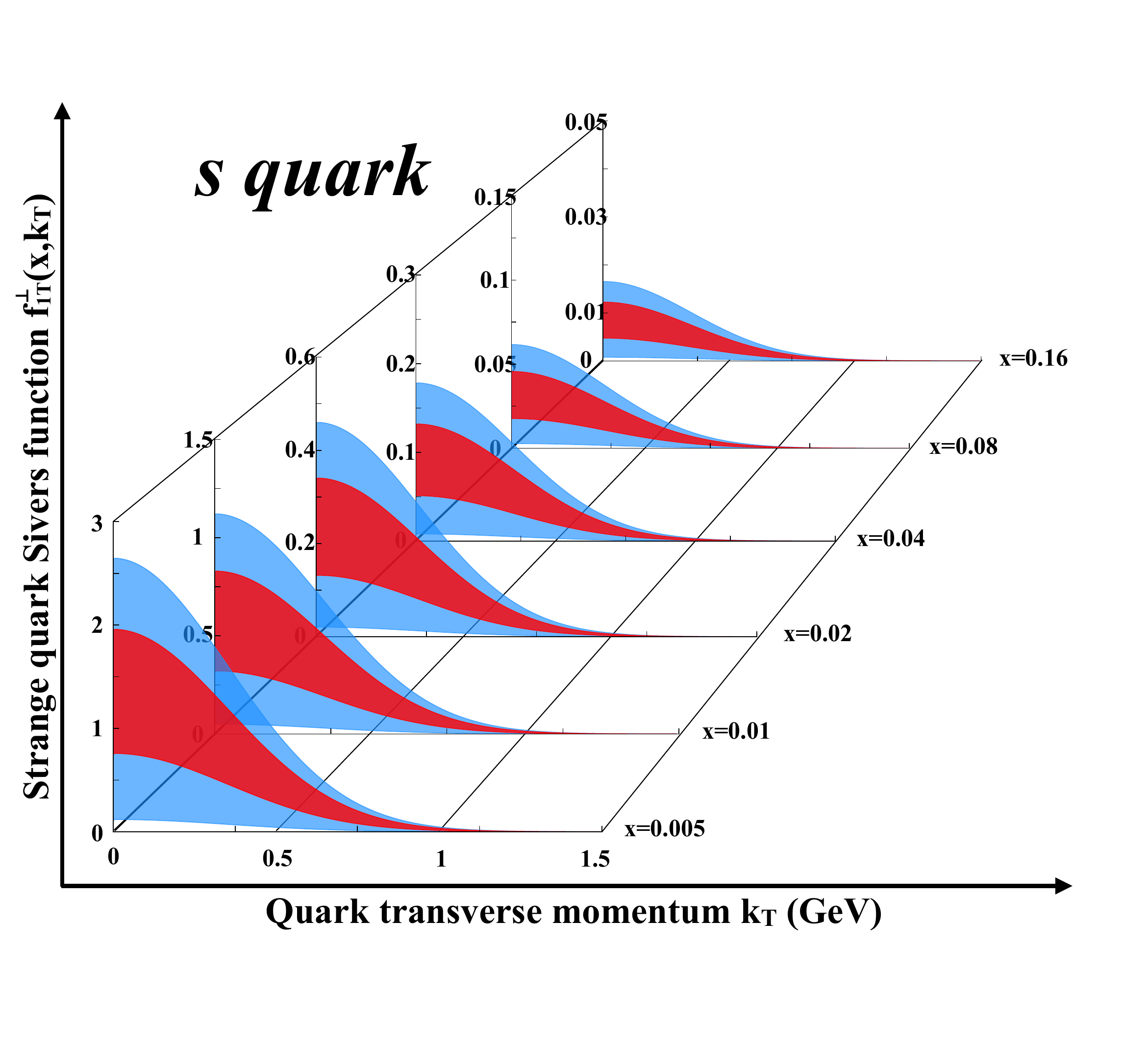}
\caption{The precision of the extraction of strange quark Sivers function. The red bands represent the accuracy by including the projected EicC data with statistical uncertainty only, and the blue bands represent the accuracy by including the projected EicC data with part of systematic uncertainties as described in the text. Integrated luminosities of $50\,\rm fb^{-1}$ for $ep$ and $50\,\rm fb^{-1}$ for $e\,^3\rm He$ are adopted in this projection.}
\label{fig:sivers-strange}
\end{figure}

We simulate SIDIS events according to EicC kinematics: $3.5\,\rm GeV$ electron beam, $20\,\rm GeV$ transverse polarized proton beam, and $40\,\rm GeV$ transverse polarized $^3$He beam serving as effective polarized neutron beam since the spin of $^3$He is mainly given by the neutron spin. Integrated luminosities are chosen as $50\,\rm fb^{-1}$ for $ep$ and $50\,\rm fb^{-1}$ for $e\,^3\rm He$ collisions. A $4\pi$ angle coverage is assumed for the acceptance. For the projection of EicC results, we select simulated SIDIS events with $Q^2>1\,\rm GeV^2$, $W>5\,\rm GeV$, $W'>2\,\rm GeV$, $0.3<z<0.7$, and the current fragmentation cut~\cite{Boglione:2019nwk}. We also require the scattered electron momentum $P_e>0.35\,\rm GeV$ and the identified final-state hadron momentum $P_h>0.3\,\rm GeV$ for detection reason. Apart from the statistical uncertainty, we consider some major systematic uncertainties, including $3\%$ relative uncertainty for beam polarizations and $5\%$ relative uncertainty for $^3$He nuclear effect. Other systematic uncertainties, {\it e.g.} detector resolution, particle identification, random coincidence, radiative corrections, are not expected to be dominant based on existing experience, although more detailed studies will be carried out with the final detector design. 
We utilize the parametrization form of the Sivers function in~\cite{Boglione:2018dqd} as the input model. In Fig.~\ref{fig:sivers-updown}, we show the results of the extraction of up and down quark Sivers functions. In Fig.~\ref{fig:sivers-strange}, we show the results of the extraction of the strange quark Sivers function. The outer light green bands represent the present accuracy from world existing SIDIS data, the inner red bands represent the accuracy including projected EicC data with statistical uncertainty only, and the blue bands represent the accuracy including projected EicC data with part of the systematic uncertainties mentioned above. As the world existing $K$ meson production data from polarized SIDIS are very limited, the current uncertainty of the strange quark Sivers function is huge, and even the sign is not yet determined. Hence the present accuracy bands are not shown in Fig.~\ref{fig:sivers-strange}. Recent theoretical study suggested an opposite sign between the strange quark and anti-strange quark Sivers functions~\cite{Dong:2018wsp} apart from the sign-flip prediction of the Sivers functions probed in SIDIS and Drell-Yan processes.
The experimental test of this prediction at EicC will enrich our knowledge about the proton spin structure, particularly in the relatively small $x$ region. We should also note that due to the large uncertainty of the limited available polarized SIDIS data one cannot apply a very flexible functional form in the global analysis. Once more precise data are available from EicC and other future experiments, we will be able to have less biased extractions of the Sivers function as well as other TMDs by using much more flexible parametrizations and more realistic estimations of the uncertainties. Results in Figs.~\ref{fig:sivers-updown} and~\ref{fig:sivers-strange} can be understood as the impact of EicC SIDIS experiments from statistics point of view since the same analysis method is applied to world data and projected data.

The EicC design  enables us to precisely measure all 18 structure functions by combining different beam polarization configurations and the separation of different azimuthal modulation terms.  Other TMDs also receive great interest. For instance, the transversity TMD, which survives after $k_T$ integration, measures the net density of transversely  polarized quarks in a transversely polarized nucleon.  The first moment of the integrated transversity distribution is the tensor charge, which is a fundamental QCD quantity defined by the matrix element of the tensor current operator. It is recognized  as a benchmark of the lattice QCD study of hadron structures. Therefore, a precise determination of the tensor charge including the flavor separation can serve as an experimental test of lattice QCD.  Due to its chiral-odd property, the transversity TMD contribution to inclusive DIS cross section is highly suppressed by the power of $m_q/Q$. In contrast,  it can be extracted from a leading power single spin asymmetry in SIDIS process, known as the Collins asymmetry, which arises from the coupling of the transversity TMD and the Collins fragmentation function. Alternatively, the transversity distribution can also be accessed by analyzing the di-hadron SIDIS events at EicC. In the present global analysis, the sea quark transversity distribution is commonly assumed to be zero. EicC as an ideal facility for the study of sea quark distributions will provide the opportunity to test this assumption.

In conclusion, EicC with  wide kinematic coverage and high luminosity has the capability to deliver the high precision experimental data. The SIDIS measurements at EicC combined with those at 12-GeV upgraded JLab focusing on the study of valence quark distributions and the recently approved high-energy EIC at BNL  will provide complementary extractions of TMDs covering the full $x$ range towards a complete three-dimensional imaging of the nucleon in the momentum space. EicC as a facility that bridges the energy gap between JLab-12GeV and the future EIC at BNL is a perfect machine to study TMD evolution effects, in particular, to constrain the non-perturbative part of the evolution kernel~\cite{Collins:2014jpa}. 

\subsection{Generalized parton distributions}

Generalized parton distributions (GPDs)  encode information on the three dimensional structure of nucleon in the joint transverse position-longitudinal momentum phase space~\cite{Mueller:1998fv,Ji:1996nm,Ji:1996ek,Radyushkin:1997ki,Diehl:2005pc}. They were initially introduced to describe the exclusive processes where an active parton participating in the hard scattering is re-absorbed into nucleon that remains intact after collisions. GPDs depend on two longitudinal momentum fractions $x$ and $\xi$ and on the squared momentum transfer
$t$ to the proton.

GPDs are widely connected to other physics quantities. In the different kinematical limits, GPDs are reduced to the normal parton PDFs and electromagnetic form factors of the nucleon. In particular, setting $\xi = 0$ and performing a Fourier transform with respect to the transverse component of $t$,  one obtains an impact parameter distribution, which describes the joint distribution of partons in their longitudinal momentum and their transverse position $b_\perp$ inside the proton~\cite{Burkardt:2000za,Burkardt:2002hr}.

One of the most important physics motivations of GPD studies is to understand nucleon spin structure. The GPDs' connection with partons angular momentum is quantified through the Ji's sum rule~\cite{Ji:1996ek}, 
\begin{equation}
J_{q,g}=\frac{1}{2}\int_{-1}^{1}dxx[H_{q,g}(x,\xi,0)+E_{q,g}(x,\xi,0)].
\label{eq:quark_angular_mom}
\end{equation}
Where $J_{q,g}$ represents the total angular momentum for quark and gluon, which can be further decomposed as,
\begin{equation}
\frac{1}{2}=J_q + J_g = \frac{1}{2}\Delta\Sigma + L_q+ J_g,
\label{eq:sin-decomp}
\end{equation}
with  $\frac{1}{2}\Delta\Sigma$, $L_q$ and $J_g$ being the quark spin angular momentum, quark orbital angular momentum and gluon total angular momentum respectively.  The quark orbital angular momentum can be extracted through the measurements of GPDs $H$ and $E$ in exclusive processes by subtracting the quark helicity contribution. It is also worth to mention that GPDs encode the rich information on the mechanical properties of nucleon internal structure~\cite{Polyakov:2002yz,Lorce:2018egm,Shanahan:2018nnv,Polyakov:2018zvc} through the gravitational form factors(GFFs), which is related to the second moment of the unpolarized GPD. These mechanical properties, such as the pressure and shear force distributions, the mechanical radius, and the mechanical stability of a particle, contain the crucial information on how the strong force inside nucleon balance to form a bound state.  However, the precise extraction of GFFs at the current facilities  remains problematic due to poor data constraints~\cite{Burkert:2018bqq,Kumericki:2019ddg,Moutarde:2018kwr,Moutarde:2019tqa}.  As GPDs play an essential role in exploring the internal nucleon structure from many aspects, the experimental studies of GPDs have been and are a cutting-edge field of high energy nuclear physics during the last two decades.

\begin{figure}[htbp]
\centering
\includegraphics[width=0.98\textwidth]{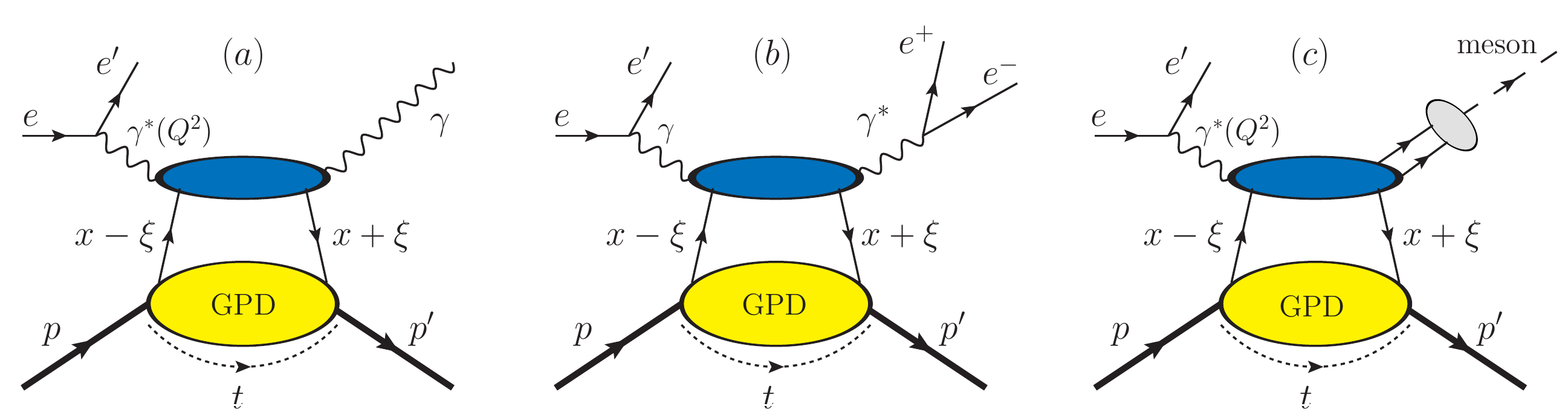}
\caption{Diagrams of various processes to study Generalized Parton Distributions (GPDs). 
(a) deeply virtual compton scattering, (b) time-like compton scattering, (c) deeply virtual 
meson production.}
\label{fig:DVCS}
\end{figure}

The main exclusive processes which allow to access to the GPDs in $ep$ collisions are deeply virtual Compton scattering (DVCS) $ep \longrightarrow ep \gamma$ (Fig. \ref{fig:DVCS}(a)), time-like Compton scattering $ep \longrightarrow epl^+l^- $ (Fig. \ref{fig:DVCS}(b)) and deeply virtual meson production(DVMP) $ep \longrightarrow ep M$ (Fig. \ref{fig:DVCS} (c))~\cite{Ji:1996nm,Ji:1996ek}.  
At the leading order,  DVCS process is described by the partonic channel
$ q \gamma^* \longrightarrow q \gamma $ where the virtual photon is provided by the electron. 
GPDs enter the  cross section of DVCS process through the  Compton form factors(CFF)  defined as (for example, the quark GPD $H^q$)~\cite{Kumericki:2007sa,Kumericki:2009uq,Guidal:2013rya,Kumericki:2016ehc},
\begin{equation}
{\mathcal H} (x_B,t,{\cal Q}^2) =
  \int_{-1}^1\!dx\, \left[\frac{1}{\xi-x- i \epsilon}-\frac{1}{\xi+x-i \epsilon} \right]
 \sum_{q=u,d,s,\cdots} e_q^2 H^q (x,\xi,t,{\cal Q}^2)\,,
\label{eq:CFF}
\end{equation}
where $\xi \approx x_B/(2-x_B)$ with $x_B$ being the Bjorken's variable. The similar relation holds for other GPDs. The precise measurements of the various angular modulations and polarization dependence of DVCS cross section at different kinematic points in ($Q^2,x_B,t$) would allow us to extract different CFFs, as each of  them has unique angular and polarization dependencies (see the reference~\cite{Diehl:2005pc} and therein). The corresponding GPDs can be subsequently constrained by the extracted CFFs.

Let us highlight some specific features of different production channels. The main limitation of DVCS process is that it is sensitive only to the sum of quark and anti-quark distributions in a particular flavor combination. In contrast, exclusive meson production offers substantial help in the separation of different quark and antiquark flavors and of gluons. For example, the valence quark and sea quark GPDs can be probed via pseudo scalar mesons ($\pi, K, \eta, ...$) production processes, whereas the vector mesons ($\rho, \phi, \omega$) production is more sensitive to sea quark and gluon GPDs. However, extracting GPDs from exclusive meson production requires the knowledge of additional non-perturbative matrix element, the meson distribution amplitude.

The precise extraction of GPDs from the measurements of exclusive processes puts the highest demands on experiments for various reasons, including the smallness of cross sections in exclusive processes, the interference with the Bethe-Heitler (BH) process, etc. The measurements of GPD-related observables in the region of moderate to large $x$ have been carried out at HERMES~\cite{Airapetian:2012pg}, COMPASS~\cite{Sandacz:2018zsy}, and  JLab~\cite{Hyde:2011ke}. However, most of these measurements have sizable statistical uncertainties and provide reasonable constraints for only one GPD, $H$. The complete and precise extraction of all GPDs requires high luminosity, detectors with full hemisphere coverage, beams with various polarization choices, and wide kinematic reach. Until now, there have been no facilities being able to meet all these demands. For  example,  the  luminosity  at  HERA  and  COMPASS is low, meanwhile COMPASS has its complication while flipping the muon beam helicity, which makes the extractions of some polarization dependent GPDs extremely difficult. On the other hand, though the luminosity at JLab-12GeV is very high, the  most DVCS events lie well below  $Q^2<10\,{\rm GeV}^2$~\cite{JLab12GeV-HallA,JLab12GeV-HallB,JLab12GeV-HallB-2,JLab12GeV-HallC}, where the various high twist effects or high order effects can play a role and complicate the  extraction of GPDs (see \emph{e.g.}~\cite{Moutarde:2013qs,Braun:2014sta,Defurne:2015kxq,Defurne:2017paw}). It is usually believed that these theoretical uncertainties can be well controlled if going to higher $Q^2$ region~\cite{Braun:2014sta}. 

\begin{figure}[htbp]
\includegraphics[width=0.55\textwidth,angle =0]{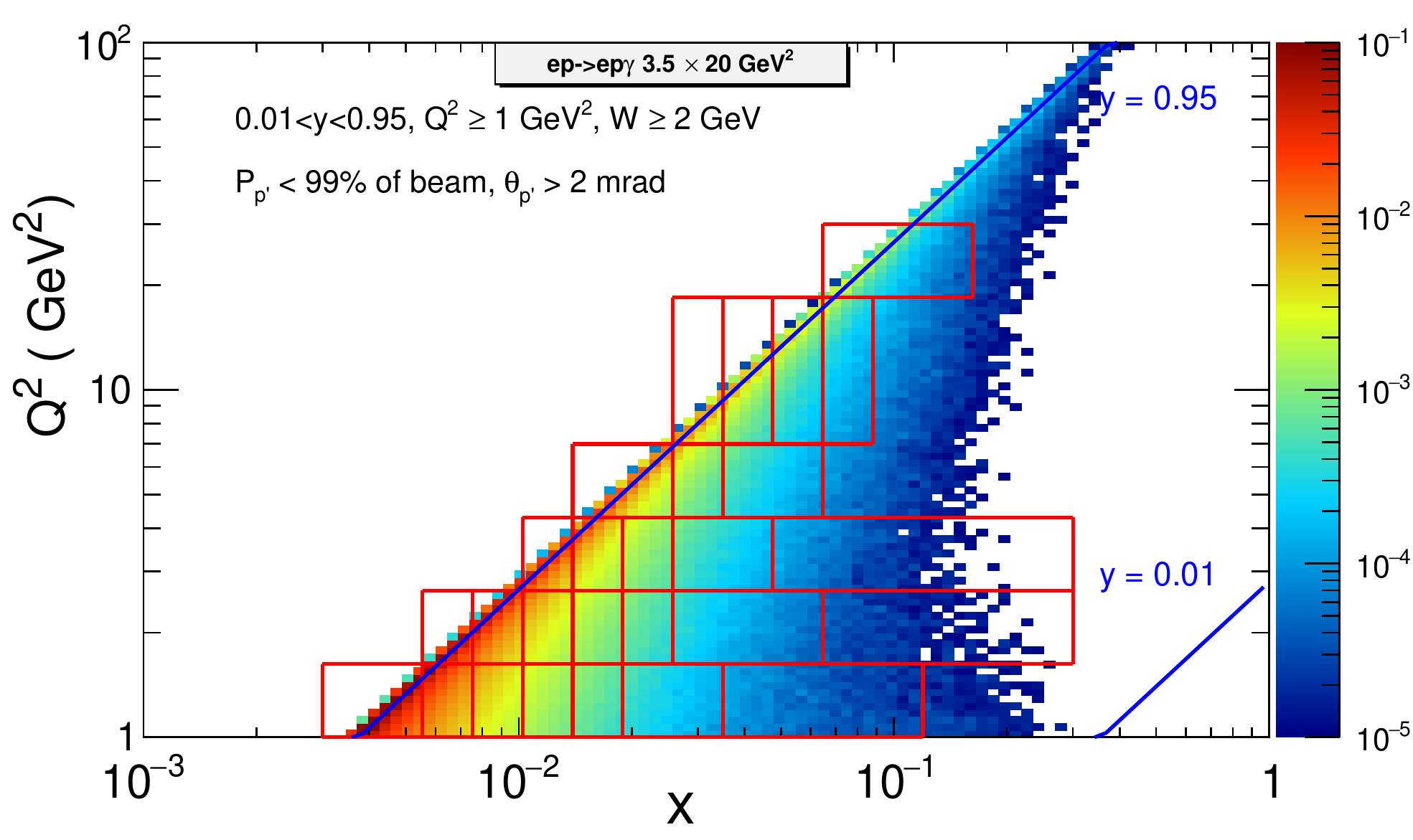}
\includegraphics[width=0.47\textwidth,angle =0]{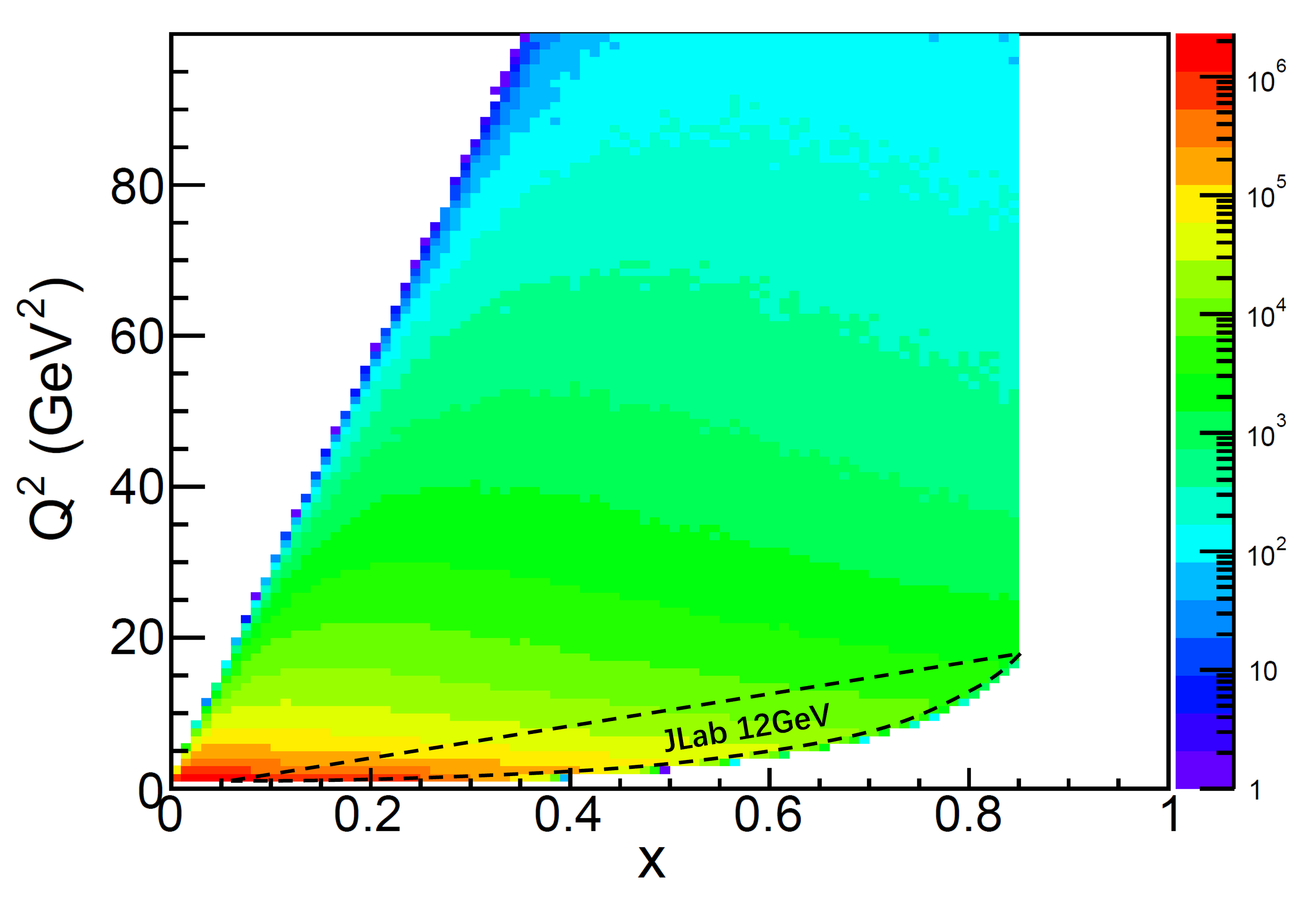}
\caption{The distributions of DVCS(left panel) and $\pi^0$ DVMP(right panel) events in the $x$ and $Q^2$ for 3.5GeV electron beams colliding with 20GeV proton beams energy at EicC. The event number with arbitrary normalization is indicated by different colors. In the left panel, the contributions from both the DVCS and the BH processes as well as their interference are included, and the used kinematic cuts and binning scheme are also shown.}
\label{GPDkin}
\end{figure}

The programs of the experimental GPDs studies  will be dramatically extended by EicC, predominantly in the sea quark kinematical region. The precise measurements of both DVCS and DVMP  can be carried out at EicC with versatile beam species/polarizations. The typical $Q^2$ range accessible with EicC is  $1$ GeV$^2 < Q^2 <30$ GeV$^2$ (see  the left panel of Fig.~\ref{GPDkin} adapted from MILOU package \cite{Perez:2004ig}). Meanwhile, the Bjorken variable $x$ can reach down to $x \approx 0.05$ when restricting  $Q^2$ to the perturbative region  $Q^2 >10$ GeV$^2$. This kinematic coverage with the current design will make EicC a unique machine to explore parton spatial imaging of the nucleon in the sea quark region. In the valence quark region, the theoretical uncertainties will be substantially reduced at EicC as compared to that at JLab-12GeV due to relatively higher $Q^2$. Moreover, EicC is also  complementary to the US EIC in studying the DVCS process.  This is because the interference contribution between the Compton and the Bethe-Heitler processes is more prominent at a lower energy machine, whereas the Compton process may dominate at EIC for  given $x$ and $Q^2$ values.

\begin{figure}[htbp]
\centering
\includegraphics[width=0.8\textwidth,angle =0]{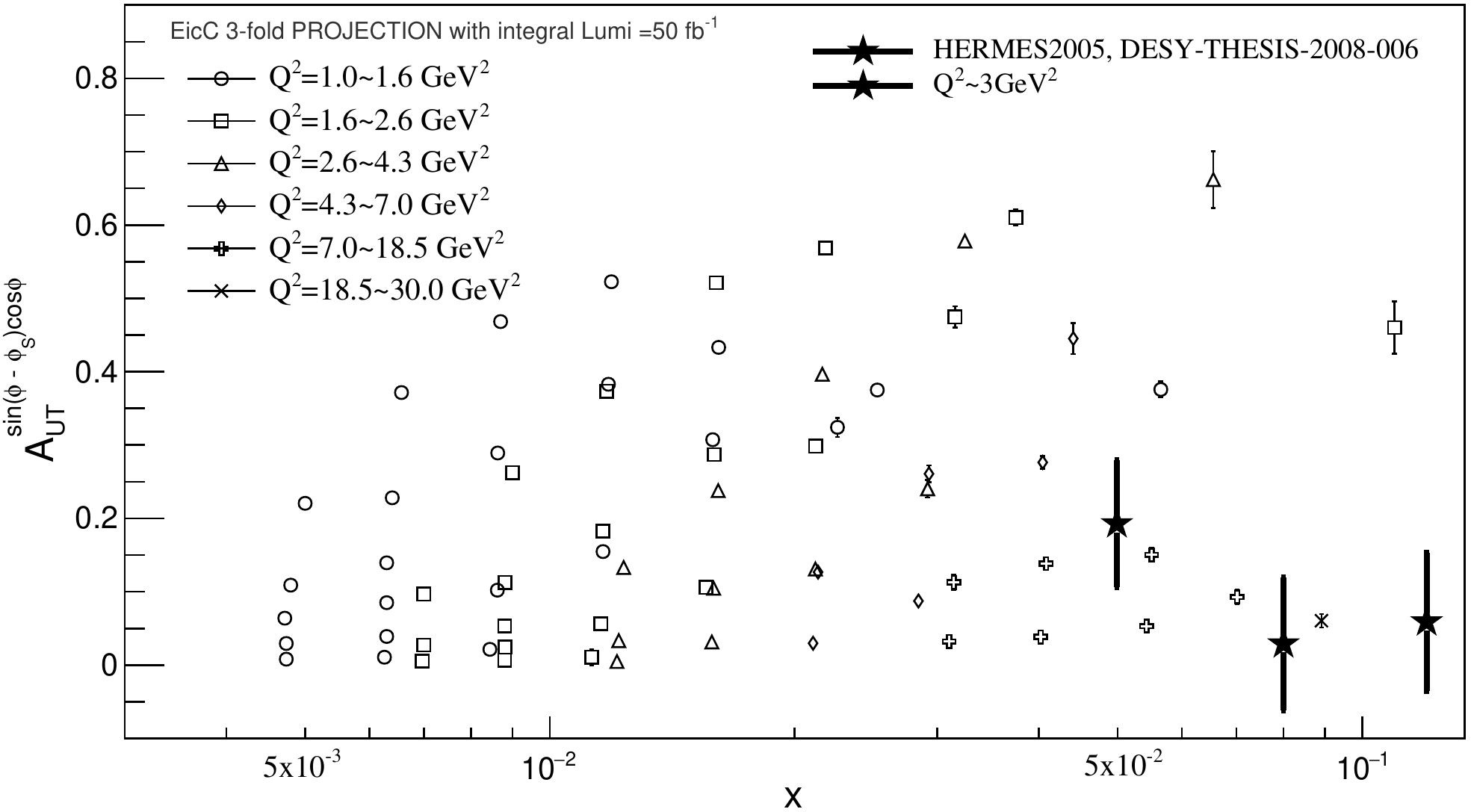}
\caption{The projected accuracy for $A_{UT}^{\sin(\phi-\phi_s)\cos\phi}$ asymmetry  in the process of DVCS off a transversely polarized proton target  at EicC in the region 1 GeV$^2<Q^2<30$ GeV$^2$. Only statistics uncertainty is included. The size of $A_{UT}$ is estimated with the Goloskokov-Kroll model~\cite{Goloskokov:2009ia,Goloskokov:2011rd,Kroll:2012sm}.
The black star is the HERMES data of $A_{UT,I}^{\sin(\phi-\phi_s)\cos\phi}$ asymmetry~\cite{Airapetian:2008aa}. The values of $|t|$ bins under the same $Q^2$ are not shown here for simplicity. }
\label{DVCSsim}
\end{figure}

As mentioned, parton orbital angular momentum  can be determined by extracting  GPD E and H from exclusive processes according to the Ji's sum rule.  However, in practice, it is extremely challenging to achieve this because it requires measuring the values of H and E for all $x$ at fixed $\xi$. Nevertheless, EicC  has great potential to advance our knowledge of GPD E and also H with a transverse polarized proton beam. Fig.~\ref{DVCSsim} displays the simulation results for the Compton scattering off a polarized proton. The EicC measurements of the DVCS transverse polarization asymmetry $A_{UT}$ with a single azimuthal modulation ${\sin(\phi-\phi_s)\cos\phi}$ have a rather small statistical uncertainties for a wide kinematic region with $|t|>0.01$ GeV$^2$. Here  the DVCS events are selected in the kinematic region   0.01 $<y<$ 0.95 and  $\gamma p$ center of mass energy W $>$ 2 GeV.

\begin{figure}[htp]
  \begin{center}
{\includegraphics*[width=0.9\textwidth,angle =0]{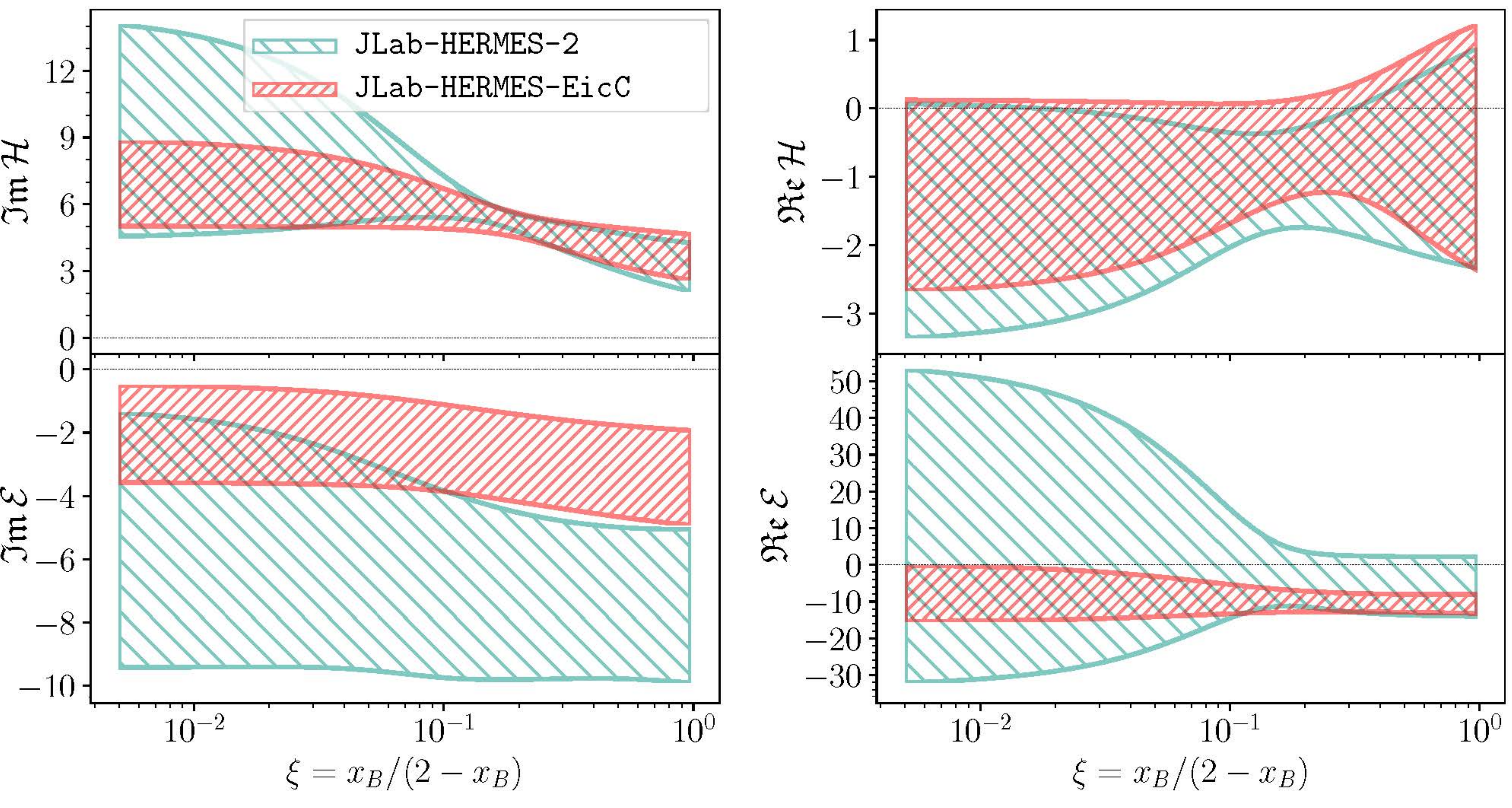}}
       \caption{
An exploratory  extraction of the real (${\mathcal Re}$) and the imaginary (${\mathcal Im}$) part of CFF ${\mathcal E}$ and ${\mathcal H}$ at $t = -0.2\,{\rm GeV}^2$ from the pseudodata generated for kinematical points shown in Fig.~\ref{DVCSsim} using  the neural network  method ~\cite{Kumericki:2011rz,Kumericki:2019ddg,Cuic:2020iwt}. 
The systematic uncertainty is not included yet.
      \label{figCPT2-2:CFF}}
  \end{center}
\end{figure}

The impact of these pseudodata on the extraction of DVCS  Compton form factors (CFF) ${\mathcal E}$ and ${\mathcal H}$ is displayed in Fig~\ref{figCPT2-2:CFF}. The approach of  global analysis utilizing the artificial neural network~\cite{Kumericki:2011rz,Berthou:2015oaw,Moutarde:2019tqa} is employed in order to reduce model dependency and propagate the uncertainties properly.  The poorly constrained  real part of $\widetilde{\mathcal E}$ and $\widetilde{\mathcal H}$ simply are assumed to be vanishing~\cite{Kumericki:2011rz,Kumericki:2019ddg,Cuic:2020iwt} in the simulation.  The light green bands, manily driven by statistical uncertainties, represent the accuracy of the existing JLab and HERMES data. The red bands show the accuracy after including the projected $A_{UT}$ data of EicC with statistical uncertainty only. One can see that the uncertainty for the extraction of the CFF ${\mathcal E}$ is  reduced in the sea quark region  once the  EicC measurements are included.
In most of the kinematic space systematic uncertainty
is at the similar level or even smaller than the statistical one. So we would still retain power of extraction of CFFs similar to what is shown on Fig 2.12, after including such level of systematic uncertainty.
  In addition to the $A_{UT}$ asymmetry,    other asymmetries, e.g. $A_{LU}$, $A_{UL}$, $A_{LL}$ and $A_{LT}$ can be precisely measured   with the different beam polarization configurations and the different azimuthal modulations at EicC as well. A combined analysis  with all these measured  modulations  would further reduce significantly the error bands shown in Fig~\ref{figCPT2-2:CFF}.  
So under the high luminosity design of EicC, the statistical uncertainties will not play essential role in the future GPD extraction and the systematic uncertainties, whose main sources are discussed in previous sections, are anticipated to be dominant, which will be under good control by facility design. 

\begin{figure}[htp]
\centering
\includegraphics[width=0.85\textwidth]{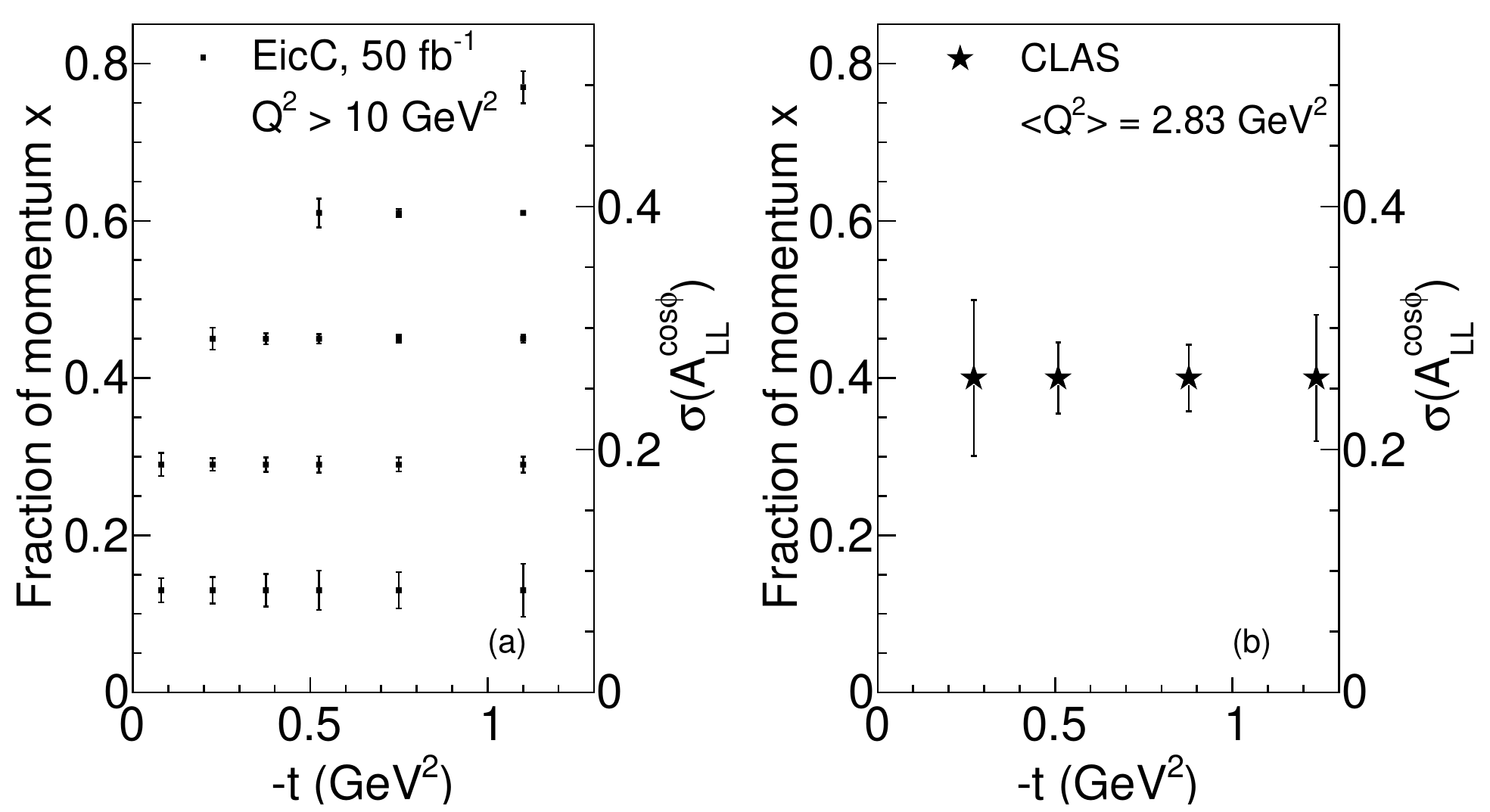}
\caption{The statistics error of the projected $A_{LL}^{\cos \phi}$ asymmetry for $\pi^0$ production in DVMP process at EicC. The CLAS data is taken from Ref.~\cite{Kim:2015pkf}. }
\label{fig:pion0-DVMP-ALLcosphi-errors}
\end{figure}

As discussed, EicC can deliver high precision data for  the DVMP process in the large $Q^2>10$ GeV$^2$ region, where the non-perturbative effects and higher-twist contributions are suppressed, so that GPDs can be extracted reliably. 
Fig.~\ref{GPDkin} (right panel) displays the expected distribution of DVMP events at EicC in  bins of $Q^2$ and $x$.  Note that DVMP cross sections rapidly decrease with increasing $Q^2$.  One sees that there are substantial event numbers in the moderate $Q^2$ region where perturbative treatment is justified. This will facilitate clearly separating different quark flavor contributions by combining with DVCS data. For instance, one can  separate up quark and down quark contributions by carrying out the measurements of  $A_{LL}^{\cos \phi}$ which arises from the coupling of the chiral-odd GPDs and the twist-3 distribution ampliutde of pion~\cite{Goloskokov:2009ia,Goloskokov:2011rd,Goldstein:2013gra,Kim:2015pkf}.  This observable could provide valuable information on transversity PDF that is hard to access in inclusive process. Despite the higher twist nature of this observable,  we found that the asymmetry is significant for the DVMP channel of $\pi^0$  production.   With the EicC  DVMP pseudo-data displayed in Fig.~\ref{GPDkin}, it is shown  in Fig.\ref{fig:pion0-DVMP-ALLcosphi-errors}  that at large $Q^2$  the statistical uncertainty of the $A_{LL}^{\cos\phi}$ in $\pi^0$ production is significantly reduced. Therefore, EicC presents an unique opportunity to study the chiral odd GPDs. 

%

In summary, judging from our simulation results presented here, EicC  will greatly advance our knowledge about the internal structure of nucleons. The combined kinematic coverage of the EicC, JLab and of EIC-US is essential for ultimately yielding the complete 3D images of proton from the large $x$ down to the saturation regime, and for much more profound understanding of the proton spin puzzle as well.

\section{Partonic structure of nucleus}
\label{PartonicNucleus}

The electron-ion collision has been recognized as an ideal process to explore the distributions of quarks and gluons inside the nucleus, as well as to study the QCD dynamics of multiple parton interactions in the nuclear medium. In this process, the electron scattering part, which can be well controlled both experimentally and theoretically, provides a high precision probe to reveal the detailed partonic structure of the nucleus which is impossible to be calculated theoretically. Besides, the nucleus can also serve as a QCD laboratory at the fermi scale to investigate the strong interactions between the energetic parton and the nuclear medium by carefully studying the so-called hadronization process which largely depends on the type of the nucleus. The detailed analysis of these nontrivial nuclear medium effects can help us to probe the fundamental differences of partonic properties in free nucleons and the nuclear medium, as well as to understand the mystery of hadronization mechanisms and the QCD confinement of quarks and gluons. 

\subsection{The nuclear quark and gluon distributions}
A full understanding of the difference between the properties of quarks and gluons inside a free nucleon and that inside a nucleon bounded within the nucleus will help us understand how the nucleus is formed at the partonic level. The longitudinal momentum distributions of quarks and gluons in a free nucleon are characterized by the usual leading twist parton distribution functions (PDFs) which have been precisely measured in the high-energy electron-proton collisions. A natural question is: how these PDFs are modified by the nuclear medium when the nucleon is bounded? To answer such a fundamental question remains one of the biggest challenges in the nuclear physics community. Due to the lack of experimental data and the limited kinematic coverage, the precision for nPDFs global extraction is far less than that for PDFs in free nucleons \cite{AbdulKhalek:2020yuc,Eskola:2016oht,Hirai:2007sx,deFlorian:2011fp,Kovarik:2015cma,Khanpour:2016pph,Walt:2019slu}. In particular, the extraction of nPDFs of sea-quarks and gluons is suffering from even much larger uncertainties. It is strongly desired to perform more high-precision measurements of conventional experimental observables as well as to explore new observables that are sensitive to the sea-quark and gluons. 

\begin{figure}[htbp]
\begin{center}
\includegraphics[width=0.8\textwidth]{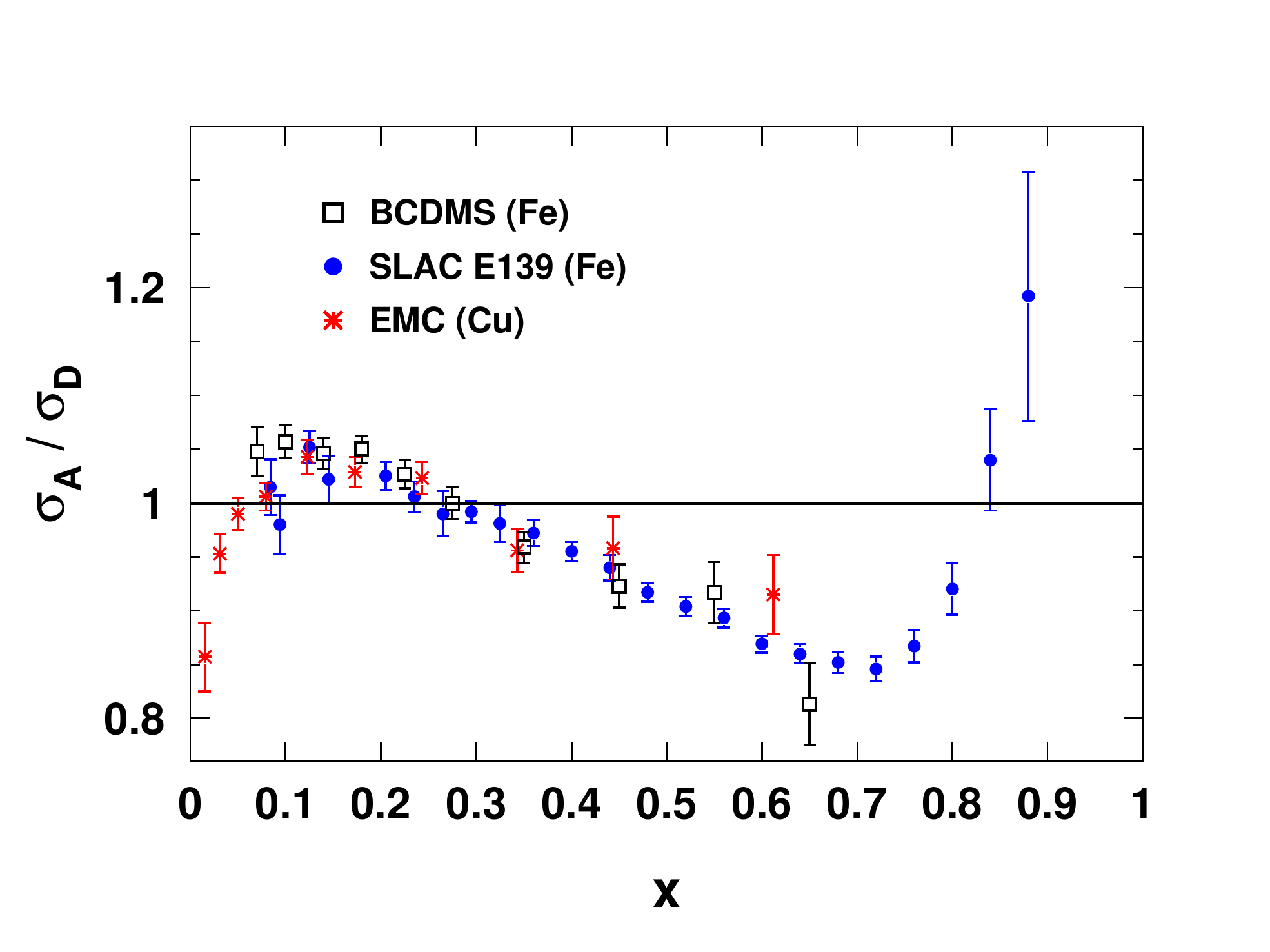}
\caption{\label{fig:EMC} The cross section ratio between electron-ion and electron-deuteron deep inelastic scattering \cite{Guzey:2012yk}.}
\end{center}
\end{figure}

In the past three decades, various experiments have confirmed that the PDFs measured in free nucleons and bounded nucleons are significantly different. Shown in Fig.\ref{fig:EMC} reveals the cross-section ratios for inclusive DIS between eA and eD collisions in terms of Bjorken x distributions.  The solid circles, the open squares, and stars correspond to the data from SLAC E139 \cite{Gomez:1993ri}, BCDMS \cite{Benvenuti:1987az}, and EMC \cite{Ashman:1992kv}, respectively.  Despite their different reactions and kinematic ranges, these data exhibit a very similar nuclear medium effect. There are four distinguishable regions ~\cite{Geesaman:1995yd}: 1. Fermi motion in $x>0.7$; 2. EMC effect in the range of $0.3<x<0.7$; 3. Anti-shadowing around $x\sim 0.1$; 4. Shadowing in $x<0.01$. 

Among them, the Fermi motion region can be interpolated as the result of the momentum distributions of nucleons bounded in the nucleus which differ among light to heavy nuclei. The so-called EMC effect, which is believed to be due to the modification of the valance quarks in the nuclear medium, has been exclusively studied since its discovery in 1980s but no satisfactory explanation has yet been reached to address its root cause. The most recent discovery of its strong correlations with the short-range corrections (SRC)~\cite{Weinstein:2010rt,Arrington:2012ax,Hen:2016kwk} sheds a light to fully unveil this mystery and provide a new way to study the nuclear structure in the partonic level \cite{Schmookler:2019nvf}. Enormous new experimental programs and theoretical calculations have been planned to continue studying this effect. On the other hand, no experimental evidence has shown that such an effect also exists in sea-quarks and gluons. Furthermore, the physics origins of the anti-shadowing and shadowing at small $x$ remain unknown due to a lack of experimental measurements and theoretical interpolations. 

\begin{figure}[htbp]
\begin{center}
\includegraphics[width=0.95\textwidth]{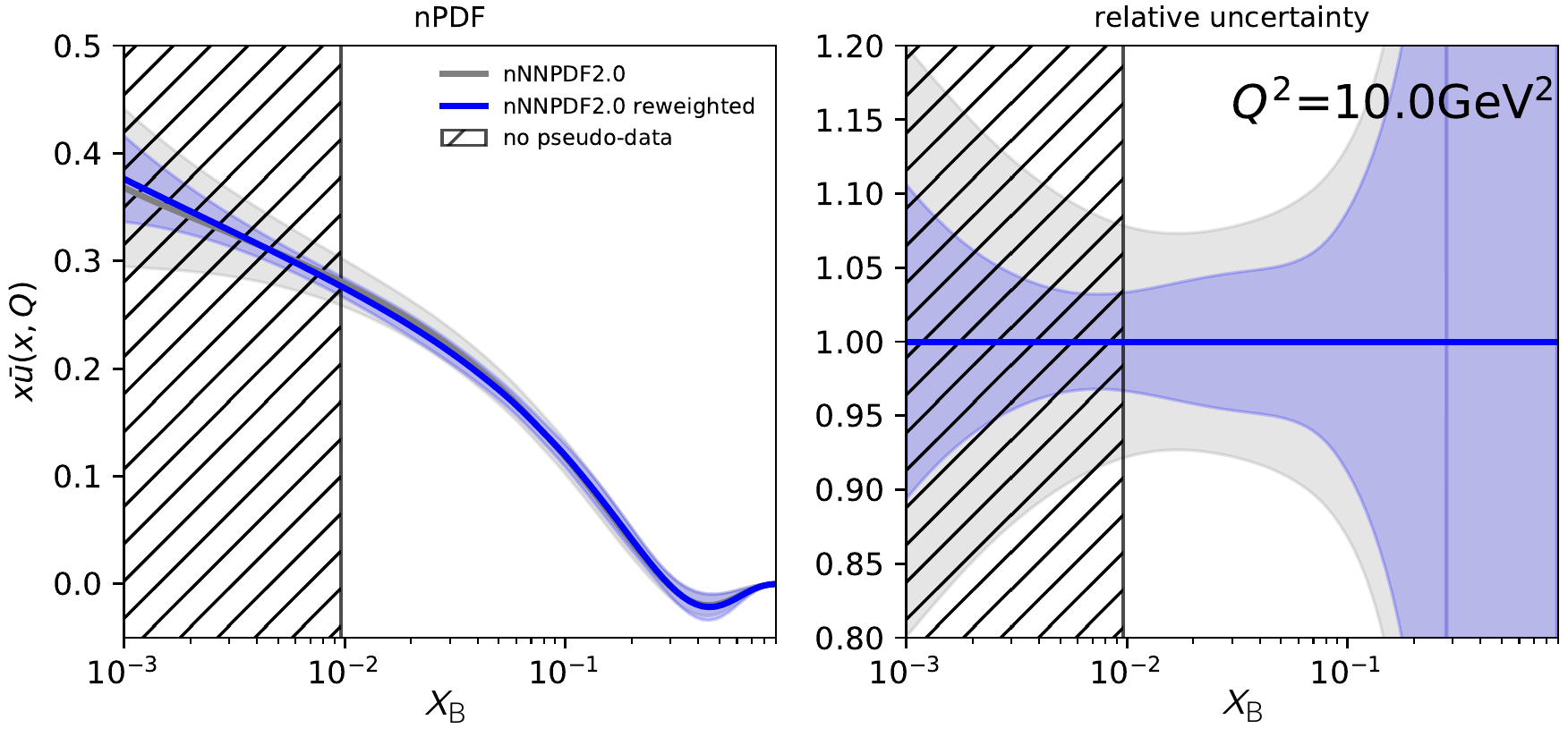}
\caption{\label{fig:ub-impact} Left: the $\bar u$ distribution in Pb at $Q^2=10$ GeV$^2$, the gray band comes from the original nNNPDF2.0 set, the blue band corresponds to the set reweighted with the EicC pseudo data based on the integrated luminosity $\mathcal{L}=0.01$ fb$^{-1}$. Right: the relative uncertainty for two sets of nPDFs.}
\end{center}
\end{figure}

\begin{figure}[htbp]
\begin{center}
\includegraphics[width=0.95\textwidth]{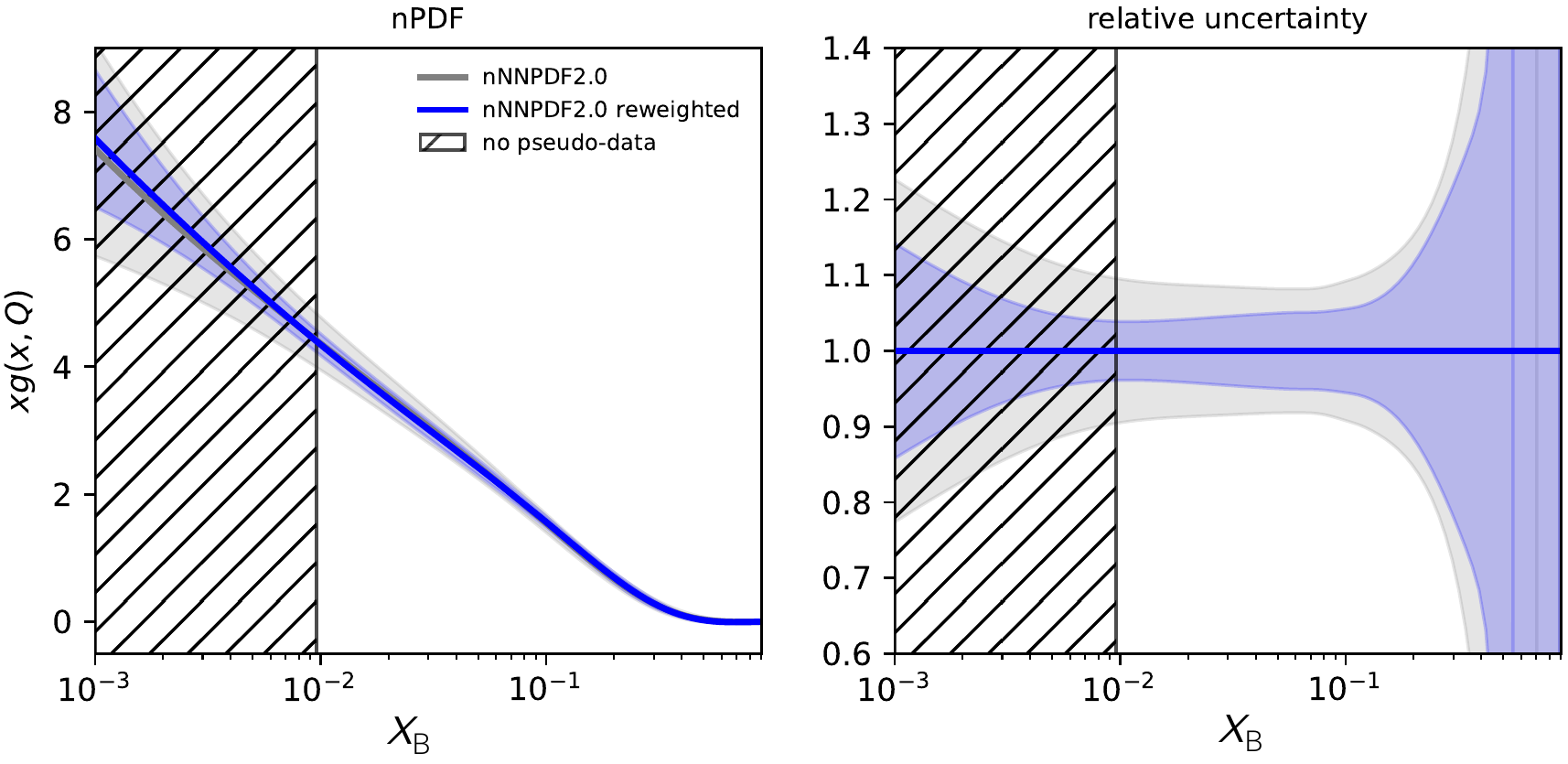}
\caption{\label{fig:g-impact} Same as Fig. \ref{fig:ub-impact}, but for gluon.}
\end{center}
\end{figure}

In the most recent nPDF determination, nNNPDF2.0, a robust quark flavor separation and a good handle of the gluon are performed \cite{AbdulKhalek:2020yuc}. However, compared to those for free nucleons large uncertainties of nPDFs remain due to the very limited existing measurements. This is also true in other global fittings such as EPPS16 \cite{Eskola:2016oht}. The future EicC will place its kinematics in the sweet spot where the nuclear medium effects of valance-quarks and sea-quarks can be extensively studied by measuring their intrinsic PDFs in bound nucleons using the eA DIS processes with a wide range of nuclei beams. Based on the projections for the EicC pseudo-data which are generated through NLO calculation of DIS cross-section using the nNNPDF2.0 nuclear PDFs, the impact of future EicC measurements on nuclear PDFs utilizing Bayesian reweighting is shown in Figs. \ref{fig:ub-impact} and \ref{fig:g-impact}. In Fig. \ref{fig:ub-impact}, the sea quark distribution in Pb at $Q^2 =10$ GeV$^2$ for both the original and reweighted nNNPDF2.0 fits are shown with uncertainty bands correspond to $90\%$ confidence level. In particular, the reduction of reweighted $\bar u$ uncertainty in the kinematic region covered by EicC, i.e. $x>0.01$, strongly indicates EicC pseudo data are adding a significant amount of new information to the global fit. A similar analysis for gluon distribution in Pb is shown in Fig. \ref{fig:g-impact}, which indicates the constraining power of EicC measurements on gluon nuclear modification. Notice that the analysis shown in Figs. \ref{fig:ub-impact} and \ref{fig:g-impact} are based on the integrated luminosity $\mathcal{L}=0.01$ fb$^{-1}$, which corresponds to only a few hours of running \footnote{Studies using EicC pseudo data with an integrated luminosity equivalent to one week or more of running have also been performed. However, the impact on the nNNPDF2.0 PDFs estimated by reweighting is so significant that the number of effective replicas with non-zero weight “surviving” the analysis reduces from an initial set of 1000 to a few dozen or less. This reflects the fact that the leap in precision between the data already included in the nNNPDF2.0 analysis and the future EicC data is too wide for reweighting techniques to return viable results and strongly suggests the need of a new fit. Similar conclusions on the reweighting procedure have been found recently in impact studies for polarized PDFs at the future US EIC using electron-helium SIDIS pseudo data \cite{Aschenauer:2020pdk}.}. Therefore, the real measurements with high precision and large coverage will provide a stringent constraint on nPDFs from the shadowing to the anti-shadowing region.      

In addition to the electrons, the detection of varied final-state hadrons, such as pions, kaons and heavier mesons, serves as the flavor-tagging to decouple the contributions from different quark flavors. The high precision data with large kinematic coverage will greatly improve the global extraction of quark nPDFs in the medium to low $x$ regions. Most importantly, EicC will, for the first time, precisely measure the undiscovered medium modification effect of sea-quarks as well as unveil the puzzle of anti-shadowing which remains largely unknown. 

On the other hand, various species of beam nuclei at EicC will also provide unique opportunities to shedding light on studying the nuclear structure at the partonic level. By detecting the outgoing protons and neutrons at the forward angles during the eA collision, the spectator-tagging DIS process serves as a powerful tool to exam the QCD origin of nucleon-nucleon interactions, such as the link between the SRC to EMC effects at $x>0.2$ region as well as the nucleon coherence phenomena at low $x$ that lead to the anti-shadowing and shadowing effects~\cite{Cosyn:2016oiq,Frankfurt:2011cs}. 

\subsection{Hadronization and parton energy loss in nuclear medium}
\label{sec:hadronization}

Precise measurements and phenomenological investigations on the SIDIS process with different nuclei in electron-ion collisions are the fundamental tools to analyze two widely discussed nuclear effects, i.e., the parton energy loss effect and hadronization in the medium ~\cite{Gyulassy:1993hr,Baier:1996sk,Gyulassy:2000er,Zakharov:1997uu,Guo:2000nz}. In a nuclear medium, the highly energetic parton generated in the hard scattering process will continuously encounter the multiple scatterings with the surrounding nucleons before it completely escapes from the nucleus or is converted inside the nucleus into charge-neutral particles, a complicated process also known as the hadronization. The collective consequence eventually leads to nontrivial phenomena of nuclear modifications, including the attenuation and broadening of the hadron spectrum in eA collisions comparing to ep collisions. These phenomena, which are very sensitive to the nuclear parton densities and QCD dynamics of multiple parton interactions in the nuclear environment, have been observed in semi-inclusive deep inelastic scatterings in HERMES \cite{Airapetian:2007vu} and Drell-Yan dilepton production in proton-nucleus collisions in FNAL-866 \cite{Vasilev:1999fa}. The available data are used to extract the transport properties of cold nuclear matter and study the parton energy loss mechanism in the nucleus \cite{Wang:2002ri,Chang:2014fba,Xing:2011fb,Arleo:2018zjw}. However, large uncertainties still exist, mainly originated from two aspects: 1. limited kinematic coverage of experimental data; 2. the assumption in energy loss calculation that the partons fragments outside the nuclear medium. To obtain more precise information about the hadronization process and the mechanism of parton energy loss in medium,
we need EicC to fill in the open window that has not been covered by existing experimental measurements.

Hadronization, as encoded in fragmentation functions, describes the process of quarks and gluons fragment into final state hadrons. In the presence of a large size nuclear medium, the hadronization dynamics will be affected and eventually leads to different hadron spectrum comparing to that in a vacuum \cite{Bialas:1980at,Akopov:2004ap}. As we know, parton energy loss effects also lead to the suppression of the hadron spectrum as functions of beam energy $\nu$ and fragmentation fraction $z_h$ in eA collisions comparing to that in ep collisions \cite{Airapetian:2007vu,Chang:2014fba}. Therefore, we can not disentangle these effects from the available experimental data, as they lead to the same phenomena but with very different mechanisms. This requires us to perform the measurement more differentially and consider as many as possible the final state identified hadrons. The high collision energy and the high luminosity, as well as the capability of identifying various hadrons in future EicC, will play a key role to differentiate the parton energy loss effect and medium hadronization effect.

\begin{figure}[htbp]
\centering
\includegraphics[width=0.45\textwidth]{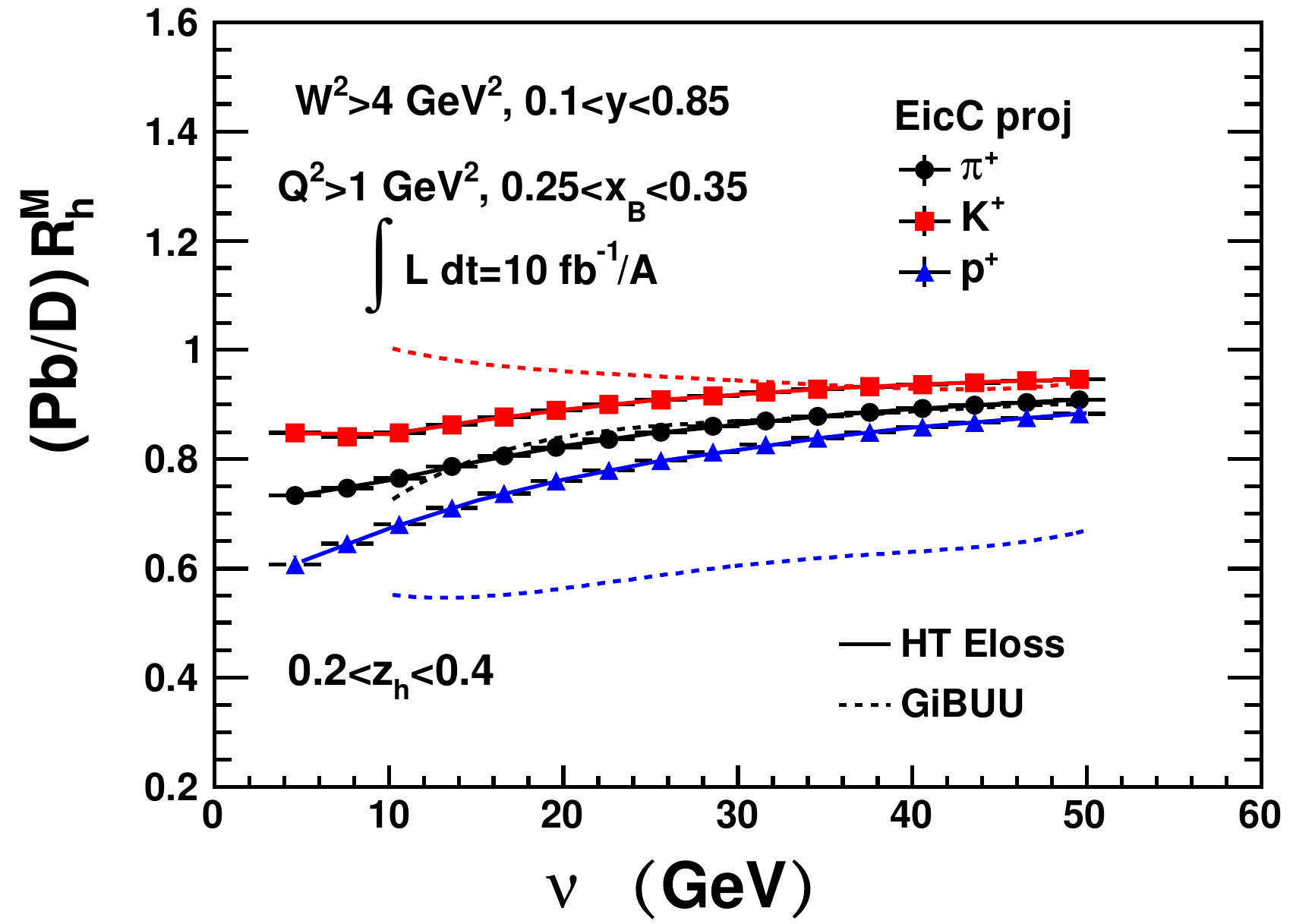}
\includegraphics[width=0.48\textwidth]{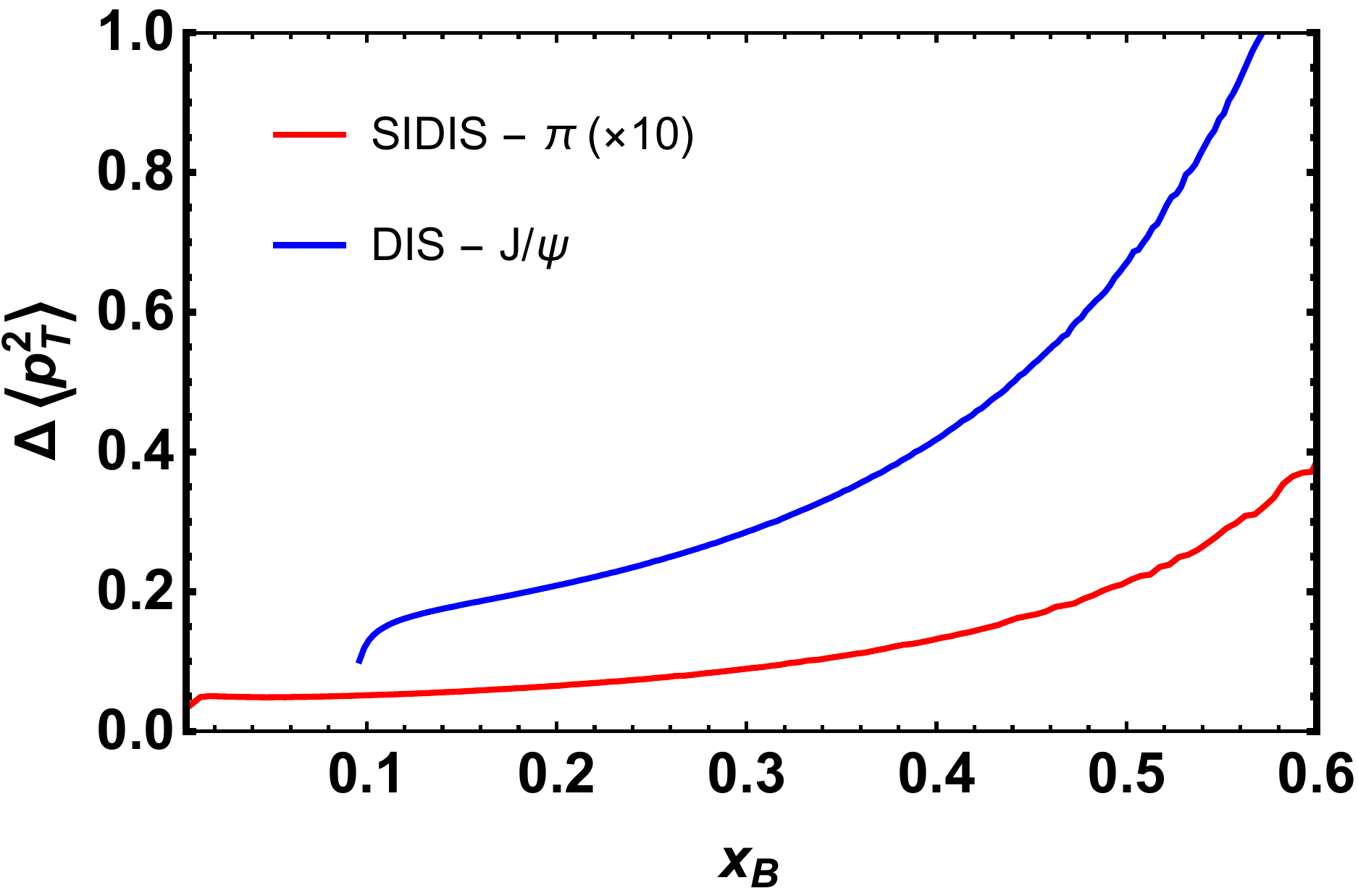}
\caption{\label{fig:RhM_z}  Left plot: the cross section ratios for $\pi^+, K^+$ and $p$ between election-ion and electron-proton collisions at EicC energy region, i.e., 3.5 GeV electron beam and 20 GeV per charge for heavy ion beam, as a function of virtual photon energy $\nu$. Right plot: the transverse momentum broadening for $\pi$ and $J/\psi$ at future EicC.}
\end{figure}

Shown in Fig. \ref{fig:RhM_z} is the comparison between predictions from parton energy loss model (solid curves) and hadron transport model (dashed curves) for the nuclear modification factors, where only events with $0.1<y<0.85,~W^2>4 \rm{GeV}^2,~Q^2>1\rm{GeV}^2$ are selected in the process when 3.5 GeV electron collides with 20 GeV (per nucleon charge) Pb beam, and various hadrons represented by different shaped points are considered in the simulations. By looking at the dependence of $R_M^h$ as a function of the virtual photon energy $\nu$, the capability of particle identification as well as the kinematic coverage in EicC will allow us to disentangle the hadronization mechanism from the parton energy loss effect as indicated by the difference between solid and dashed curves. Though the two models give very similar nuclear modification effect for $\pi^+$ production, enormous differences for $p$ and $K^+$ are predicted. These differences can be identified in EicC considering its high luminosity $50 fb^{-1}$, which leads to invisible statistical uncertainty as shown in Fig. \ref{fig:RhM_z}.

The transverse momentum broadening effect is very sensitive to the QCD dynamics of multiple parton interactions in the nuclear environment and the nuclear medium transport property. It has been extensively studied in heavy-ion collisions, see for example \cite{Baier:1996sk, Burke:2013yra, Ru:2019qvz}. Similarly, we can also use this observable to probe the fundamental properties of the nuclear medium in eA collisions. Comparing to pA collisions, eA collisions is much cleaner due to the absence of the strong interaction between the beam electron and the target nucleus. Based on the assumption that the partons hadronize outside the nuclear medium, we show in Fig. \ref{fig:RhM_z} the transverse momentum broadening for light hadron (red curve) and $J/\psi$ (blue curve), which can be used to probe the jet transport parameters for quark jet and gluon jet, respectively. Notice that the available measurements on the gluon jet transport parameter are very limited, and EicC can make a significant contribution to this subject.  

\section{Exotic hadronic states}
\label{ExoticHadrons}

Hadron spectroscopy started a new era in 2003 when the $D_{s0}^*(2317)$, $D_{s1}(2460)$ and $X(3872)$\footnote{It is denoted as $\chi_{c1}(3872)$ according to its quantum numbers in the latest version of Review of Particle Physics (RPP) by the Particle Data Group~\cite{Zyla:2020}.} 
were discovered at the $B$ factories.
Since then many new hadron resonances or resonant structures were discovered at various experiments all over the world. 
In particular, most of them contain at least one heavy (charm or bottom) quark, and have properties at odd with expectations from quark model. The meson states discovered in the heavy-quarkonium mass region are called $XYZ$ states, see Table~\ref{tab:XYZ} for a list. 
Notable examples include the $X(3872)$, $Z_c(3900)$, $Z_c(4020)$ and others. In 2015 and 2019, the LHCb Collaboration discovered pentaquark candidates with hidden charm, $P_c(4312)$, $P_c(4440)$ and $P_c(4457)$. 
The charged heavy-quarkonium like $Z_c$ and $Z_b$ states as well as these $P_c$ states are clearly beyond the scope of the conventional quark model for mesons and baryons, and thus excellent candidates of exotic multiquark states.
Understanding the nature of these structures has been the main concern for hadron spectroscopy, and is a challenge that needs to be solved toward revealing the mystery of how massive hadrons emerge from the interaction between quarks and gluons.

Various models were proposed to explain (some of) these observations, including multiquark states, hadronic molecules, hybrid states, mixing of different components and non-resonant effects such as kinematical singularities and  interference.
These investigations were witnessed by a large number of review articles in the past few years, see Refs.~\cite{Jaffe:2004ph,Swanson:2006st,Voloshin:2007dx,Klempt:2007cp,Klempt:2009pi,Brambilla:2010cs,Chen:2016qju,Hosaka:2016pey,Lebed:2016hpi,Esposito:2016noz,Guo:2017jvc,Dong:2017gaw,Ali:2017jda,Olsen:2017bmm,Karliner:2017qhf,Yuan:2018inv,Kou:2018nap,Cerri:2018ypt,Brambilla:2019esw,Liu:2019zoy,Yamaguchi:2019vea,Guo:2019twa,Zyla:2020} emphasizing on various aspects of these new resonant structures.
Many of the observed structures need to be confirmed by other experiments, and most of the theoretical models also predicted light-flavor and/or heavy-quark partner states of the observed ones. Thus, in order to understand the pattern behind the messy spectrum of these new hadrons and to be able to classify them into a clear picture, which can in turn give important hints towards understanding the confinement mechanism, more experimental measurements are urgently needed. 

\subsection{Status of hidden-charm and hidden-bottom hadron spectrum}

\begin{figure}[tbp]
\centering
\includegraphics[width=0.95\textwidth]{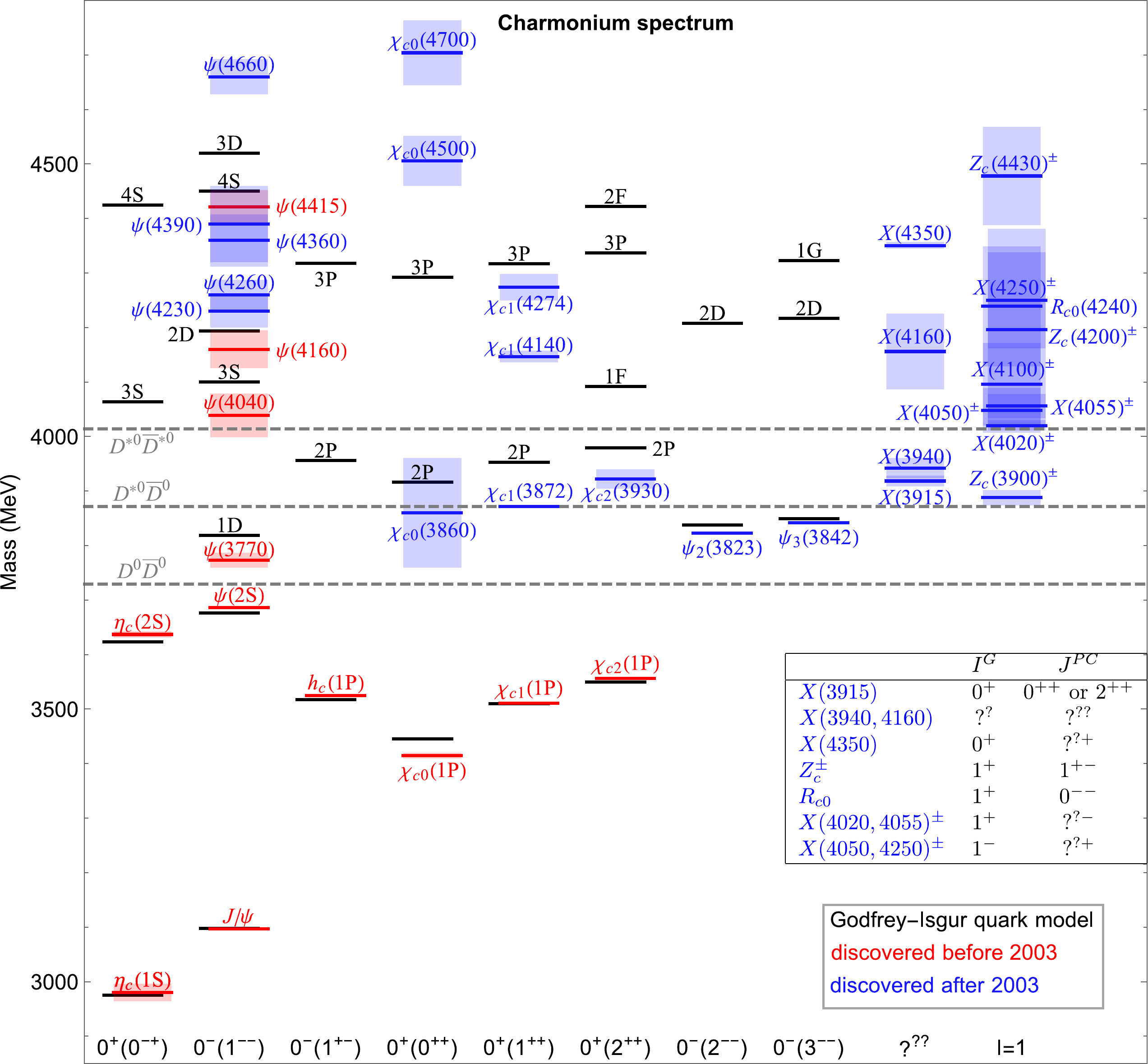}
\\
\includegraphics[width=0.95\textwidth]{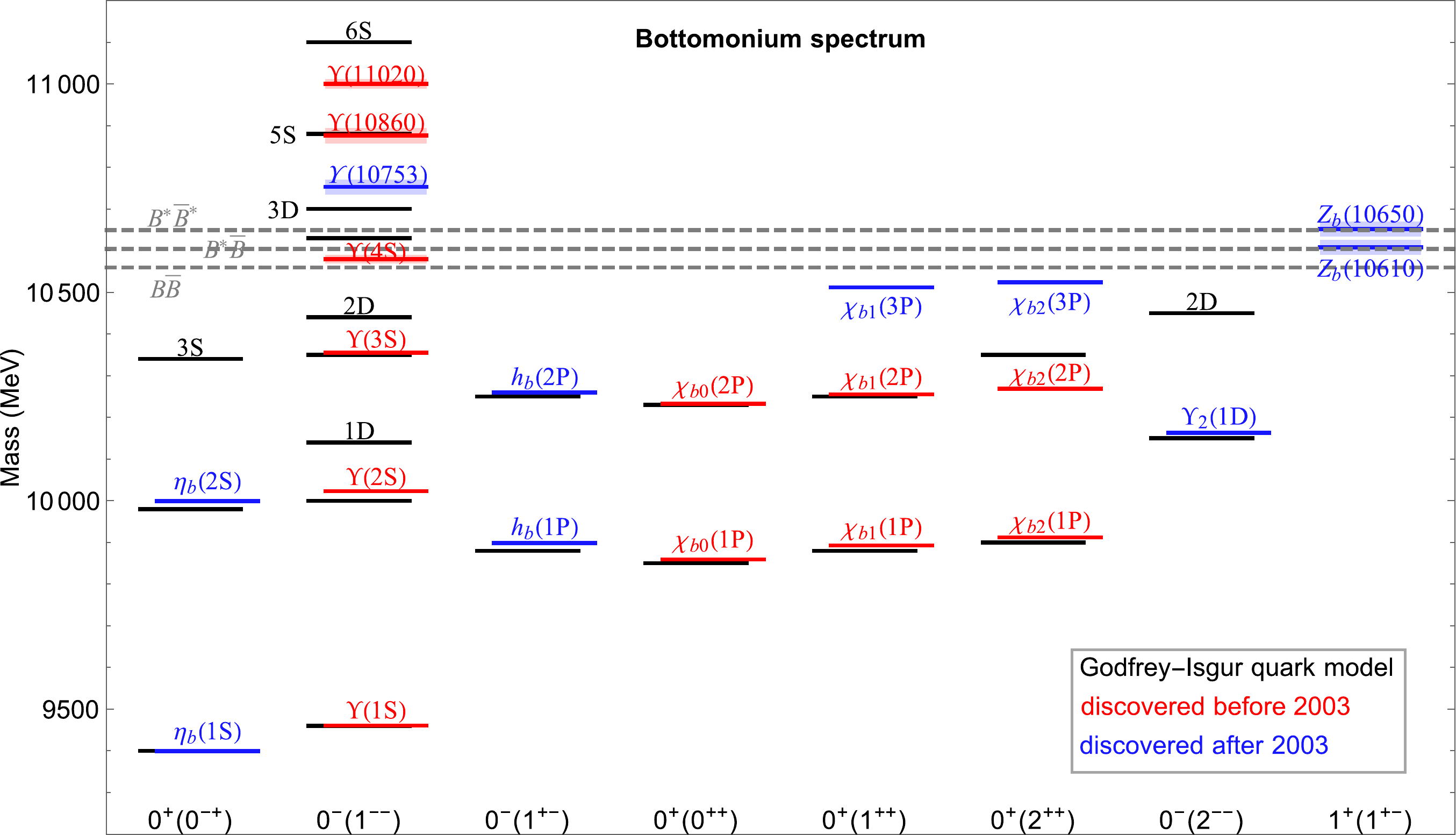}
\caption{\label{fig:sec2_4-QQspec} Comparison of the mass spectra of the observed heavy quarkonia and quarkonium-like states~\cite{Zyla:2020} with those predicted by the Godfrey--Isgur quark model~\cite{Godfrey:1985xj,Barnes:2005pb}. The states are listed according to their quantum numbers $I^G(J^{PC})$. The states with quantum numbers not fully determined are listed in the column ``$?^{??}$'' with the exception of $\Upsilon(10753)$ which is listed in the $0^-(1^{--})$ bottomonium column though its $I^G$ have not been fixed. 
For the experimentally observed states, the shaded areas indicate the central values of the widths of the observed states.}
\end{figure}

\begin{table}[tbp]
\begin{threeparttable}
  \caption{\label{tab:XYZ} Structures observed experimentally since 2003 in the charmonium mass region and their production and decay processes. 
  We group the mesonic structures into three blocks: those with $PC=--$ quantum numbers are given in the second block; the isospin-1 structures are given in the third block; the others are given in the first block.
  Their quantum numbers can be found in Fig.~\ref{fig:sec2_4-QQspec}. 
  The exotic baryon candidates are listed in the fourth block.
  Here we use the nomenclature of RPP~\cite{Zyla:2020}, according to which states are named according to their quantum numbers. For more information of the $XYZ$ states and their properties and the original experimental references, see Ref.~\cite{Zyla:2020}. }
\centering
\begin{tabular}{l l l}
\hline \hline
     $XYZ$ & Production processes
            & Decay channels 
                 \\\hline
    $\chi_{c0}(3860)$ & $e^+e^-\to J/\psi X$, $\gamma\gamma \to X$\tnote{a} & $D\bar D$, $\gamma\gamma$\tnote{a} \\
           $\chi_{c1}(3872)$ & $B \to K X/ K\pi X,e^+e^-\to\gamma X$,  & {$\pi^+\pi^-J/\psi, \omega J/\psi, D^{*0}\bar{D}^0, D^0\bar D^0\pi^0,$} \\
    & $pp/p\bar p$ semi-inclusive, &
             $ \pi^0\chi_{c1}, \gamma J/\psi, \gamma \psi(2S)$  \\ 
    & $\gamma^* N \to X\pi^\pm N$\tnote{b} & 
             \\  
    $X(3915)$ & $B \to K X,\,\gamma\gamma\to X, e^+e^-\to \gamma X$ & $\omega J/\psi, \gamma\gamma$ \\ 
    $\chi_{c2}(3930)$ & $\gamma\gamma\to X$, $pp$ semi-inclusive & $D\bar D, \gamma\gamma$ \\ 
   $X(3940)$ &  $e^+e^- \to J/\psi + X$ & $D {\bar D}^*$\\ 
    $\chi_{c1}(4140)$ & $B \to K X$, $p\bar p$ semi-inclusive\tnote{c} & $\phi J/\psi$ \\
  $X(4160)$ &   $e^+e^- \to J/\psi + X$ & $D^* {\bar D}^*$\\ 
    $\chi_{c1}(4274)$ & $B \to K X$ & $\phi J/\psi$ \\
    $X(4350)$ & $\gamma\gamma \to X$ & $\phi J/\psi, \gamma\gamma$\\
    $\chi_{c0}(4500)$ & $B \to K X$ & $\phi J/\psi$ \\
    $\chi_{c0}(4700)$ & $B \to K X$ & $\phi J/\psi$ \\
   \hline\hline
   $\psi_2(3823)$ & $B\to K \psi_2, e^+e^-\to \pi\pi \psi_2$ & $\gamma\chi_{c1}$ \\
   $\psi_3(3842)$ & $pp$ semi-inclusive & $D\bar D$ \\
  $\psi(4230/4260)$ & $e^+e^- \to Y, e^+e^- \to Y\gamma_{\rm ISR}$ & $\pi\pi J/\psi, \pi\pi \psi(2S), \chi_{c0}\omega, h_c\pi\pi,$ \\
  & & $D\bar D^*\pi, \gamma \chi_{c1}(3872), J/\psi K\bar K$ \\
   $\psi(4360)$ & $e^+e^- \to Y, e^+e^- \to Y\gamma_{\rm ISR}$ & $\pi \pi \psi(2S)$, $\pi\pi\psi_2(3823)$\tnote{d}, $D_1(2420)\bar D$\tnote{d} \\
   $\psi(4390)$ & $e^+e^- \to Y$ & $\pi \pi h_c, \pi\pi\psi(3770)$\tnote{d} \\
  $\psi(4660)$ & $e^+e^- \to Y\gamma_{\rm ISR}$ & $\pi \pi \psi(2S), \Lambda_c\bar \Lambda_c, D_s^+ D_{s1}(2536)^-$ \\\hline\hline
  $Z_c(3900)^\pm$ & $e^+e^-\to \pi Z_c,$ & $\pi J/\psi, D\bar D^*$ \\
  & $b$-hadron semi-inclusive decays & \\
  $X(4020)^\pm$ & $e^+e^-\to \pi Z_c$ & $\pi h_c, D^*\bar D^*$ \\
    $X(4050)^\pm$ & $B \to K  Z_c$ & $\pi^\pm \chi_{c1}$\\
    $X(4055)^\pm$ & $e^+e^-\to\pi X$ & $\pi^\pm \psi(2S)$\\
    $X(4100)^\pm$ & $B \to K  Z_c$ & $\pi^\pm \eta_{c}$\\
   $Z_c(4200)^\pm$ & $B \to K Z_c$ & $\pi^\pm J/\psi$ \\
   $R_{c0}(4240)^-$ & $B \to K  R_{c0}$  & $\pi^- \psi(2S)$ \\
   $X(4250)^\pm$ & $B \to K  Z_c$ & $\pi^\pm \chi_{c1}$ \\ 
    $Z_c(4430)^\pm$ & $B \to K Z_c$ & $\pi^\pm J/\psi,\pi^\pm \psi(2S)$ \\ \hline\hline 
    $P_c(4312)^+$ & $\Lambda_b\to K^- P_c$\tnote{e} & $p J/\psi$ \\
    $P_c(4380)^+$ & $\Lambda_b\to K^- P_c$ & $p J/\psi$ \\
    $P_c(4440)^+$ & $\Lambda_b\to K^- P_c$ & $p J/\psi$ \\
    $P_c(4457)^+$ & $\Lambda_b\to K^- P_c$ & $p J/\psi$ \\
    \hline\hline
\end{tabular}
\vspace{-2mm}
\begin{tablenotes}
\begin{minipage}[t]{14.5cm}
  \noindent\item[a] From the analysis of Ref.~\cite{Guo:2012tv}.
\noindent\item[b] It is likely a different state. It was reported by the COMPASS Collaboration in muoproduction~\cite{Aghasyan:2017utv}; however, the $\pi\pi$ invariant mass spectrum does not agree with that coming from a $\rho$, and a negative $C$-parity is preferred.
\noindent\item[c] Not seen in $\gamma\gamma\to J/\psi \phi$ and $e^+e^-\to \gamma J/\psi \phi$.
\noindent\item[d] Noted as ``possibly seen" in RPP.
\noindent\item[e] No signal of $P_c$ was seen in $\gamma p\to J/\psi p$~\cite{Ali:2019lzf}.
\end{minipage}
\end{tablenotes}
\end{threeparttable}
\end{table}

In Fig.~\ref{fig:sec2_4-QQspec}, we show the spectrum of the charmonium(-like) and bottomonium(-like) states listed in RPP~\cite{Zyla:2020}.
The hidden-charm structures that were reported in various experiments since 2003 are also listed in Table~\ref{tab:XYZ}, together with their production processes and observed decay channels. The nomenclature of the latest RPP is used in the figure and table and will be used in the following discussion. 

One sees that the charmonium-like structures were observed mainly in three types of processes, and many of them were only seen in one particular production process as well as in only one particular final state.
Thus, one immediate question is whether they can be found in other processes and, in particular, in other types of experiments such as the electron-proton/electron-ion collisions. Let us first discuss the experiments that have already significantly contributed to the field.

Weak decays of the ground state bottom-hadrons contributed to the observation of hidden-charm states more than other processes. The decays happen mainly via the singly-CKM suppressed $b\to c\bar c s$ at the quark level. The main experiments are the $B$ factories Belle, BaBar and the LHCb experiment at LHC. There are two main processes. 
First, the three-body decays of $B$ mesons with the final state being a kaon and a pair of charmed mesons or a charmonium and light mesons. The maximal mass of a charmonium-like state that can be produced in this way is the mass difference between the $B$ and $K$ mesons, which is about 4.8~GeV. In fact, the highest charmonium-like structure reported so far is the $X(4700)$ observed by the LHCb Collaboration~\cite{Aaij:2016iza}, close to this bound. 
Second, the three-body decays of the $\Lambda_b$, such as the $\Lambda_b^0\to J/\psi p K^-$, which is the process where the hidden-charm pentaquark candidates $P_c(4312)$, $P_c(4380)$, $P_c(4440)$ and $P_c(4457)$ were discovered at LHCb~\cite{Aaij:2015tga,Aaij:2019vzc}. The hidden-charm pentaquark state that can be produced through such a process needs to have a mass lower than the mass difference between $\Lambda_b$ and $K$, which is about 5.1~GeV. 
For these weak decay processes, the mass of the initial particle is fixed. Furthermore, the final states always involve at least three particles, which complicates the data analysis and may bring ambiguities due to insufficient treatment of cross channels and three-body final state interaction.
   
Vector charmonium(-like) states are mostly easily studied in $e^+e^-$ collisions. They have the same $J^{PC}$ quantum numbers as a virtual photon so that they can be directly produced in $e^+e^-$ collisions; by emitting a photon to adjust the energy to the interested region, they can also be produced using the initial-state radiation (ISR) process, whose cross section, however, is smaller by two orders of magnitude because of the suppression of $\alpha$. The main experiments include BESIII, CLEO-c, Belle and BaBar. Thus, more vector states have been observed than other quantum numbers, and they are normally observed in more channels as well due to the high luminosity of the $e^+e^-$ machines.
There is also one new vector state in the bottomonium mass region, the $\Upsilon(10753)$, reported by Belle~\cite{Abdesselam:2019gth}.
The structures with other quantum numbers need to be produced through the decays of higher states or two-photon collisions, and thus have much smaller production rates or beyond the energy reach of BESIII.

Some of the structures were also produced in the prompt production processes at hadron colliders, and observed semi-inclusively. The main experiments are CDF and D0 at Tevatron and CMS, ATLAS and LHCb at the LHC. Such processes have much larger cross sections than those via virtual photons. However, the large energy and large strong-interaction cross sections also imply huge backgrounds, which lowers the detection efficiency. The types of final state particles that can be efficiently detected are usually restricted to charged light hadrons and muons, and soft photons are hardly detected.

\subsection{Exotic hadrons at EicC}
\label{sec:exotic_eicc}

For the study of exotic hadrons, each of the experiments has its advantages and limitations.
Different kinds of experiments complement to one another, and they are needed to establish a more complete picture of the heavy-flavor hadron spectroscopy.
Even in the charm sector, the so-far collected information is not enough to build up a clear picture for all of these new structures.
Furthermore, although it is expected that there should be analogues of resonances with open or hidden charm in the bottom sector, far fewer states with bottom have been observed due to the limitations of the current experiments.
Let us take heavy quarkonia as an example. 

In Fig.~\ref{fig:sec2_4-QQspec}, we present a comparison of the observed heavy quarkonia and quarkonium-like states with the predictions of the $Q\bar Q$ mass spectrum predicted in the Godfrey--Isgur quark model~\cite{Godfrey:1985xj,Barnes:2005pb}.
It is clear that all of the $XYZ$ structures are located above or at least very close to the open-charm thresholds, while there are only a couple of analogous states in the bottomonium sector.
The messy situation of the charmonium(-like) states nicely illustrates how little we understand the confinement aspect of QCD.
Although the importance of the open-charm coupled channels was already noticed in the seminal Cornell model~\cite{Eichten:1978tg,Eichten:1979ms}, their role in forming the observed spectrum is still far from being understood.
The highly excited states close or above the open-flavor thresholds contain important information about the long-distance interaction between heavy quarks and about how the light degrees of freedom come into play their role. Thus, a detailed study of these states is highly valuable for understanding confinement. 

Furthermore, no matter how the hidden-charm meson spectrum emerges, one would expect to have an analogous spectrum for hidden-bottom mesons as well as in hidden-charm and hidden-bottom baryonic sectors. Especially, if the coupled channels are crucial to form the spectrum, phenomena similar to those of the $XYZ$ states would repeat in these sectors.

The EicC energy region covers all these interesting physics. In the following, let us briefly discuss a few topics on heavy-flavor hadron spectroscopy that EicC can significantly contribute to. 

\begin{figure}[t]
    \centering
    \includegraphics[width=0.95\textwidth]{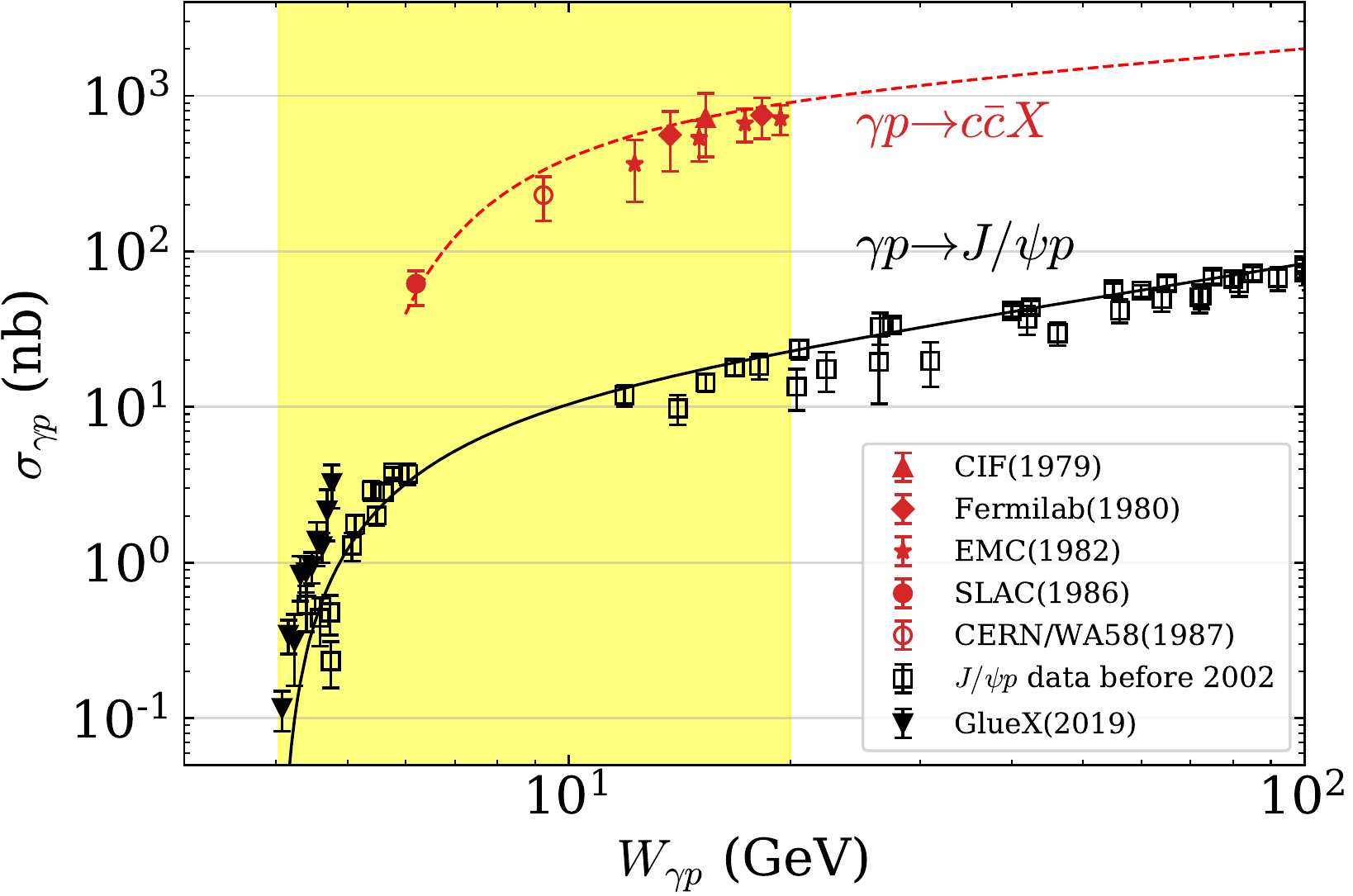}
    \caption{The dependence of the photoproduction cross sections on the $\gamma p$ c.m. energy for the exclusive $\gamma p\to J/\psi p$  and semi-inclusive $\gamma p\to c\bar c X$ processes.~\cite{Gittelman:1975ix,Camerini:1975cy,Binkley:1981kv,Denby:1983az,Frabetti:1993ux,Adloff:2000vm,Chekanov:2002xi,Ali:2019lzf,Atiya:1979hx,Clark:1980ed,Aubert:1982tt,Adamovich:1986gx}. 
    The EicC energy coverage is denoted by the shaded area. Here, $X$ denotes the all particles that are not detected and should not be confused with the $X$ charmonium-like states. }
    \label{fig:sec2_4_xsect_ccbar}
\end{figure}
\begin{itemize}
    \item \noindent{\bf Charmonium(-like) states}
\end{itemize}

  The photoproduction cross section of the exclusive process $\gamma p\to J/\psi p$ is at the order of 10~nb for the c.m. energy of the $\gamma p$ within 10 to 20~GeV, which is the energy region of EicC. The cross section of the semi-inclusive production of $c\bar c X$ is larger by almost two orders of magnitude, see Fig.~\ref{fig:sec2_4_xsect_ccbar}. 
  The data shown in this figure include the exclusive $J/\psi$ production data from Refs.~\cite{Gittelman:1975ix,Camerini:1975cy,Binkley:1981kv,Denby:1983az,Frabetti:1993ux,Adloff:2000vm,Chekanov:2002xi} ($J/\psi p$ data before 2002), \cite{Ali:2019lzf} (GlueX), and the semi-inclusive $c\bar c X$ data from Refs.~\cite{Atiya:1979hx} (CIF), \cite{Clark:1980ed} (Fermilab), \cite{Aubert:1982tt} (EMC), and \cite{Adamovich:1986gx} (SLAC). The data were fitted using parametrization origined from the vector-meson dominance model of Ref.~\cite{Gryniuk:2016mpk}. 
  The cross section for the electroproduction process is about two orders of magnitude smaller due to an additional factor of electromagnetic coupling $\alpha$. Considering an integrated luminosity of 50~fb$^{-1}$, one may estimate that the $J/\psi$ events produced from the exclusive process is about  $\mathcal{O}(5\times10^6)$. 
  Because almost all excited charmed mesons (baryons) will decay into $D$ ($\Lambda_c$) and their antiparticles, one can expect that there must be many more $D$ and $\Lambda_c$ events.
  Therefore, in addition to the hidden-charm channels, the $XYZ$ charmonium-like states, including the highly excited ones beyond the capability of BESIII and JLab or those that cannot be produced through the $B$ meson decays, can be studied through open-channel final states.  
  As a benchmark, the production of the $\chi_{c1}(3872)$ and $Z_c(3900)$ are simulated and will be discussed in Section~\ref{sec:2-xection}. 

\begin{itemize}
    \item \noindent{\bf Hidden-charm pentaquarks}
\end{itemize}

  So far, the only observations of hidden-charm pentaquarks came from LHCb: $P_c(4312)$, $P_c(4380)$, $P_c(4440)$ and $P_c(4457)$~\cite{Aaij:2015tga,Aaij:2019vzc}.\footnote{The $P_c(4380)$ here is a broad structure introduced to improve the fitting quality in the 2015 LHCb analysis~\cite{Aaij:2015tga}, and it is not needed to fit the updated $J/\psi p$ invariant mass distribution~\cite{Aaij:2019vzc}. However, there is a hint for the existence of a narrow $P_c(4380)$~\cite{Du:2019pij} in the new LHCb data.} 
  In fact, the existence of narrow hidden-charm baryon resonances, as hadronic molecules of a pair of charm meson and charm baryon, have been predicted to exist in the mass region above 4~GeV~\cite{Wu:2010jy,Wang:2011rga,Yang:2011wz,Wu:2012md,Xiao:2013yca,Uchino:2015uha,Karliner:2015ina}. As mentioned above, similar to the existence of many hidden-charm $XYZ$ states, there should also be lots of hidden-charm baryonic excited states. Searching for them and verifying the LHCb observations will provide valuable inputs to understanding the spectroscopy of excited hadrons.
  The nonobservation of the $P_c$ states at the GlueX experiment~\cite{Ali:2019lzf} indicates that the branching fractions of the $P_c$ states into $ J/\psi p$ to be small (for a combined analysis of the GlueX and LHCb measurements, see Ref.~\cite{Cao:2019kst}). Then the dominant decay modes of the $P_c$ should be the open-charm channels, including the $\bar D^{(*)}\Lambda_c$ and $\bar D^{(*)}\Sigma_c$~\cite{Lin:2017mtz,Lin:2019qiv,Du:2019pij,Dong:2020nwk}.
  Therefore, at EicC, the $P_c$ need to be searched for in exclusive processes with the final states being not only the $J/\psi N$, but also the open-charm $\bar D^{(*)}\Lambda_c$ and $\bar D^{(*)}\Sigma_c$ channels~\cite{Huang:2016tcr,Wu:2019adv}. 
  Semi-inclusive processes of these processes will also be a crucial part as they have much larger cross sections. Pentaquarks with both hidden charm and hidden (or open) strangeness can also be searched for in analogous processes. 
  For an estimate of the semi-inclusive production rates in the hadronic molecular model of the $P_c$ states, see the next subsection.

  From the above discussions, one sees that an efficient detection of the $D/\bar D$ and $\Lambda_c$ particles is essential for the study of the hidden-charm mesons and baryons. From RPP~\cite{Zyla:2020}, one finds that the most important decay channels of the $D^+$ are $K^-2\pi^+$ [$(9.38\pm0.16)\%$] and $K_S^0\pi^+\pi^0$ [$(7.36\pm0.21)\%$], those for the $D^0$ are $K^-\pi^+\pi^0$ [$(14.4\pm0.5)\%$] and $K^-\pi^+$ [$(3.950\pm0.031)\%$], and those for the $\Lambda_c^+$ are $\Lambda\pi^+\pi^0$ [$(7.1\pm0.4)\%$] and $pK^-\pi^+$ [$(6.28\pm0.32)\%$]. Thus, both the charged and neutral pions and kaons need to be efficiently detected.
  Once one of the open-charm final state particles is reconstructed, the events for the other one can be selected from the missing mass spectrum in the relevant energy region. In this way, searching for hidden-charm states in the open-charm final states is promising.

\begin{itemize}
    \item \noindent{\bf Bottom hadrons}
\end{itemize}

\begin{figure}[t]
    \centering
    \includegraphics[width=0.95\textwidth]{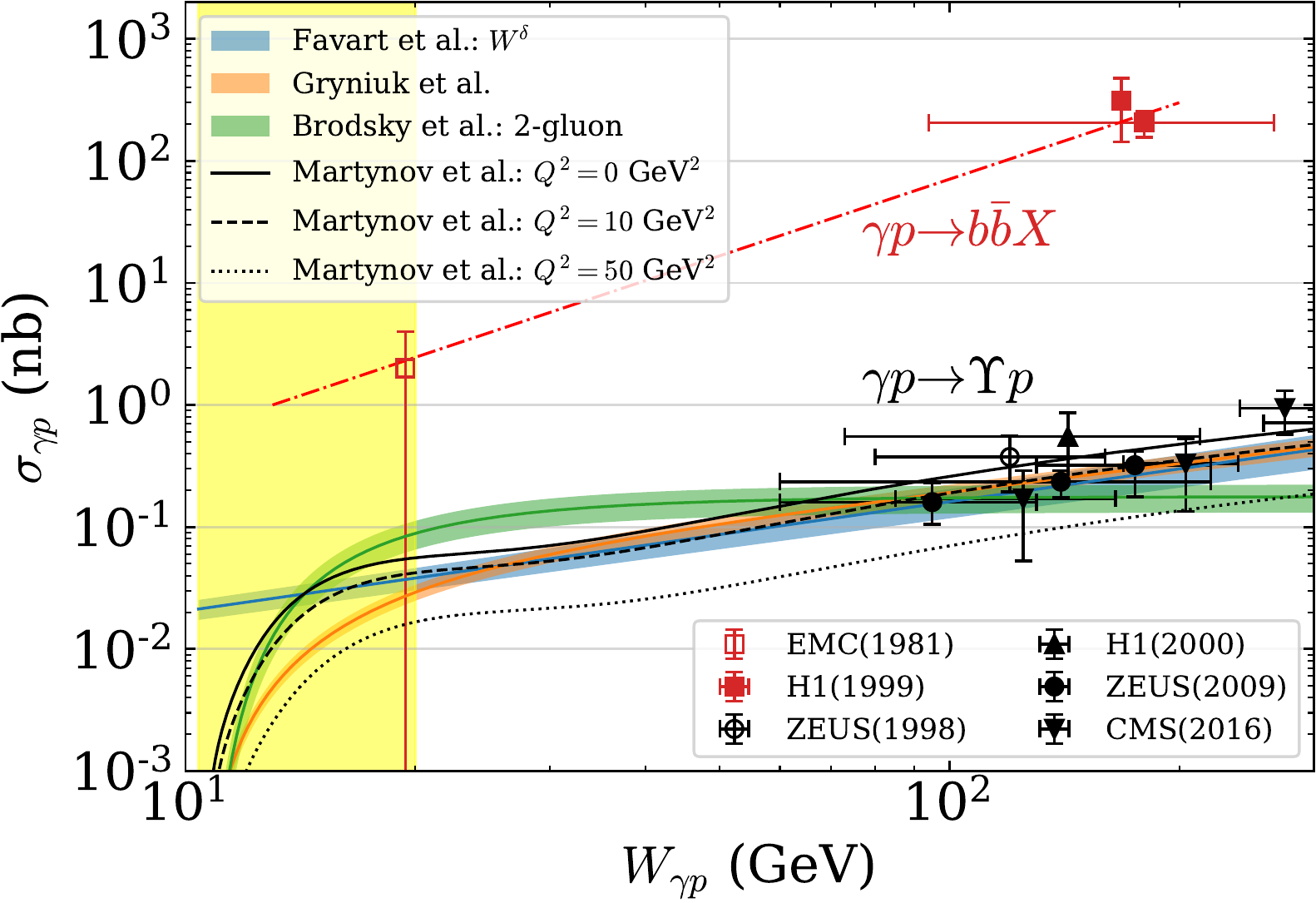}
    \caption{The dependence of the photoproduction cross sections on the $\gamma p$ c.m. energy for the exclusive $\gamma p\to \Upsilon p$  and semi-inclusive $\gamma p\to b\bar b X$ processes. The EicC energy coverage is denoted by the shaded area. }
    \label{fig:sec2_4_xsect_bbbar}
\end{figure}

  In Fig.~\ref{fig:sec2_4_xsect_bbbar}, we show the cross sections for the exclusive photoproduction of the $\Upsilon$ and for the semi-inclusive $b\bar b$. 
  The shaded area in corresponding to the EicC energy region covers the hidden-bottom hadron masses. 
  The exclusive data are taken from Refs.~\cite{Breitweg:1998ki,Chekanov:2009zz} (ZEUS), \cite{Adloff:2000vm} (H1), and \cite{CMS:2016nct} (CMS); the semi-inclusive data are taken from Refs.~\cite{Aubert:1981gx} (EMC) and \cite{Adloff:1999nr} (H1).
  The models used to fit the data include the empirical formula for the deeply-virtual meson production (DVMP) model~\cite{Favart:2015umi} (Favart {\it et al.}), the 2-gluon exchange model~\cite{Brodsky:2000zc} (Brodsky {\it et al.}), the parametrization~\cite{Gryniuk:2016mpk} (Gryniuk {\it et al.}), and the dipole Pomeron model ($Q^2 =$0, 10, 50~GeV$^2$)~\cite{Martynov:2001tn,Martynov:2002ez} (Martynov {\it et al.}).
  One sees that for the c.m. energy in the range between 15 and 20~GeV, the photoproduction cross section for $\Upsilon p$ is of $\mathcal{O}(10~\text{pb})$; thus, the corresponding electroproduction $e^- p \to e^- \Upsilon p$ cross section should be of $\mathcal{O}(0.1~\text{pb})$. Considering an integrated luminosity of 50~fb$^{-1}$, the event number of $\Upsilon$ that can be produced through the exclusive $\Upsilon p$ process is of $\mathcal{O}(10^4)$, consistent with the simulation in Sec.~\ref{sec:protonmass} and the estimate in Ref.~\cite{Xu:2020uaa}. 
  The cross section for the semi-inclusive $b\bar b +\text{anything}$ is two orders of magnitude higher. Thus, millions of bottom mesons $B$ and $\Lambda_b$ can be produced. 
  If these bottom hadrons can be efficiently detected, the EicC will be able to contribute to the study of excited bottom hadrons. Although the ground state bottom hadrons are more difficult to be detected than their charmed cousins, their life times are much longer, making the secondary decay vertices useful in detecting them. Hidden-bottom pentaquark states that are expected to be decay into $\Upsilon N$, $\Lambda_b B^{(*)}$ and $\Sigma_b B^{(*)}$ final states may also be searched for in processes similar to those for the searching of the $P_c$ states.

\medskip

Next, let us briefly discuss the advantages of EicC in the study of hadron spectroscopy.
Comparing with experiments utilizing  electron-positron collisions and the $B/\Lambda_b$ decays, one special feature of the EicC is that it has different kinematics which can avoid the ambiguity of interpreting resonance signals induced by the so-called triangle singularity.
Triangle singularity is a type of kinematical singularity, which occurs because of the simultaneous on-shellness of three intermediate particles, and can produce peaks mimicking the behavior of resonances (for a recent review, see Ref.~\cite{Guo:2019twa}). 
For instance, the triangle singularity mechanism have been constructed for producing resonance-like signals for the prominent multiquark candidates $P_c(4450)$~\cite{Guo:2015umn,Liu:2015fea}, $Z_c(3900)^\pm$~\cite{Wang:2013cya,Wang:2013hga,Pilloni:2016obd} (see also Refs.~\cite{Szczepaniak:2015eza,Albaladejo:2015lob,Gong:2016jzb,Guo:2020oqk}), $Z_c(4200,4430)^\pm$~\cite{Nakamura:2019btl} and $X(4050,4250)^\pm$~\cite{Nakamura:2019emd}.
This means that it is essential to distinguish the signals from resonances from those from kinematical singularities. 
Triangle singularity is very sensitive to the involved kinematical variables such as masses and energies. The examples mentioned in the above were all reported in the decays of bottom hadrons ($B$ or $\Lambda$), which have fixed masses, or $e^+e^-$ collisions. The photoproduction or electroproduction processes at EicC have completely different kinematical regions from these experiments; the photon is space-like or nearly on shell, making the occurrence of triangle singularities in the exclusive processes impossible. Furthermore, the dependence of the signals on the energy and $Q^2$ can also be measured. 
In addition, the double polarized beams facilitate the EicC to determine the quantum numbers, such as spin and parity, of hadron resonances. 
In comparison with hadron colliders, the EicC has a better signal to noise ratio. 
Furthermore, the EicC covers all the mass regions for charmonium, bottomonium, $P_c$, $P_b$ and excited heavy hadrons.
As a result, the study of exotic mesons and baryons will be one of the foci of EicC.

A search for the photoproduction of $Z_c(3900)$~\cite{Adolph:2014hba} and $\chi_{c1}(3872)$~\cite{Aghasyan:2017utv} has been performed by  COMPASS with muon beam, giving valuable input for the simulation of EicC. Within similar range of c.m. energy, EicC can search for these and also other states with more than one order higher luminosity. Furthermore, with solid angle coverage, especially hadron PID in the forward angle and good vertex detector for decay topology, the acceptance and reconstruction efficiency of EicC are expected to be significantly increased, so the discovery potential of exotica states will be exploited. 

\subsection{Cross section estimates and simulations}
\label{sec:2-xection}

In this subsection, more quantitative estimates of the production rates for a selected list of exotic hadron candidates and simulations of the $ep\rightarrow e \chi_{c1}(3872)p$, $ep\rightarrow e Z_c^+(3900)n$, $ep\rightarrow e P_c \rightarrow e J/\psi p$, and $ep\rightarrow e P_b \rightarrow e \Upsilon p$ processes are reported. For more model estimates of the exclusive productions of hidden-charm and bottom exotic hadrons, we refer to Refs.~\cite{Liu:2008qx,He:2009yda,Galata:2011bi,Lin:2013mka,Lin:2013ppa,Huang:2013mua,Adolph:2014hba,Wang:2015jsa,Wang:2015lwa,Kubarovsky:2015aaa,Karliner:2015voa,Huang:2016tcr,Blin:2016dlf,Aghasyan:2017utv,Meziani:2016lhg,Joosten:2018gyo,Paryev:2018fyv,Wang:2019krd,Goncalves:2019vvo,Wang:2019zaw,Cao:2019gqo,Wu:2019adv,Winney:2019edt,Xie:2020niw,Paryev:2020jkp,Yang:2020eye,Albaladejo:2020tzt}.

\begin{figure}[t]
\centering
\includegraphics[width=0.48\textwidth,angle =0]{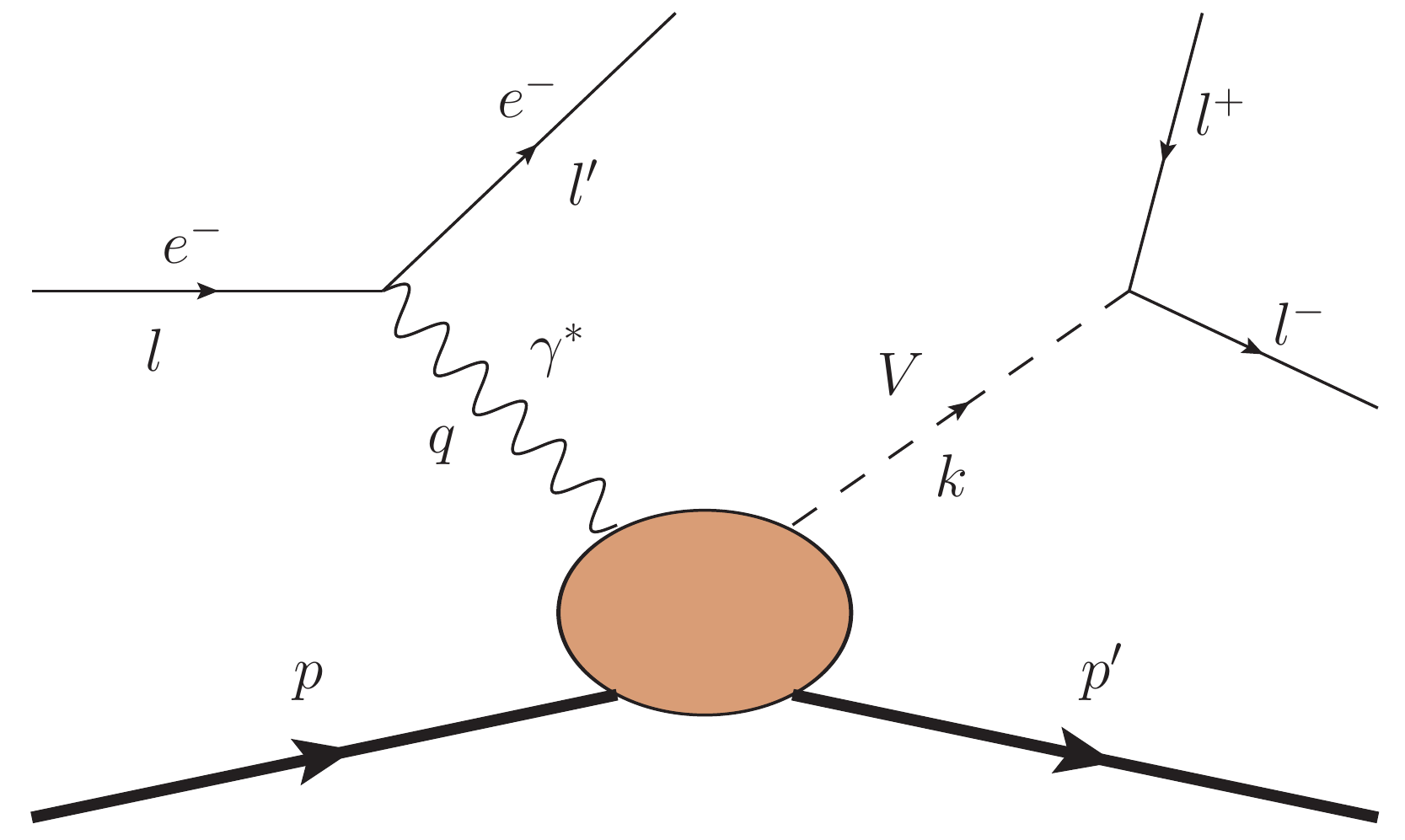}
\includegraphics[width=0.48\textwidth,angle =0]{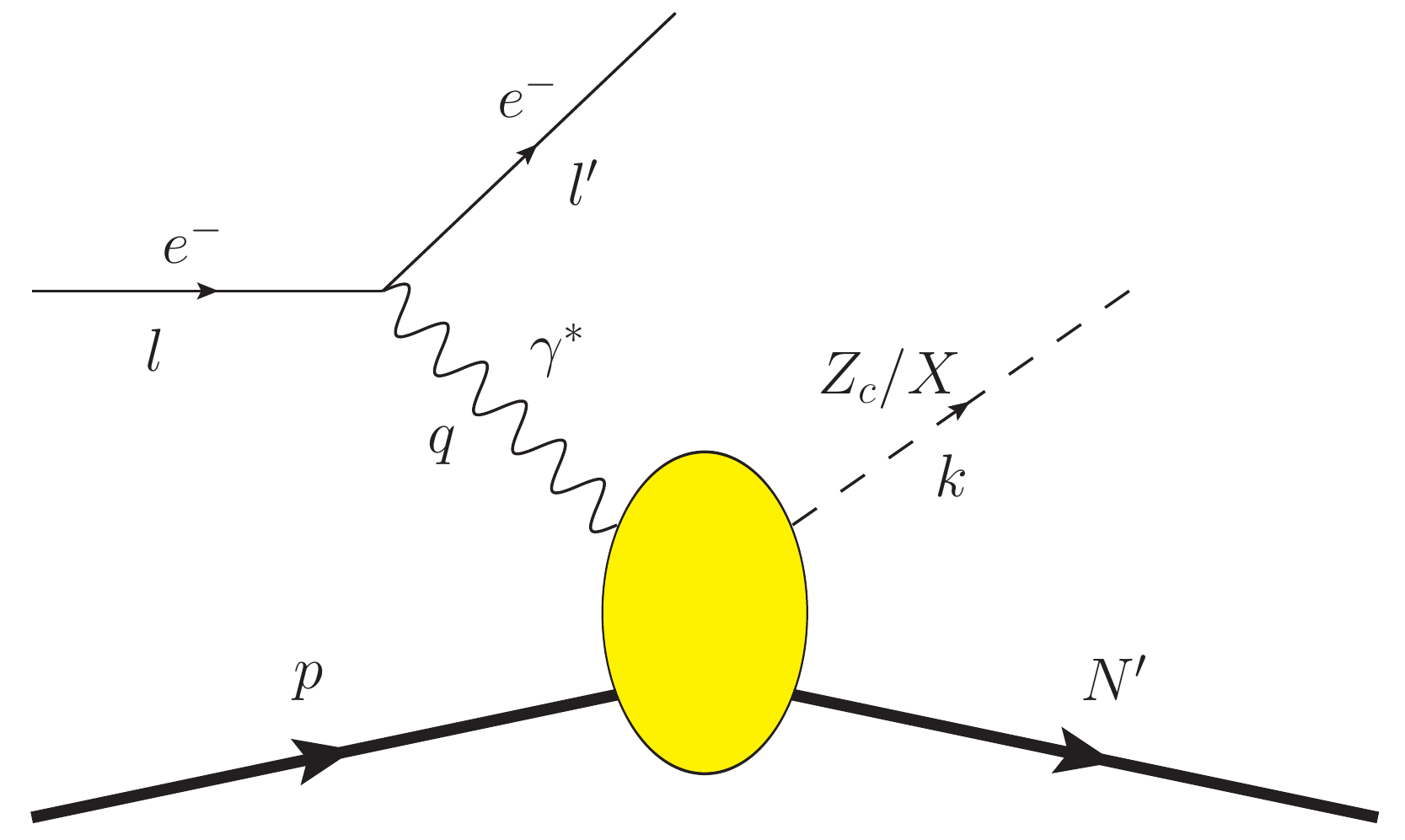}
\vspace*{-0.cm}
\caption{\label{fig:Zc_X_production_feyman_diagram} Left: The $ep \to epV \to ep l^+ l^-$ process. Right: The $ep\rightarrow e \chi_{c1}(3872)p$ and $ep\rightarrow e Z_c^+(3900)n$ process.}
\end{figure}

\begin{table}[tb]

  \caption{\label{tab:Exotic_simulation} Estimated event numbers that can be collected at EicC assuming an integrated luminosity of 50 fb$^{-1}$. The lepton pairs $l^+l^-$ denote both $\mu^+ \mu^-$ and $e^+e^-$. The event numbers are estimated using the assumed detection efficiencies listed in the third column, which are expected to be higher in the middle rapidity than that in the forward region. }
\centering

\begin{tabular}{c | c | c |  c}
\hline \hline
     Exotic states & Production/decay processes & Detection efficiency & Expected events
    \\\hline
     \multirow{3}{*}{$P_c(4312)$}   &   $ep \to e P_c(4312)$       &   \multirow{3}{*}{$\sim$30\% } &   \multirow{3}{*}{ 15$-$1450 } \\
    &  $P_c(4312) \to p  J/\psi$   & &\\
    &  $J/\psi \to l^+l^-$   & & \\\hline

   \multirow{3}{*}{$P_c(4440)$}   &   $ep \to e P_c(4440)$         &   \multirow{3}{*}{$\sim$30\% } &   \multirow{3}{*}{ 20$-$2200 } \\
    &  $P_c(4440) \to p  J/\psi$   & &\\
    &  $J/\psi \to l^+l^-$   & & \\\hline

   \multirow{3}{*}{$P_c(4457)$}   &   $ep \to e P_c(4457)$         &   \multirow{3}{*}{$\sim$30\% } &   \multirow{3}{*}{ 10$-$650 } \\
    &  $P_c(4457) \to p  J/\psi$   & &\\
    &  $J/\psi \to l^+l^-$   & & \\\hline

    \multirow{3}{*}{$P_b(\text{narrow})$}   &   $ep \to e P_b(\text{narrow})$    &   \multirow{3}{*}{$\sim$30\% } &   \multirow{3}{*}{ 0$-$20 } \\
    &  $P_b(\text{narrow}) \to p  \Upsilon$   & &\\
    &  $\Upsilon \to l^+l^-$   & & \\\hline

     \multirow{3}{*}{$P_b(\text{wide})$}   &   $ep \to e P_b(\text{wide})$       &   \multirow{3}{*}{$\sim$30\% } &   \multirow{3}{*}{ 0$-$200 } \\
    &  $P_b(\text{wide}) \to p  \Upsilon$  & &\\
    &  $\Upsilon \to l^+l^-$  & & \\\hline

    \multirow{3}{*}{$\chi_{c1}(3872)$}   &   $ep \to e\chi_{c1}(3872)p$      &   \multirow{3}{*}{$\sim$50\% } &   \multirow{3}{*}{ 0$-$90 } \\
    &  $\chi_{c1}(3872) \to \pi^+\pi^- J/\psi$  & &\\
    &  $J/\psi \to l^+l^-$   & & \\ \hline

   \multirow{3}{*}{$Z_c(3900)^+$}   &   $ep \to eZ_c(3900)^+n$      &   \multirow{3}{*}{$\sim$60\% } &   \multirow{3}{*}{ 90$-$9300 } \\
    &  $Z_c^+(3900) \to \pi^+  J/\psi$   & &\\
    &  $J/\psi \to l^+l^-$   & & \\\hline

    \hline\hline
\end{tabular}

\end{table}

\begin{figure}[tb]
\centering
\includegraphics[width=0.99\textwidth,angle =-0]{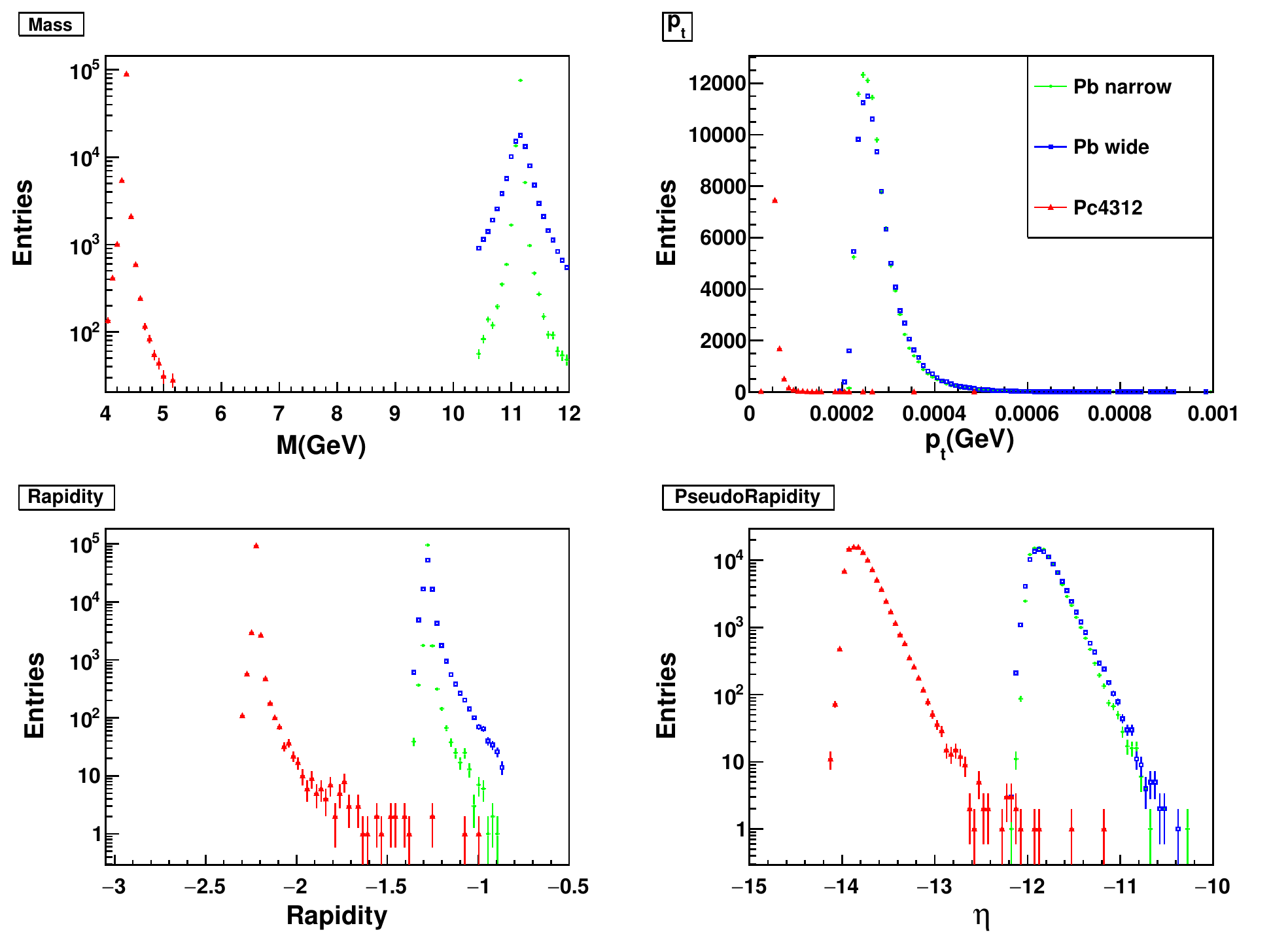}
\vspace*{-0.cm}
\caption{\label{fig:PbPc}The distributions of invariant masses, transverse momenta, pseudo-rapidities and rapidities of the $P_c$ and $P_b$ states.
}
\end{figure}

\begin{figure}[tb]
\centering
\includegraphics[width=0.99\textwidth,angle =-0]{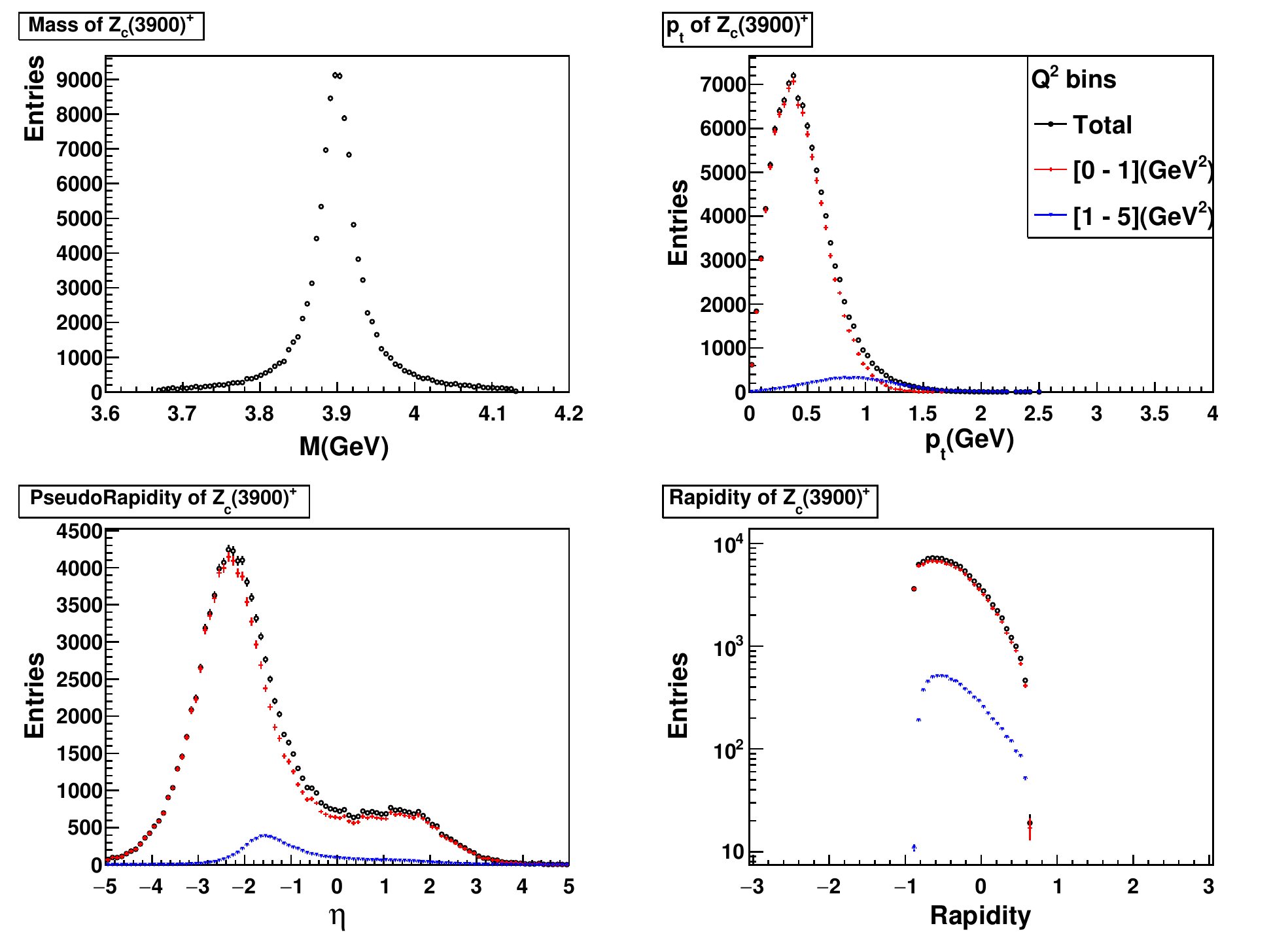}
\vspace*{-0.cm}
\caption{\label{fig:Zc3900_Q2_bin_Zc_distribution}
The distributions of invariant masses, transverse momenta, pseudo-rapidities and rapidities of the $Z_c(3900)^+$ in different $Q^2$ ranges.
}
\end{figure}
Motivated by the heavy quark flavor symmetry for the potential between heavy mesons and baryons, analogues of the pentaquark candidates $P_c$ in the bottom sector, labeled as $P_b$ here, are expected to exist~\cite{Wu:2010rv,Lin:2018kcc,Karliner:2015voa,Karliner:2015ina,Yang:2018oqd,Anwar:2018bpu,Huang:2018wed,Huang:2015uda,Shen:2017ayv,Xiao:2013jla}.
A resonant state $P_b$ coupling to $\Upsilon p$ with a mass around 11.12 GeV and a width ranging from tens of MeV to 300 MeV is predicted by nearly all models.
We simulate the exclusive electroproduction of two $P_b$ at EicC with a width of 30 MeV ($P_b({\rm narrow})$) and 300 MeV ($P_b({\rm wide})$), respectively, together with the three narrow $P_c$ states with the resonance parameters reported by LHCb.

The process is depicted in the left panel of Fig.~\ref{fig:Zc_X_production_feyman_diagram}, where $V$ represents the $J/\psi$ ($\Upsilon$) and the $P_c$ ($P_b$) couples to the $J/\psi N$ ($\Upsilon N$) in the $s$-channel.
The non-resonant background is modeled by Pomeron-exchange to be discussed in Sec.~\ref{sec:protonmass}.
The $P_c/P_b$ states are produced from the interaction between the virtual photon, emitted from the electron beam, and the proton beam in the $s$-channel. 
The upper limit of production rates at EicC can be determined by the upper limit of cross section of $\gamma p \to P_c \to J/\psi p$ measured by the GlueX Collaboration~\cite{Ali:2019lzf}, by properly including the photon flux and $Q^2$ dependence of amplitudes.
According to the analysis of the LHCb data, the lower limit of the branching ratio $P_c \to J/\psi p$ is around
0.5\%~\cite{Cao:2019kst}, and we assume the same lower limit for the $P_b \to \Upsilon p$.
The lower limit of production rates at EicC is obtained by the vector meson dominance (VMD) model, see, e.g.~\cite{Cao:2019gqo,Yang:2020eye}.

After taking into account the detection efficiency and the dilepton decay rates of the $J/\psi$ and $\Upsilon$, the expected production rates of these exotic states at EicC are shown in Table~\ref{tab:Exotic_simulation}. Here the detection efficiencies in the third column are estimated from the simulated distribution of final-state particles. The production rates on light nuclei are supposed to be larger~\cite{Paryev:2018fyv,Paryev:2020jkp}, which however will have lower luminosity. The distributions of the invariant mass spectra, the transverse momenta, the pseudo-rapidities, and the rapidities of the $P_b/P_c$ states are presented in Fig.~\ref{fig:PbPc}. 
They are obviously characterized by the $s$-channel resonances, with small transverse momenta and narrow ranges of pseudo-rapidity and rapidity, apparently different from the non-resonant Pomeron-exchange contribution which is smoothly spanned in the full range, as demonstrated in detailed investigation~\cite{Yang:2020eye,Xie:2020niw}. It is suggested to extract the pentaquark signal from large non-resonant contribution with a proper kinematic cut~\cite{Yang:2020eye}.

As discussed above, the cross section of the open charm channel $\bar D^{(*)}\Lambda_c$ is expected to be much bigger than the that of $J/\psi p$, so is the case of the open bottom channel $\bar B^{(*)}\Lambda_b$ in comparison with the $\Upsilon p$. In particular, the branching fractions of the $P_c$ and $P_b$ states into open-flavor channels are expected to be at least one-order-of-magnitude larger than those of the $J/\psi p$ and $\Upsilon p$.
In addition, as mentioned in Section~\ref{sec:exotic_eicc}, the open-charm ground state hadrons could be reconstructed at the level of 10\%. Thus, it is optimistic that the $P_c$ states can be studied in detail through processes  $ep\to J/\psi p$ and $e\bar D^{(*)}\Lambda_c$ at EicC. If the open-bottom hadrons can also be efficiently reconstructed, the hypothesized $P_b$ states may also be sought at EicC.

In addition to the $P_c$ and $P_b$ states, the exclusive productions of the $\chi_{c1}(3872)$ and $Z_c(3900)$ are also simulated.
The  $\chi_{c1}(3872)$ is arguably the most interesting charmonium-like state. It mass is $(3971.69\pm0.17)$~MeV, coinciding with the $D^0\bar D^{*0}$ threshold exactly, and width is smaller than 1.2 MeV~\cite{Zyla:2020}.\footnote{The mass quoted in RPP is from averaging previous experiments. Recently, the LHCb Collaboration reported precise determinations of the mass and width~\cite{Aaij:2020qga,Aaij:2020xjx}. In particular, a Flatt\'e analysis, which is more proper than the Breit-Wigner one for near-threshold states, is performed in Ref.~\cite{Aaij:2020qga}. } The quantum numbers are  $1^{++}$~\cite{Aaij:2013zoa}.
It is the first discovered and most studied exotic meson candidate in the charmonium region~\cite{Choi:2003ue}.
The mass of $Z_c(3900)^\pm$ is (3888.4$\pm$2.5) MeV with a width of (28.3$\pm$2.5)~MeV~\cite{Zyla:2020}. It was discovered in the $J/\psi \pi^{\pm}$ spectrum by BESIII~\cite{Ablikim:2013mio} and Belle~\cite{Liu:2013dau} and confirmed by other experiments~\cite{Xiao:2013iha,D0:2019zpb}.
The BESIII collaboration identified its spin-parity as $1^+$ and isospin as 1~\cite{Collaboration:2017njt,Ablikim:2015tbp}.
As a charged state decaying into a charmonium, the quark content of $Z_c(3900)$ contains at least $u\bar{d}c\bar{c}$, making it a prominent candidate of tetraquarks or molecular states.

The right panel of Fig.~\ref{fig:Zc_X_production_feyman_diagram} shows the $ep\rightarrow e
\chi_{c1}(3872)p$ and $ep\rightarrow e Z_c^+(3900)n$ processes with the yellow ellipse representing the $t$-channel exchange.
The $\gamma p\rightarrow \chi_{c1}(3872) p$ can proceed through the exchange of vector mesons (e.g. $\rho$, $\omega$ and $J/\psi$) in the $t$-channel. The $\gamma p\rightarrow Z_c^+(3900) n$ can proceed through charged mesons (e.g. $\pi^+$ and $a_0^+$)~\cite{Liu:2008qx} or mesonic Regge trajectories~\cite{Galata:2011bi} in the $t$-channel.
Around the averaged c.m. energy $W_{\gamma p} = 13.7$~GeV, which is in the EicC energy region, the COMPASS Collaboration has searched for the $Z_c(3900)$, and the nonobservation sets an upper limit for the cross section:
$\mathcal{B}(Z_c(3900)^\pm\to J/\psi \pi^\pm) \sigma(\gamma N\to Z_c(3900)^\pm N) <52$~pb at the 90\% confidence level~\cite{Adolph:2014hba}. 
In a similar energy region, the COMPASS Collaboration has also obtained an upper limit for the $\chi_{c1}(3872)$ at the same confidence level: $\mathcal{B}(\chi_{c1}(3872)\to J/\psi \pi \pi) \sigma(\gamma N\to \chi_{c1}(3872) N) <2.9$~pb~\cite{Aghasyan:2017utv}.
A parametrization of these information with a reasonable formula is used as input to the eSTARlight generator~\cite{Lomnitz:2018juf} to obtain the production rates at EicC. The resulting estimated event numbers are shown in Table~\ref{tab:Exotic_simulation}. In particular, the upper bounds for the production of the $\chi_{c1}(3872)$ and $Z_c(3900)^+$ are constrained by the COMPASS measurements. The lower bounds are very roughly estimated by reducing these values by two orders of magnitude, as inferred from the difference between the lower and upper bounds of the $P_c$/$P_b$ production rates.

It is expected that a considerable amount of $Z_c^+(3900)$ could be observed while the events of $\chi_{c1}(3872)$ are lower. The distributions of the invariant mass distribution, the transverse momentum, the pseudo-rapidity and the rapidity for the $Z_c^+(3900)$ are shown in Fig.~ \ref{fig:Zc3900_Q2_bin_Zc_distribution} for different $Q^2$ regions. The red cross histogram is the contribution from $0<Q^2<1$~GeV$^2$, the blue star one is that from $1<Q^2<5$ GeV$^2$, and the black circle one is the overall contribution. It is seen that most of the events are within the low $Q^2$ range.

The above simulations show that the exotic hadrons produced at EicC are close to the middle rapidity range.
This is beneficial for detection. Thus, EicC provides an excellent platform to study the nature of known, but barely understood, charmonium-like states and search for new states.

\begin{figure}[tb]
\centering
\includegraphics[width=0.5\textwidth]{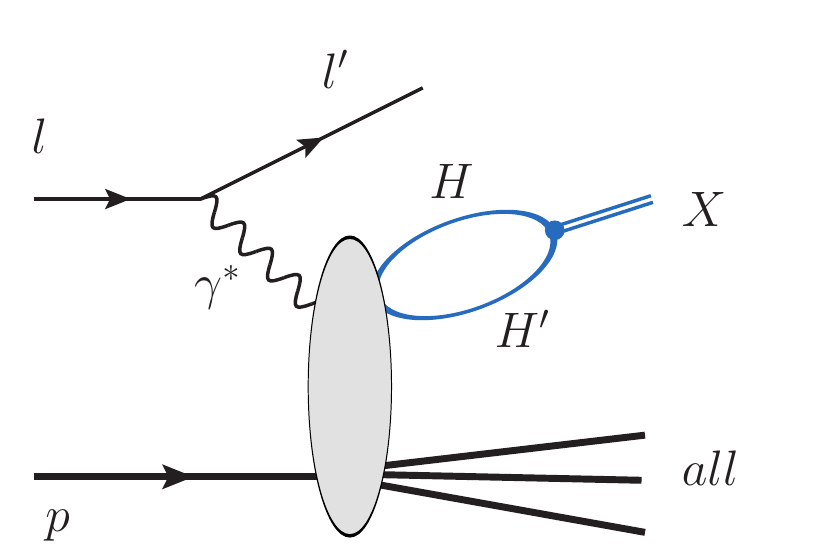} 
\caption{\label{fig:2_semidiagram}The mechanism considered  in Ref.~\cite{Yang:2020} for the semi-inclusive production of exotic hadrons (denoted as $X$) which couple strongly to a pair of hadrons ($H$ and $H'$) in lepton-proton collisions.}
\end{figure}

In addition to the exclusive processes discussed so far, exotic hadrons can be searched for in semi-inclusive processes. In particular, for those hidden-flavor exotic hadrons which couple strongly to a pair of heavy hadrons, such as the $\chi_{c1}(3872)$ and the $Z_c(3900)$, one can achieve an order-of-magnitude estimate of the production cross sections following the method of Refs.~\cite{Bignamini:2009sk, Artoisenet:2010uu, Guo:2013ufa, Guo:2014sca, Albaladejo:2017blx}. The production can be factorized into a short-distance part and a long-distance part. At short distances the hadron pairs of interest are produced, which can be simulated using the Pythia event generator~\cite{Sjostrand:2006za}. The hadron pairs then merge to form the exotic hadrons which couple to them strongly, and the long-distance piece can be computed at the hadronic level. 
The mechanism is shown in Fig.~\ref{fig:2_semidiagram}. This mechanism, when applied to hadron colliders, can produce cross sections for the prompt production of the $\chi_{c1}(3872)$ if the momentum integration range for the hadron-hadron Green's function extends up to a few hundreds of MeV~\cite{Artoisenet:2010uu, Guo:2014sca, Albaladejo:2017blx}.

\begin{figure*}[tb]
\centering
\includegraphics[width=0.495\textwidth]{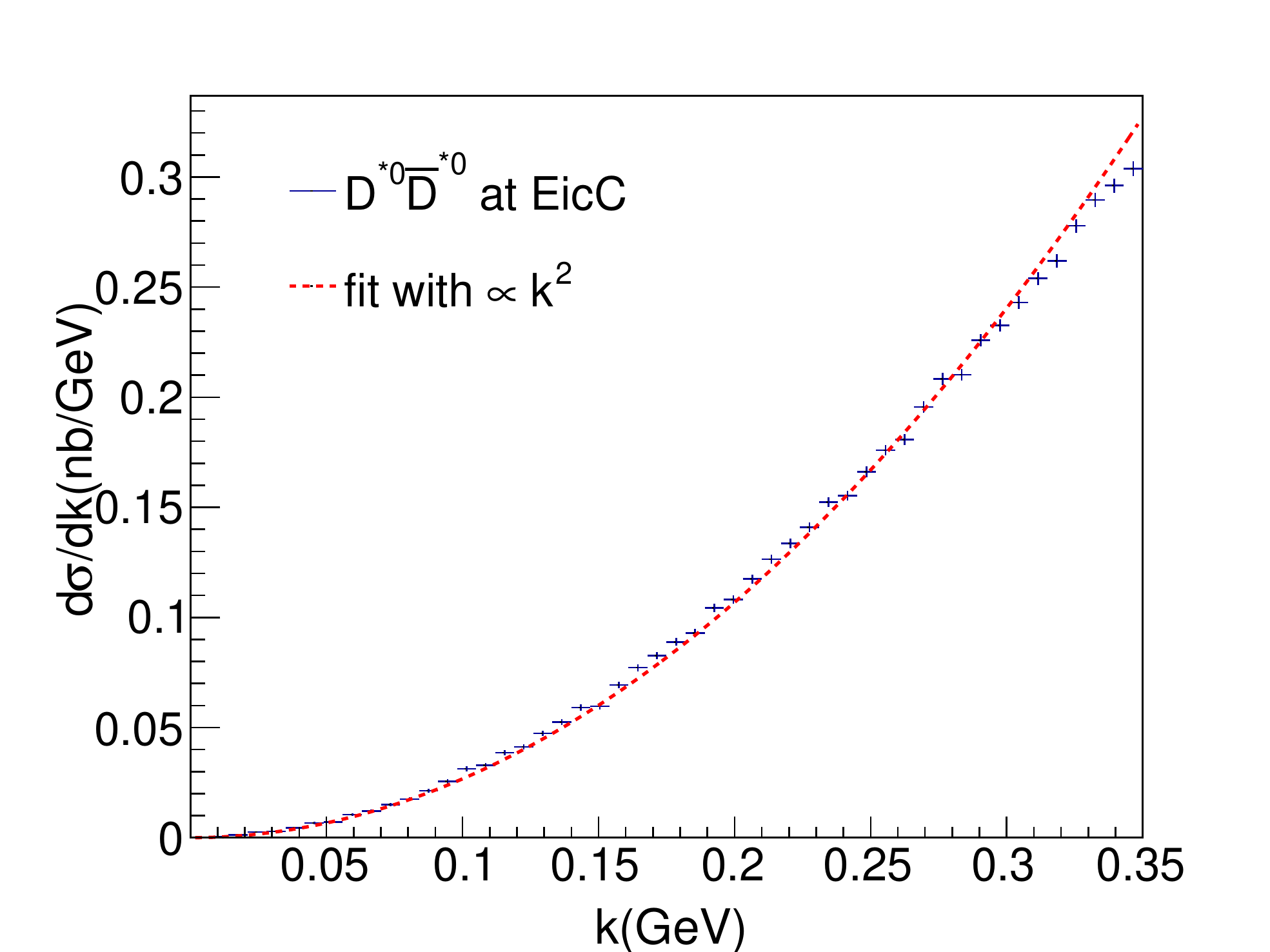} \hfill
\includegraphics[width=0.495\textwidth]{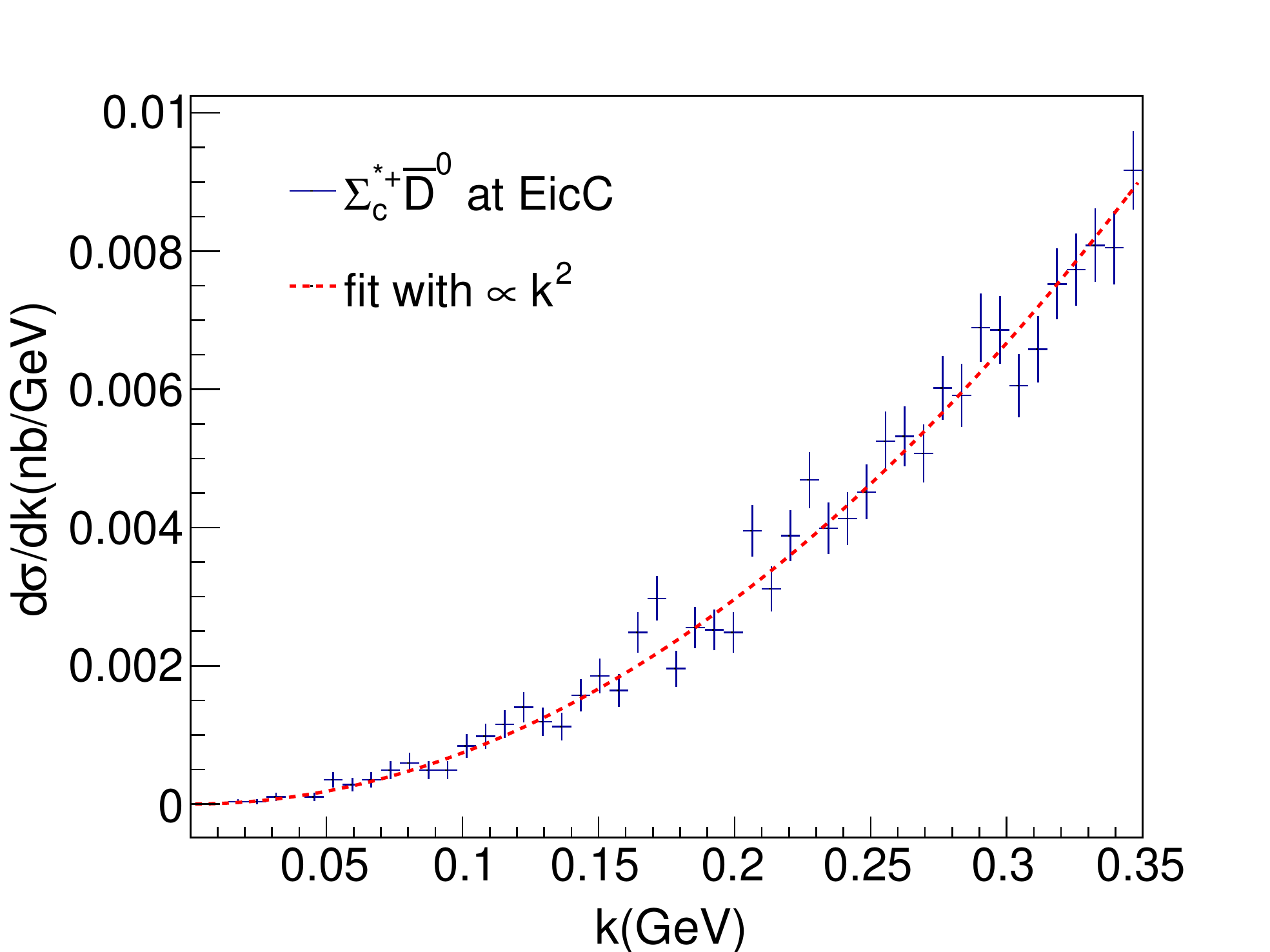} 
\caption{\label{fig:2_semiinclusive_pythia} 
Differential cross sections ${d\sigma}/{ dk}$ (in units of nb/GeV) 
for the semi-inclusive production of $D^{*0}\bar{D}^{*0}$ and $\Sigma_c^{*+}\bar{D}^0$ through electron-proton scattering~\cite{Yang:2020}, where $k$ is the c.m. momentum of the open-charm hadrons. The histograms are obtained from the Pythia event generator while the curves are  fitted according to the momentum dependence $k^2$. The electron and proton energies are set to 3.5 and 20 GeV, respectively. }
\end{figure*}
As an example, in Fig.~\ref{fig:2_semiinclusive_pythia} we show the differential cross sections generated using Pythia~\cite{Sjostrand:2006za} for the semi-inclusive productions of the $D^{*0}\bar D^{*0}$ and $\Sigma_c^{*+}\bar D^0$ pairs in electron-proton collisions with the electron and proton beam energies set to 3.5 and 20~GeV, respectively. The distribution can be well fitted by a $k^2$ dependence with $k$ the c.m. momentum of the open-charm hadrons.
The $XHH'$ coupling in Fig.~\ref{fig:2_semidiagram} can be extracted from measurements or evaluated in phenomenological models. In particular, for the hadronic molecular model, the coupling is connected to the binding energy (see~\cite{Guo:2017jvc}).
The loop in Fig.~\ref{fig:2_semidiagram} is ultraviolet divergent, and the divergence in principal needs to be absorbed into the multiplicative renormalization of the short-distance part. For an order-of-magnitude estimate, the loop integral is evaluated using a Gaussian regulator with a cutoff $\Lambda$ of 0.5 and 1~GeV. 
We list rough estimates for the production cross sections of the $\chi_{c1}(3872)$, $Z_c(3900)^{+,0}$, $X(4020)^0$ and the $P_c$ states observed by LHCb in Table~\ref{tab:sec2_4_xections_semiinclusive}.
The estimates for four more $P_c$ states predicted in the hadronic molecular model using heavy quark spin symmetry~\cite{Xiao:2013yca,Liu:2019tjn,Du:2019pij} are also presented with masses and couplings taken from Ref.~\cite{Du:2019pij}.
Considering an integrated luminosity of 50~fb$^{-1}$, this leads to $\mathcal{O}(10^5)$ events for each of the $P_c$ states, and  $\mathcal{O}(10^6)$ for the $\chi_{c1}(3872)$ and  $\mathcal{O}(10^7)$ events for $Z_c$ states.
Notice that neither branching fractions of further decays nor the detection efficiency is taken into account here. For more details, we refer to Ref.~\cite{Yang:2020}.

To conclude this section, it is promising that EicC can contribute significantly to the study of heavy-flavor exotic hadrons, in particular to charmonium-like states and hidden-charm pentaquarks, and thus contribute to the understanding of how the emergent hadron spectrum is formed by the nonperturbative strong interaction.

\begin{table}[tb]
\caption{Rough estimates of integrated cross sections (in units of pb) at EicC for the semi-inclusive production of a few selected states in the hadronic molecular (HM) model~\cite{Yang:2020}, where $\Lambda$ refers to the cutoff in the Gaussian regulator of the two-hadron Green's function. The processes are $e^-+p\to
\text{HM}$+all, where
$\text{HM}=\chi_{c1}(3872)$, $Z_c(3900)^{0/+}$, $X(4020)^0$, and seven $P_c$ states. The energy configuration considered here is $E_e=3.5$~GeV and $E_p=20$~GeV. The branching fractions of further decays and the detection efficiency are not yet considered here.
}
\label{tab:sec2_4_xections_semiinclusive}
\centering
\begin{tabular}{lccc}
\hline\hline
  & Channel  & $\Lambda = 0.5$~GeV & $\Lambda=1.0$~GeV   \\
\hline
$\chi_{c1}(3872)$     & $D\bar{D}^{*}$ & 21 & 89 \\
$Z_c(3900)^0$ & $(D\bar{D}^{*})^0$ & $0.3\times10^3$ & $1.1\times10^3$ \\
$Z_c(3900)^+$ & $(D\bar{D}^{*})^+$ & $0.4\times10^3$ & $1.3\times10^3$ \\
$X(4020)^0$   & $(D^{*}\bar{D}^{*})^0$ & $0.1\times10^3$ & $0.5\times10^3$ \\
$P_c(4312)$   & $\Sigma_c\bar{D}$ & 0.8 & 4.1 \\
$P_c(4440)$   & $\Sigma_c\bar{D}^{*}$ & 0.7 & 4.7 \\
$P_c(4457)$   & $\Sigma_c\bar{D}^{*}$ & 0.5 & 1.9 \\
\hline
$P_c(4380)$   & $\Sigma_c^{*}\bar{D}$ & 1.6 & 8.4  \\
$P_c(4524)$   & $\Sigma_c^{*}\bar{D}^{*}$ & 0.8 & 3.9  \\
$P_c(4518)$   & $\Sigma_c^{*}\bar{D}^{*}$ & 1.2 & 6.9  \\
$P_c(4498)$   & $\Sigma_c^{*}\bar{D}^{*}$ & 1.2 & 9.8  \\
\hline\hline
\end{tabular} 
\end{table}

\section{Other important exploratory studies}
\subsection{Proton mass}
\label{sec:protonmass}

The major part of the mass of observable matter is carried by the nucleons (neutrons and protons) that constitute all the atomic nuclei in the Universe.  Nucleons themselves are constituted from the gluons and quarks of QCD.  So a key piece of the puzzle surrounding the origin of mass lies with understanding how the proton's mass emerges from the QCD Lagrangian, expressed in terms of light valence-quarks, massless gluons and the interactions between them.  The current-quark masses are produced by the Higgs mechanism; but based on those masses, a straightforward application of notions from relativistic quantum mechanics delivers a proton mass that is two orders-of-magnitude too small.  Plainly, the source of the proton's mass is far more subtle.  Consider, therefore, a sum rule connected with the trace of QCD's energy-momentum tensor (EMT) \cite{Shifman:1978zn}:
\begin{equation}
\label{mpsumrule}
m_p = H_m + H_a\,,\quad
H_m =\langle p|m\bar{\psi}\psi|p\rangle\,, \quad
H_a = \langle p| [\gamma_m m \bar{\psi}\psi + \beta(g) (E^2-B^2)] | p\rangle\,,
\end{equation}
where: $\langle p | p \rangle = 1$; and $m$ represents the light-quark current masses, whose values are of the order $2$-$4\,$MeV at a renormalisation scale $\zeta = 2\,$GeV.  The first of the two terms in Eq.\,\eqref{mpsumrule} is that seeded by the Higgs mechanism.  The second term is more interesting in many ways: it is the trace anomaly, whose origin is intimately connected with the need to regularize and renormalize any quantum field theory defined in four dimensional spacetime.  (These and related issues are discussed further in Sec.\,2.6.1.)

The $H_m$ term in Eq.\,\eqref{mpsumrule} is unambiguous; but there are many ways to decompose and rearrange $H_a$ \cite{Roberts:2016vyn, Lorce:2017xzd}.  One popular approach is to focus on the energy of a proton at rest and write \cite{Ji:1994av, Ji:1995sv, Yang:2018nqn}:
\begin{equation}
\label{mpsumrule2}
m_p = H_m + H_a = H_m + H_q + H_g + \tfrac{1}{4} H_a\,,
\end{equation}
where
\begin{equation}
H_q = \langle p|\psi^\dagger(-i\textbf{D} \cdot \alpha )\psi|p\rangle\,,\quad \quad
H_g=\langle p|\tfrac{1}{2}(E^2+B^2)|p\rangle
\end{equation}
are, respectively, the quark and gluon kinetic energies.
Contemporary lattice-QCD calculations reveal \cite{Yang:2015uis, Abdel-Rehim:2016won, Bali:2016lvx} that only about 9\% of $m_p$ is generated by the $u$-, $d$-, $s$-quark contributions in $H_m$.  However, even this contribution is nontrivial because it is built from products of the Higgs-generated current-quark masses and enhancement factors that express nonperturbative QCD dynamics, \emph{viz}.\ $H_m \sim m_{\rm Higgs} \times \langle p |\bar\psi\psi|p\rangle_{\rm np\,QCD}$.  The remaining 91\% is essentially dynamical.  It can be considered as the contribution generated by strong QCD forces through the trace anomaly; or in the second decomposition, broken into identified pieces that include those from quark and gluon kinetic energies.  Nonetheless, no matter how one chooses to cut the pie, a very large fraction of the proton's mass emerges as a dynamical consequence of strong interactions within QCD.

As noted above and described in more detail below, the appearance of a nonzero contribution to the trace of the QCD EMT, call it $\Theta$, is a quintessentially quantum field theoretical effect.  Empirically, the expectation value of $\Theta$ is large in almost every hadron state.  The exception is the chiral-limit pseudoscalar meson, for which the expectation value is expected to vanish.  Providing a mathematically rigorous proof that these outcomes are truly dynamical consequences of QCD would constitute a solution to one of the seven Millennium Prize Problems \cite{millennium:2006}.  Modern progress in theory is seemingly bringing such a proof within reach; and numerical simulations of lattice-regularized QCD are beginning to yield quantitative results for the independent terms in Eq.\,\eqref{mpsumrule2}.

Given the fundamental importance of the scale set by $m_p$, much attention is now focused on the experimental confirmation of Eq.\,\eqref{mpsumrule}.  In this context it has been argued, using the operator product expansion and low-energy theorems \cite{Kharzeev:1995ij,Kharzeev:1998bz,Boussarie:2020vmu}, that the heavy-vector-meson--proton scattering amplitude near threshold is dominated by $H_a$ and sensitive to the $H_m$ correction. Recent theoretical study also suggests that this process at large photon virtualities $Q^2$ can be used to extract the gluon part of the proton gravitational form factor and sensitive to the trace anomaly effect at subleading level\cite{Boussarie:2020vmu}. A preliminary analysis of GlueX data on $J/\psi$  photoproduction in Hall-D at JLab is broadly consistent with the prediction.  However, a range of theoretical issues complicate the interpretation of such $J/\psi$ measurements in this way and substantially more work is needed before firm conclusions can be drawn.

On the other hand, the case for a connection between the $\Upsilon p$ near-threshold scattering amplitude and the proton mass sum rule is theoretically much cleaner.  Experimentally, this system is inaccessible at JLab; and existing measurements at other facilities are restricted to $W\gtrsim 90\,$GeV, as evident in Fig.\,\ref{fig:sec2_4_xsect_ccbar}, which is far above threshold.  Consequently, the EicC can here contribute uniquely, being able to explore collision energies $W < 20\,$GeV.

\begin{figure}[htbp]
\includegraphics[width=0.495\textwidth,angle =0]{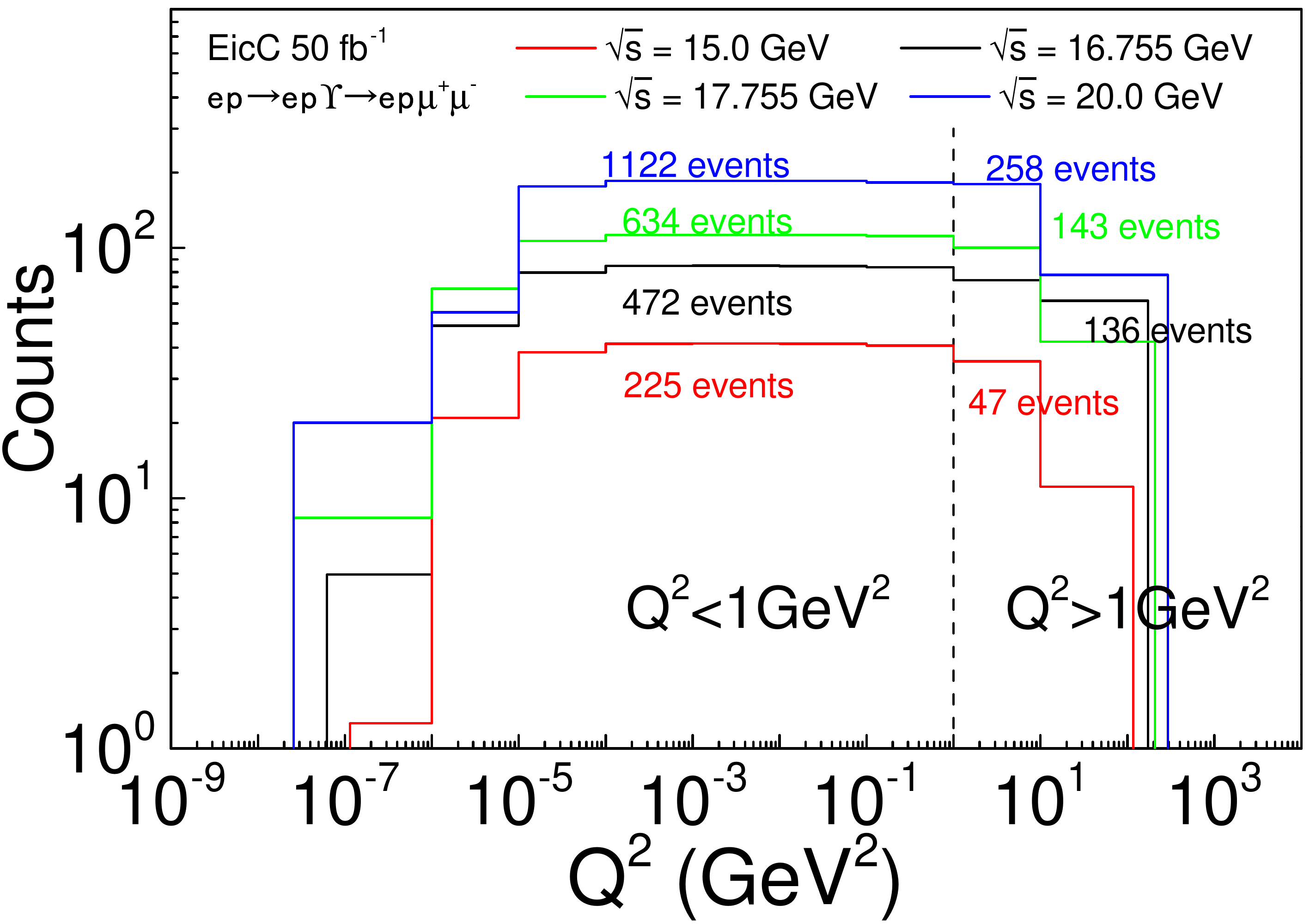}
\includegraphics[width=0.495\textwidth,angle =0]{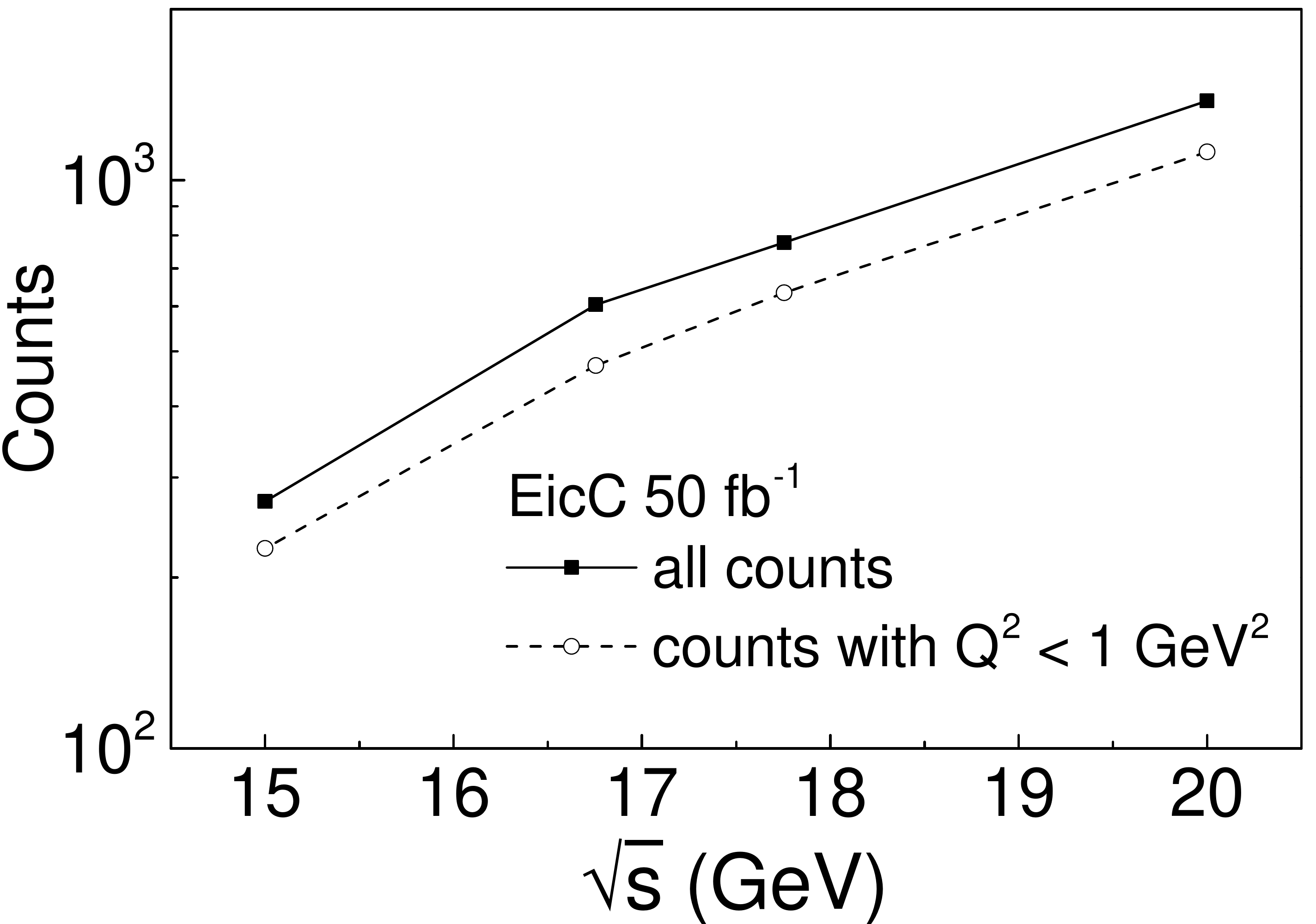}
\caption{\label{fig:UpsilonQ2} 
Distribution of $ep \to ep \Upsilon \to ep \mu^+ \mu^-$ events, assuming $50\,{\rm fb}^{-1}$ integrated luminosity. The $ep$ central energy $\sqrt{s} = 16.755\,$GeV in the figure corresponds to that of EicC.  Cases with central energy$\sqrt{s} = 15.0$, $17.755$ and $20.0\,$GeV are also illustrated, but only $\sqrt{s} = 16.755$ GeV is expected for a running to accumulate 50 fb$^{-1}$ luminosity. \emph{Left panel} -- number of events as a function of photon virtuality,$Q^2$; and \emph{right panel} -- event number as a function of $\sqrt{s}$.}
\end{figure}

\begin{figure}[htbp]
\includegraphics[width=0.95\textwidth,angle =0]{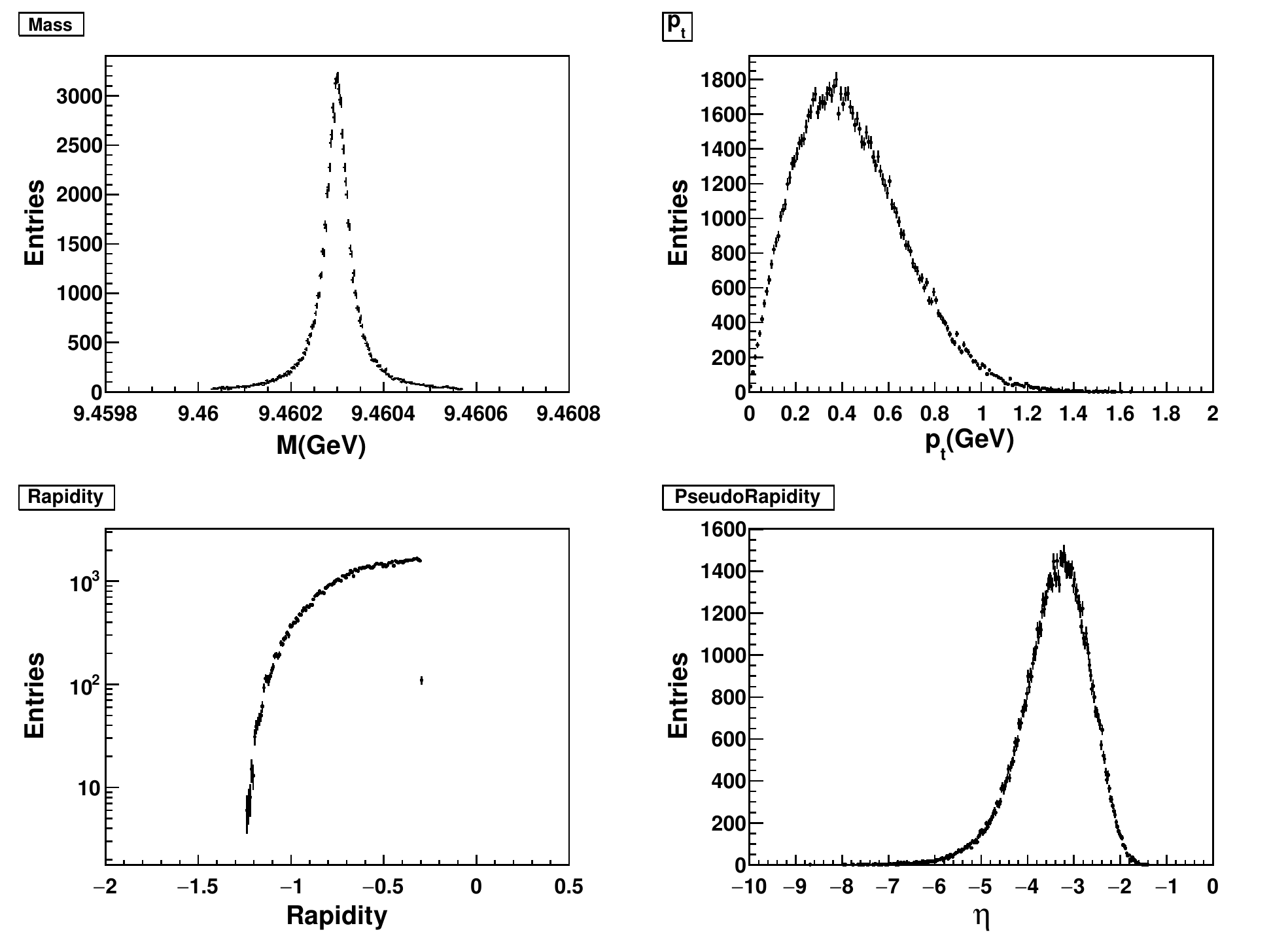}
\caption{\label{fig:UpsilonBin}
Reconstruction profiles in the distributions of mass, transverse momentum, rapidity and quasi-rapidity, for the final state $\Upsilon$ in the reaction $ep \to ep \Upsilon \to ep \mu^+ \mu^-$.}
\end{figure}

Using the reaction $ ep \to ep \Upsilon \to ep l^+ l^-$, as pictured in Fig.\,\ref{fig:Zc_X_production_feyman_diagram}
where $V=\Upsilon$ and the orange area represents a $t$-channel Pomeron \cite{Laget:1994ba,Martynov:2002ez}, the distribution of events achievable with EicC is shown in Fig.\,\ref{fig:UpsilonQ2}. It seems that the cross section estimation within Pomeron exchange under a proper consideration of phase space \cite{Martynov:2002ez} is indeed consistent at the order-of-magnitude level with other models as shown in Fig. \ref{fig:sec2_4_xsect_bbbar}, in which dipole Pomeron model is used as the input of simulation. Evidently, EicC could produce around 600 events under the proposed design, $80$-$85$\% of which lie in the $Q^2<1\,$GeV$^2$ region, with more than $90$\% at $Q^2<10\,$GeV$^2$.  Moreover, if the two decay channels $ \Upsilon \to \mu^ + \mu^-$, $ e^+ e^-$ are detected simultaneously, the number of reconstructed events is even larger. Fig.~\ref{fig:UpsilonBin} displays the anticipated reconstruction profiles of the $\Upsilon$ in the distributions of mass, transverse momentum, rapidity and quasi-rapidity. Though the detector reconstruction efficiency would be lower than that of the resonant process in Tab. \ref{tab:Exotic_simulation}, one can confidently assume a value of $20$\%, given that the final states are all charged particles. So it is feasible to investigate detailedly the $t$-dependent cross sections at EicC.

This discussion demonstrates that EicC can deliver precision in the study of $\Upsilon$ production and potentially thereby open a window onto the origins of the proton's mass.  As a significant collateral benefit, $\Upsilon$ production is an important background that must be understood when analysing data taken in searches for the hidden-bottom five-quark state, $P_b$.

\subsection{Structure of light pseudoscalar mesons}
\label{StructureNGmodes}
Theoretical imperatives for investigating and revealing the structure of light pseudoscalar mesons are detailed in Sec.\,2.6.  The case has many facets because the pion is the lightest known hadron and it has a unique and crucial position in nuclear and particle physics \cite{Horn:2016rip}.  For example, the pion is the closest approximation to a Nambu-Goldstone (NG) boson in Nature.  It is massless in the absence of Higgs-boson couplings into QCD and remains unusually light when those couplings are switched on.  In addition, this light pion is essential to the formation of nuclei, carrying the strong force over length-scales large enough to enforce stability against electromagnetic repulsion between the protons within a nucleus.  Thus, understanding pion structure is of the utmost importance.  Two clear paths are available: namely, the measurement of pion elastic form factors and of pion structure functions.

Existing empirical knowledge of pion structure is poor.  Elastic form factor measurements do not extend beyond $Q^2=2.45\,$GeV$^2$ \cite{Volmer:2000ek, Horn:2006tm, Tadevosyan:2007yd, Horn:2007ug, Huber:2008id, Blok:2008jy} and existing structure function measurements are more than thirty years old \cite{Badier:1980jq, Badier:1983mj, Betev:1985pg, Falciano:1986wk, Guanziroli:1987rp, Conway:1989fs}.  The kaon situation is worse; and that is unsatisfactory for many more reasons.  Largest amongst them being that the standard model of particle physics has two sources of mass:
explicit, generated by Higgs boson couplings;
and emergent, arising from strong interaction dynamics, responsible for the $m_N \sim 1\,$GeV mass-scale that characterises nuclei, and the source for more than 98\% of visible mass.
Emergent hadronic mass 
is dominant for all nuclear physics systems; but the Higgs mechanism applies modulations that are critical to the evolution of the Universe, \emph{e.g}.\ CP-violation, discovered in neutral kaon decays \cite{Christenson:1964fg}.  Thus, knowledge of kaon structure is necessary because it provides a window onto the interference between Higgs boson effects and 
emergent mass \cite{Aguilar:2019teb, Roberts:2020udq}.

The impediment to experimentally mapping the structure of light pseudoscalar mesons is simply explained. These systems are unstable, they decay quickly: so, how can one build a target?  One answer is to use the Drell-Yan process at high-energy accelerators.  This is the mode exploited thirty years ago \cite{Badier:1980jq, Badier:1983mj, Betev:1985pg, Falciano:1986wk, Guanziroli:1987rp, Conway:1989fs}.  Another approach is to measure the leading neutron in high-energy $ep$ collisions \cite{Chekanov:2002pf, Aaron:2010ab}.  In these processes, the range of light-front momentum fraction, $x$, has been somewhat limited and existing errors are large.

\begin{figure}[htbp]
\begin{tabular}{ccc}
\includegraphics[clip, width=0.4\textwidth]{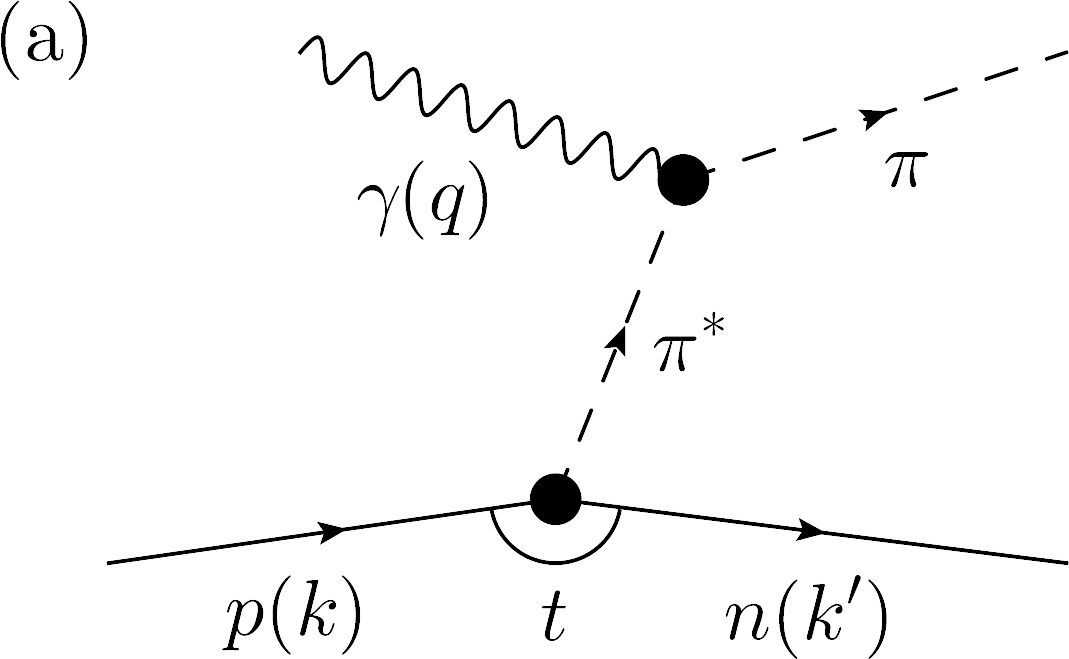} & \hspace*{2em} &
\includegraphics[clip, width=0.4\textwidth]{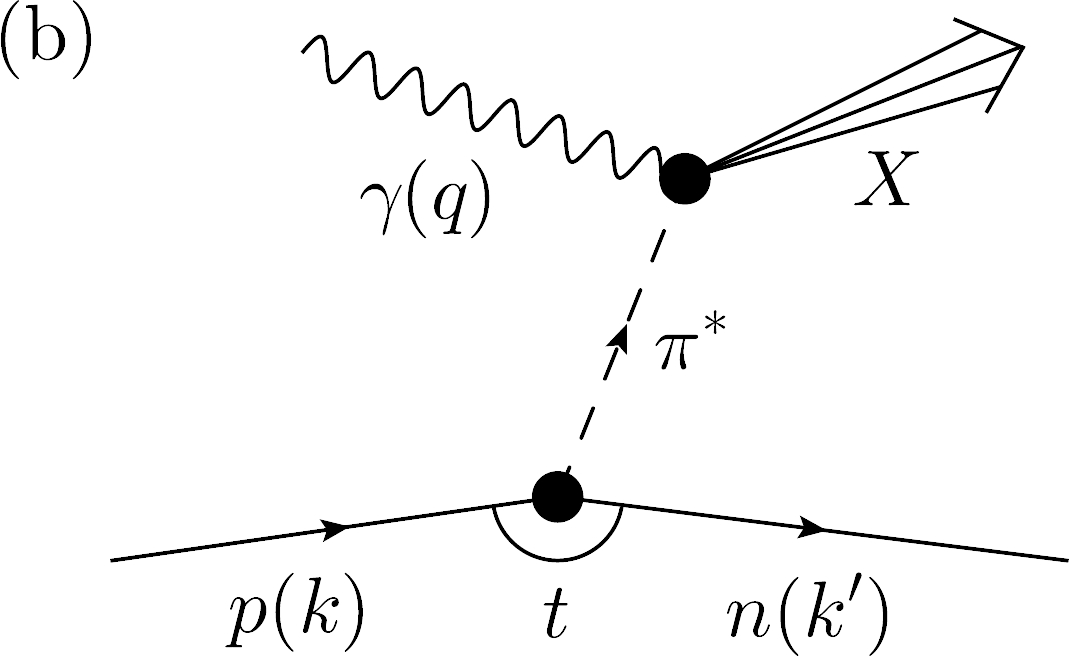}
\end{tabular}
\caption{\label{figSullivan}
Sullivan processes: a nucleon's pion cloud provides access to (a) pion elastic form factor and (b) pion parton distribution functions. The intermediate pion, $\pi^\ast(P=k-k^\prime)$ with the Mandelstam variable $t = P^2$, is off-shell.}
\end{figure}

At high-luminosity facilities, the Sullivan process, illustrated in Fig.\,\ref{figSullivan}, becomes a very good method for for gaining access to meson targets. The approach capitalizes on the feature that a proton is always surrounded by a meson cloud and can sometimes be viewed as a correlated $\pi^+ n$ $(K^+ \Lambda)$ system.  Theory predicts \cite{Qin:2017lcd} that such processes provide reliable access to a pion target on $-t\lesssim 0.6\,$GeV$^2$; and for the kaon, on $-t\lesssim 0.9\,$GeV$^2$.  The 12\,GeV-upgraded JLab facility will exploit these reactions to extend the $Q^2$ reach of existing $\pi$ and $K$ form factor measurements \cite{E1206101, E12-07-105} and the $x$-range of available data on the pion structure function \cite{Keppel:2015}.  Efforts will also be made to measure the kaon structure function \cite{C12-15-006A}.  Given the wider kinematic range of EicC, its capacity to achieve these goals will be far greater.
At the EicC, the scattering angle and energy of the final state baryons can be measured precisely, to tag the Sullivan process and to measure the invariant mass of the virtual light mesons. Technologies of the far-forward hadron calorimeter and trackers could be learned from projects at HERA (H1 and ZEUS), and from similar experiments at the US-EIC. The final state neutron at the EicC has energy around 15 GeV, a theta angle around 25 mrad, and $p_T$ $<$ 0.5 GeV. Assuming the energy resolution 
to be 35\%/$\sqrt{(E/GeV)}$ and the spatial resolution to be 1 cm for the far-forward neutron detector,  
one can estimate the resolution for the invariant mass of the virtual meson, $\frac{\sigma_{|m^2(\pi^*)|}}{|m^2(\pi^*)|}$, to be around 27\% at the EicC.

\begin{figure}[htbp]
\centering
\includegraphics[width=0.5\textwidth]{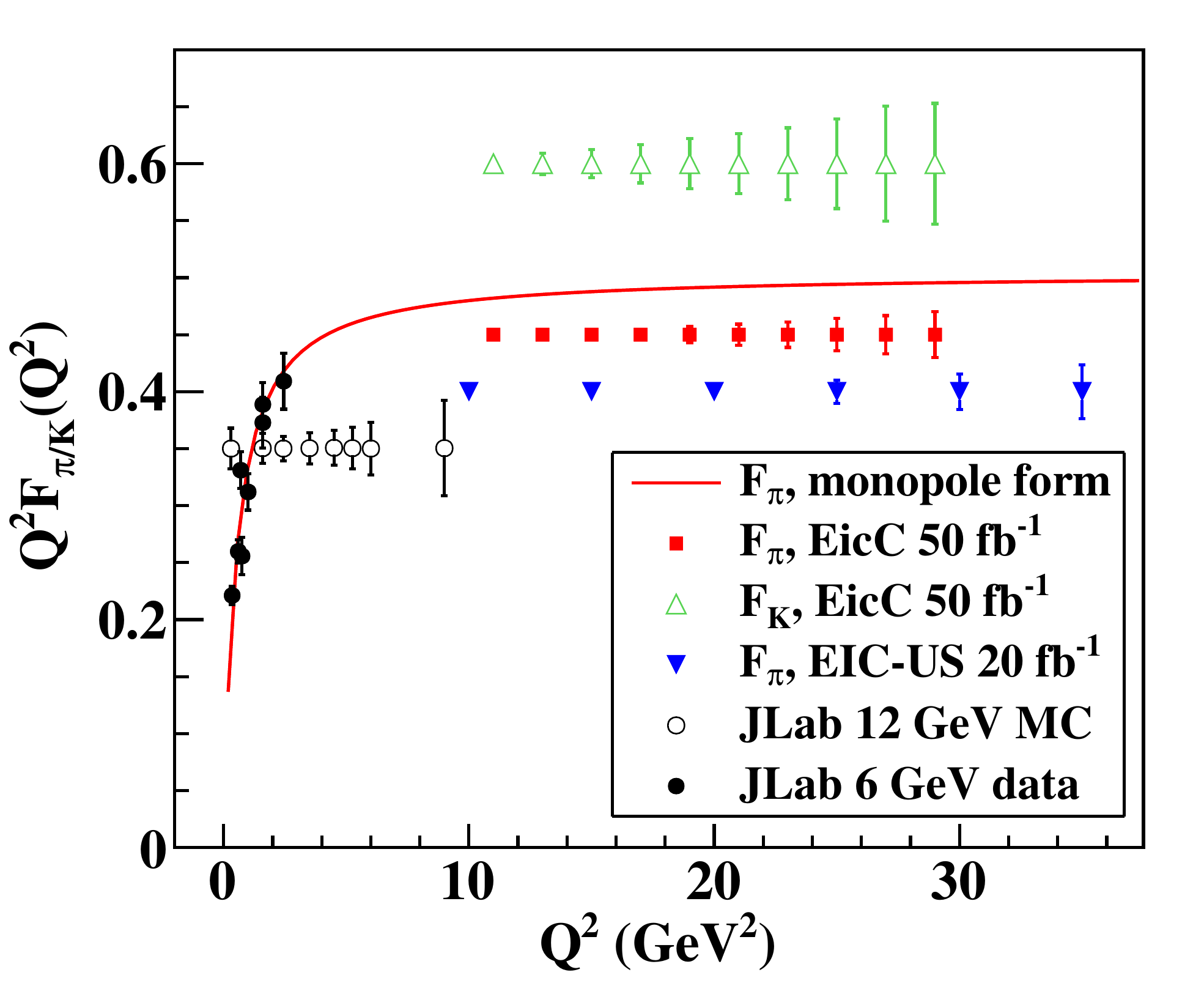}
\caption{
Projected statistical errors on EicC pion form factor measurements compared with those at the US EIC \cite{Aguilar:2019teb}
and JLab 12 GeV program \cite{E1219006}.   Also depicted: extant JLab data obtained by the F$_{\pi}$-Collaboration \cite{Horn:2006tm, Tadevosyan:2007yd, Horn:2007ug, Huber:2008id, Blok:2008jy}; and statistical error projections for kaon form factor measurements at EicC.
}
\label{fig:FormFactorSimulation}
\end{figure}

Application and experience have delivered a reliable approach to obtaining the pion elastic form factor, $F_\pi(Q^2)$, from the Sullivan process \cite{Guidal:1997by, Vanderhaeghen:1997ts, Choi:2015yia, Perry:2020hli}.  The longitudinal electroproduction cross-section is expressed such that $F_\pi(Q^2)$ is the only unknown, which can be determined by comparison between data and the model.  The $\pi^+/\pi^-$ production ratio extracted from electron-deuteron beam collisions in the same kinematics as charged pion data from electron-proton collisions is used to ensure that the longitudinal cross-section has been isolated \cite{Horn:2016rip}.  Fig.\,\ref{fig:FormFactorSimulation} shows that a high-luminosity, high-energy EicC can deliver precise results on pion and kaon electromagnetic form factors, providing detailed maps on the domain $Q^2/{\rm GeV}^2 \in [10,30]$, which has the highest physics discovery potential.

The estimates in Fig.\,\ref{fig:FormFactorSimulation} were prepared using the following cuts: $|t|<0.6\,$GeV and $W/{\rm GeV}\in (2,10)$.  Separating the data into ten bins on $Q^2/{\rm GeV}^2 \in [10,30]$, the statistical errors on $F_\pi(Q^2)$ data are competitive with all other existing proposals.  Regarding the size of the estimated statistical errors, the same is true for the EicC kaon form factor measurement.

Fig.\,\ref{figSullivan}(b) shows that a Sullivan process can also be used to measure $\pi$ and $K$ structure functions.  Compared with elastic form factor measurements, the cross-section is much larger because the meson target is shattered.  Deep inelastic $e\pi$ interactions are ensured by selecting events in which the transverse momentum of the tagged final-state neutron is small and its longitudinal momentum exceeds 50\% of that of the incoming proton.

For kaon structure functions, the tagged outgoing system is the $\Lambda$-baryon.

\begin{figure}[htbp]
\centering
\includegraphics[width=0.8\textwidth]{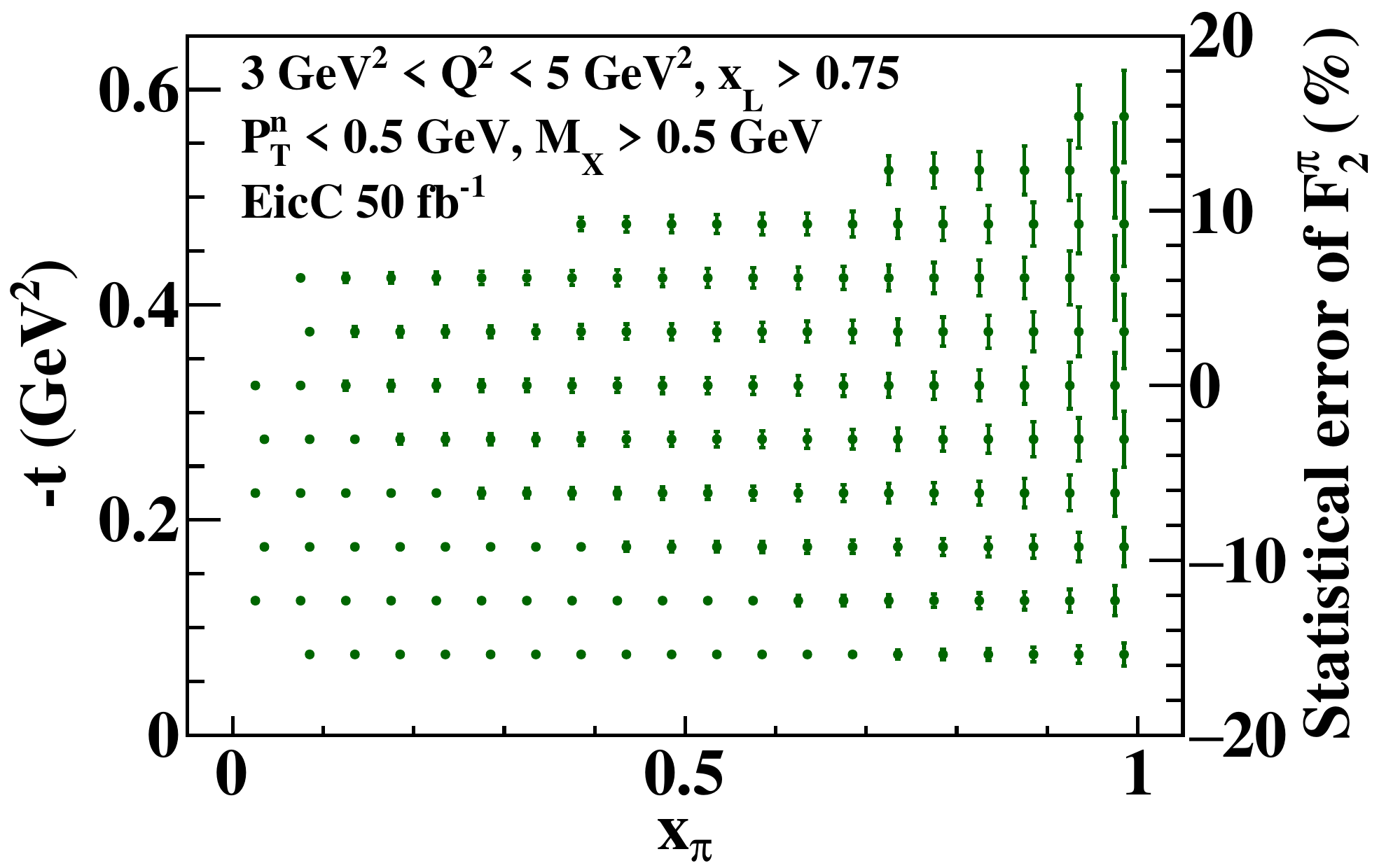}
\caption{
Projected statistical uncertainty on pion structure function in a certain $Q^2$ bin as 
a function of the four-momentum-transfer $-t$ and the Bjorken-x of pion.
The cuts for producing the projection are also shown in the plot. In order to select
the scattering between electron and virtual pion, a cut on
$x_L>0.75$ is made \cite{Aaron:2010ab}. Moreover, the detection efficiency for the final
state neutron is assumed to be $50\%$.
The measurement at EicC could be performed on a large kinematic domain with uncertainty that is uniformly $\lesssim 5$\%.}
\label{fig:NeutTagged-DIS-Domain-A}
\end{figure}

The projected statistical precision of an EicC pion structure function measurement is sketched in Fig.\,\ref{fig:NeutTagged-DIS-Domain-A}.  It assumes roughly one year of running and is based on a Monte-Carlo simulation of leading-neutron tagged DIS with the pion valence-quark distribution function taken from Ref.\,\cite{Gluck:1999xe}.
Fig.\,\ref{fig:NeutTagged-DIS-Domain-A} indicates that EicC can map the domain $x_\pi \in (0.01,0.95)$ with precision, yielding data that could be crucial to resolving the pion structure function controversy described in Sec.~\ref{sec:2.6.3.2}.

\subsection{Intrinsic charm}

Great progress has been made in the literature in understanding the fundamental structure of the nucleon in recent years. However, we still know little about the properties and distributions of the heavy quarks in nucleon. As a potential constituent of the nucleon, information for those heavy-flavor content plays an important role in testing Standard Model (SM), and in searching for new physics beyond the SM.

Heavy content in nucleon is predicted theoretically in QCD theory. In the light-cone framework, the wave function of a proton can be expanded in terms of superposition of Fock states as following,
\begin{eqnarray}\label{Fexpan}
|p\rangle = c_1 |uud \rangle + c_2 |uudg \rangle + c_3 |uud {\color{red} c\bar{c}} \rangle + \cdots.
\end{eqnarray}
Here, $|F\rangle$ are the Fock states, and the coefficients $c_j$ ($j=1,2, \cdots$) are proportional to the wave function amplitudes of Fock components~\cite{Lepage:1979zb, Lepage:1980fj, Brodsky:1997pd}. It indicates that there are heavy charm quarks arise in the proton, even the states such as $|uudc\bar{c}\rangle$ are extremely rare. On the view point of QCD theory, the extra ${\color{red} c\bar{c}}$ pair in the proton can be generated in two distinct processes involving in perturbative and nonperturbative effects.

One probable charm content arisen in a proton is the ``extrinsic charm (EC)". In this case, a gluon radiates from the valence quark in the proton, and then splits to a $c\bar{c}$ pair associated with large transverse momentum. The gluon must be hard enough in order to produce a heavy $c\bar{c}$ pair. The process is therefore governed by the QCD evolution corresponding to the short distance effects. In this scenario, the charm and anti-charm quarks have the same significant features in the proton. The EC behaves like a sea quark and is generally softer than the gluon by a factor of $(1-x)$, where $x$ is the momentum fraction of EC in the proton. The parton distribution function (PDF) describes the extrinsic charm density in the proton, which is related to the momentum fraction $x$ and the factorization scale $\mu_F$. Using an initial form at a specific scale, such as the charm quark mass $m_c$, the EC PDF can be evolved to any factorization scale $\mu_F$ with the help of the Dokshitzer-Gribov-Lipatov-Altarelli-Parisi (DGLAP) evolution equations~\cite{Gribov:1972ri, Altarelli:1977zs, Dokshitzer:1977sg, Field:1989uq}. In practice, the initial form of the EC PDF could be modeled, whose parameters can be fixed from a comprehensive global analysis of hard scattering data generated from a variety of fixed-target or collision experiments. The extrinsic charm distribution is assumed to be concentrated in the small region of momentum faction, e.g. $x \in [0,0.1]$, since it drops down quickly at $x>0.1$. The EC component is usually neglected in early studies on the processes with small and moderate center-of-mass energy (CME) in collisions, especially for the hadronic colliders where the small $x$ events are generally not recorded. With much higher CME of colliders today, the EC becomes more and more important in high energy processes.

Another probable charm content arisen in a proton is the ``intrinsic charm (IC)'', which is quite different from the extrinsic one and has strong hints from experimental observations. In year 1979, the production of charmed particles was reported by the CERN and ACCDHW Collaborations via $pp \to D^+ (\Lambda_c^+)X$ with the center-of-mass energy $\sqrt{s}=53$ and 63 GeV, respectively~\cite{Drijard:1978gv, Giboni:1979rm, Lockman:1979aj, Drijard:1979vd}. The cross section was measured to be $0.1\sim 0.5\,\rm mb$, which is larger than theoretical predictions ($10\sim 50\,\rm \mu b$~\cite{Georgi:1978kx}) by about one order of magnitude. Moreover, the $D^+$ was found to be generated abundantly in the forward region, which is hard to be explained by simply using the ordinary extrinsic charm. In order to narrow the gap between experimental data and theoretical predictions, the idea of IC was proposed by S. J. Brodsky {\it et al.} in Ref.\cite{Brodsky:1980pb, Brodsky:1981se, Brodsky:2015fna}. Theoretically, this fluctuation of the proton can be achieved in two ways, one is the interaction between valance quarks and multiple soft gluons, another one is vacuum polarization which is usually extremely rare and can be neglected. The intrinsic charm generated in this manner exhibits remarkable differences from the extrinsic charm, and the probability of finding IC in a proton is proportional to ${\cal O}(1/m_c^2)$. In contrast to the EC, the IC can be generated at or even below the energy scale of the heavy quark mass threshold. The intrinsic charm and anti-charm are not necessary to have the same distributions in the proton. The IC has a ``valence-like'' character, which has small contributions in low momentum fraction $x$ and dominates in the relatively large $x$ region. The IC distribution is non-perturbative, and many models have been suggested in the literature, whose inputs can be fixed by global fitting of various known experimental data.

Some previously done measurements could indicate the existence of IC in the nucleon. In 1980s, the charm structure function $F_2^c(x,\mu)$ was measured by the European Muon Collaboration (EMC)~\cite{Aubert:1981ix, Aubert:1982tt}. An enhancement at large $x$ beyond theoretical expectation was reported by EMC, and such a gap could be compensated by taking the IC into consideration~\cite{Harris:1995jx}. The inclusive photon production with heavy-flavor $Q$ jets in hadronic collisions would also provide valuable information for the IC distributions in the nucleon. The partonic process $g+Q \to \gamma+Q+X$ gives the main contributions to the photon events, which depends strongly on the heavy contents in an incident hadron. Data from the D\O\ experiment at the Tevatron~\cite{Abazov:2009de} shows disagreement between the prediction without IC and the measurement for $\gamma+c+X$ in large transverse momentum of the photon. By including IC contributions, the discrepancy is reduced but still prominent in the larger $p_T^{\gamma}$ region, in particular when $p_T^{\gamma}$ is larger than 80 GeV.


Up to now, a definite conclusion on the existence of IC is still missing. Additional experiments are necessary for investigating the IC. The measurement of inclusive charmed hadron production at hadron colliders is an additional promising way to investigate the IC~\cite{Aitala:2000rd, Aitala:2002uz}. The doubly heavy baryons (especially $\Xi_{cc}$) production associated with initial charm quark at the parton level will be suitable to investigate the intrinsic content, since the IC impacts significantly on these production channels. The $\Xi_{cc}$ production in hadronic collisions involve three typical mechanisms, i.e. the gluon-gluon fusion $(g + g)$, the gluon-charm collision $(g + c)$, and the charm-charm collision $(c + c)$, at the proton-proton (or proton-antiproton) colliders. Conventionally, contributions from the gluon-gluon fusion are expected to be dominant in the hadronic production of $\Xi_{cc}$. However, other production mechanisms may also have sizable contributions. For the $(g + c)$ and $(c + c)$ production mechanisms, the initial $c$ quark can be either extrinsic or intrinsic in the incident protons. Because the proportion of the IC components in the nucleon is small, which is only up to $\sim 1\%$, the intrinsic charm usually gives no significant contribution in most of the high-energy processes. However, in specific kinematic regions, the IC may lead to unexpected conspicuous consequences in the $\Xi_{cc}$ production.

The hadronic production of $\Xi_{cc}$ baryon was investigated at the LHC, the Tevatron, and the Fixed-target Experiments at hadron collider (FixExp@HC)~\cite{Chang:2006xp, Chen:2019ykv, Chen:2014hqa}. Using the generator GENXICC~\cite{Chang:2007pp, Chang:2009va, Wang:2012vj,Wu:2013pya}, one may simulate the hadronic production of $\Xi_{cc}$ with both extrinsic and intrinsic charm being considered. Because the intrinsic component at large $x$ will decrease at the high factorization scale, the $(g+g)$ channel becomes dominant in the $\Xi_{cc}$ hadroproduction with high CME. The IC mechanisms shall have a significant impact on the hard processes with moderate factorization scales at those colliders with a relatively lower center-of-mass energy. Therefore, colliders with high luminosity at lower center-of-mass energy, such as the fixed-target experiment (like After@LHC~\cite{Brodsky:2012vg, Hadjidakis:2018ifr, Lansberg:2012wj, Lansberg:2012sq, Lansberg:2013wpx} operating at a center-of-mass energy $\sqrt{s}=115\;\rm GeV$) and future electron-ion colliders (e.g., EIC US~\cite{Accardi:2012qut} and EicC, etc.), would be ideally suited to discover or constrain the intrinsic content in nucleon.


The EicC will provide a brand new mode to study the production of doubly charmed baryon. Two important subprocesses occur in the electron-nucleus ($e$-N) collision with exchange of a virtual photon between the electron and nucleus, i.e., $\gamma + g \to \Xi_{cc}+\bar{c}+\bar{c}$ and $\gamma + c \to \Xi_{cc}+\bar{c}$, which are classified by the virtuality $Q^2$ of the photon. Numerically, we observe that the intrinsic charm enhanced the total cross sections by nearly 3 times to that without intrinsic components with $A_{\rm in}=1\%$ in the $e$-Au collision mode at the EicC. Moreover, at
the instantaneous luminosity of $2.0 \times 10^{33}\;\rm cm^{-2}s^{-1}$, 
the estimated number of $\Xi_{cc}$ events in one year is about $4.0\times10^5$ by adopting the Non-relativistic Quantum Chromodynamics~\cite{PhysRevD.51.1125}.
It makes the precise investigation of the properties of the doubly charmed baryon accessible at the EicC.

The observation of the doubly charmed baryon $\Xi_{cc}^{++}$ had been reported by the LHCb collaboration~\cite{Aaij:2017ueg} in 2017. However, the $\Xi_{cc}^+$ with a similar production rate but a bit shorter lifetime than $\Xi_{cc}^{++}$, has not been discovered yet. The shorter lifetime means more loss events of $\Xi_{cc}^+$ before tested by the detectors, which requires more efforts in experimental measurements. Recently, the LHCb collaborations reported zero results of searching for $\Xi_{cc}^+$~\cite{Aaij:2019jfq} again with higher integrated luminosity than that in 2013 by an order of magnitude. This is in contradiction with the observation of $\Xi_{cc}^+$ baryon by the SELEX in 2002~\cite{Mattson:2002vu} and in 2005~\cite{Ocherashvili:2004hi}. This contrary may be interpreted by the difference of kinematic cuts between those two experiments. The LHCb experiment may lose more events in the small $p_t$ region than that in the SELEX, which are related closely to the intrinsic content in a nucleon. Therefore, more different kinds of experiments, such as the After@LHC, the EicC, and the LHeC, etc., are needed for clarifying the puzzle in searching for the $\Xi_{cc}^+$. Conversely, experimental studies on $\Xi_{cc}$ production at those colliders could help to further progress in verifying the existence of IC.

In conclusion, different high-energy colliders (such as the After@LHC, the EicC, etc.), are expected to increase the probability for the discovery of IC. Moreover, those different experimental platforms could provide important cross-checks to ensure the correctness of experimental measurements and theoretical analysis. And it shall shed light on the intrinsic heavy mechanism and fundamental structure of the nucleon.


\section{QCD theory and phenomenology}
\subsection{Synergies}
High level theory and phenomenology are required in order to inspire, guide, and capitalize on the wide ranging program of EicC experiments.  Rigorous QCD theory must be complemented by insightful phenomenology so that every discovery opportunity can be seized.  The lattice formulation maintains the closest connections with the QCD Lagrangian.  Here, steady progress continues and testable predictions are being delivered.  Tight links with QCD are also provided by continuum Schwinger function methods, e.g.\ the Dyson-Schwinger equations (DSEs).
With the added flexibility of continuum methods, DSEs provide access to a wider range of observables.  Thus, with their complementary content, lattice QCD and DSEs have a natural synergy that can be exploited for the benefit of EicC physics.

\subsection{Lattice QCD}
Lattice QCD is a fundamental method to study strong interactions non-perturbatively from first principles. It directly computes the QCD path integral in a discretized finite-volume Euclidean space-time, and then the finite lattice spacing and volume provide natural ultraviolet and infrared cut-offs. The input parameters in lattice QCD 
are the quark masses and the coupling constant, which is related to the lattice spacing by the renormalization group.
Today, with the help of powerful computing facilities, lattice QCD can provide important information on many aspects of EicC physics and comparisons with experimental results.

\subsubsection{Nucleon spin structure}
As discussed in Sec.\,\ref{OneDSpinStructure}, the proton's spin can be decomposed into the helicity and orbital angular momentum of \mbox{quarks} and gluons. The contribution from quark helicity is the sum of quark polarization along the proton spin direction, which can be obtained from lattice QCD by computing the first Mellin moment of the polarized parton distribution function of quarks. There have been many lattice calculations, see Refs.~\cite{Lin:2017snn,Aoki:2019cca} for reviews.  Recent lattice results~\cite{Alexandrou:2017oeh,Liang:2018pis,Lin:2018obj} are mutually consistent and agree well with global fits; and strange flavor results are more precise than the global fits.  Both experiment~\cite{deFlorian:2014yva} and lattice calculation~\cite{Yang:2016plb} indicate that the gluon helicity contributes considerably to proton spin, but improved lattice QCD calculations are needed in order to deliver precise predictions~\cite{Yang:2019dha}.

The QCD energy-momentum tensor admits several expressions for the quark and gluon orbital angular momentum contribution to the proton spin.  
With the Jaffe-Manohar decomposition \cite{Jaffe:1989jz}, both the orbital and total angular momentum of quarks and gluons can be defined in a gauge invariant manner with appropriate light-cone gauge links. On the other hand, gauge invariant and frame independent total angular momentum of quarks and gluons can be defined through the symmetrized energy-momentum tensor based on Ji's decomposition~\cite{Ji:1996ek}, expressed in Eq.\,(\ref{eq:quark_angular_mom}).  The difference between the two is that the quark-gluon interaction term is allocated to the gluon (quark) orbital angular momentum in the Jaffe-Manohar (Ji) decomposition.
The gauge invariant total angular momenta of quarks and gluons have been calculated in lattice QCD~\cite{Deka:2013zha, Alexandrou:2017oeh,Alexandrou:2020sml}: the gluon angular momentum result needs improved precision. Exploratory lattice studies have been made for the quark orbital angular momentum and preliminary results have been obtained~\cite{Engelhardt:2019lyy, Engelhardt:2017miy}.

Lattice QCD is playing a valuable role in the study of nucleon spin structure, with important recent progress.
Quark helicity results for different flavors have reached 10\% precision, and preliminary estimates have been obtained for the gluon helicity, quark orbital angular momentum and quark (gluon) total angular momentum. It is expected that lattice QCD will be able to provide more accurate and extensive information on nucleon spin structure.

\subsubsection{Proton mass decomposition}
From the sum rule of the QCD energy-momentum tensor, Eq.\,\eqref{mpsumrule}, the invariant mass of proton can be decomposed into quark mass term and the trace anomaly~\cite{Shifman:1978zn}. The quark mass term is a product, involving the chiral condensate and quark current masses; hence, it is directly related to the mass generated by the Higgs mechanism and scale invariant. This part only contributes about 9\% of the total proton mass~\cite{Yang:2018nqn} for three light flavors; so most of the proton mass can be considered as arising from the trace anomaly~\cite{Shifman:1978zn}.

In the chiral limit, the pion matrix element of the trace anomaly is zero, whereas it is nonzero for the nucleon, yielding the entire nucleon mass. As yet, there is no direct calculation of the trace anomaly using lattice QCD, but the proton mass sum rule has been used to predict the trace anomaly contribution to $m_p$~\cite{Yang:2018nqn}. A calculation of the trace anomaly contribution to $m_p$ is being pursued using lattice QCD; and success will enable a comparison between QCD theory and the EicC measurements described in Sec.\,\ref{sec:protonmass}.

Using one definition of the quark kinetic energy \cite{Ji:1994av}, its contribution to $m_p$ can be obtained from the quark light-front momentum fraction, i.e.\ the first moment of the unpolarized quark distribution functions; similarly, for the gluon kinetic energy contribution. Lattice results for the momentum fraction~\cite{Yang:2018nqn} agree with phenomenological analyses~\cite{Lin:2017snn}; but uncertainties still need to be further suppressed, especially for the sea quarks and gluons.

\subsubsection{1-D and 3-D structure of nucleons}
Most lattice QCD calculations of PDF-related quantities have been limited to the first few PDF moments, mainly due to the fact that light-cone PDFs are not directly accessible in Euclidean space-time.
However, novel theoretical developments in recent years~\cite{Ji:2013dva,Ma:2014jla,Radyushkin:2016hsy} are enabling lattice QCD to directly calculate PDFs, including 1-D Bjorken-x dependent PDFs, generalized parton distribution functions(GPDs), and also transverse moment dependent parton distribution functions(TMDs)

Large momentum effective theory (LaMET) is one approach \cite{Ji:2020ect}.  It uses an equal time correlation function to reach a light cone correlation function using standard methods of effective field
theory matching and running.  Namely, from the equal time correlations that can be computed in lattice QCD, one obtains the so-called quasi-PDFs; and when the nucleon momentum is large enough, quasi-PDFs can be matched perturbatively to the physical PDFs, with computable large-momentum power corrections.

LaMET has been employed in some lattice QCD calculations, with notable progress on the non-singlet quark PDFs ($u(x)-d(x)$), including unpolarized, polarized and transverse distributions. The latest results agree with global fits~\cite{Lin:2018qky,Alexandrou:2018pbm,Alexandrou:2018eet}. The small $x$ region is challenging for laMET because it requires simulations with high nucleon momenta, which is difficult in lattice QCD. At present, lattice QCD can compute valence-quark PDFs on $x\gtrsim 0.1$ and provide some constraints at small-$x$ for the sea quarks via PDF moments~\cite{Liang:2019xdx}.
%
LaMET can also be used to compute GPDs, e.g.\ the pion's unpolarized quark GPD has been explored~\cite{Chen:2019lcm} and this may be extended to the nucleon case.

\begin{figure}
\centering
\includegraphics[width =0.49 \textwidth]{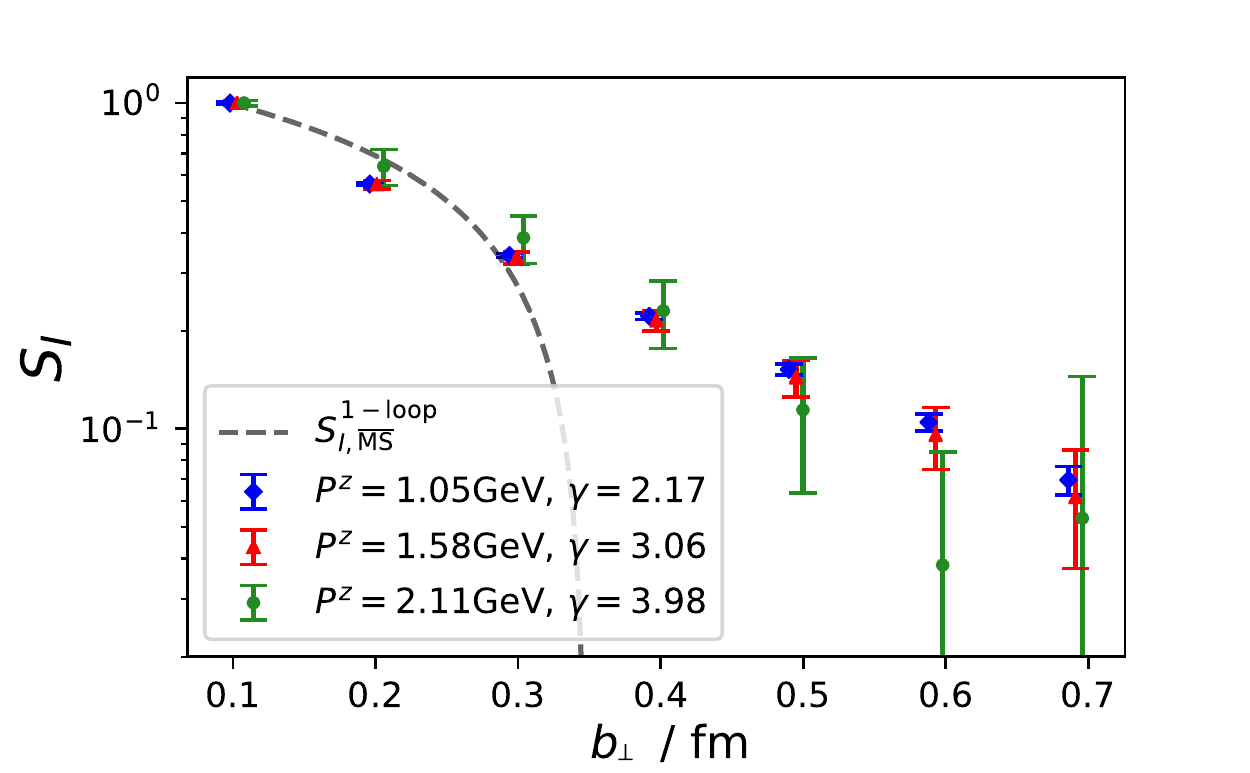}
\includegraphics[width =0.49 \textwidth]{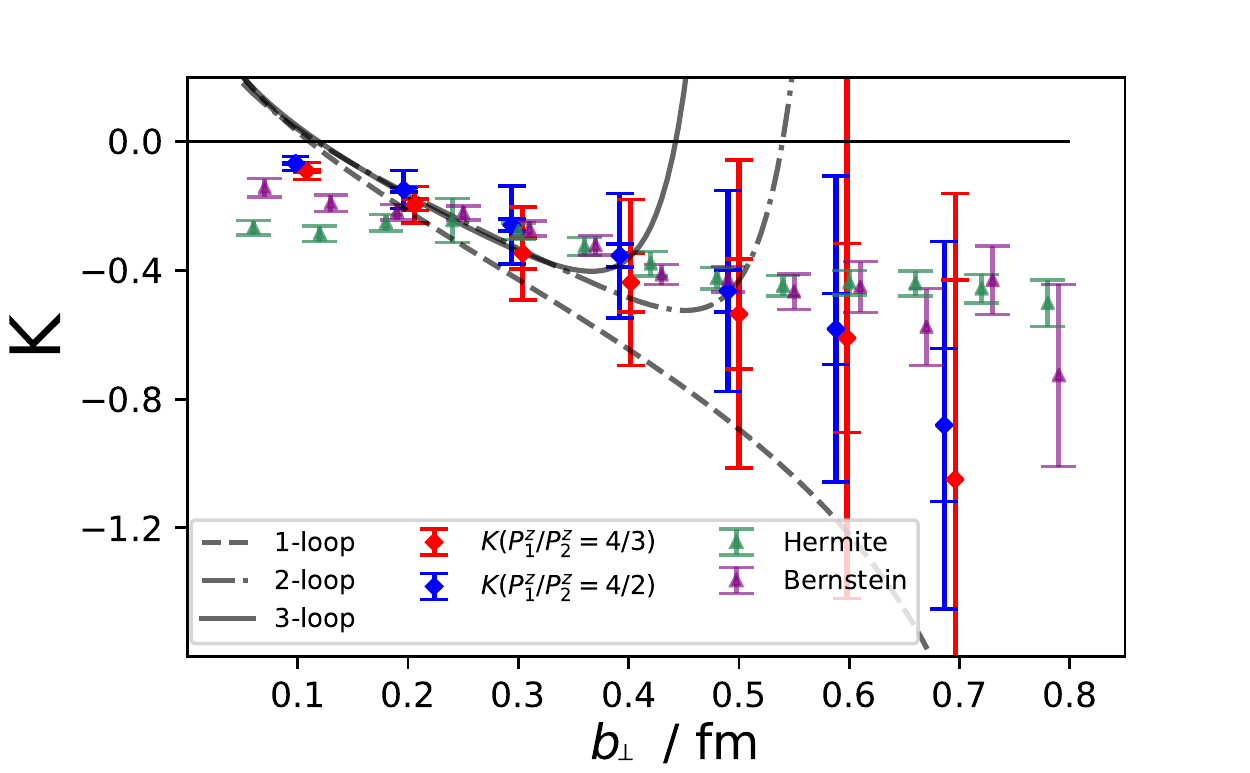}
\caption{Lattice results for the TMD soft function: intrinsic part (left panel)~\cite{Zhang:2020dbb} and rapidity dependent part (right panel)~\cite{Shanahan:2020zxr,Zhang:2020dbb}.  For the rapidity dependent part, the results from Ref.~\cite{Zhang:2020dbb} (red/blue points) are consistent with those from Ref.~\cite{Shanahan:2020zxr} (green/brown points). The lattice calculation reaches much larger transverse separation $b_{\perp}$ than perturbative calculations.}
\label{Fig:LatticeSF}
\end{figure}

TMD calculations using lattice QCD are more complicated than those of PDFs and GPDs. An additional soft function is required to factorize processes involving small transverse momentum, like Drell-Yan production and semi-inclusive DIS. This function involves two light-like gauge links along reversed light-cone directions; hence, cannot be simulated directly in Euclidean space. Recently, it was argued that this difficulty can be overcome~\cite{Ji:2019ewn}; and as shown in Fig.~\ref{Fig:LatticeSF}, several lattice QCD attempts at compuation of the soft function have been carried out~\cite{Shanahan:2020zxr,Zhang:2020dbb}.  This opens a path to prediction of TMD-related quantities using lattice QCD.

\subsubsection{Partonic structure of the nucleus}
Nucleons within the nucleus appear to interact weakly with each other via long-range forces and the binding energy per nucleon is small in comparison with the nucleon mass; yet, as illustrated in Fig.\,\ref{fig:EMC}, various experiments have indicated that the PDFs of bound nucleons are different from those in free space \cite{Benvenuti:1987dv, Ashman:1992kv, Gomez:1993ri}.  This highlights the importance understanding nuclear structure from the gluon-quark level.  Naturally, for quarks and gluons contained within nucleons and nuclei, the non-perturbative nature of the bound-state problem makes the theoretical study very difficult. Lattice QCD simulations can shed light on this problem.

Lattice calculation at physical kinematics are very challenging owing to the signal-to-noise problem.  In a first step, Ref.~\cite{Detmold:2020snb} reported a study of the PDFs of ${}^3\text{He}$, extracting the first Mellin moment of the unpolarized isovector quark PDFs at an unphysical quark mass corresponding to $m_\pi\sim800$ MeV. The ratio of the quark momentum fraction in ${}^3\text{He}$ to that in a free nucleon was found to be consistent with unity. Although no EMC effect was observed, this study together with an earlier lattice analysis of the ${}^4\text{He}$ binding energy~\cite{Yamazaki:2009ua} show that 
lattice methods are reaching a level of practicable maturity for very light nuclei. Calculation precision can be controlled to a level of few percent by using a relatively heavy pion mass; and even in this unphysical realm, results from the HALQCD collaboration suggest that some nuclear physics remains~\cite{Iritani:2018zbt}.

Further reduction of both statistical and systematic uncertainties requires more effort from the lattice QCD community. One can anticipate that, with the continuing development of computer hardware and software, lattice calculations may come to play a unique role in hunting the EMC effect.

\subsubsection{Exotic hadrons}
The methods to study hadron spectroscopy in lattice QCD are relatively mature.  One computes correlation functions of operators with the desired quantum numbers and obtains the spectrum from their time dependence. The interpolating operators encode the hadron structure information. Usually the operators are constructed by a quark--anti-quark pair (meson) or three quarks (baryon). To explore the structure of exotic hadrons, such as those discussed in Sec.\,\ref{ExoticHadrons}, one needs to build operators that express the exotic's likely composition, e.g.\ multi-quark states, hybrids and glueballs, etc. Most hadrons are unstable resonances, which appear as poles of hadron scattering amplitudes. Again, owing to the Euclidean space-time used in lattice QCD, real-time dependent matrix elements related to the scattering processes cannot be computed directly. One way to circumvent this problem is to use the finite volume method developed by M.~L\"uscher~\cite{Luscher:1990ux}, by which the scattering information can be extracted from the energies of analogous systems in a finite box.
This approach offers a path forward for lattice QCD in calculations of the structure and properties of exotic states.

EicC is ideal for the study of heavy flavor hadrons. Many charmed exotic hadrons have been observed in experiments, e.g.\ the $XYZ$ particles and the pentaquark candidates $P_c$s, but the structure of these states are not known yet.
Some studies on $XYZ$ particles have been performed in Lattice QCD~\cite{Liu:2012ze,Prelovsek:2014swa,Chen:2014afa,Ikeda:2016zwx}. However, the results are generally contaminated by systematic uncertainties, which come from finite lattice spacing, finite volume, unphysical light quark mass, ignoring coupled channel effects, and so on.
At the same time, it is expected that the bottom counterpart of the charm exotic states should also exist; yet the number of observed exotic bottom hadrons is fewer than in the charm sector.
%
With continuing steady improvement, lattice QCD will be able to contribute to resolving the puzzles associated with heavy flavor hadrons.

\subsection{Continuum theory and phenomenology}
\subsubsection{Mass and matter}
Sec.\,\ref{sec:protonmass} highlights that the masses of the neutron and proton, the kernels of all visible matter, are roughly 100-times larger than the Higgs-generated masses of the light $u$- and $d$-quarks. In contrast, Nature's composite Nambu-Goldstone bosons are (nearly) massless.  In these states, the strong interaction's $m_N \approx 1\,$GeV mass-scale is effectively hidden. Furthermore, in the chiral limit, when the Higgs-generated current masses in QCD's Lagrangian are omitted, the $\pi$ and $K$ mesons  are exactly massless and perturbation theory suggests that strong interactions cannot distinguish between quarks with negative or positive helicity. Such a chiral symmetry would have many consequences, but none of them is realised in Nature.  Instead, the symmetry is broken by interactions. Dynamical chiral symmetry breaking (DCSB) entails that the massless quarks in QCD's Lagrangian acquire a large effective mass \cite{Bhagwat:2003vw, Bowman:2005vx, Bhagwat:2006tu} and ensures that the interaction energy between those quarks cancels their masses exactly so that the chiral-limit pion is massless \cite{Maris:1997hd, Qin:2014vya, Binosi:2016rxz}.

DCSB underpins the notion of constituent-quark masses and, hence, sets the characterising mass-scale for the spectrum of mesons and baryons constituted from $u$, $d$ quarks and/or antiquarks.   Moreover, restoring the Higgs mechanism, then DCSB is responsible for, \emph{inter alia}: the measured pion mass ($m_\pi \approx 0.15 \,m_N$); and the large mass-splitting between the pion and its valence-quark spin-flip partner, the $\rho$-meson ($m_\rho > 5 \,m_\pi$).  There are many other corollaries, extending also to the physics of hadrons with strangeness, wherein the competition between dynamical and Higgs-driven mass-generation has numerous observable consequences.  The competition extends to the charm and bottom quark sectors; and much can be learnt by tracing its evolution.

Such phenomena, their origins and corollaries, entail that the question of how did the Universe evolve is inseparable from the questions of how does the $m_N \approx 1\,$GeV mass-scale that characterizes atomic nuclei appear; why does it have the observed value; and, enigmatically, why does the dynamical generation of $m_N$ have no apparent effect on the composite Nambu-Goldstone bosons in QCD, \emph{i.e}.\ whence the near-absence of the pion mass?  A decisive challenge is to determine whether the answers to these questions are contained in QCD, or whether, even here, the Standard Model is incomplete.

\begin{figure}[htbp]
\centering
\includegraphics[width=0.6\textwidth]{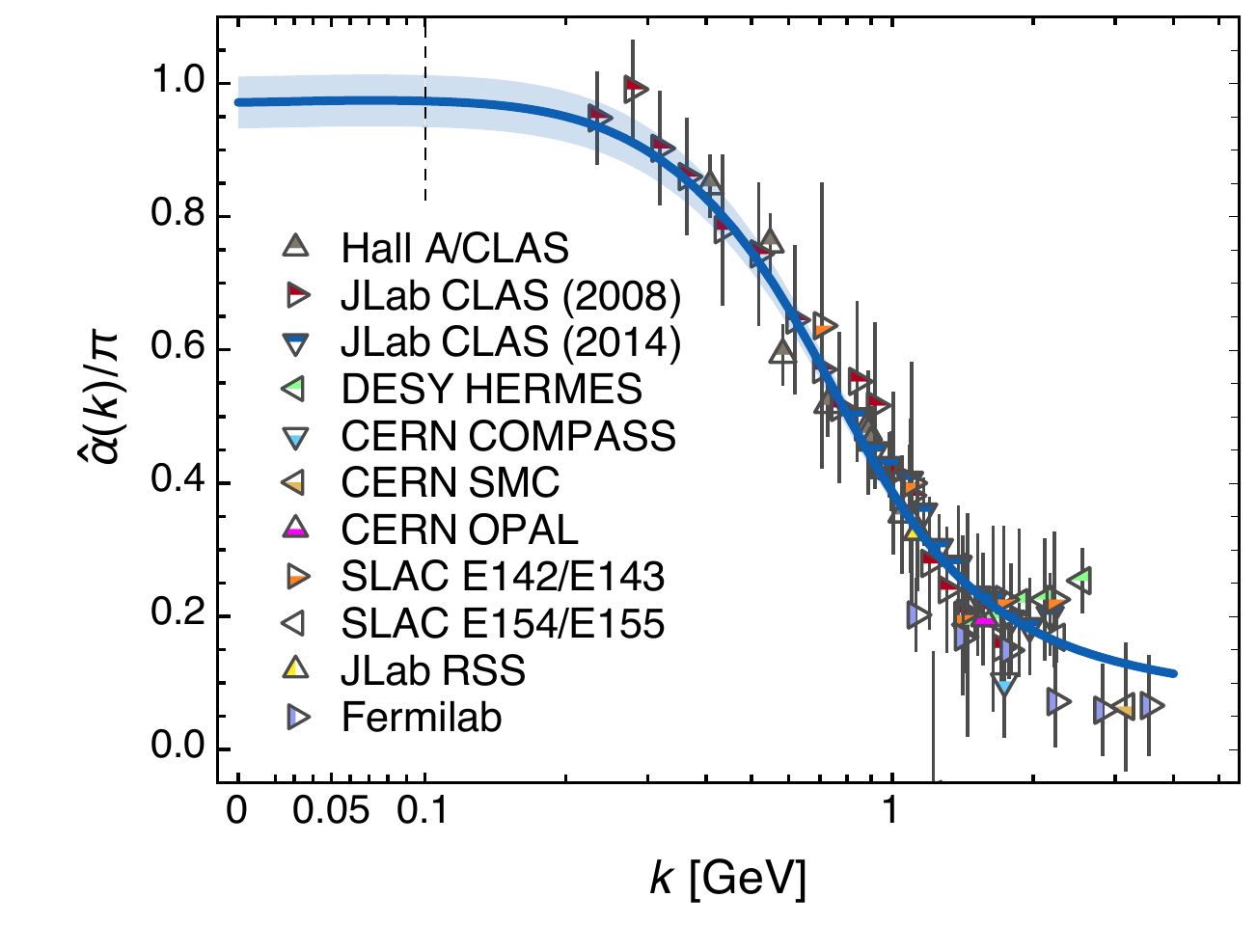}
\caption{
QCD's process-independent running-coupling, $\hat{\alpha}(k^2)/\pi$, obtained by combining the best available results from continuum and lattice analyses \cite{Cui:2019dwv}.
%
%
World data on the process-dependent charge, $\alpha_{g_1}$, defined via the Bjorken sum rule, are also depicted.  (Sources are detailed elsewhere \cite{Cui:2019dwv}.
Image courtesy of D.\,Binosi.)
}
\label{FigwidehatalphaII}
\end{figure}

These questions are being addressed using DSEs, which provide a symmetry-preserving approach to solving the continuum hadron bound-state problem \cite{Roberts:2015lja, Horn:2016rip, Eichmann:2016yit, Burkert:2019bhp, Qin:2020rad}.  
%
In connection with the emergence of hadronic mass, 
the framework has delivered a significant advance with the prediction of a process-independent QCD effective charge \cite{Cui:2019dwv}.  Depicted in Fig.\,\ref{FigwidehatalphaII}, $\hat\alpha(k)$ saturates at infrared momenta, $\hat\alpha(0)/\pi=0.97(4)$, owing to the emergence of a gluon mass-scale: $m_0/{\rm GeV} = 0.43(1)$.  These and other features of $\hat\alpha(k)$ suggest that QCD is rigorously defined; if so, then it is unique amongst known four-dimensional quantum field theories.  Numerous consequences can be tested with EicC experiments, a few of which are subsequently described.

\subsubsection{1-D hadron structure\label{sec:2.6.3.2}}
%
\hspace*{-0\parindent}\emph{Meson Form Factors}.\\
The best known and most rigorous QCD predictions are those made for the electromagnetic form factors of pseudoscalar mesons, \emph{e.g}.\ the pion and kaon.  These hadrons are abnormally light; yet, their properties provide the cleanest window onto the emergence of mass within the Standard Model  \cite{Roberts:2019ngp}.  This connection is expressed most forcefully, in the behaviour of meson form factors at large momentum transfers.  On this domain, QCD relates measurements simultaneously to low- and high-energy features of QCD, \emph{viz}.\ to subtle features of meson wave functions and to the character of quark-quark scattering at high-energy \cite{Lepage:1979zb, Efremov:1979qk, Lepage:1980fj}. These relationships are expressed concretely in advanced DSE analyses, with predictions that expose the crucial role of 
emergent mass \cite{Gao:2017mmp}.

Throughout the modern history of nuclear and particle physics, much attention has focused on finding evidence for power-law scaling in experimental data.  This is an important step; but it should be remembered that QCD is not found in scaling laws.  Instead, since quantum field theory requires deviations from strict scaling, then QCD is to be found in the existence and character of scaling violations.  Considering meson elastic electromagnetic form factors, theory predicts that (\emph{i}) scaling violations will become apparent at momentum transfers $Q^2\gtrsim 10$\,GeV$^2$ \cite{Gao:2017mmp} and (\emph{ii}) the magnitude of any given form factor on a sizeable domain above $Q^2= 10$\,GeV$^2$ is determined by the physics of emergent mass. 
Hence, experiments focused in this area are of the greatest importance; and as discussed in connection with Fig.\,\ref{fig:FormFactorSimulation}, EicC can here make crucial contributions.

\medskip

\hspace*{-\parindent}{\emph{Meson Structure Functions}.}\\
At a similar level of rigor is the QCD prediction for the behavior of meson structure functions. The momentum distributions of light valence quarks within the pion have the following behaviour at large-$x$ \cite{Ezawa:1974wm, Farrar:1975yb, Berger:1979du}: ${\mathpzc u}^{\pi}(x;\zeta \approx m_0) \sim (1-x)^{2}$.
The most recent measurements of ${\mathpzc u}^\pi(x;\zeta)$ are thirty years old \cite{Badier:1980jq, Badier:1983mj, Betev:1985pg, Falciano:1986wk, Guanziroli:1987rp, Conway:1989fs}; and conclusions drawn from those experiments have proved controversial \cite{Holt:2010vj}.  For example, using a leading-order (LO) perturbative-QCD analysis, the E615 experiment \cite{Conway:1989fs} reported: ${\mathpzc u}_{\rm E615}^{\pi}(x; \zeta_5=5.2\,{\rm GeV}) \sim (1-x)^{1}$, in striking conflict with the expected behavior.  Subsequent calculations \cite{Hecht:2000xa} confirmed the QCD prediction and eventually prompted reconsideration of the E615 analysis, demonstrating that E615 data may be viewed as consistent with QCD \cite{Wijesooriya:2005ir, Aicher:2010cb}.  However, uncertainty over ${\mathpzc u}^{\pi}(x)$ remains because more recent analyses of available data have failed to consistently treat higher-order effects and, crucially, modern data are lacking.

\begin{figure}[htbp]
\begin{tabular}{ccc}
\includegraphics[clip, width=0.46\textwidth]{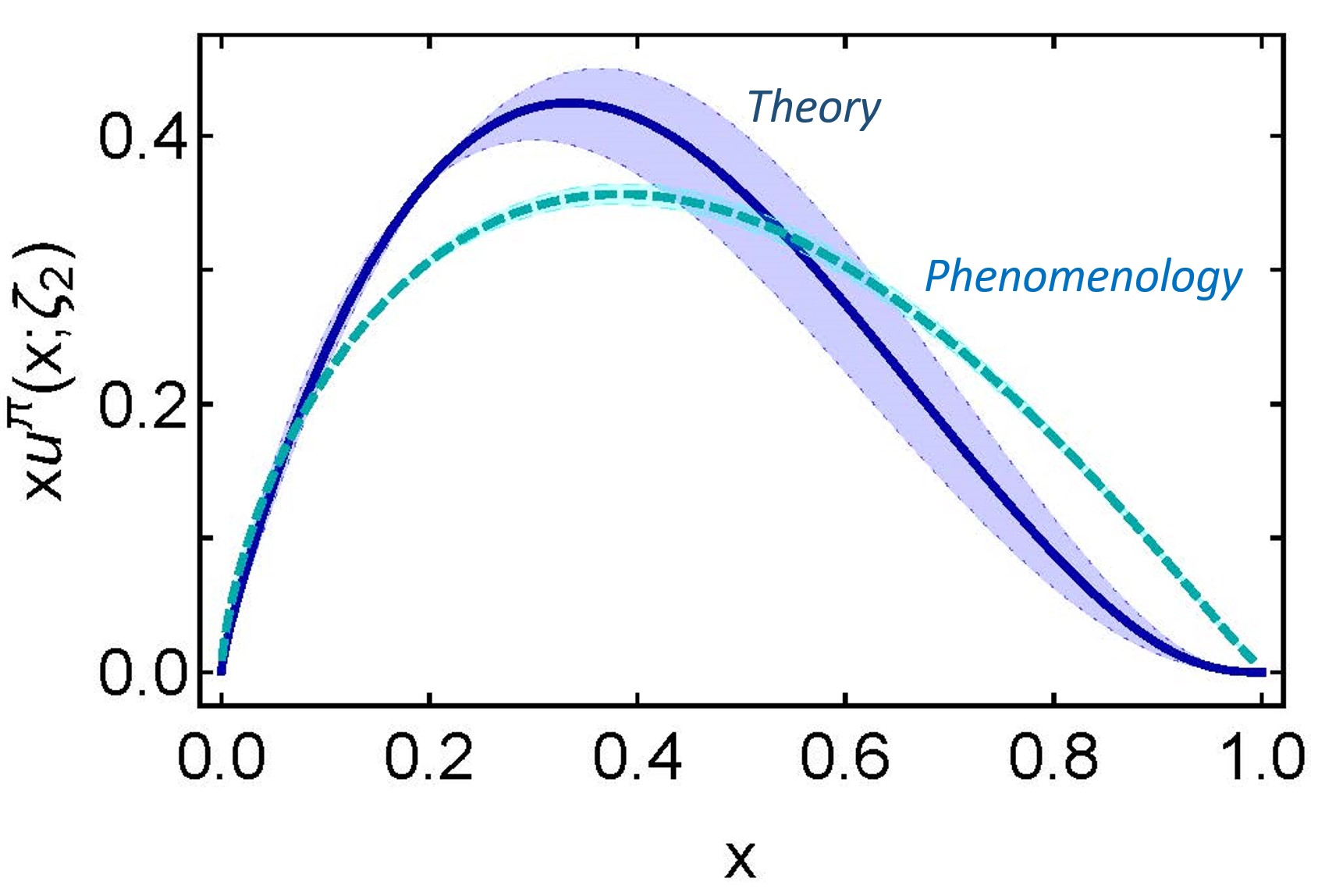} & \hspace*{1em} &
\includegraphics[clip, width=0.46\textwidth]{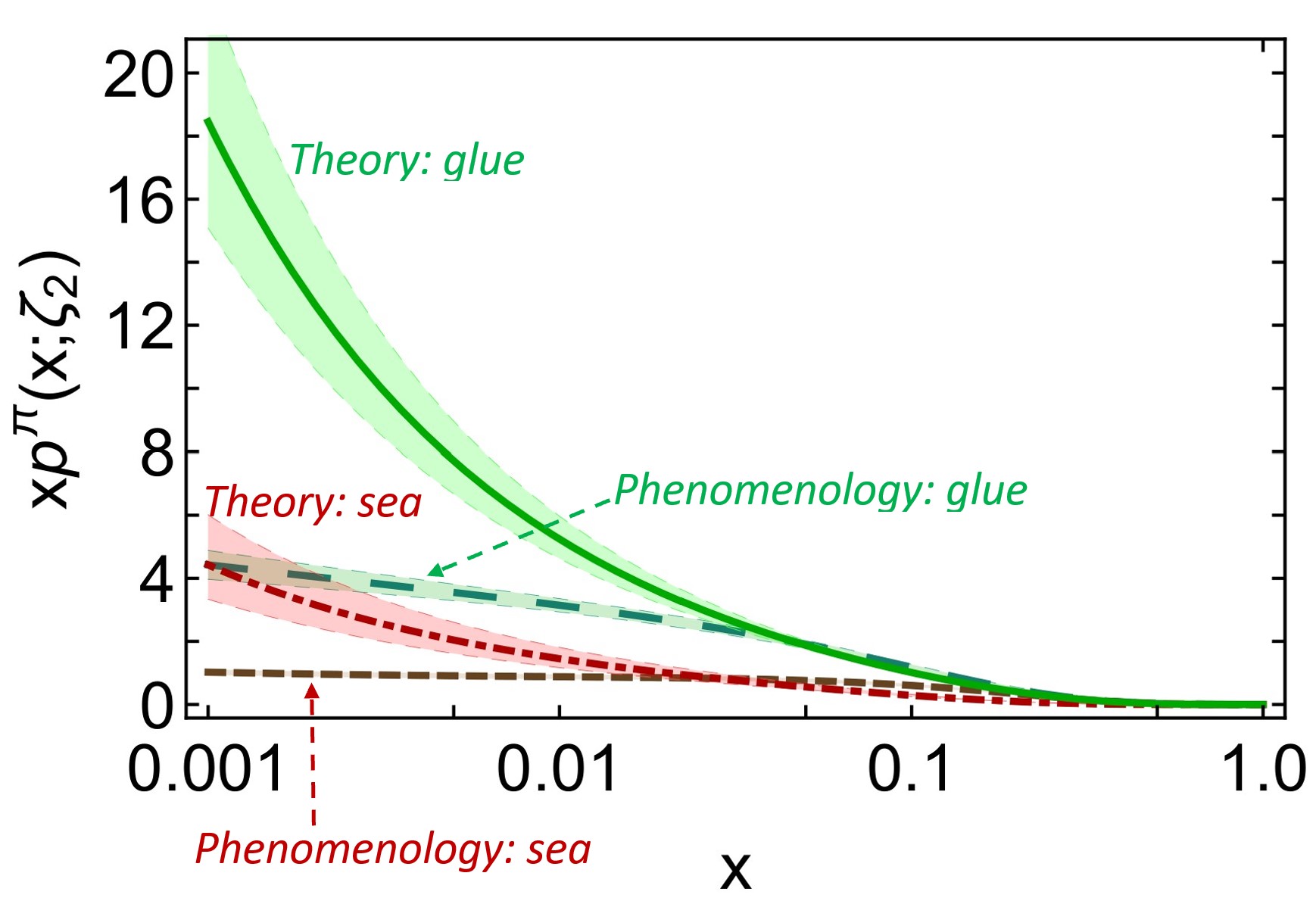}
\end{tabular}
\caption{\label{figF12}
\emph{Left panel}.
Blue solid curve -- theory prediction for ${\mathpzc u}^\pi(x;\zeta_2=2\,{\rm GeV})$ \cite{Ding:2019qlr, Ding:2019lwe};
and cyan short-dashed curve  -- phenomenological result from Ref.\,\cite{Barry:2018ort}.
\emph{Right panel}. Theory predictions for the pion's glue and sea-quark distributions \cite{Ding:2019qlr, Ding:2019lwe}: green solid curve -- ${\mathpzc g}^\pi(x;\zeta_2)$; and red dot-dashed, ${\mathpzc S}^\pi(x;\zeta_2)$.  The associated momentum fractions are:
$\langle 2 x {\mathpzc q}^\pi(x;\zeta_2) \rangle = 0.48(4)$,
$\langle x {\mathpzc g}^\pi(x;\zeta_2) \rangle = 0.41(2)$,
$\langle x {\mathpzc S}^\pi(x;\zeta_2) \rangle = 0.11(2)$.
For comparison, phenomenological results from Ref.\,\cite{Barry:2018ort}:
$p=\,$glue -- dark-green long-dashed; and $p=\,$sea -- brown dashed.
(The bands around the theory curves express the impact of the uncertainty in $\hat\alpha(k^2=0)$, Fig.\,\ref{FigwidehatalphaII}.  Normalisation: $\langle x[2 {\mathpzc u}^\pi(x)+g^\pi(x)+S^\pi(x]\rangle=1$.)
}
\end{figure}

Pressure is also being applied by modern advances in theory.  Lattice QCD is beginning to yield results for the pointwise behaviour of the pion's valence-quark distribution \cite{Chen:2018fwa, Oehm:2018jvm, Karthik:2018wmj, Sufian:2019bol}.
Furthermore, continuum analyses \cite{Ding:2019qlr, Ding:2019lwe} have yielded the first parameter-free predictions for the valence, glue and sea distributions within the pion; and revealed that, like the pion's leading-twist parton distribution amplitude (PDA) \cite{Chang:2013pq}, the valence-quark distribution function is hardened as a direct consequence of emergent mass. 


The valence, glue and sea distributions from Refs.\,\cite{Ding:2019qlr, Ding:2019lwe} are drawn in Fig.\,\ref{figF12}.  Also shown are the phenomenological extractions from Ref.\,\cite{Barry:2018ort}.  Even though the valence distribution fitted in Ref.\,\cite{Barry:2018ort} yields a momentum fraction compatible with the theory prediction, its $x$-profile is very different.  
(Ref.\,\cite{Barry:2018ort} did not include the higher-order corrections discussed in Ref.\,\cite{Aicher:2010cb}.)
It is significant, therefore, that the continuum result for ${\mathpzc u}^\pi(x;\zeta_5)$ matches that obtained using lattice QCD \cite{Sufian:2019bol}.
Consequently, the Standard Model prediction: ${\mathpzc u}^{\pi}(x;\zeta =m_0) \sim (1-x)^{2}$, is stronger than ever before;
and as demonstrated by Fig.\,\ref{fig:NeutTagged-DIS-Domain-A}, EicC design specifications would enable it to deliver clear experimental validation.

If the pion's valence-quark distributions are controversial, then its glue and sea distributions can be described as uncertain or worse.  Fig.\,\ref{figF12}\,--\,right panel compares the predictions from Refs.\,\cite{Ding:2019qlr, Ding:2019lwe} with fits from a combined analysis of $\pi$-nucleon Drell-Yan and HERA leading-neutron electroproduction data \cite{Barry:2018ort}.
The gluon distribution predicted in Refs.\,\cite{Ding:2019qlr, Ding:2019lwe} and that fitted in Ref.\,\cite{Barry:2018ort} are markedly different on $x\lesssim 0.05$; and both glue DFs in Fig.\,\ref{figF12} disagree with those inferred previously \cite{Gluck:1999xe, Sutton:1991ay}.  These remarks highlight that the pion's gluon content is empirically uncertain.  Thus, new measurements which are directly sensitive to the pion's gluon content are necessary.  Prompt photon and $J/\psi$ production could address this need \cite{Denisov:2018unj, Chang:2020rdy}.
The sea DFs in Fig.\,\ref{figF12} have different profiles on the entire $x$-domain.  Hence, it is fair to describe the sea-quark distribution as empirically unknown.  This problem could be addressed with DY data obtained from $\pi^\pm$ beams on isoscalar targets \cite{Londergan:1995wp, Denisov:2018unj}.  Equivalent tagged DIS measurements at the EicC could also provide this information.
%
Evidently, precision measurements that are sensitive to meson glue and sea-quark distributions are highly desirable.  Here, too, EicC can have a significant impact.

\medskip

\begin{figure}[htbp]
\begin{tabular}{ccc}
\includegraphics[clip, width=0.46\textwidth]{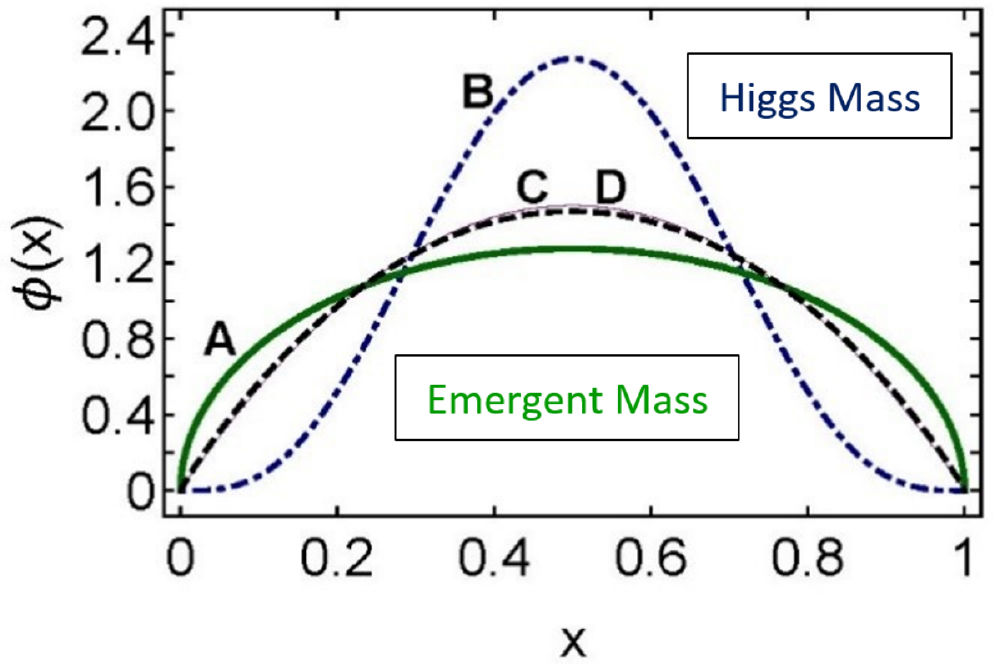}
&  &
\includegraphics[clip, width=0.46\textwidth]{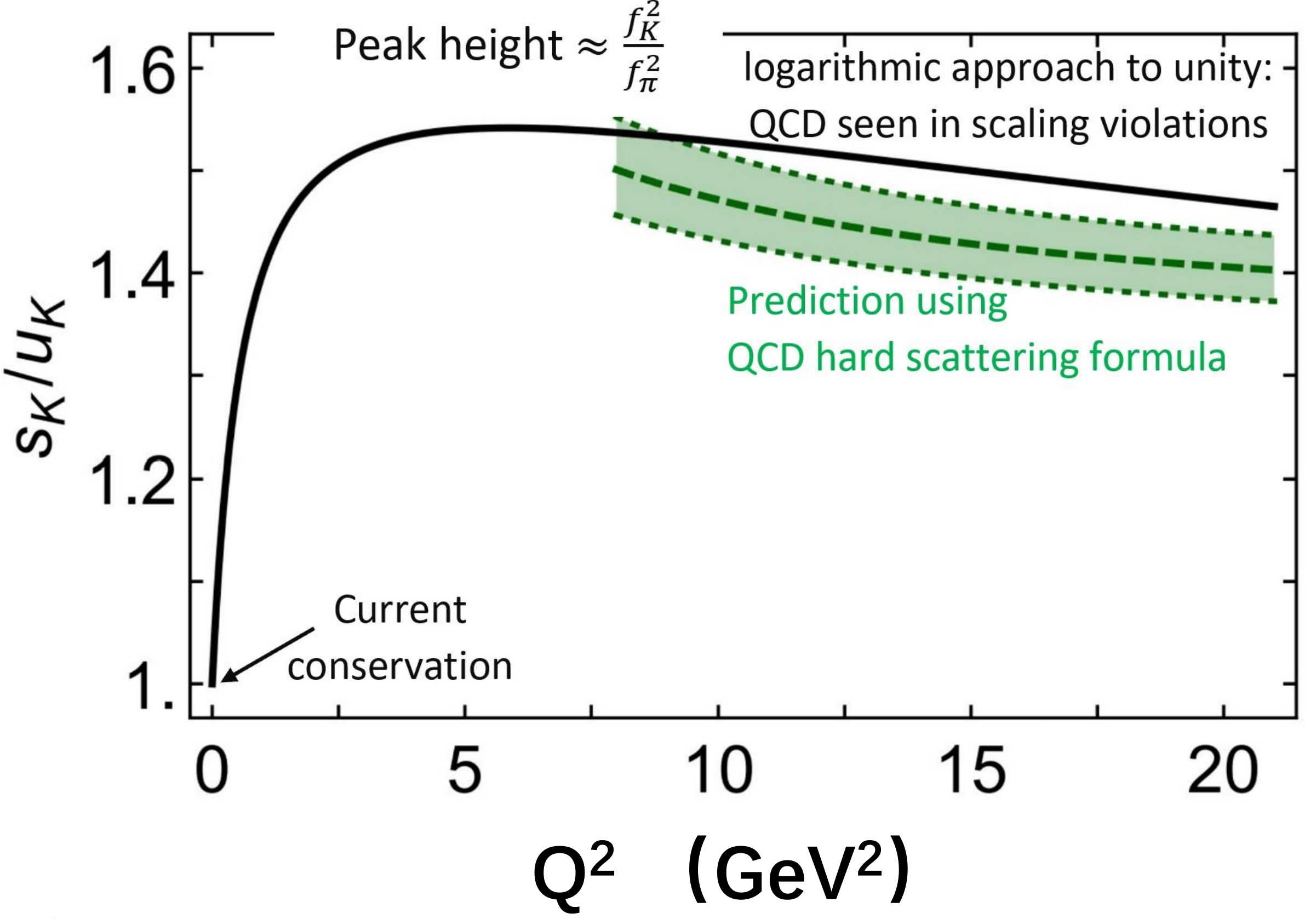}
\end{tabular}

\caption{\label{figsquark}
\emph{Left panel} -- Twist-two parton distribution amplitudes at $\zeta=2 \,$GeV$=:\zeta_2$. \textbf{A} solid (green) curve – pion $\Leftarrow$ emergent mass generation is dominant; \textbf{B} dot-dashed (blue) curve – $\eta_c$ meson $\Leftarrow$ Higgs mechanism is the primary source of mass generation;  \textbf{C} solid (thin, purple) curve -- asymptotic profile, $\varphi_{\rm as}(x)$; and \textbf{D} dashed (black) curve – ``heavy-pion'', \emph{i.e}.\ a pion-like pseudo-scalar meson in which the valence-quark current masses take values corresponding to a strange quark $\Leftarrow$ the boundary, where emergent and Higgs-driven mass generation are equally important.
\emph{Right Panel} -- Solid curve: ratio of strange-to-normal matter in the $K^+$; green dashed curve and band: results obtained using the QCD hard-scattering formula \cite{Lepage:1979zb, Efremov:1979qk, Lepage:1980fj}.
}
\end{figure}

\hspace*{-\parindent}{\emph{Balancing Emergent Mass and the Higgs Mechanism}.}\\
The ability of form factor and structure function measurements to expose emergent hadronic mass 
is enhanced if one includes similar kaon data because theory has revealed that $s$-quark physics lies at the boundary between dominance of strong (emergent) mass generation and weak (Higgs-connected) mass \cite{Roberts:2019ngp}, as highlighted in Fig.\,\ref{figsquark}-left panel.  Hence, comparisons between distributions within systems constituted solely from light valence quarks and those associated with systems containing $s$-quarks are ideally suited to exposing measurable signals of emergent mass 
in counterpoint to Higgs-driven effects.

One example can be found in the contrast between the parton distribution functions of the pion and kaon, first theory predictions for which are now available \cite{Cui:2020dlm, Cui:2020piK}.  Another is illustrated in Fig.\,\ref{figsquark}\,--\,right panel, which displays a flavor-separation of the charged-kaon, $K^+$, elastic electromagnetic form factor: $s_K:=F_K^{\bar s}$ is the $s$-quark contribution to the form factor and $u_K:=F_K^u$ is the analogous $u$-quark term.
The rate of growth from $Q^2=0$, the peak height and location, and the logarithmic decay away from the peak are all expressions of emergent mass 
and its modulation by Higgs-boson effects \cite{Chen:2020ijn}.
%
%
%
%
%
%

Analogous predictions exist or are being completed for nucleon form factors, following the approach in Ref.\,\cite{Wang:2018kto}, and parton distribution functions \cite{Roberts:2013mja}.  Simulations indicate that the planned EicC could be used to validate the connection between emergent mass 
and these observables; and also many others that are members of the same class.

\subsubsection{Meson fragmentation functions}

Along with the promised rewards described in Sec.\,\ref{ThreeDnucleons}, many new challenges are faced in extracting 3-D images from new-generation experiments.  Phenomenological models of a wide variety of parton distribution functions will be crucial.  They can provide guidance on the size of the cross-sections to be measured and the best means by which to analyse them \cite{Berthou:2015oaw}.  On the other hand, as experiences with meson structure functions have shown, in order to profit fully from such experiments, one must use computational frameworks that can reliably connect measurements with qualities of QCD.  Here, continuum calculations can provide valuable insights \cite{Mezrag:2014jka, Mezrag:2016hnp, Chouika:2017rzs, Xu:2018eii, Shi:2018zqd}.

Regarding TMDs, there is an additional complication.  Namely, every cross-section that can yield a given hadronic TMD involves a related parton fragmentation function (PFF) \cite{Field:1977fa}, the structure of which must be known in detail.  Therefore, the future of momentum imaging depends critically on making significant progress with the measurement and calculation of PFFs.  Notwithstanding these demands, there are currently no realistic computations of PFFs.  Even a formulation of the problem remains uncertain.  Here are both a challenge and an opportunity for EicC.

If EicC can provide precise data on quark fragmentation into a pion or kaon, it will deliver results that directly test those aspects of QCD calculations which incorporate and express emergent phenomena, \emph{e.g}.\ confinement, DCSB, and bound-state formation.  The fact that only bound-states emerge from such processes is one of the cleanest available manifestations of confinement.  As a collider, EicC measurements will potentially have enormous advantages over earlier and existing fixed-target experiments.  By capitalising on energy range, versatility, and detection capabilities in the collider environment, EicC should be able to first cleanly single out the pion or kaon target and subsequently the fragmentation process tag, thereby delivering an array of information that will best test theory aimed at the calculation and interpretation of PFFs. 


\chapter{Accelerator conceptual design}
To achieve the proposed scientific goals in chapter 2, the EicC project will construct a high performance polarized electron-proton collider which can reach the luminosity of $2.0 \times 10^{33}$ $\mathrm{cm^{-2} s^{-1}}$ at the center-of-mass energy of 16.76 GeV while ensuring an average polarization for electrons and protons of about $80\%$ and $70\%$, respectively. For convenience, unless 
stated, the following polarization all represents the average polarization. Moreover, the center-of-mass energy of the EicC has a flexible range from 15 GeV to 20 GeV in order to serve different experimental purposes. To keep a balance between the physics goals and the overall cost, we will take full advantage of the existing HIAF accelerator complex and its ancillary facilities. Therefore, only a new figure-8 ion collider ring, a polarized electron injector, and a racetrack electron collider ring will be constructed. This chapter is organized as follows:
In Section~\ref{sec:3.1}, we present a comprehensive overview of the EicC accelerator facility, including its design goals, main layout, key accelerator parameters, and potential technical challenges. 
Section~\ref{sec:3.2} will mainly discuss the basic operation modes of the ion accelerator complex and the electron accelerator complex of the EicC. The following section~\ref{sec:3.3} to section~\ref{sec:3.5} will provide a detailed description of three key conceptual design ingredients for the EicC accelerator facility, i.e. beam cooling, beam polarization, and interaction regions (IRs). Last but not the least, several topics of the technical pre-research in the accelerators will be presented in Section~\ref{sec:3.6}.

\section{\label{sec:3.1}Overall design and key parameters}

The various scientific goals pose the various requirements that must be met in the EicC accelerator facility. Taking the proton beam as the reference beam (also for the following sections), these requirements are explained 
as follows:

\begin{enumerate}

\item The center-of-mass energy should range from 15 GeV to 20 GeV, which corresponds to the energy of 2.8 GeV to 5.0 GeV for electron beams in the electron collider ring and the energy of 19.08 GeV for proton beams in the ion collider ring. The center-of-mass energy of other ion beams can be obtained according to the maximum momentum rigidity which is 86 $\mathrm{T\cdot m}$.

\item The luminosity should reach up to  $2.0 \times 10^{33}$  $\mathrm{cm}^{-2} \mathrm{s}^{-1}$ and should be optimized mainly at the center-of-mass energy of 16.76 GeV.

\item At the interaction point (IP), the electron beams should be polarized longitudinally with the  polarization of $80\%$, while the proton beams should be polarized both longitudinally and transversely with the polarization of $70\%$. Besides the polarized proton beams, the ion accelerator complex should also provide polarized deuterium ($^2 \mathrm{D}^{+}$) beams, polarized Helium-3 ($^3\mathrm{He}^{2+}$) beams as well as unpolarized heavy ion beams.

\item The design of the IRs should fully accommodate the design of the EicC detectors of which a large number of components, such as the polarimetry, luminosity monitors, forward detectors, and so on, will be integrated into the beamlines. 

\end{enumerate}

\begin{sidewaysfigure}
\centering
\includegraphics[width=25cm]{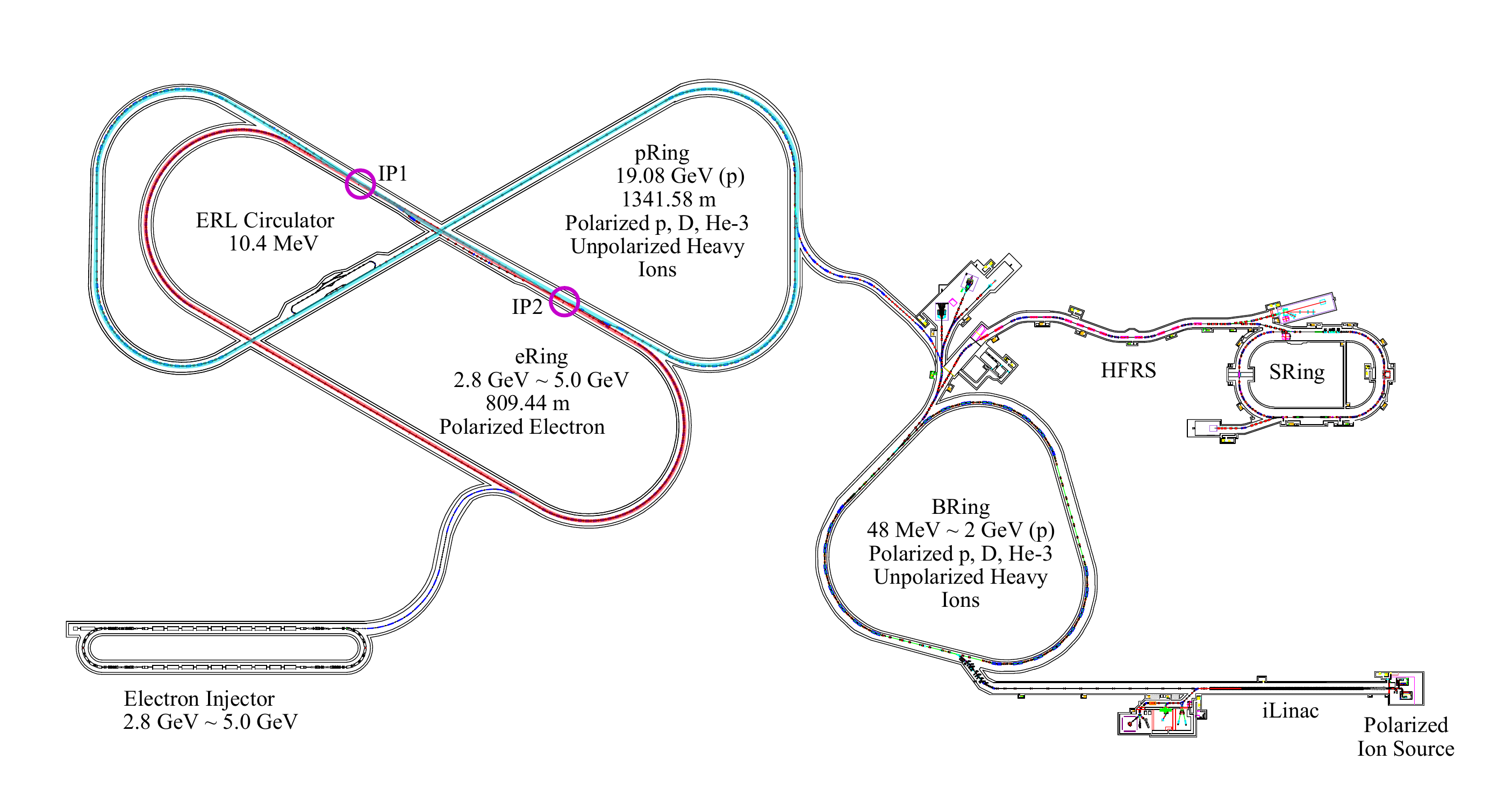}
\caption{\label{fig:accelerator:3.1} The layout of EicC accelerator facility.}
\end{sidewaysfigure}

As shown in the Fig.\ref{fig:accelerator:3.1}, the EicC was carefully designed to meet all requirements listed above. 
The ion accelerator complex will fully utilize the existing HIAF complex which helps us to largely reduce the construction cost. 
On top of that, a new ion collider ring (pRing) will be constructed. 
The high-intensity heavy-ion booster ring (BRing) in the HIAF can produce proton beams of energy up to 9.3 GeV, which covers the injection energy of 2 GeV in the pRing. Furthermore, in the BRing, there will be sufficient free space reserved for the installation of a DC electron cooler for beam cooling and the Siberian snake for spin preservation. Therefore, the BRing in the HIAF is planned to be upgraded and fully employed as the booster in the EicC ion accelerator complex. To avoid the depolarization resonances during the acceleration process, a figure-8 ring will be adopted for the pRing. This design ensures the spin tune of the pRing is always zero and makes it possible to achieve the high polarization of proton beams during the acceleration process with a large energy range. Meanwhile, it also allows us to control the polarization direction of proton beams more efficiently. The figure-8 ring design is one of the essential technical solutions to achieve the high polarization in the EicC ion accelerator complex. Besides the polarization design, it is also important to achieve high luminosity in the EicC. To this end, a multistage beam cooling scheme is proposed. Firstly, the pre-cooling of proton beams will be performed in the BRing. Secondly, a high energy bunched beam electron cooler based on an energy recovery linac (ERL) will be installed in the pRing and employed to cool the proton beams after they are accelerated to the energy of 19.08 GeV. The beam cooling will be employed to suppress the intra-beam scattering effect over the whole collision process, which is the key to achieve the required luminosity.

The EicC electron accelerator complex mainly consists of an electron injector and an electron collider ring (eRing). The electron injector is a superconducting radio frequency (SRF) linac which is considered to be the best choice to deliver the electron beams with the energy range from 2.8 GeV to 5.0 GeV. Since the eRing adopts the full energy injection scheme which means there is no acceleration process, and the electron beams will experience the self-polarization process, the eRing is designed as a racetrack ring which is widely used in the electron facilities. Meanwhile, the polarization direction of the electron beams produced by the electron injector should be matched with the eRing.

Based on the layout of the EicC accelerator facility, two IRs are available at two intersections of the long straight sections of the pRing and the eRing, as shown in the Fig.~\ref{fig:accelerator:3.1}. 
These two identical IRs will allow us to install two independent detector systems for different scientific goals simultaneously or the same physics goal but with different experimental techniques for cross-validation. 
However, the current design only considers one detector to be installed in one of these IRs, with the other one reserved for the future upgrade

The luminosity is one of the most important characteristics of a collider. 
Different from other EIC designs in the world which commonly choose high collision frequency and low bunch intensity, 
the EicC accelerator facility will adopt the unique scheme of low collision frequency and high bunch intensity to achieve a fourfold increase of luminosity to reach the required luminosity of $2 \times 10^{33}$ $\mathrm{cm}^{-2} \mathrm{s}^{-1}$, satisfying the physics goals of the EicC project. This is an essential part of the luminosity design of the EicC. For the collision between a proton with constant current $I_p$ and an electron beam with constant current $I_e$, the luminosity can be expressed as

\begin{gather}
\label{eq:3.1}N_p f_c \propto I_p \equiv \mathrm{const},\\
\label{eq:3.2}N_e f_c \propto I_e \equiv \mathrm{const},\\
\label{eq:3.3}L = \frac{N_p N_e f_c}{4 \pi \sigma_x \sigma_y } H \propto \frac{I_p I_e}{\sigma_x \sigma_y f_c}.
\end{gather}
Here $N_p$ and $N_e$ are the numbers of particles in one proton bunch and electron bunch, respectively. $f_c$ is the collision frequency. $\sigma_x$ and $\sigma_y$ are the transverse beam size in horizontal and vertical directions, respectively. $H$ is the hourglass factor, which remains constant when the ratio of the $\beta$ function to the bunch length is fixed. Although there is a positive correlation between $\sigma_x$ or $\sigma_y$ and the number of particles in the bunch, it can be proven that the influence of $\sigma_x$ and $\sigma_y$ on the luminosity is less than that of $f_c$. Therefore, a decrease of $f_c$ can increase the luminosity. In fact, constant average beam current and constant intra-beam scattering in the bunch is a major factor of collision lifetime, i.e.
\begin{equation}
    \frac{1}{\tau} \propto \frac{N_p}{\varepsilon_x \varepsilon_y \sigma_{\delta} \sigma_s},
\end{equation}
from which it can be derived that if the bunch intensity is increased by $m$ times, the increase of the horizontal emittance $\varepsilon_x$, the vertical emittance $\varepsilon_y$, the longitudinal bunch length $\sigma_{\delta}$ and the momentum spread $\sigma_s$ reads
\begin{equation}\label{eq:3.4}
    \varepsilon_x \propto m^{\frac{1}{3}},\ \varepsilon_y \propto m^{\frac{1}{3}},\ \sigma_{\delta} \propto m^{\frac{1}{6}},\ \sigma_s \propto \beta_{x,y}^{*} \propto m^{\frac{1}{6}}.
\end{equation}
Substituting the Eq.\ref{eq:3.4} to the Eq.\ref{eq:3.3}, the relationship between the luminosity and the beam intensity is
\begin{equation}\label{eq:3.5}
    L \propto \sqrt{m},
\end{equation}
indicating that the luminosity increases by $\sqrt{m}$ times while the bunch intensity increases by $m$ times.

Based on the optimized luminosity scheme introduced above, the main parameters of the EicC accelerator facility are generated, as listed in the Tab.\ref{tab:accelerator:3.2} in which proton beams are used as the reference beam and the optimization is performed at the center-of-mass energy of 16.76 GeV. Furthermore, several technical limitations are considered in the Tab.\ref{tab:accelerator:3.2}, as is shown below.
\begin{enumerate}
  \item The power density of the synchrotron radiation in the eRing should be less than $20$ $ \mathrm{kW/m}$.
  \item The beam-beam interaction parameter of proton beams is less than 0.03.
  \item The beam-beam interaction parameter of electron beams is less than 0.1.
\end{enumerate}

To meet the scientific goals, the parameters of collisions between electron beams and several heavy ion beams are also shown in the Tab.\ref{tab:accelerator:3.3}. In the following sections of this chapter, the design and the implementation of each parameter in the Tab.\ref{tab:accelerator:3.2} will be discussed in detail.

\begin{table}
    \caption{\label{tab:accelerator:3.2} Main parameters for the EicC.}
       \hspace{0.2cm}
        \centering
        \begin{tabular}{lcc}
          \toprule
          \specialrule{0em}{3pt}{3pt}
          Particle &
          $\rm{e}$ &
          $\rm{p}$ \\
          \specialrule{0em}{3pt}{3pt}

          \midrule
          Circumference(m) &
          809.44 &
          1341.58\\
          
          Kinetic energy(GeV) &
          3.5 &
          19.08\\

          Momentum(GeV/c) &
          3.5 &
          20\\

          Total energy(GeV) &
          3.5&
          20.02\\

          CM energy(GeV) &
          \multicolumn{2}{c}{16.76}
          \\
          
          $\rm{f_{collision}(MHz)}$&
          \multicolumn{2}{c}{100}\\

          Polarization &
          80\%&
          70\%\\

          $\rm{B\rho(T\cdot m)}$ &
          11.7 &
          67.2\\

          Particles per bunch$(\times \rm{10^9})$ &
          170&
          125\\

          $\rm{\varepsilon_x, \varepsilon_y(nm\cdot rad, rms)}$ &
          60/60&
          300/180\\

          $\rm{\beta^*_x/\beta^*_y (m) }$&
          0.2/0.06 &
          0.04/0.02\\

          Bunch length(m, rms) &
          0.02&
          0.04\\

          Beam-Beam Parameter $\rm{\xi_y}$ &
          0.09/0.05&
          0.004/0.004\\

          Laslett tune shift &
          -&
          0.09\\

          Energy loss per turn(MeV)&
          0.32&
          -\\

          Total SR power(MW)&
          0.86&
          -\\

          SR linear power density(kW/m)&
          3.3&
          -\\

          Current(A) &
          2.7&
          2\\

          Crossing angle(mrad) &
          \multicolumn{2}{c}{50}
          \\

          Hourglass &
          \multicolumn{2}{c}{0.78}
          \\

          Luminosity at nucleon level ($\rm{cm^{-2}s^{-1}}$)  &
          \multicolumn{2}{c}{$\rm{2.0\times 10^{33}}$}
          \\

          \bottomrule
          \end{tabular}
\end{table}

\begin{sidewaystable}
\caption{\label{tab:accelerator:3.3} Main parameters for the EicC.}
   \hspace{0.2cm}
    \centering
    \resizebox{\textwidth}{!}{
    \begin{tabular}{lccccccccc}
      \toprule
      \specialrule{0em}{3pt}{3pt}
      Particle &
        $\rm{e}$ &
        $\rm{d}$ &
        $\rm{^{3}He^{++}}$ &
        $\rm{^{7}Li^{3+}}$ &
        $\rm{^{12}C^{6+}}$ &
        $\rm{^{40}Ca^{20+}}$ &
        $\rm{^{197}Au^{79+}}$ &
        $\rm{^{208}Pb^{82+}}$ &
        $\rm{^{238}U^{92+}}$ \\
      \specialrule{0em}{3pt}{3pt}
      \midrule
      Kinetic energy(GeV/u) &
        3.5 &
        12.00 &
        16.30 &
        10.16 &
        12.00 &
        12.00 &
        9.46 &
        9.28 &
        9.09    \\
      Momentum(GeV/c/u) &
        3.5 &
        12.90 &
        17.21 &
        11.05 &
        12.90 &
        12.90 &
        10.35 &
        10.17 &
        9.98  \\
      Total energy(GeV/u) &
        3.5&
        12.93 &
        17.23 &
        11.09 &
        12.93 &
        12.93 &
        10.39 &
        10.21 &
        10.02 \\
      CM energy(GeV/u) &
         -&
         13.48&
         15.55&
         12.48&
         13.48&
         13.48&
         12.09&
         11.98&
         11.87\\
        $\rm{f_{collision}(MHz)}$&
        -&
        499.25&
        499.82&
        498.79&
        499.25&
        499.25&
        498.54&
        498.47&
        498.39\\
        
      Polarization &
         80\%&
         Yes&
         Yes&
         No&
         No&
         No&
         No&
         No&
         No\\
      $\rm{B\rho(T\cdot m)}$ &
        11.67&
        86.00&
        86.00&
        86.00&
        86.00&
        86.00&
        86.00&
        86.00&
        86.00\\
      Particles per bunch$(\times \rm{10^9})$ &
        40&
        6.1&
        3.0&
        2.04&
        1.00&
        0.30&
        0.07&
        0.065&
        0.055\\
      $\rm{\varepsilon_x, \varepsilon_y(nm\cdot rad, rms)}$ &
        20&
        100/60&
        100/60&
        100/60&
        100/60&
        100/60&
        100/60&
        100/60&
        100/60\\
      $\rm{\beta^*_x/\beta^*_y (m) }$&
        0.2/0.06 &
        0.04/0.02&
        0.04/0.02&
        0.04/0.02&
        0.04/0.02&
        0.04/0.02&
        0.04/0.02&
        0.04/0.02&
        0.04/0.02\\
      Bunch length(m, rms) &
        0.01&
        0.015&
        0.015&
        0.02&
        0.015&
        0.015&
        0.02&
        0.02&
        0.02\\
      Beam-Beam Parameter $\rm{\xi_y}$ &
        0.007 &
        0.002&
        0.002&
        0.002&
        0.002&
        0.002&
        0.002&
        0.002&
        0.002\\
      Laslett tune shift &
        -&
        0.07&
        0.06&
        0.04&
        0.06&
        0.06&
        0.06&
        0.06&
        0.06\\
      Current(A) &
        3.3&
        0.49&
        0.48&
        0.49&
        0.48&
        0.48&
        0.44&
        0.43&
        0.40\\
      Crossing angle(mrad) &
        \multicolumn{9}{c}{50}
        \\
      Hourglass &
        -&
        0.94&
        0.94&
        0.92&
        0.94&
        0.94&
        0.92&
        0.92&
        0.92\\
      Luminosity at nucleon level ($\rm{cm^{-2}s^{-1}}$)  &
        -&
        $\rm{8.48\times 10^{32}}$&
        $\rm{6.29\times 10^{32}}$&
        $\rm{9.75\times 10^{32}}$&
        $\rm{8.35\times 10^{32}}$&
        $\rm{8.35\times 10^{32}}$&
        $\rm{9.37\times 10^{32}}$&
        $\rm{9.22\times 10^{32}}$&
        $\rm{8.92\times 10^{32}}$\\
      
      \bottomrule
      \end{tabular}
      }%
\end{sidewaystable}

\section{Accelerator facilities}\label{sec:3.2}

\begin{figure}
\centering
\includegraphics[width=15cm]{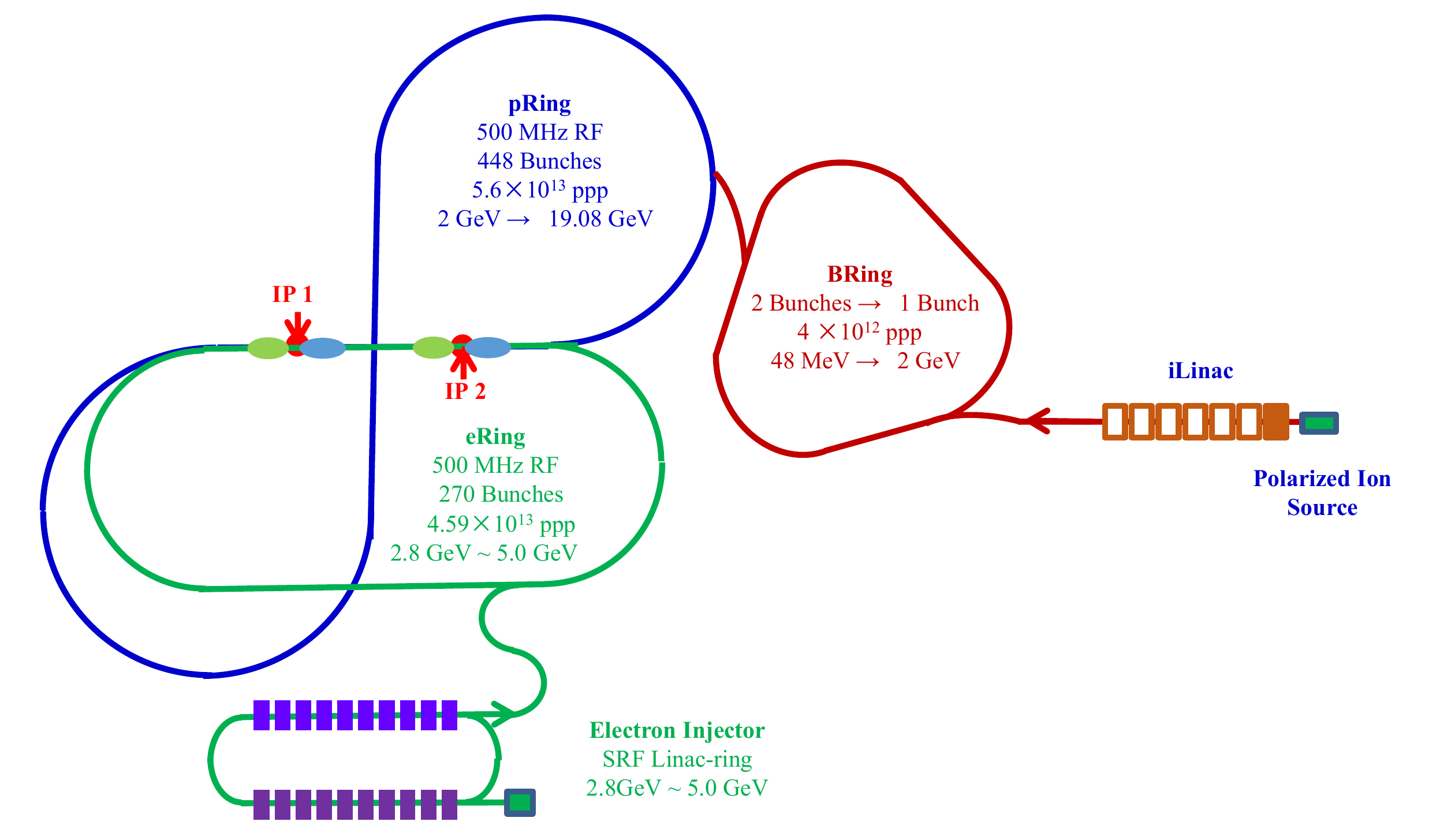}
\caption{\label{fig:accelerator:3.2} The beam path of the EicC accelerator facility.}
\end{figure}

The EicC mainly consists of an ion accelerator complex and an electron accelerator complex, as shown in the Fig.\ref{fig:accelerator:3.1}. The Fig.\ref{fig:accelerator:3.2} illustrates the operation mode of the two accelerator complexes. There are a lot of differences between the ion accelerator complex and the electron accelerator complex, not only on the complicated beam manipulations but also on the key design and technical challenges, all of which will be described in detail in the following subsections. In Section~\ref{sec:3.2.1}, the ion accelerator complex of the EicC will be described, mainly including the acceleration scheme of the pRing, the collective effects, and so on. The details of the electron accelerator complex, including some key designs and beam manipulations of the eRing, will be presented in Section~\ref{sec:3.2.2}. The topic of beam cooling and beam polarization will be discussed in Section~\ref{sec:3.3} and Section~\ref{sec:3.4}, respectively. The design of the IRs, which is one of the most important accelerator designs in the EicC accelerator facility, will be presented in Section~\ref{sec:3.5}.

\subsection{Ion accelerator complex\label{sec:3.2.1}}

In the proposed ion accelerator complex of the EicC accelerator facility, the existing Ion Linac (iLinac) and the BRing of the HIAF will be employed respectively as the ion injector and the booster to provide injection beams for the pRing. The iLinac will be operated in pulsed mode, in which the polarized proton beams can be accelerated to the energy of 48 MeV, and then injected into the BRing with a matched polarization direction. In the BRing, a two-plane painting injection scheme will be adopted for the beam accumulation with coasting beams to increase the beam intensity. In this scheme, a tilted electrostatic septum will be employed for painting simultaneously in the horizontal and vertical directions during the injection. As a result, the beam intensity will be increased by 100 times and as high as $4 \times 10^{12}$ $ \mathrm{ppp}$. The proton beams of the energy of 48 MeV in the BRing will be captured into two bunches and accelerated to the energy of 2 GeV. During the acceleration, the Siberian snakes will be used to maintain the polarization of the proton beams\cite{alekseev2003polarized}, since the beams will cross several depolarization resonances.

The proton beams of the energy of 2 GeV which reach the extraction energy of the BRing in the ion accelerator complex will experience the first stage of beam cooling provided by a DC electron cooler. There are several advantages of performing beam cooling at the energy of 2 GeV. Firstly, the beam cooling effect is still strong around such energy while the space charge effect is relatively weak. Secondly, as a well-developed technology widely used for the beam cooling, the DC electron cooler can effectively reduce the technical difficulties and the construction costs. At this stage, to obtain a more efficient beam cooling, a longitudinal bunch rotation at the energy of 2 GeV will be performed before the beam cooling to minimize the bunch momentum spread as much as possible. After the bunch rotation, the bunched proton beams will be debunched into coasting beams to improve the beam cooling efficiency. After the beam cooling, the coasting proton beams in the BRing will be bunched once again into one bunch and compressed to match the injection settings of the pRing. The polarization direction of the proton beams extracted will get matched to the polarization direction of the pRing by using a spin rotator.

The pRing will accelerate the injected proton beams of the energy of 2 GeV to reach the energy of 19.08 GeV. For such a wide energy range, the Siberian snakes are unable to keep the polarization of the proton beams. Therefore, the pRing will be designed as a figure-8 shape, in which the spin tune purely contributed by the pRing itself remains zero. The figure-8 design, along with a solenoid in the long straight sections, ensures that the proton beams will not cross any depolarization resonances during acceleration. Moreover, such a design will facilitate the control of the beam polarization direction. The bucket-to-bucket injection scheme will be adopted for the injection of 14 bunches from the booster BRing to the pRing. The beam intensity will increase to $\rm{5.6 \times 10^{13}}$ $\mathrm{ppp}$, equivalent to an average beam current of 2 A. After the injection, the beams will be split into 448 bunches step by step and finally accelerated to the energy of 19.08 GeV. The collision frequency introduced by these 448 bunches is 100 MHz, which meets the requirements of the luminosity design.

To increase the luminosity, the pRing will install a 500 MHz RF system to shorten the bunch length. After the proton bunches are accelerated to the energy of 19.08 GeV, a bunch rotation will be performed. And during the bunch rotation, the bunch length will be sharply shortened while the momentum spread will be increased, which is quite different from the bunch rotation in the BRing. Along with the bunch rotation, the 500 MHz RF system will be turned on to make the bunch length as short as possible while keeping particle numbers in the bunch unchanged, so that the requirements of the luminosity design will be satisfied.

The longitudinal manipulation of the proton beams is followed by the second stage of beam cooling which is supported by high energy bunched beam electron cooler based on an energy-recovery linac (ERL). The intra-beam scattering effect will be also suppressed during the whole collision process by this cooler to improve the luminosity life.

The luminosity optimization scheme of low collision frequency, high beam intensity could bring strong single-bunch collective effects. Therefore, besides the threshold values of the average current, and beam-beam parameters, the impedance limitations introduced by the single-bunch collective effects are also significant in the collider rings since it determines the feasibility of luminosity optimization scheme in the pRing. In principle, the longitudinal microwave instability, which is caused by the longitudinal broadband impedance, is one of the most concerned longitudinal collective effects. When the particle numbers in the bunch exceed the corresponding limits, the longitudinal microwave instability will lead to an increase in the momentum spread, which can further induce bunch lengthening and weaken the luminosity. The typical growth time of the instability is shorter than one synchrotron period. In the pRing, it has been calculated that the longitudinal broadband impedance ($Z^{\parallel}/n$) should be lower than 87.4 $\Omega$, which is achievable in practice. In the transverse planes, the transverse mode coupling instability caused by the transverse broadband impedance is the most likely to occur among the single-bunch collective effects. When the bunch intensity exceeds the threshold, particles in the bunch will be lost quickly. Calculation results have shown that the transverse broadband impedance threshold in the pRing is 30.3 $\mathrm{M\Omega/m} $, which is also achievable in practice. From the view of the pRing, the luminosity optimization scheme with low collision frequency, the high beam intensity is feasible for the EicC accelerator facility.

One of the biggest challenges in the optical design of colliders is the correction of the large chromaticity introduced by the very small $\beta$ function at the IPs, with the requirement of large dynamic aperture. As shown in the  Fig.\ref{fig:accelerator:3.1}, there are four arc sections in the pRing. These arc sections are connected via two short straight sections and two long straight sections. The long straight sections are employed for the IPs, the beam cooling sections, the spin rotators as well as the RF devices. Accelerator control devices, such as the chromaticity correction sextupole magnets and so on, are placed on the short straight sections and the arc sections. The chromaticity correction scheme based on sextupole magnets on the arc sections and the straight sections with dispersion will be employed in the pRing. Each arc section consists of eight FODO cells with $90^{\circ}$-phase advance per FODO cell, where 12 sextupole magnets will be installed. Dispersion exist in the short straight sections, and the optical parameters are designed symmetrically to the center of the short straight section. By doing this, the larger $\beta_y$-value appears at the position $\pm \pi/2$ from the center, which can enhance the capability of chromaticity correction contributed by the sextupole magnets here. The chromaticity correction scheme, which is based on 52 sextupole magnets in total placed on the arc sections and the short straight sections, can correct the chromaticity of the pRing to zero, as well as keep the dynamic aperture larger than $8 \sigma$. This scheme satisfies the requirement of the EicC accelerator facility.

\subsection{Electron accelerator complex}\label{sec:3.2.2}

A recirculating superconducting radio frequency (SRF) linear accelerator will be employed as the injector of the electron accelerator complex of the EicC accelerator facility. The recirculating SRF linac is considered to be the best choice for the electron injector since it has the advantages not only of linear accelerators, i.e. high accelerating gradient, greater compactness but also of circular accelerators, i.e. high efficiency and low cost. The polarized electron beams generated from the polarized photocathode electron gun will get matched to the electron injector, and then accelerated to the extraction energy range from 2.8 GeV to 5.0 GeV via passing through the RF cavities several turns to provide electron beams for the full energy injection scheme in the eRing. In this process, the polarized electron beams will be bunched into micro bunches with the bunch length of picoseconds. Taking into account the limitations of beam power and beam dump, the recirculating SRF linac can provide the electron beams of micro-Ampere beam intensity. Furthermore, the polarization direction of the electron beams extracted from the electron injector should get matched to the eRing, which can be achieved via a spin rotator.

Because of the full energy injection scheme and the self-polarization effect in the electron beams, the eRing is designed to be racetrack-shaped as many other electron colliders. The RF system of the eRing will adopt the frequency of 500 MHz to generate 1350 stable buckets. The polarized electron beams will be injected into 270 stable buckets, forming into 270 equally spaced electron bunches, to meet the collision frequency of 100 MHz in the luminosity design. Moreover, as the transverse and longitudinal beam sizes will decrease rapidly because of the damping effects introduced by the synchrotron radiation, the full energy injection scheme will be adopted to reach the bunch intensity of $1.70 \times 10^{11}$ ppb required by the luminosity design. After beam accumulation is finished, the electron bunches in the eRing will be reinforced or replaced by the recirculating SRF linac on line to maintain the luminosity.

In addition to the limitations of average beam current, synchrotron radiation power density, and beam-beam interaction parameters as listed in the Tab.\ref{tab:accelerator:3.2}, the luminosity optimization scheme with low collision frequency, high beam intensity could also bring longitudinal microwave instability and transverse mode coupling instability. This is similar to the case in the pRing. It can be calculated theoretically that the thresholds of the longitudinal and the transverse broadband impedance ($Z^{\parallel}/n$ and $Z^{\bot}$) in the eRing are $0.040$  $\Omega$ and $0.259$ $\mathrm{M} \Omega /\mathrm{m}$, respectively. The thresholds are considered to be reachable, if we take into account the case in the KEKB electron-positron collider in Japan with the corresponding thresholds of $0.012$ $\Omega$ and $0.235$ $\mathrm{M} \Omega/\mathrm{m}$. Therefore, the luminosity optimization scheme of low collision frequency, the high beam intensity is feasible in the eRing.

Since the eRing is designed to be racetrack-shaped, the chromaticity correction scheme base on the sextupole magnets on the arc sections will be adopted. Specifically, there are two arc sections in the eRing, as shown in the Fig.\ref{fig:accelerator:3.1}. Each of them consists of 20 FODO cells with $120^{\circ}$ phase advance. Two pairs of FODO cells at both ends of one arc section are used for optical matching, while the other 16 FODO cells are employed for chromaticity correction. Each three of the 16 FODO cells form a super periodical structure to cancel out the non-linear effects caused by the sextupole magnets within the arc sections. By using such a chromaticity correction scheme, the chromaticity can be corrected to zero while keeping the dynamic aperture larger than $20 \sigma$, which satisfies the accelerator design requirements.

\section{Beam cooling}\label{sec:3.3}

\begin{figure}
    \centering
\includegraphics[width=15cm]{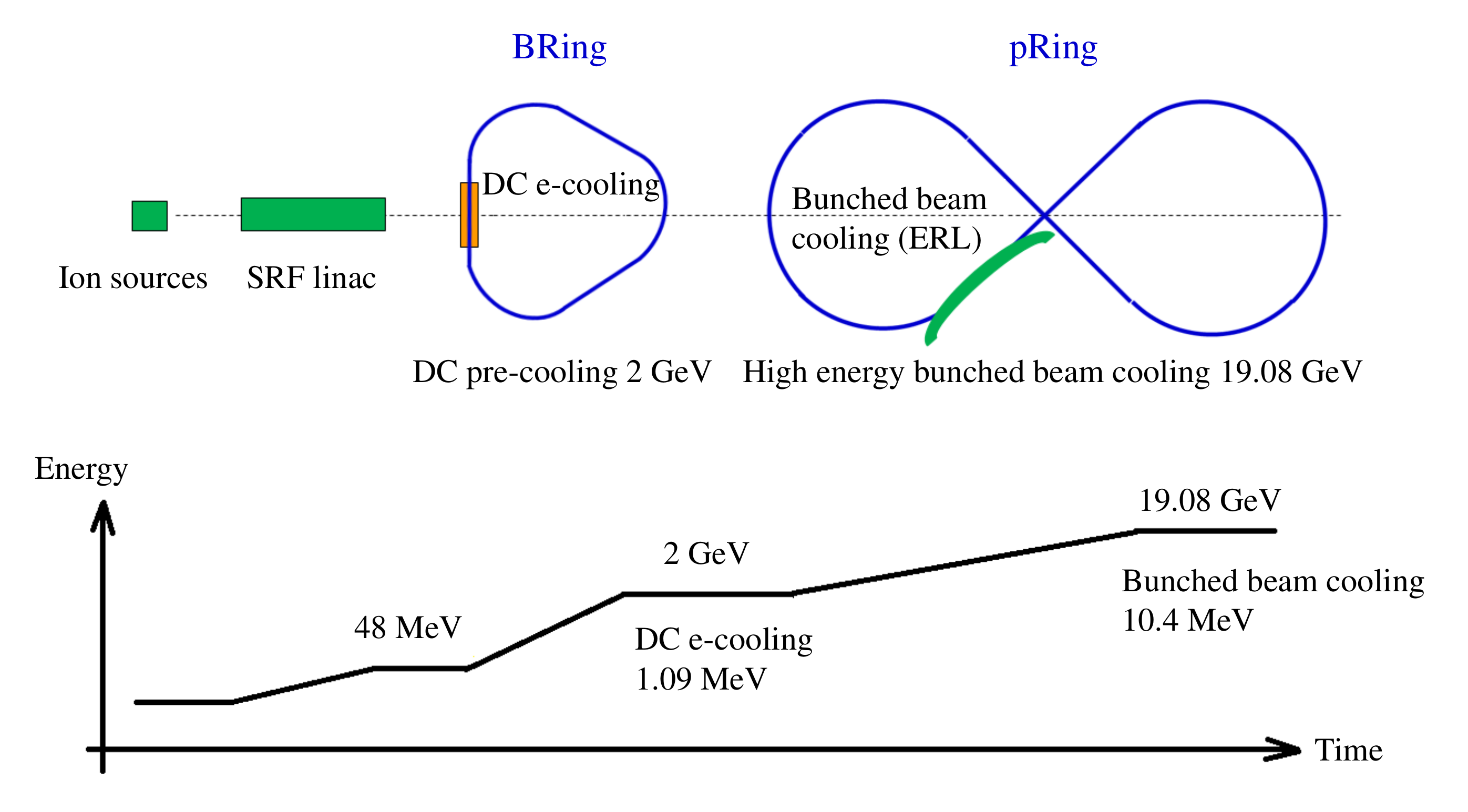}
    \caption{The layout of the staged electron cooling scheme for the EicC accelerator facility.}
    \label{fig:accelerator:3.3}
\end{figure}

High luminosity, which is the primary goal of an electron-ion collider, is thus the most essential parameter for the design of the EicC accelerator facility. It can be proven from the luminosity formula that an efficient way to increase the luminosity is to decrease the six-dimensional emittance of ion beams. Since there is no synchrotron radiation damping effect in the heavy ion synchrotrons, an external cooling mechanism is required to reduce the ion beam emittance. For decades, now electron cooling has become one of the most effective and well-developed methods to reduce the ion beam emittance. Since electron cooling has a greater effect on the ion beams of low energy and low emittance, the EicC accelerator facility will adopt the staged electron cooling scheme to shorten the cooling time and improve the cooling efficiency. In the first cooling stage, an electron cooler\cite{mao2015electron}, based on conventional electrostatic high voltage acceleration, will be installed in the BRing to reduce the transverse emittance and the momentum spread of the medium-energy ion beams to reach the design value. In the second cooling stage, a high energy bunched beam electron cooler, based on an energy recovery linac (ERL), will be installed in the pRing to further reduce the transverse emittance and the momentum spread of the high-energy ion beams. The intra-beam scattering effect in the collision will be suppressed by this cooler to keep the emittance and the momentum spread to be the design value, which can ensure high luminosity and long luminosity life required by the scientific goals. Due to a reduced emittance after the low-energy cooling in the first stage, the cooling time for the high-energy beam can also be largely reduced, leading to a shortened total cooling time and enhanced cooling efficiency. The staged electron cooling scheme for the ion beams of the EicC accelerator facility is shown in the Fig.\ref{fig:accelerator:3.3}.

%
%

The Tab.\ref{tab:accelerator:3.4} lists the details of the staged beam cooling scheme in the EicC accelerator facility. Taking the polarized proton beam as an example, it will be first accelerated to the energy of 48 MeV by the iLinac, then injected into the BRing for further accumulation, and accelerated rapidly to 2 GeV via the RF cavities. The proton beam will be debunched into a coasting beam after the momentum spread is reduced by a bunch rotation manipulation. The coasting beam will be cooled via DC electron beams of the energy of 1.09 MeV which are produced from a conventional electrostatic electron cooler. The cooling process can reduce the emittance and the momentum spread of the proton beam to the design value. After that, the high intensity polarized proton beam in the BRing will be injected to the pRing for further accumulation and acceleration. When the proton beam is accelerated to reach the energy of 19.08 GeV, the second cooling stage, which is based on high energy, high-intensity and high-quality electron bunches produced by an energy recovery linac (ERL), will be performed to cool the beam again and suppress the emittance growth and the bunch lengthening caused by intra-beam scattering effects over the whole collision process. For other ion beams with lower energy and higher cooling efficiency, the staged electron cooling process requires a shorter cooling time compared to proton beams.

\begin{table}
\centering
\caption{The staged electron cooling scheme of the EicC accelerator facility.}
\label{tab:accelerator:3.4}
\begin{tabular}{p{0.09\textwidth} p{0.08\textwidth} p{0.3\textwidth} p{0.12\textwidth} p{0.12\textwidth} p{0.1\textwidth}}
\hline
 & Position & Function & Proton energy(GeV) & Electron energy(MeV) & Cooler\\
\hline
Phase 1 & BRing & Reduction of beam emittance &2 & 1.09 & DC \\
Phase 2 & pRing & Beam  emittance reduction and intra-beam scattering suppression & 19.08 & 10.4 & Bunched beam (ERL) \\
\hline
\end{tabular}
\end{table}

\begin{figure}
    \centering
    \includegraphics[width=15cm]{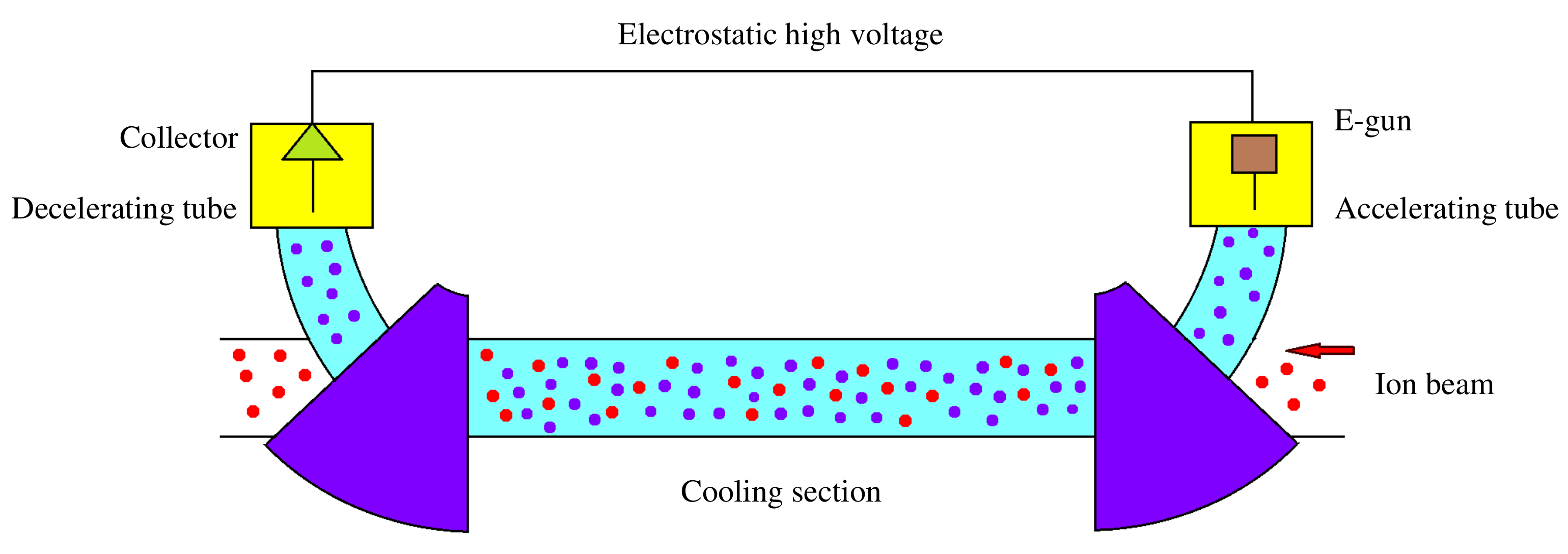}
    \caption{Layout of the low energy DC electron cooler.}
    \label{fig:accelerator:3.4}
\end{figure}

The DC electron cooler in the BRing consists of an electron gun, an accelerating section, a cooling section, a decelerating section, a collector, several solenoids, and several correctors, as shown in the Fig.\ref{fig:accelerator:3.4}.
The electron beams emitted from the cathode of the electron gun can be extracted to the accelerating section via a potential difference introduced at the anode. After acceleration via electrostatic high voltage to obtain the same average speed as the ion beams, the electron beams will be transported to the cooling section, in which they will interact with the ion beams and absorb part of the heat from the ion beams via the Coulomb interaction. Then the electron beams will pass through the decelerating section and be collected in the collector. The repeated electron beams with low temperatures can finally reduce the emittance and the momentum spread of the ion beams to the design value. Nowadays the technology of the DC electron cooling has already been well-developed and there are many DC electron coolers around the world, with the energy ranging from a few tens of KeV to a few MeV. For instance, in the Recycler Ring, Fermi-Lab, USA, the energy of the electron beam is as high as 4.3 MeV. In the COSY electron cooler in the J\"ulich, Germany, the electron beam reaches the energy of 2 MeV. And also the first RF linac-based electron cooler (bunched beam cooling) with an electron energy of 1.6 and 2.0 MeV was successfully commissioned at RHIC. For the EicC accelerator facility, the energy of the electron beam is designed to be 1.09 MeV for the DC electron cooler, which is technically achievable.

\begin{figure}
    \centering
\includegraphics[width=15cm]{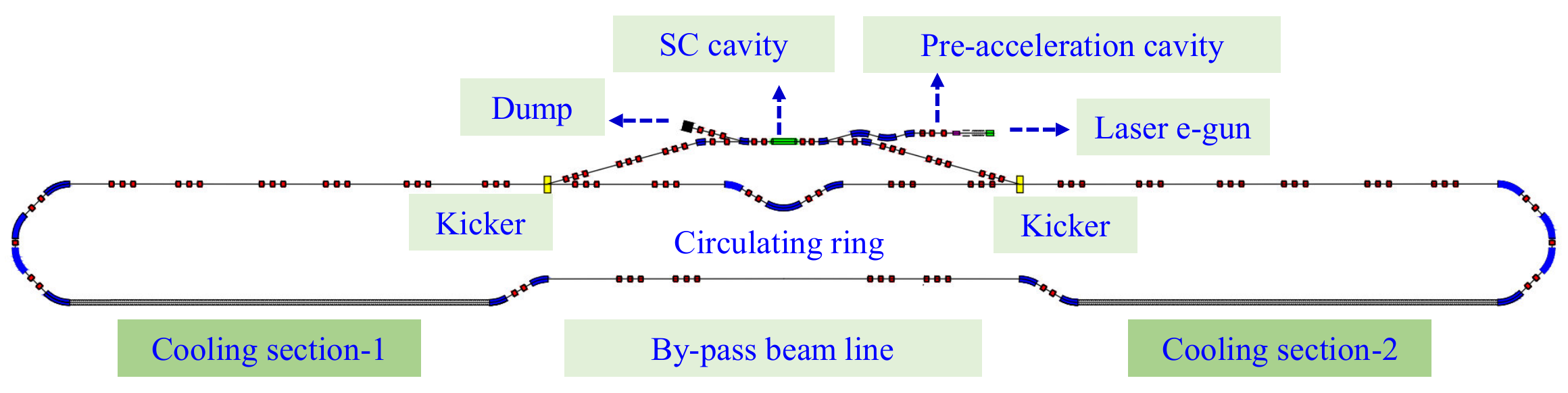}
    \caption{The high energy bunched beam cooler based on the ERL.}
    \label{fig:accelerator:3.5}
\end{figure}

In the pRing, beam cooling requires electron beams with a maximum energy of 10.4 MeV. Since the conventional DC electron cooler are unable to accelerate the electron beams to this energy, it is necessary to employ the electron beams accelerated by RF cavities, which is high energy bunched beam electron cooling. However, the specific electron beams cannot be used in the beam cooling for a very long time and should be always replaced by new electron beams, because the electron cooling requires the electron beams of very high quality (low emittance, low energy spread, high beam intensity). Moreover, the discarded electron beams should be collected to avoid radiation protection issues, since there is almost no energy loss during beam cooling. The quality of the electron beams in the synchrotrons hardly satisfies the requirement of the beam cooling. It is limited by the equilibrium conditions of the synchrotron radiation effects. The electron linear accelerator can accelerate, transport beams effectively, and keep the high beam quality. However, the high power level of the RF cavity increases the cost of construction and operation. Meanwhile, the collection of such high power electron beams can cause severe issues of radiation protection, such as neutron activation, and environmental pollution. In comparison, the energy recovery linac (ERL) cannot only accumulate the electron beam intensity as effectively as a synchrotron but also keep high beam quality as the linear accelerators, which satisfies the requirement of the high energy bunched beam cooling. The electron beams in the ERL can be sent back into the RF cavities with a decelerating phase, where the power of the high energy electron beams can be transferred to the power of the microwave acceleration field which can be used to accelerate newly injected electron beams. The energy recovery from the discarded electron beams in the ERL cannot only largely reduce the technical difficulty and the costs of the RF power, but also decrease the power deposition in the beam collector, solving the possible problems of the radiation protection and environmental pollution caused by the high power electron beam. The ERL-based high energy bunched beam electron cooler employed in the EicC accelerator facility is shown in the Fig.\ref{fig:accelerator:3.5}. It consists of a photocathode electron gun, merger line, superconducting RF cavities, arc sections, two 25-meter-long cooling sections, beam matching sections, circulating sections, and collectors.

The high intensity, high-quality electron bunches generated from the photocathode will be firstly accelerated via the pre-accelerating cavity to 2 MeV, then transported into the main accelerating section of the ERL through the merger line while keeping the initial emittance and energy spread. The electron bunches are further accelerated by the superconducting RF cavities to reach the energy of 10.4 MeV. The electron bunches are transported through the arc sections to the cooling sections to travel along with the ion beams. Then, the electron bunches interact with the ion bunches within the cooling sections in the strong solenoid magnetic field. After that, the electron bunches are stored in the circulating ring until making 53 bunches in the circulating ring, with the frequency between the bunches being 100 MHz. Then, the first stored electron bunch is kicked out of the circulating ring by an ultra-fast kicker cavity, with its energy recovered in the superconducting RF cavities. Finally, the residual power can be deposited in the collectors. The newly-injected electron beams will be accelerated by the recycled energy in the RF cavities and replace the bunch which had been kicked out of the circulating ring every 6.25 MHz. This process will be repeated until the emittance and the momentum spread of the ion beams meet the design requirements. The ERL and its circulating ring are shown in the Tab.\ref{tab:accelerator:3.5}.

\begin{table}
\caption{The main parameters of ERL and circulating ring.}
\label{tab:accelerator:3.5}
 \centering
\begin{tabular}{p{0.28\textwidth}lp{0.28\textwidth}l}
\hline
Gun type & SRF & RMS energy spread & $5 \times 10^{-4}$ \\
PRF & 6.25 MHz & Cooling energy & 10.4 MeV\\
Main RF frequency & 700 MHz & Cooling section length & 50 m \\
Charge per bunch & 4 nC & Cryostat number & 1 \\
Injection energy & 2 MeV & Cryostat length & 1.5 m \\
Beam current in ERL & 25 mA & Beam current in circulating ring & 400 mA \\
RMS bunch length in the cooling sections & 150 ps & Circumference of circulating ring & 159 m \\
Transverse RMS normalized emittance & 2.5 $\pi \cdotp \mathrm{mm}\cdotp \mathrm{mrad} $ & Bunch frequency in circulating ring & 100 MHz\\
\hline
\end{tabular}
\end{table}

In recent years, compared to many other types of high energy particle accelerators, the ERL has been paid more attention and experienced rapid development and wide application, since it has the advantages of less power consumption, high beam quality, and so on. The ERLs are being considered to replace the conventional linear or ring accelerator for the application of the free-electron laser (FEL), the synchrotron radiation sources, the colliders, as well as the electron coolers. Up to now, there exist several established ERL facilities around the world, including the Novosibirsk ERL at BINP in Russia, the CEBAF-ER and the IR ERL in the Jefferson laboratories (JLab) in the USA, the S-DALINAC ERL at Technical University Darmstadt in Germany and the ERL of the High Energy Accelerator Research Organization (KEK) in Japan, and so on. BNL has also tried to build test ERL and CeC ERL for high-energy beam cooling before. With the technology development of the photocathode gun and the superconducting radio frequency (SRF), the difficulties for the construction of the high energy, high-intensity ERL has been reduced a lot. There are several ERLs under construction or proposed, including the bERLinPro of the Helmholtz-Zentrum Berlin(HZB) and the MESA at the Johannes Gutenberg-Universit\"at Mainz in Germany, Cornell-BNL-ERL-Test-Accelerator (CBETA), the light source in the Cornell University, the US-EIC in the BNL and the LHeC in CERN. Some of them will be employed as high energy bunched beam electron cooler, which can lay a solid foundation for the construction of the high energy bunched beam electron cooler based on the ERL in the EicC accelerator facility while largely reducing the technical risk of the beam cooling in the project.

\section{Beam polarization}\label{sec:3.4}

As a dual-polarized electron-ion collider, the beam polarization is another important part of the design of the EicC accelerator facility, besides the luminosity and the beam cooling which are typical designs of a collider. The physics goals of the EicC project pose the following requirements on the beam polarization design of the EicC accelerator facility.

1. The polarized electron beams will collide with the polarized proton beams, the polarized deuterium beams, and the polarized Helium-3 beams, respectively. The required polarization here is about $80\%$ for the electron beams, $70\%$ for the proton beams. Other ion beams are non-polarized.

2. At the IPs, the polarization direction of the electron beams should be longitudinal, while for the proton beams, the deuterium beams, and the Helium-3 beams, the polarization direction could be arbitrary.

3. The measuring errors of the polarization for the proton beams, the deuterium beams, and the Helium-3 beams should be less than $5\%$, while for the electron beams it is less than $2\%$.

\begin{figure}
    \centering
\includegraphics[width=15cm]{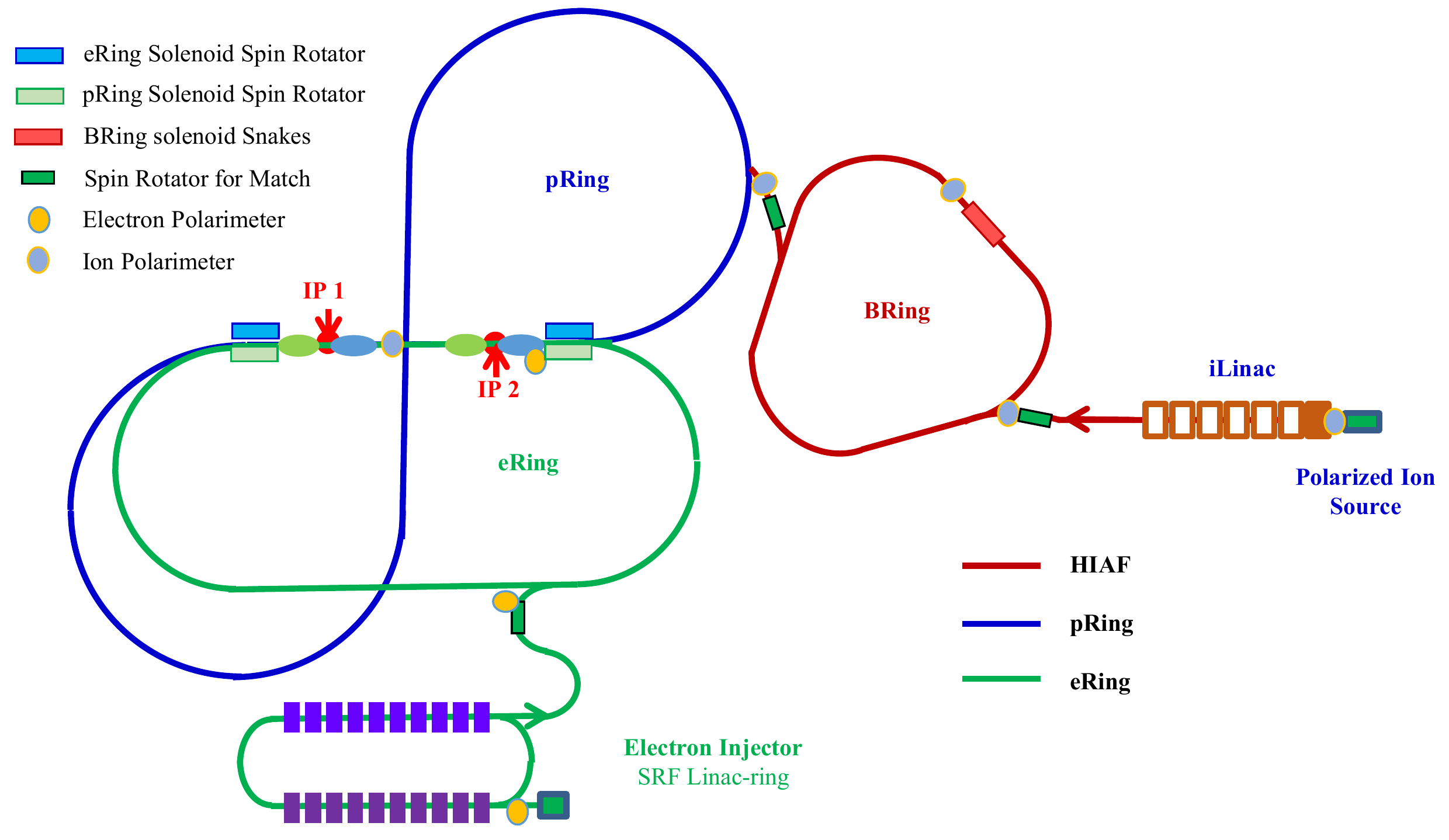}
    \caption{The polarization design of the EicC accelerator facility.}
    \label{fig:accelerator:3.6}
\end{figure}

For these requirements and goals listed above, the scheme of the beam polarization is designed and optimized based on the layout in the Fig.\ref{fig:accelerator:3.1} and the operation mode in the Fig.\ref{fig:accelerator:3.2}, including polarization control and polarization measurement, which is shown in the Fig.\ref{fig:accelerator:3.6}.

\subsection{Ion polarization}\label{sec:3.4.1}

For the ion accelerator complex, the atomic beam polarized ion source (ABPIS) can produce both polarized and non-polarized proton beams, while other polarized ion beams are similar to the proton beams. There are four successive processes to transform hydrogen atoms into polarized proton beams, i.e. the dissociation, the separation in a sextupole magnet, the transition in a weak RF field, and ionization. Since the dissociated hydrogen atom has an isotropic spin distribution, the hydrogen atoms of high energy level and with an electron spin of $1/2$ can be selected and guided into the next devices. During the transition, an atom with a proton spin of $-1/2$ will jump selectively into the atomic state of $1/2$ proton spin, forming into high intensity and high polarized atomic beams. Furthermore, with the selection of the polarization direction of the weak RF field, the polarization direction of the atomic beams in the transition module can be controlled to match the polarization directions of the following beamlines and accelerators. Proton beams of high intensity and high polarization can thus be produced via the ionization of such atomic beams. The ABPIS is one of the most widely-employed ion sources and can generate the proton beams of polarization as high as $90\%$\cite{zelenski2010review}.

The polarized proton beams extracted from the polarized ion source will be measured via a Lamb-shift polarimeter. The Lamb-shift polarimeter scans the magnetic field in the rf spin filter to quench the different meta-stable hydrogen atoms and measures the Lyman - $\alpha$ photons from the quenching downstream. The meta-stable hydrogen atoms neutralized from the polarized protons will be quenched at special spin filter magnetic field values and emit Lyman - $\alpha$ photons. The polarized proton beams will be accelerated in the iLinac to reach the energy of 48 MeV needed by the injection of the BRing, and finally injected to the BRing via a beamline for injection. During this process, the polarization of the proton beams remains unaffected since there is no spin resonance. However, to measure the polarization direction of the beams injected into the BRing, a polarimeter will be still installed in the beamline for the polarization direction matching to the BRing. The polarimeter mainly records the counting rate of the Coulomb elastic scattering between the polarized proton beams and the target at different angles, from which the polarization at a certain angle can be calculated.

High intensity polarized proton beams of the energy of 48 MeV will be accelerated to 2 GeV in the BRing. During the acceleration, the polarization of the beam will significantly decline because the spin tune will experience several depolarization resonances. To maintain high polarization, two Siberian snakes will be installed at the two ends of the electron cooling section of the BRing to change the spin tune of the ring. The Siberian snakes, i.e. the solenoids whose magnetic fields are synchronized with the beam energy, can keep the spin tune always to be $1/2$ which is far from the depolarization resonances. In this way, the problem of the decline of the polarization of the proton beams caused by the depolarization resonances can be solved. Besides, a polarimeter will be installed, which can be used for the on-line measurement of the polarization direction and the polarization of the beams and provide useful information for adjustment and optimization of the polarization-related parameters of the BRing. The polarimeter should operate at the energy range from 48 MeV to 2 GeV in which a high counting rate should remain and the requirement of high-precision control for the BRing should be met.

\begin{figure}
    \centering
\includegraphics[width=6cm]{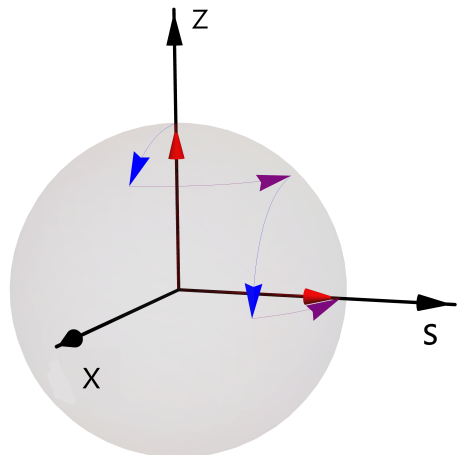}
    \caption{The spin rotation in a spin rotator.}
    \label{fig:accelerator:3.7}
\end{figure}

The polarized proton beams of 2 GeV energy will be extracted via a fast extraction scheme from the BRing and injected into the pRing after measuring and matching the polarization direction. The polarization matching can be implemented via a spin rotator. Based on a set of magnets consisting of solenoid-horizontal bending dipole magnet-solenoid-horizontal bending dipole magnet, the polarization direction can be guided arbitrarily. This is one of the most essential polarization control devices in the EicC accelerator facility. The polarization direction control scheme is illustrated in the Fig.\ref{fig:accelerator:3.7}, taking as the example of a vertical polarization rotating to the longitudinal polarization. In the Fig.\ref{fig:accelerator:3.7}, the blue line denotes the rotation of polarization direction contributed by the solenoids, and the purple line represents the rotation from the horizontal bending dipole magnets. The initial and final polarization directions are shown in red. When passing through a solenoid, the polarization direction of the polarized beam will rotate with a certain angle with the beam direction as the axis. In the horizontal bending dipole magnets, the polarization direction will rotate around the vertical direction. With once more such rotations in another solenoid and horizontal bending dipole magnet, the polarization direction of the beam can be transferred from the vertical direction to the horizontal direction. Similarly, the polarization direction of the beam will rotate to an arbitrary direction by this spin rotator with carefully-chosen rotation angles in the solenoids and the bending dipole magnets. Moreover, the polarization direction can be well-controlled in a wide energy range while leaving the closed orbit unchanged, because the four spin angles provide sufficient polarization control variables, which makes it possible to provide multiple solutions for the adjustment of a certain polarization direction. With the help of the spin rotator, the polarization direction of the extracted high intensity polarized proton beams can be matched with one of the arc sections in the pRing, thus avoiding the depolarization caused by the mismatch of the polarization direction. Another polarimeter will be installed near the injection section of the pRing to ensure an efficient and complete matching of the polarization direction during the injection.

After injected into the pRing, the polarized proton beams will be further accelerated to the energy of 19.08 GeV. During the acceleration, the proton beams will experience several depolarization resonances, which means it is very difficult to maintain the polarization of the proton beams in the conventional racetrack-shaped accelerators. Furthermore, for the large energy range in the pRing, there are quite a lot of technical challenges in Siberian snakes consisting of solenoids. However, the alternative Siberian snakes consisting of dipole magnets will cause the closed orbit distortion. These make it difficult to avoid the depolarization resonances by only using the Siberian snakes. To solve the issues, the pRing of the EicC accelerator facility is designed to be a novel figure-8 structure, in which the spin precession in the arc sections of one side can be always exactly canceled by the ones of the other side, keeping the total spin tune in the ring to be zero. The spin precession angle of the proton along the reference orbit is G$\gamma$ times the bending angle. The total bending angle of figure-8 ring is zero, so the spin tune will be zero independent of the beam energy for the convenience of spin control and the spin tune is far away from all intrinsic depolarization resonances. Meanwhile, a small solenoid added in the long straight section to introduce additional spin precession angle can ensure the spin tune to be far away from the imperfection depolarization resonances. It can be shown that in the optimized condition the solenoid at the IPs can also meet requirements for adjusting the spin tune, which is a novel scheme for the ion accelerator complex design in the EicC project.

Due to the arbitrary polarization direction of the proton beams needed at the IPs, two spin rotators will be placed on either side of the long straight section containing the IPs and the arc sections at both ends of the long straight section. As discussed above, each spin rotator can perform polarization direction rotation to generate the arbitrary direction needed by the experiments. Moreover, by using the spin rotator, the polarization direction will get matched to the direction of the arc sections before the proton beams enter the arc sections upstream of the IPs. With this scheme, the requirements of the arbitrary polarization directions and the polarization matching can be satisfied easily.

A polarimeter will also be installed in the pRing for the online measurement of the polarization direction and the polarization of the proton beams at the IPs, as well as providing measuring data of the polarization for various physics experiments. The measuring accuracy of the polarimeter should be as high as $5\%$, which meets the physics experiment requirements.

\subsection{Electron polarization}\label{sec:3.4.2}

In the electron accelerator complex, the polarized electron beams can be produced via the photocathode polarized electron gun. Since the energy band of the electron excitation in the cathode plated by Cs and GaAs can overlap with the energy band of the vacuum, the electrons will become free electrons when transiting to the excitation energy band. With a laser of a specified energy and polarization direction, the electrons of the specified spin state will transit to the excitation energy band, forming highly-polarized electron beams. The polarization direction of the electron beams can be controlled by changing the polarization direction of the laser. Up to now, the polarization of the beam generated from the polarized electron gun can be as high as $90\%$\cite{tsentalovich2019high}, which can meet the requirement of the EicC project. Before electron beams are injected into the electron injector, their polarization characteristics will be measured via a polarimeter. This polarimeter can be also employed to adjust the parameters of the photocathode polarized electron gun according to the requirements of physics experiments and accelerator operations.

After the polarized electron beams are injected into the electron injector, it will be accelerated to reach the energy range from 2.8 GeV to 5.0 GeV. During the acceleration, the depolarization resonances will not occur since the electron beam travels through different arc sections in the different turns, causing a lack of resonances between the periodic transverse and longitudinal motion, as well as the spin periodic precession\cite{higinbotham2009electron}. Therefore, it is not necessary to place any polarization preservation devices in the electron injector.

The polarized electron beam will be accelerated to its maximum energy in the electron injector and then injected into the eRing. During this process, the polarization direction of the electron beam should get matched to one of the arc sections of the eRing. The polarization matching can be realized via a spin rotator in the ion accelerator complex.

The polarization direction and the polarization of the beam will be measured before the beam is injected into the eRing. It is different from the case in the ion accelerator complex or in the low energy electron accelerators, where the internal-target measuring scheme is adopted. The differential Compton cross section is a function of the initial electron and photon polarizations\cite{fano1949remarks}. The polarization direction and the polarization of the electron beams can be obtained indirectly via detecting the backscattered photons after the scattering of circularly polarized photons on polarized electrons. This is a non-interception measurement scheme that will not affect the beam quality.

The polarized electron beams should preserve the polarization of $80\%$ during the whole collision. The eRing is designed to be racetrack-shaped as other facilities according to the experiment results of the existing facilities in the world\cite{johnson1983beam}. Such a design has the advantage that the depolarization effects can be canceled by the self-polarization effect introduced by the synchrotron radiation of the electron beams, resulting in high equilibrium polarization at the level of $80\%$ during the whole collision.When the beam energy lies nearly on the depolarization resonance energy and it is not enough for the self-polarization effect to cancel the depolarization effects, the spin tune can be moved away from the depolarization resonance via a solenoid at the IR. At the IR, the polarization direction of the electron beam is longitudinal, so the spin tune is moved via the solenoid while the polarization direction remains longitudinal. When colliding with the polarized proton beams, the polarization direction of the polarized electron beams should be longitudinal, but the polarization direction matching to the arc sections of the eRing is transverse. Therefore, it is necessary to place two spin rotators between the long straight sections containing the IPs and the arc sections to satisfy the matching of the polarization direction and the requirement of the physics goals. The spin rotator here is similar to the one in the ion accelerator complex. The online measurement of the polarization direction and the polarization in the eRing, as well as its beamline for injection, is also designed based on the 
Compton backscatter, only with an accuracy of $2\%$ required by the physics experiments.

\section{Design of the interaction regions (IR)}\label{sec:3.5}

To achieve the physics goals, one full-acceptance detector will be built at one IR for detection and identification of reaction products, such as charged particles, i.e. electrons $e$, muon $\mu$, $\pi$-meson $\pi$, k-meson $K$, proton $p$ and so on, as well as neutral particles, i.e. photon $\gamma$, neutron $n$, etc. Another IR will be reserved for the future upgrade. This detector consists of a central detector with end caps and forward detectors. The central detector with end caps will be placed around a superconducting solenoid and can be divided into two parts: the barrel part and the end cap part on both sides, which is used to detect the reaction products in a large angular range. It consists of a series of detectors, including vertex detectors, tracking detectors, time-of-flight (TOF) detectors, Cherenkov detectors, electromagnetic calorimeters, and hadron calorimeters. Free space with a length of 8 m is required for the installation of the central detector with end caps. The forward detectors will be mainly used for the detection of the final-state particles emitted from the reaction in a small or ultra-small angle. Those detectors are far from the IP and dipole magnets are required to separate the particles. The particle detection of the electron forward detectors is realized with the help of dipole magnets and several long drift sections after the first quadrupole magnet. For the ion forward detectors, a dipole magnet should be installed in front of the first quadrupole magnet to make it possible to identify the reaction products at a small angle. Furthermore, other dipole magnets and long drift sections will be placed behind the first quadrupole magnet to allow the identification of the reaction products in ultra-small angles. Compared to conventional colliders, such specifications from the detectors raise many more requirements of the interaction region design of the EicC accelerator facility.

Typically, a conventional collider poses the following requirements about the layout of the interaction region and its optics design.
\begin{enumerate}
  \item The beam transverse size or the $\beta^{*}$ at the IP should be as small as possible.
  \item The devices in the interaction region should meet the required minimum installation length of the central detector.
  \item The $\beta$ function in the first quadruple magnet in the interaction region should be not too large.
  \item The optics design in the interaction region should satisfy the chromaticity correction of the collider ring.
  \item The cross angle at the IP should be larger than the minimum value required for the bunch separation, and smaller than the maximum acceptable value decided by the crab cavity.
  \item In the optics design of the interaction region, the size of each magnet should not exceed the space limit introduced by the detectors, and there should be no interference among the magnets.
\end{enumerate}

Furthermore, in order to identify nearly $100\%$ of the reaction products with higher resolution, the full-acceptance detector poses more requirements on the design of the interaction region in the EicC accelerator facility.
\begin{enumerate}
  \item In the interaction region of the eRing, a large deflection of electron beams should be avoided to reduce the impact of the synchrotron radiation background generated by the electron beams themselves on the detectors in the interaction region.
  \item The interaction region of the pRing should be designed as close as possible to the arc sections to reduce the hadron background produced by the collision among the proton beams and the residual gas molecules which could affect the detectors in the interaction region.
  \item In the pRing, a dipole magnet should be placed downstream of the IP to improve the detecting resolution of the reaction products at small angles.
  \item In the pRing, a set of dipole magnets should be placed after the first focusing quadrupole magnet to improve the detecting resolution of the reaction products at ultra-small angles.
  \item In the eRing, at least one set of dipole magnets should be placed after the first focusing quadrupole magnet to improve the detecting resolution of the reaction products at small angles \cite{abeyratne2012science}.
  \item In the interaction region, the transverse aperture of the magnets should be larger than the clear zone, which is at least ten times of the transverse RMS beam size, to let the debris at larger scattering angles in the small-angle reaction products pass through the vacuum pipe of the magnets and reach the forward detectors downstream of the IP.
\end{enumerate}

\begin{figure}
    \centering
\includegraphics[width=15cm]{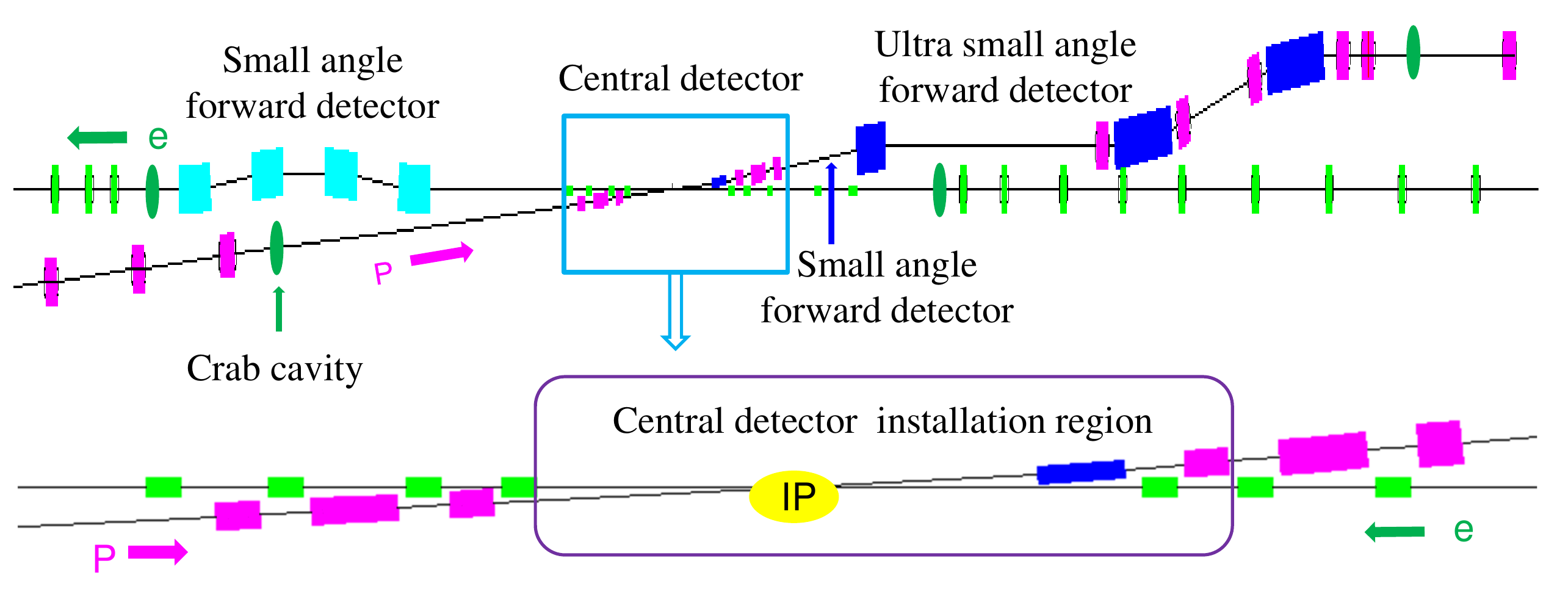}
    \caption{The interaction region of the EicC accelerator facility.}
    \label{fig:accelerator:3.8}
\end{figure}

Taking each item listed above into consideration, the overall design of the interaction region is shown in the Fig.\ref{fig:accelerator:3.8}. Since there are significant differences between electron beams and proton beams, the detectors in the different collider rings have different demands on the layout of the interaction region. So, the interaction region of the EicC accelerator facility is designed to be asymmetric. In the pRing, the interaction region is close to the upstream arc sections to reduce the hadron background caused by the interaction between the residual gas molecules and the proton beams, improving the resolution of the detectors. In the eRing, a long straight section will be placed upstream of the interaction region, keeping the interaction region away from the arc sections. Meanwhile, synchrotron radiation absorbers will be installed. These schemes can reduce the synchrotron radiation background, further improving the resolution of the detectors.

The interaction region mainly consists of two parts, i.e. the straight section with few devices close to the IP for the installation of the central detector with end caps, and the beamline that is relatively distant from the IP and contains dipole magnets and drift lines for the installation of the forward detectors. The installation of the central detector with end caps requires an 8-meter free space. To follow the magnetic field limits of the magnets and make full use of the installation space in the interaction region, all of the magnets are designed to be superconductive for increasing the length of the free space and save the installation space so that all the final state particles except the reaction products at the small and very small angles can be detected. Based on the superconducting magnets, the crossing angle at the IP is chosen to be 50 mrad, which can achieve a fast separation between the electron beams and the ion beams and suppress the long-range beam-beam interactions, so as to reduce the limit of the length of the straight section at the IP. And this design can optimize the minimum $\beta$ function at the IP, the maximum $\beta$ function at the first focusing quadrupole magnet, and the dipole magnetic field required by the forward detectors for small-angle reaction products.

The $\beta$ function values at the IP in the pRing are 0.04 m in the horizontal direction and 0.02 m in the vertical direction respectively, which can satisfy the luminosity design. And the maximum $\beta$ function value in the first focusing quadrupole magnet is around 1000 m. It makes sure that this superconducting quadrupole magnet can be installed in the central detector. Furthermore, in order to achieve higher luminosity, the dispersion at the IP should be zero to further reduce the beam transverse size. However, there is a large dispersion on both sides of the IP to improve the detecting resolution. A dispersion suppression section, which consists of two dipole magnets and two quadrupole magnets, will be installed upstream of the IP to make sure that the dispersion is zero at the IP while the dispersion at the forward detectors is large enough to provide high detecting resolution.

The $\beta$ function values at the IP in the eRing are 0.2 m in the horizontal direction and 0.06 m in the vertical direction, which satisfies the design of the luminosity. Furthermore, compared to the case of the pRing, the first focusing quadrupole magnet in the eRing is closer to the IP, in order to separate the position at which the beam size is the largest in the eRing from the corresponding position in the pRing. It makes it easier to separate the beams at the IP. In the eRing, the maximum value of $\beta$ function in the first focusing quadrupole magnet is about 280 m as it is much closer to the IP. However, the fact that the first focusing quadrupole magnet in the eRing is much closer to the IP introduces a huge advantage. Because the aperture required by electron beams is much smaller, this quadrupole magnet can reduce the required installation space, which makes it possible for the central detector with end caps to detect all the reaction products more effectively and to avoid breaking the full-acceptance feature of the central detector.

A superconducting dipole magnet with 1-meter length, 2.1 T maximum magnetic field, and 30 mrad deflection angle will be placed downstream of the IP in the pRing. The dipole magnet will be able to provide large dispersion to enhance the detecting resolution of the forward detectors installed behind it for small-angle products. Furthermore, the dipole magnet can separate the beams to make it more flexible for the optics design of the two collider rings. Behind the superconducting dipole magnet, there is a straight section that contains only focusing quadrupole magnets, which can be used for the separation of the small-angle reaction products and the installation of the small-angle forward detectors. The quadrupole magnet is followed by a dipole magnet providing a large deflecting angle to deflect the beam to the direction which is parallel with the eRing. The distance between the beamlines in the two collider rings is about 1 meter. Along with the following long straight section, this large dipole magnet can improve the detecting resolution for the ultra-small-angle reaction products. Besides, the neutral particles will be not deflected when passing through this dipole magnet. To detect the neutral particles produced from the reaction, the detectors can be put in the direction of the extension line of the tilting straight section. Based on a system consisting of two dipole magnets, the forward reaction products of small angles and ultra-small angles in the ion beams can be detected to almost $100\%$, which ensures the full-acceptance feature of the detector.

In the eRing, a set of carefully-designed beamline segments are placed downstream of the IP, for detecting the small-angle forward reaction products. With the horizontal bending dipole magnets, the electron beams will be deflected to a tilted direction with respect to the eRing, and then deflected back to the parallel direction. Two sets of dipole magnets (4 magnets in total) are placed symmetrically to cancel the dispersion generated by themselves and generate a smooth decrease of the $\beta$ function, which makes it convenient for the optics parameter matching upstream and downstream. The dipole magnet closest to the IP is chosen to be the first dipole magnet of such a set of four. It is followed by a long straight section, which can provide plenty of drifting space for the reaction products in the electron beams so that the reaction products can be observed by the small-angle forward detectors. The straight section between the second and the third dipole magnet is parallel to the long straight line of the interaction region. A polarimeter, based on the Compton backscatter of the electron beams, will be placed in the extension line of the straight section. Four dipole magnets will form a deflecting structure, which ensures the full-acceptance feature of the detector in the electron beams.

To obtain higher luminosity with a large crossing angle at the IP, crab cavities will be placed with a phase shift of $\pi/2$ from the IP, which perform a transverse bunch rotation before the beams entering the IP. A head-to-head collision between the proton beams and the electron beams at exactly the IP can be achieved by this scheme. After the collision, other crab cavities will rotate the bunch to its initial transverse state for matching other single beam dynamics in the collider rings. The crab cavities have been employed successfully in the KEKB of the KEK in Japan\cite{akai2005new}. The frequency and voltage are crucial parameters for crab cavities. For the pRing, the frequency and voltage of the crab cavities are selected to be 207 MHz and 14.4 MV, respectively. With proton bunch length of 0.04 m, the phase width is chosen as about $\pm 30^ \circ$, which can fulfill the requirements of complete crabbing. As for the crab cavities in the eRing, the same frequency used in the pRing is chosen but with a lower voltage, 3.6 MV, resulting from lower energy of electron beams.

\section{Pre-research on key technologies}\label{sec:3.6}

For the design presented above, the pre-research of the EicC accelerator facility and testing of several related key technologies will be carried out on the existing facility HIAF. It mainly includes the atomic beam polarized ion sources (ABPIS), the photocathode polarized electron gun, the high energy bunched beam electron cooler based on the energy recovery linac (ERL), the Siberian snake, the spin rotator, a verification facility of a figure-8 ring, and an elastic scattering polarimeter.

The atomic beam polarized ion source (ABPIS) is the essential device for the production of the polarized proton beams (as well as the lighter polarized heavy ion beams). The study of the maximum polarization and the maximum intensity of such beams is one of the most important topics for the pre-research of the EicC accelerator facility. In the pre-research, the polarized ion source will be installed in the HIAF for online testing to find out if it meets the requirement of the EicC project.

The technology of the photocathode polarized electron gun is relatively well-developed. However, there are still several subjects that need to be investigated thoroughly, such as the control of the polarization direction, the optimization of the beam intensity, and the photocathode lifetime. And it is necessary to study the related parameters of the photocathode polarized electron gun, in order to determine the injection and accumulation schemes in the eRing. The photocathode polarized electron gun for the pre-research of the EicC accelerator facility will be tested independently since there is no electron accelerator in the HIAF.

The high energy bunched beam electron cooler based on the energy recovery linac (ERL) is an indispensable device to achieve the required collision luminosity and the collision lifetime in the EicC accelerator facility and will be developed in the stage of the pre-research. A verification facility of the high energy bunched beam electron cooler will be installed in the BRing of the HIAF. The pre-research of the high energy bunched beam electron cooler mainly includes three aspects. The first aspect is about the development of high- quality energy recovery linac (ERL). Compared to those energy recovery linacs which are not used for electron cooling, the ERL employed for the EicC accelerator facility poses higher requirements on the electron beam quality, because the electron beams should be kept sufficiently cold. In the pre-research, a prototype of ERL with low energy will be built to provide electron beams for the experiment of the high energy bunched beam electron cooling on the BRing of the HIAF, in which key technologies and experience about the cavity design of the ERL, as well as the energy recovery, can be developed and improved. The second aspect is about the design and implementation of a circulating ring. Since the electron beams generated from the electron gun cannot reach the beam current required by the high energy bunched beam electron cooling, they need to be recirculated in the circulating ring for 16 turns so that to reduce the ERL beam current by 16 times. The third aspect is the development of the ultra-fast kicker cavity. In the high energy bunched beam electron cooler based on the ERL, the electron bunches should be either injected from the main accelerating section to the circulating ring during the accumulation or extracted to the main accelerating section for energy recovery. It is crucial that the kicker cavity is able to deflect several electron bunches or even one electron bunch. There are many technical challenges for the design and production of such kicker cavities, which need to be examined and verified carefully in the pre-research. Overall, the high energy bunched beam electron cooler based on the ERL and developed in the pre-research stage of the EicC project is expected to be able to perform high energy bunched beam electron cooling experiments and online testing in the BRing, which will lay a solid technical foundation for the development of higher energy bunched beam electron cooler in the EicC accelerator facility.

The Siberian snake is a key device to avoid the depolarization resonances in the BRing. Technologically, it will be taken as a solenoid. Compared to dipole magnets, the solenoid-shaped Siberian snake will not affect the closed orbit of the BRing. However, its magnetic field ramping rate can be far lower than the one of dipole magnets. The BRing will be operated in a rapid cycling mode, in which the ramping rate of the magnetic field of the dipole magnets is about 12 T/s. There still exist quite a lot of technical difficulties in the synchronization of the solenoid magnetic field with proton beam energy, which needs to be studied and optimized in detail. The solenoid-shaped Siberian snake for the EicC accelerator facility will be firstly designed and built in the BRing. After the polarized ion source developed in the pre-research is installed, an integration testing of the BRing performance will then be conducted to check whether the beam intensity and the polarization of the BRing fulfill the criteria of the EicC project. So the booster of the EicC accelerator facility, i.e. the BRing, can be completed during the pre-research stage.

The spin rotator is an essential device for the adjustment of the polarization at the IP, as well as for the polarization direction matching between beamlines and accelerators of the EicC accelerator facility.
There are three most important characteristics of the spin rotator, i.e. functioning well for different energies, rotating the polarization direction to arbitrary direction, and introducing no influence of the closed orbit. A spin rotator for the EicC accelerator is composed of two solenoids and two dipole magnets, with the sequence of the solenoid-dipole magnet-solenoid-dipole magnet. It is important to verify that four parameters of magnetic fields for an arbitrary given rotation angle can be always specified. To this end, a spin rotator will be designed and developed at the injection energy of the BRing, which can be used to perform online testing of the spin rotator design in the beamline for the injection of the BRing of the HIAF. Based on these studies, the design of the spin rotator will be further improved and eventually applied to the EicC accelerator facility.

There has been no synchrotron with a figure-8 structure in the world until now, and the property of a zero pure spin tune in such type of synchrotrons has not yet been verified. There is some free space reserved for the mirror ring next to the SRing in the design and construction process of the HIAF, which makes it possible to upgrade the SRing into a figure-8 shaped synchrotron with less cost to test the effect of the depolarization resonances on such synchrotrons. If the upgrade is completed smoothly during the pre-research stage and the main features (such as the maintenance of polarization and so on) are verified, the high-precision spectrometer ring SRing will become the first figure-8 synchrotron that can store high intensity polarized proton beams in the world. This work, which thus plays a crucial role in the whole pre-research of the EicC accelerator facility, will support the construction of the pRing with fruitful technical experience.

Since there is no electron accelerator in the HIAF project, it is impossible to perform the testing of the polarization direction and polarimeter based on the Compton backscatter. For this reason, the test of the polarization direction and polarimeter based on the internal target could be firstly performed. Typically, the polarimeter adopts the internal polarized gaseous target, in which the polarization in every specified direction is measured via recording the counting rate of the angle distribution of the Coulomb scattering, in which the direction with maximum polarization represents the polarization direction of the beams. In this measuring scheme, the transport of the low energy beam will be greatly affected, so it will stop the beams and is not an online measuring method. However, for the beams of high energy, the measurement will end up with less effect on the beam transport and become the online measuring method. The polarimeter is one of the most key devices for the polarization direction and polarization control of the polarized ion source and the photocathode polarized electron gun. It is also the only equipment that can be employed to perform the polarization matching between accelerators and beamlines, as well as to measure the depolarization resonances in the synchrotrons. In the pre-research of the EicC project, several internal target polarization direction and polarimeters will be built and installed at the exit of the polarized ion source, at the exit of the photocathode polarized electron gun, at the exit of the spin rotator in the beamline for the injection of the BRing, in the BRing and in the high-precision spectrometer ring SRing and so on, to check whether the technical goals of the pre-research of the EicC accelerator facility is fully satisfied.

In conclusion, the pre-research of the EicC accelerator facility is crucial to ensure the design, construction, commissioning, and operation of the EicC accelerator facility to be finished smoothly and successfully in the future. With that, the key technical barriers in the EicC project will be overcome, and their corresponding solutions and schemes will be verified.


\chapter{Detector conceptual design}

Driven by the physics program of EicC, a conceptual design for a general purpose spectrometer is presented in this chapter. The physics program includes investigating the nucleon spin structure, the nucleon 3-D structure with respect to TMDs and GPDs, and the study of exotics, etc., as detailed in previous chapters. 

\begin{figure}[tbh]
\centering
\includegraphics[width=0.95\textwidth,angle =0]{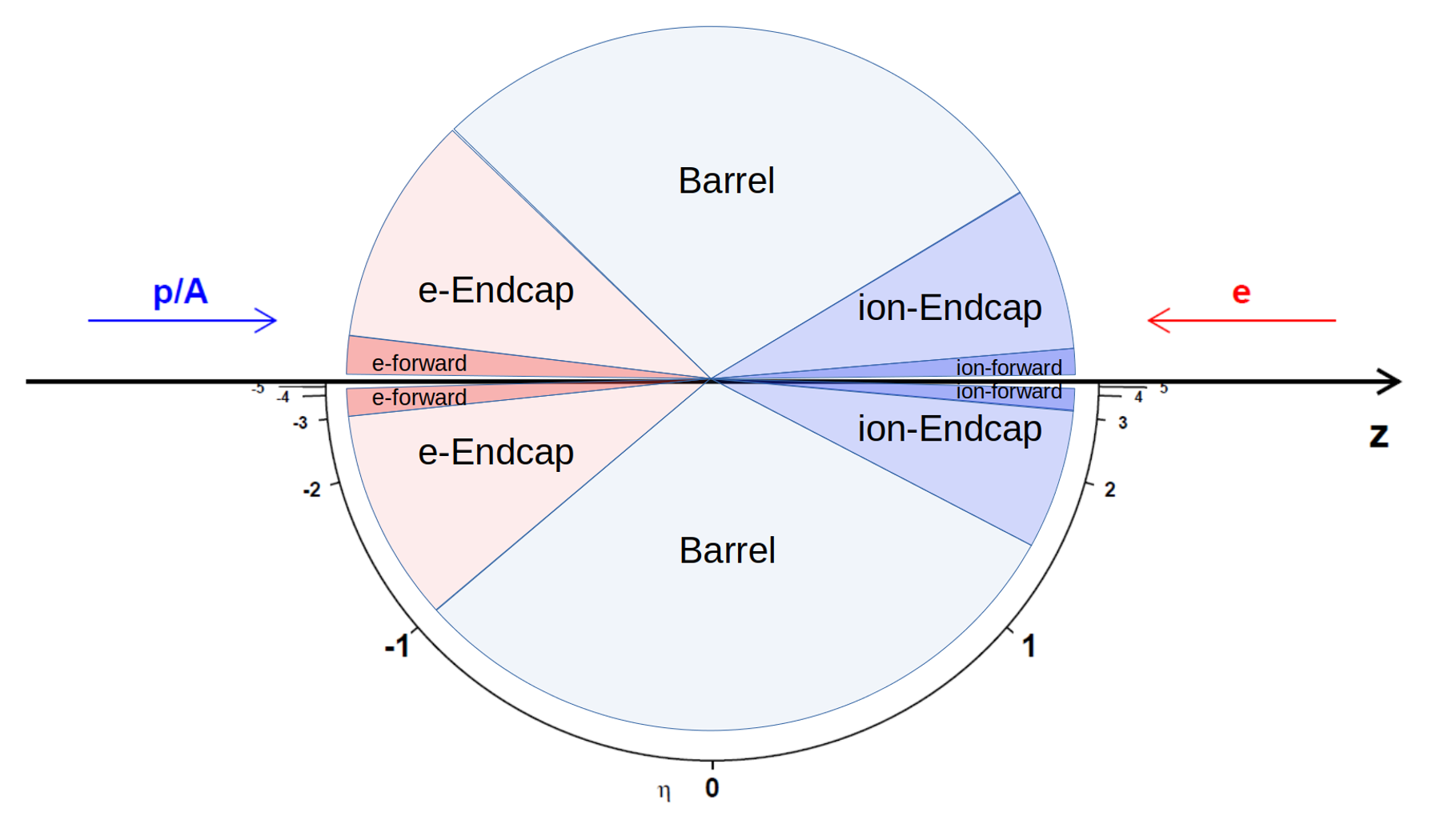}
\vspace*{-0.cm}
\caption{\label{fig:chp11_coor_def}Illustration of the coordinate system for EicC.
}
\end{figure}

Figure~\ref{fig:chp11_coor_def} shows the definition of the coordinate system, where the electron beam points into the negative $z$ direction. The pseudo-rapidity axis is shown as the half-circle. The acceptance of the detector segments is only meant for illustration.

\section{Detector performance requirements}

Based on the EicC baseline design, a PYTHIA~\cite{Sjostrand:2000wi} simulation is performed. In the EicC baseline design, an electron beam energy is 3.5 GeV, proton beam energy of 20 GeV, results in a center of mass energy of 16.7 GeV and a cross-section of 20.8 $\mu$b. The luminosity is expected to be $L = 4\times10^{33}cm^{-2}s^{-1}$  with an interaction rate of 83.2 kHz. The PYTHIA simulation shows the final state particles are concerntrated near $\eta$ = 1, with a particle density of $dN/d\eta dt = 8 \times 10^4/s$. This yields a moderate event rate that has to be considered in the design of detectors and the data acquisition system. 

\begin{figure}[tb]
\centering
\includegraphics[width=1\textwidth,angle =0]{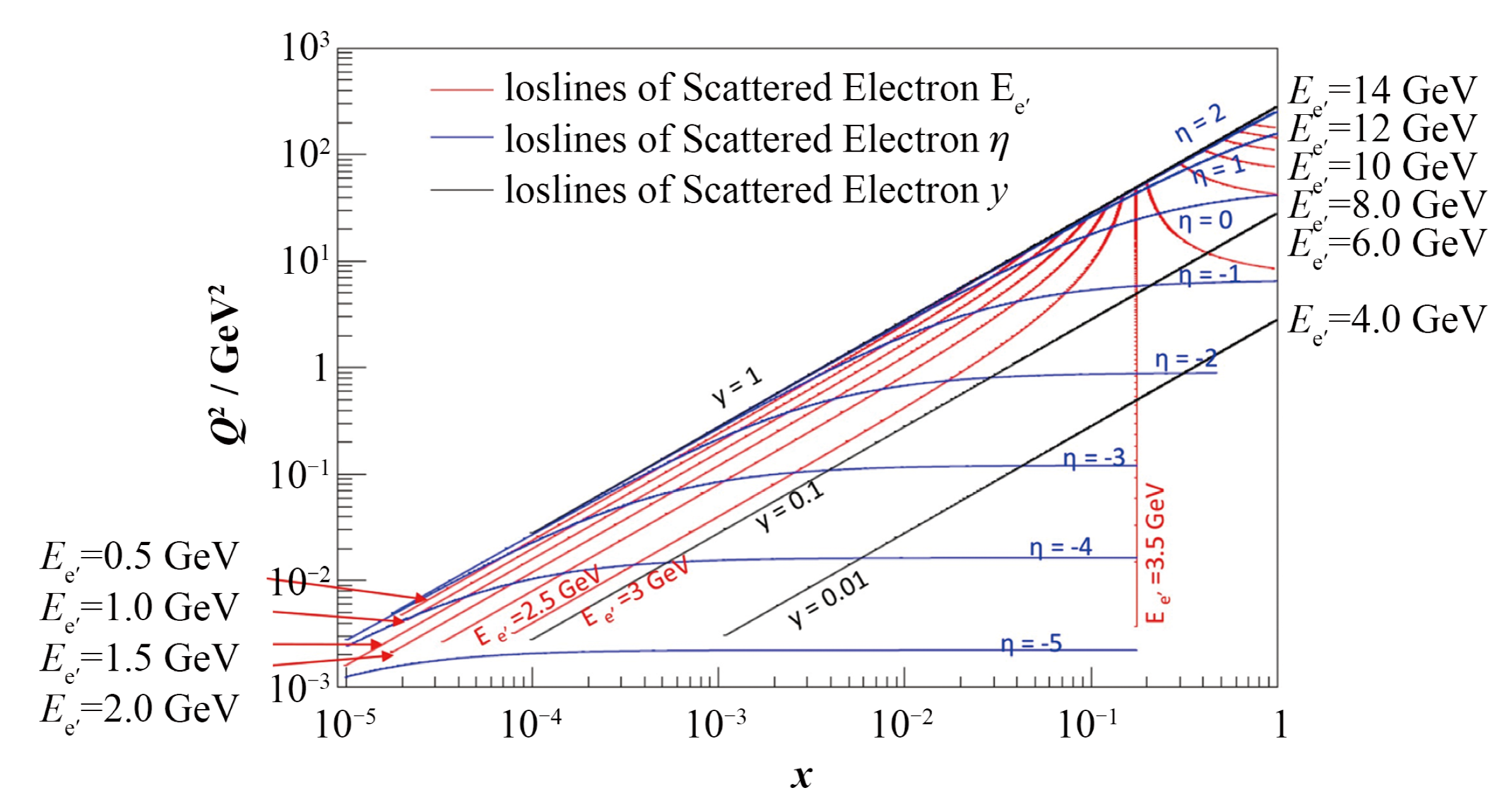}
\vspace*{-0.cm}
\caption{\label{fig:chp11_kinematics}Isolines of the scattered electron energy, pseudo-rapidity and inelasiticity.
}
\end{figure}

For electron-ion collisions, the detection of the scattered electrons play an important role in most of the physics programs. As shown in Fig.~\ref{fig:chp11_kinematics}, the isolines of energy, pseudo-rapidity, and inelasticity of the scattered electrons are drawn in the $x$-$Q^2$ space. Here $x$ is the Bjorken variable and $Q^2$ is the momentum transfer. The red lines show the iso-lines of the energy of the scattered electrons. Inelastic scattering of electrons off protons can be interpreted as elastic scattering of electrons off partons inside the proton. For $x= 0.175$ , which means that the parton carries the same momentum of 3.5~GeV as the electron, elastic scattering will result in constant 3.5~GeV momentum of the scattered electron. This leads to the vertical line at $x=0.175$, separating two groups of isolines. On the left-hand side, with $x<0.175$, the low momentum parton cannot change the trajectory of the electron significantly, which results in low $Q^2$ and very forward going electrons at large rapidity. In contrast, on the right-hand side of the vertical line, with $x>0.175$, the high momentum parton can boost the electron to a very high momentum. The blue lines in the figure show the pseudo-rapidity isolines. At very small momentum transfer $Q^2$, the scattered electron is expected to be in the extreme forward direction (see the low $Q^2$ region with $\eta = -5$ as an example). This analysis shows instructive information about the kinematics for $ep$ collisions.

\subsection{Distributions of the final state particles}

\begin{figure}[tbp]
\centering
\includegraphics[width=1\textwidth,angle =0]{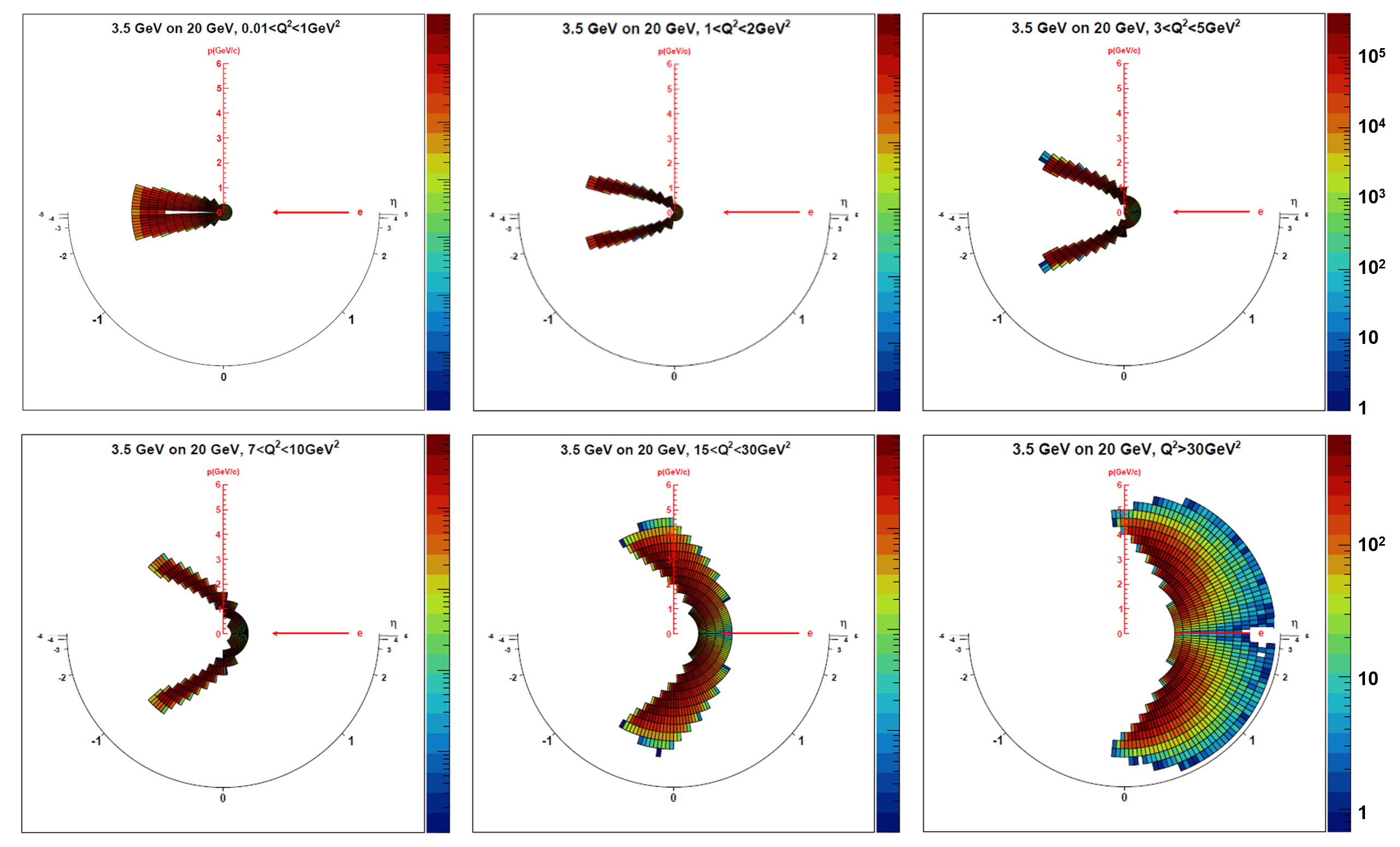}
\vspace*{-0.cm}
\caption{\label{fig:chp11_phsp_2D_Q2_bins}Kinematics of the scattered election at various $Q^2$ bins.
}
\end{figure}

Among the final state particles, the scattered electrons provide crucial information to most of the physics programs, in particular those focusing on the processes of DIS, SIDIS, DVCS, and so on. We have shown very instructive information from a general behavior analysis in Fig.~\ref{fig:chp11_kinematics}.  Here, with a detailed study, more information about the kinematics of the scattered electrons can be obtained. In Fig.~\ref{fig:chp11_phsp_2D_Q2_bins}, distributions of the scattered electrons are shown for various $Q^2$ bins. In these plots, the color scale reflects the cross-section. The distance from the center point denotes the magnitude of the scattered electron momentum, and the direction reflects the pseudo-rapidity. The events were generated with Pythia~6~\cite{Sjostrand:2006za}. Comparing the plots with various $Q^2$ bins, it is noticed that with increasing $Q^2$, the scattered electrons are less boosted to negative pseudo-rapidity. For physics requiring $Q^2$ larger than 1 GeV$^2$ (such as SIDIS or DVCS), a detector coverage of $\eta > -2$ is sufficient for the scattered electron.

\begin{figure}[htbp]
\centering
\includegraphics[width=0.9\textwidth,angle =0]{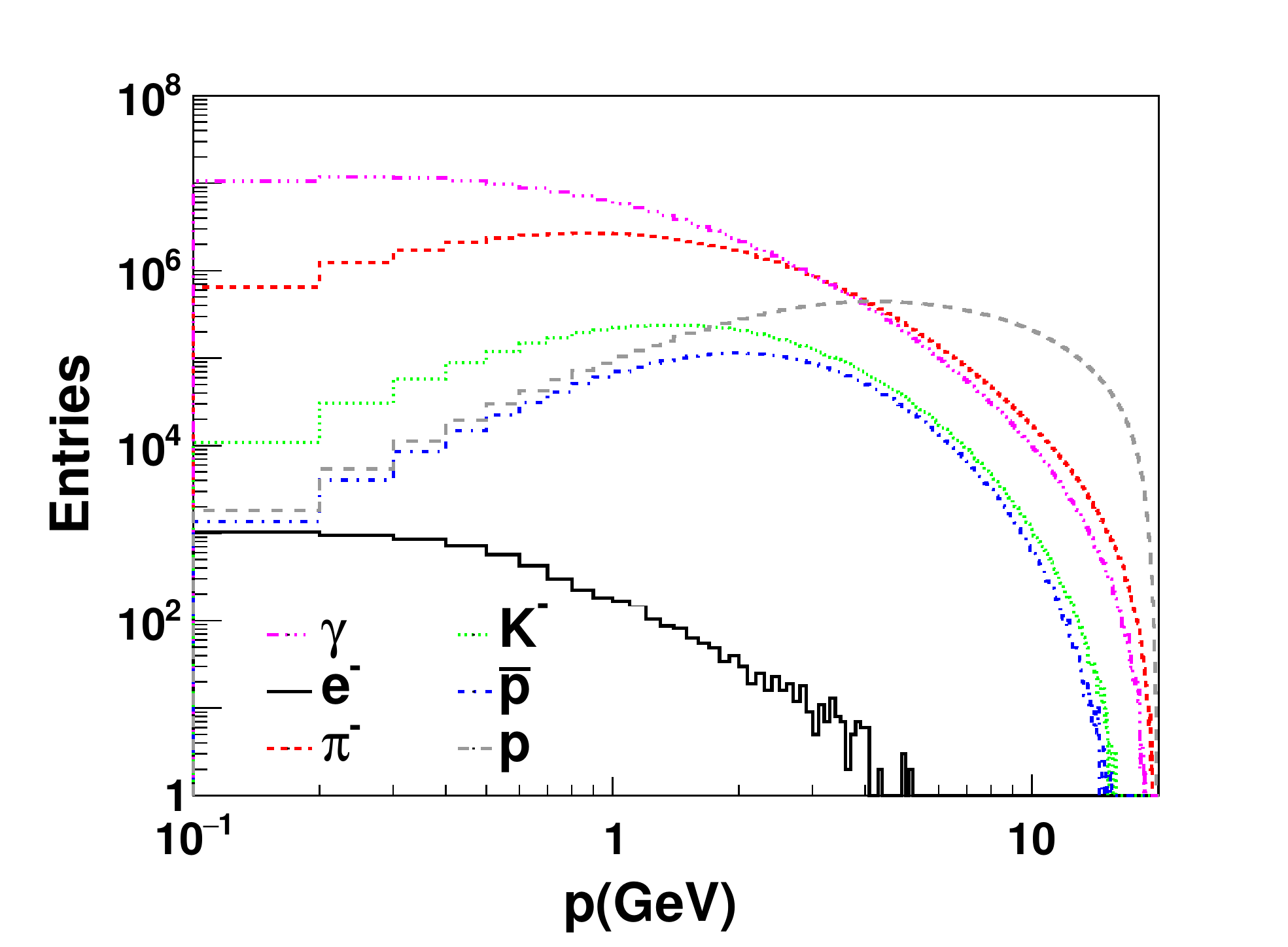}
\vspace*{-0.cm}
\caption{\label{fig:chp11_hadron_momentum_eta_2_3}
Momentum distributions for the final state $\pi^{-}$(red), $K^{-}$(green), $\bar{p}$(blue), proton(grey), $e^-$(black), and $\gamma$(purple) at a given pseudo-rapidity bin $2<\eta<3$.
}
\end{figure}

In addition to scattered electrons, the other final state particles are also important and have been studied. In Fig.~\ref{fig:chp11_hadron_momentum_eta_2_3}, the momentum distributions for the final state $\pi^{-}$(red), $K^{-}$(green), $\bar{p}$(blue), proton(grey), $e^-$(black), and $\gamma$(purple) at a given pseudo-rapidity bin are shown. By comparing the yields of these particles, it is noticed that the number of  final state pions is about 1-2 orders of magnitude larger than those of kaons and anti-protons. The momenta of hadrons in other $\eta$ regions are also investigated. At a pseudo-rapidity smaller than 1, the momenta of final state hadrons are expected to be smaller than 6~GeV, while at large pseudo-rapidity ($\eta > 2$), the final state hadrons have momenta up to 15~GeV. Thus, the detection and particle identification of the final state hadrons needs to be considered at various pseudo-rapidity regions.

\begin{figure}[tbh]
\centering
\includegraphics[width=0.95\textwidth,angle =0]{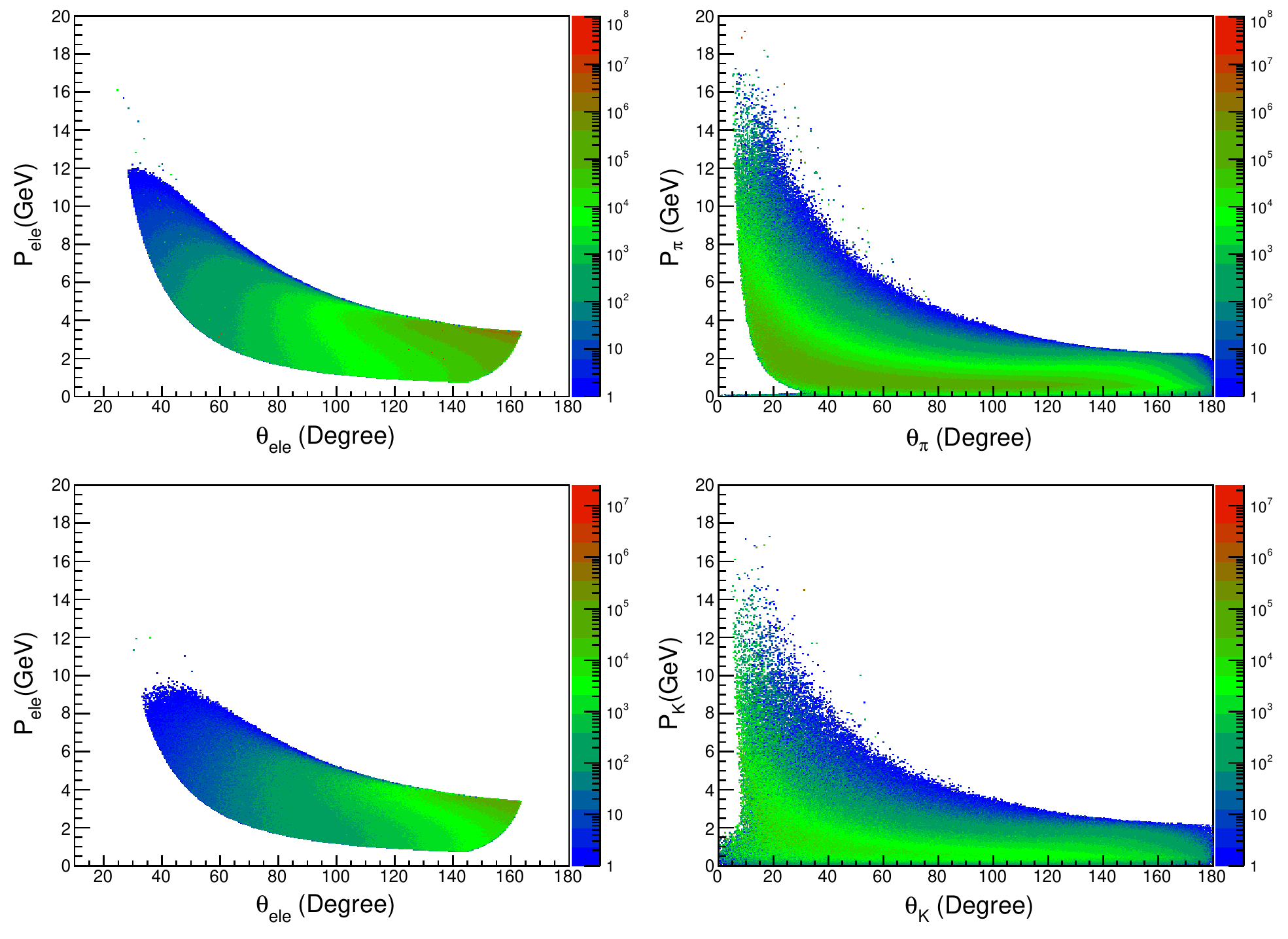}
\caption{\label{fig:chp11_SIDIS}
Distribution of the scattered electrons and charged pions (the upper two plots) and Kaons (the lower two plots) in SIDIS processes. $Q^2 > 1$~GeV$^2$ is applied. 
}
\end{figure}

\begin{figure}[tbh]
\centering
\includegraphics[width=0.99\textwidth,angle =0]{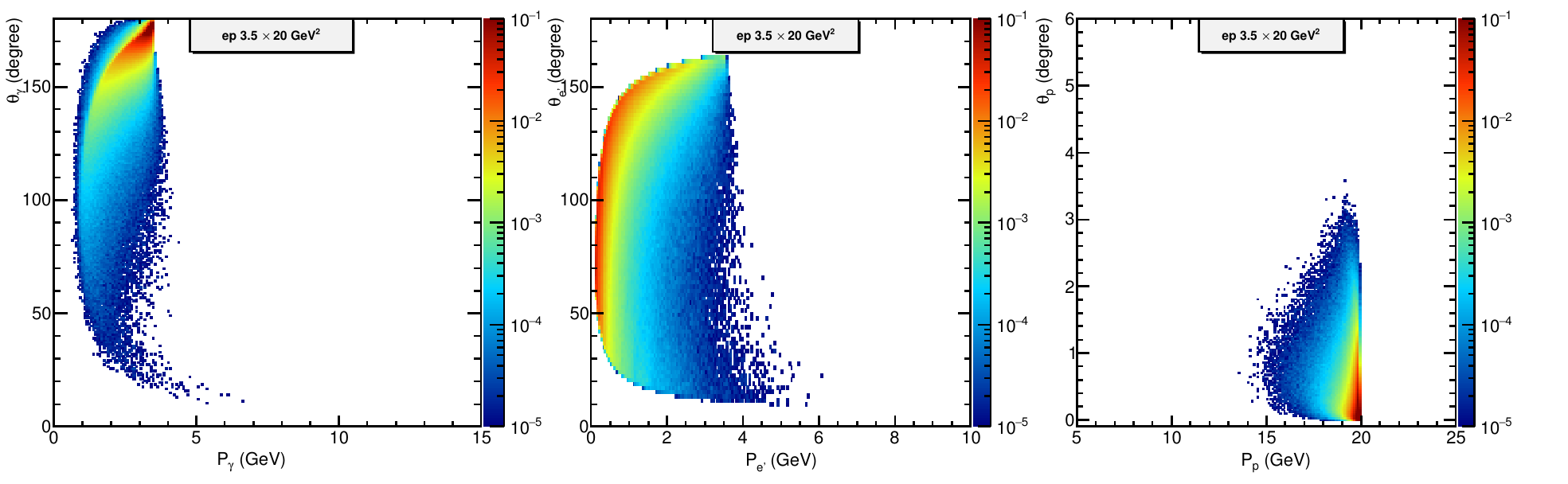}
\vspace*{-0.cm}
\caption{\label{fig:chp11_DVCS}
Distributions of the scattered electrons, protons and photons in the DVCS process.
}
\end{figure}

In addition to the above study of events produced with the Pythia generator, dedicated studies are performed for, e.g., the study of SIDIS or DVCS processes, which provide direct constraints for the detector design. Figure~\ref{fig:chp11_SIDIS} shows the distribution of scattered electrons, charged pions and kaons in the SIDIS processes. From these plots, high momenta larger than 15 GeV for the pions and kaons at small angles in the ion forward direction are observed. This is consistent with the previous Pythia simulation. These SIDIS processes set very basic requirements for the central detector. Figure~\ref{fig:chp11_DVCS} shows the distributions of the scattered electrons, protons, and photons in the DVCS process. Here, the detection of extremely forward going protons in the DVCS process needs special consideration. A dedicated device, Roman Pot~\cite{Minafra:2015gfa}, can be installed to detect these small-angle protons. Furthermore, the final state neutrons are important for some physics programs, such as the meson structure function measurements, in which a neutron is found at extreme forward angles near the proton beam. This indicates that a special neutron detection is needed for EicC to carry out this physics program. These small-angle neutrons can be detected with zero-degree calorimeters after the analysing dipole in the far ion-forward region. In the meson structure section, the Kaon structure function is mentioned. With the Sullivan process, the Kaon structure function can be measured, in which a forward Lambda baryon needs to be reconstructed. With the current detector design, the proton and pion from Lambda decay could be partly detected by the forward detectors. Detailed studies on the detection efficiency and background level are needed to provide guidance to optimize the detector design and to make clear statement on the physics potential.

In EicC, exotic hadron production is one of the important physics highlights. Figure~\ref{fig:PbPc} shows the distributions of various pentaquarks in EicC. To reconstruct these pentaquarks, the final state particles, such as protons and lepton pairs from $J/\psi$ ($\Upsilon$), or charged pions, kaons, photons for reconstruction of open-charm (open-bottom) hadrons, are under investigation.

\subsection{Luminosity and polarization measurements}

For a collider such as EicC, the precise knowledge of the beam luminosity is essential for any type of cross-section measurements. For electron-proton collisions, e.g. at HERA, the luminosity can be continuously monitored via the bremsstrahlung process, $ep\rightarrow ep\gamma$. This is a pure QED process where the cross-section is large and well known (~0.2\%). The luminosity can be quickly and precisely measured from the photon rate at very small angles by using a calorimeter system installed next to the beamline in the electron  direction. The cross-section of the bremsstrahlung process also depends on the beam polarization. 

The precise investigation of nucleon structure in the sea quark region, including one-dimensional and three-dimensional tomography of the nucleon spin-flavor structure in both the momentum and spatial spaces is one of the featured physics highlights of EicC. For any of these spin related measurements, the final results always need to be normalized to the beam polarization. Since an unprecedented statistical precision is expected from the proposed high luminosity of EicC, the precision of the beam polarization measurement becomes very demanding. For the electron beam, the polarization can be measured from different QED processes, e.g., the electron-photon Compton scattering and the electron-electron M\o{}ller scattering. The spin dependent cross-section and the analyzing power of these processes can be precisely obtained from QED calculations. Together with the known polarization of the polarimeter targets, the beam polarization can thus be extracted. Taking CEBAF at Jefferson Lab, for instance, there are three types of electron beam polarimeter which are Mott~\cite{Steigerwald:2001mt}, M\o{}ller~\cite{Hauger:1999iv}, and Compton~\cite{MAGEE2017339}, respectively, providing polarization measurements as precise as 1\% at different positions along the beamline. The Compton polarimeter uses a highly polarized and high power laser as the scattering target which is noninvasive to the electron beam so that it can continuously monitor the beam polarization. The Mott and M\o{}ller polarimeters use gold and iron foils, respectively, as the scattering targets which have very high rates and are able to perform precise measurements in a short period of time. 

For the proton beam, the polarization can be measured from elastic proton-proton or proton-nucleus scattering, where large transverse spin asymmetries are expected. For example, at RHIC two types of polarimeter are used to perform proton beam polarization measurements. One of them, the $p$-Carbon polarimeter~\cite{NakagawaAIP}, uses carbon fibers as the scattering target with large scattering rates, periodically measuring the polarization. Another one is the H-jet polarimeter~\cite{PhysRevD.79.094014}, which  continuously performs noninvasive polarization measurements, just analogous to the Compton polarimeter for the electron beam. The proton beam polarization at RHIC can be measured with a precision at the $\sim$3\% level.

\section{Detector conceptual design}

\begin{figure}[htbp]
\centering
\includegraphics[width=1\textwidth,angle =0]{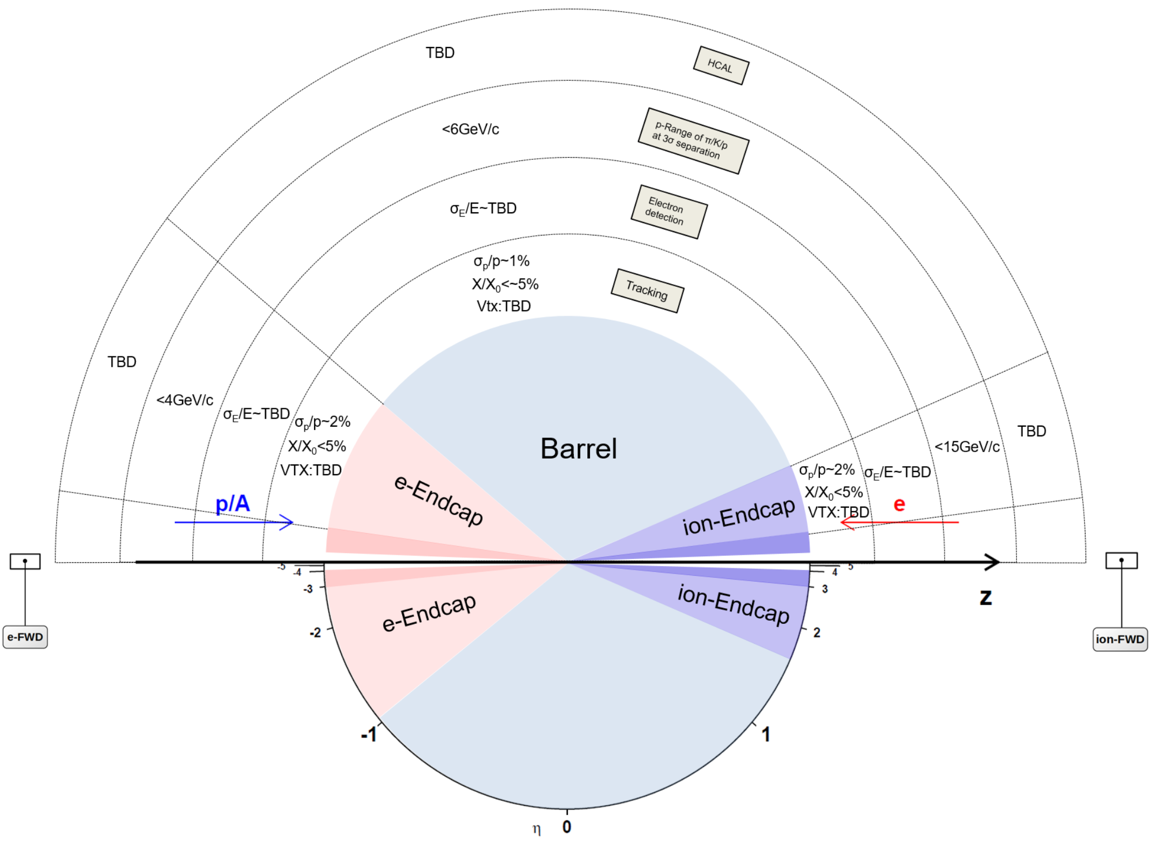}
\vspace*{-0.cm}
\caption{\label{fig:chp11_det_reqirement_summary}Physics requirements for EicC.
}
\end{figure}

In the previous section, we showed the requirements derived from the key physics programs at EicC. A summary table is given in Fig.~\ref{fig:chp11_det_reqirement_summary}, in which the physics requirements are listed in terms of momentum or energy reach at different pseudo-rapidity coverage. For example, the hadron separation power at 4~GeV is sufficient for the $e$-Endcap region. However, hadrons with momenta up to 15~GeV are expected for the ion-Endcap region. These differences indicate that different detection techniques need to be adopted in the detector design. 

As a high luminosity machine, EicC could reduce the statistical uncertainty down to a few percent for many measurements. To cope with the small statistical uncertainty, we need a matching systematic error, which requires a good detector. For example, a tracking detector with a tracking resolution of a few percent is necessary. In Fig.~\ref{fig:chp11_det_reqirement_summary}, a momentum resolution of 1\% for the central coverage and 2\% for small angles are marked based on experiences from similar experiments. For the first conceptual design, we divide the EicC detector into the central detector and forward detectors. The central detector consists of the barrel part and two endcaps, and it will be constructed around a solenoid. Four main detection components include:

\begin{enumerate}
\item[(1)] Vertex detector, for detecting the primary and secondary vertices. MAPS~\cite{Contin:2016jff} based vertex detectors have been used in many experiments and can be considered for EicC.
\item[(2)] Tracking detector, for the momentum reconstruction of charged particles. TPC~\cite{DUPRE201890}, GEM~\cite{SAULI20162}, or Straw Tube detectors~\cite{Erni:2013ita} can be considered for EicC.
\item[(3)] Particle identification detectors, such as Time-of-Flight (TOF) detectors and Cerenkov detectors.
\item[(4)] Calorimeter, including electro-magnetic calorimeter and hadron calorimeters.
\end{enumerate}

\begin{figure}[tb]
\centering
\includegraphics[width=0.9\textwidth,angle =0]{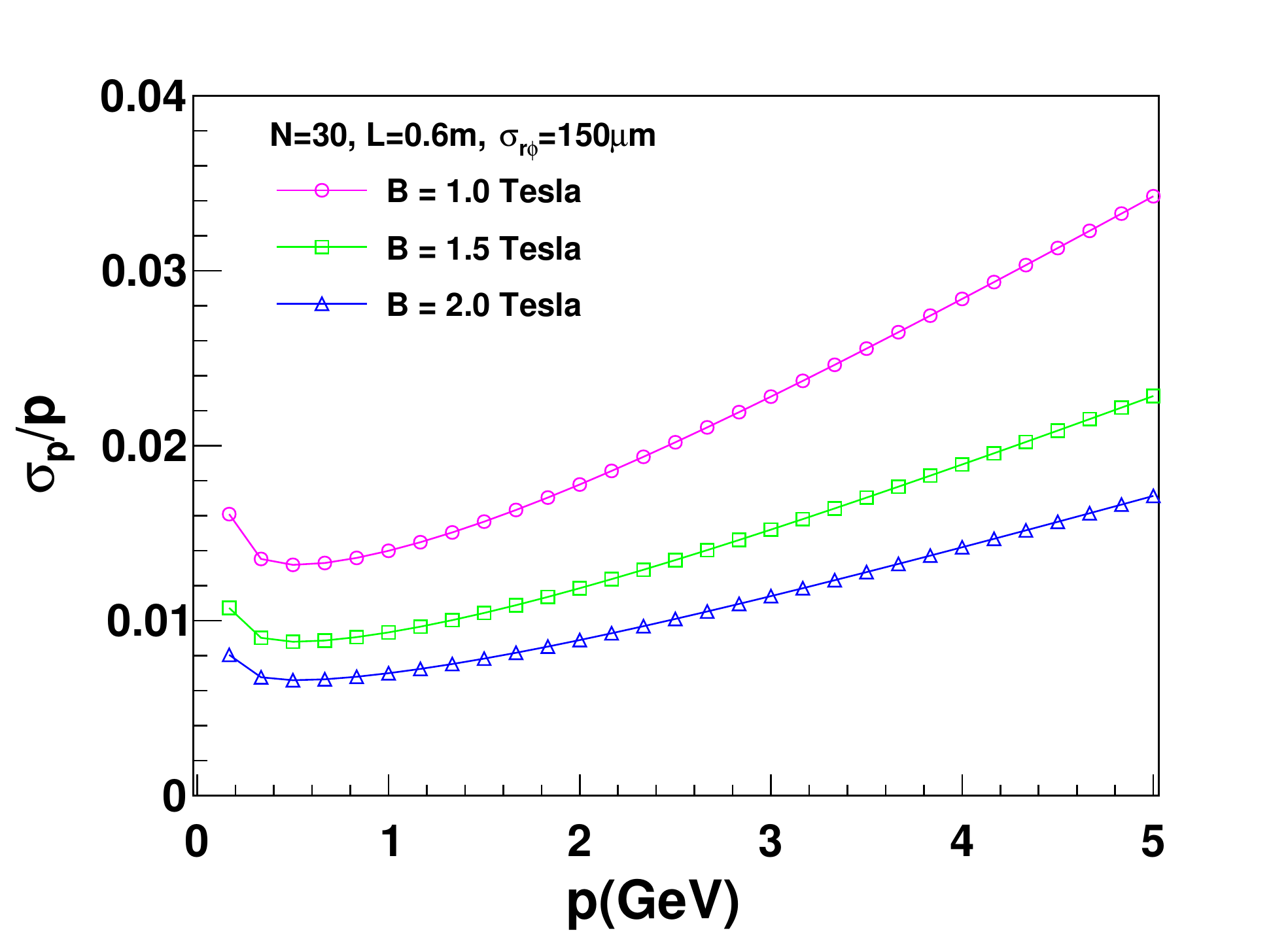}
\vspace*{-0.cm}
\caption{\label{fig:chp11_det_mom_res}Momentum resolution for tracking detectors at different magnetic fields.
}
\end{figure}

For the tracking detector, a momentum resolution of roughly 1\% at 1 GeV with  energy deposition ($dE/dx$) measurement is widely used in many experiments~\cite{ABLIKIM2010345,VanHaarlem:2010yq}. For the central rapidity coverage, the tracking detector is usually installed inside a solenoid magnet. As one of the key components of the central spectrometer, the solenoid magnet provides the bending power for charged particles inside a tracking detector. Figure~\ref{fig:chp11_det_mom_res} shows the impact of different magnet fields to the momentum resolution with the other parameters fixed. A solenoid magnet of 1 to 2~Tesla, well used in many experiments~\cite{ABLIKIM2010345,SMYRSKI201285} with adequate power to allow for a good momentum resolution and less challenging in manufacturing, is fine for EicC. For the small-angle coverage in the central spectrometer, certain layers of position-sensitive detectors can be adopted, such as the GEM disks or silicon detectors.

\begin{figure}[tb]
\centering
\includegraphics[width=0.85\textwidth,angle =0]{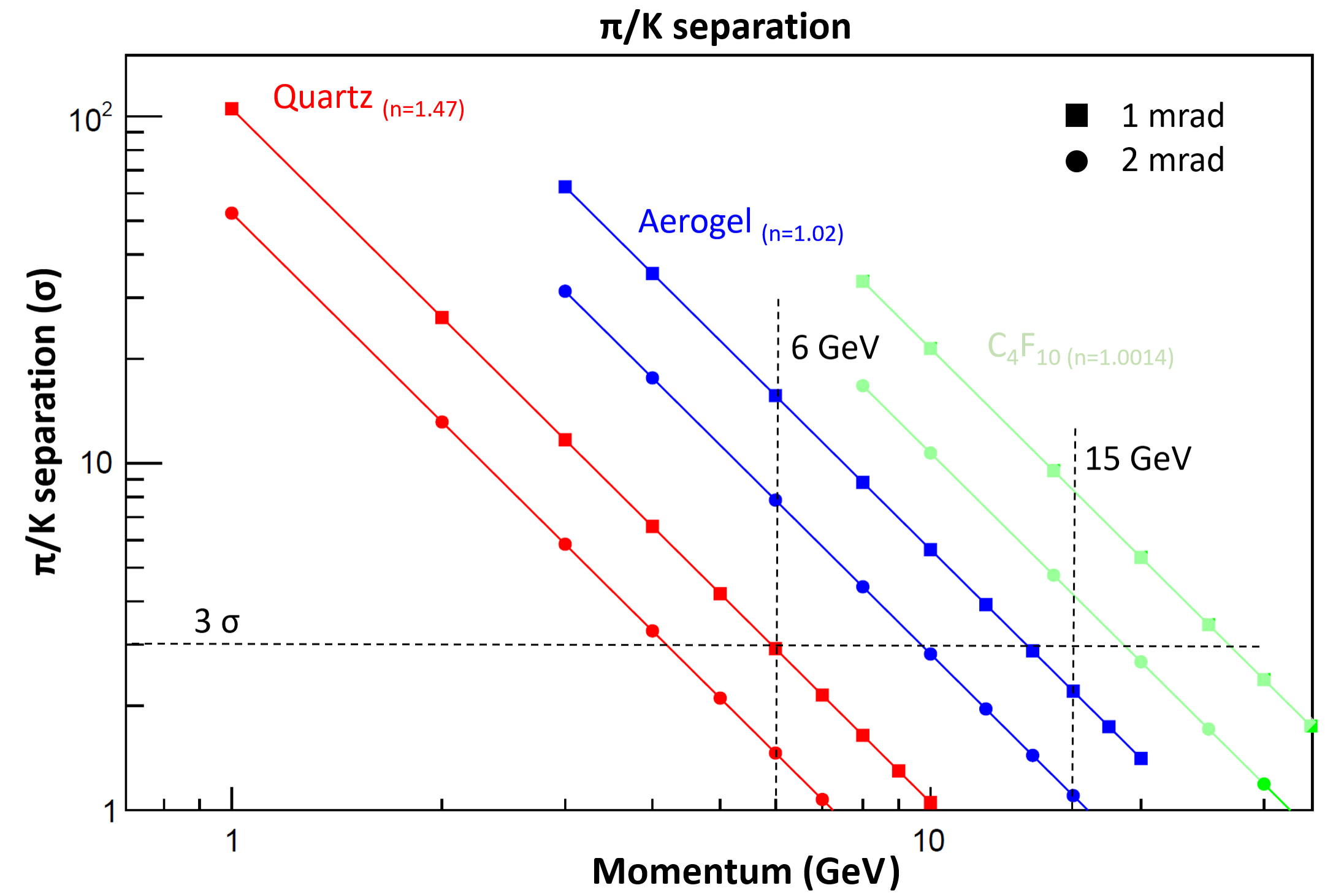}
\vspace*{-0.cm}
\caption{\label{fig:chp11_det_RICH} $\pi/K$ separation power with a Cherenkov detector of different refractive indices.
}
\end{figure}

One challenge for the EicC detector is the hadron identification, especially in the ion forward direction where the hadrons may have large momenta up to 15~GeV. A series of detectors may serve the particle identifications. The energy deposition ($dE/dx$) measurements from tracking detectors could provide particle identification in the region of low momenta around a few hundred MeV. A TOF detector with a time resolution of a few tens of picoseconds and a flight distance of about 2-3 meters (the forward region) could extend the particle identification up to a few GeV. To reach large momenta up to 15~GeV, a Cherenkov detector~\cite{Barion:2020iqw} with a small refractive index ($n = 1.02$), such as the aerogel RICH (Ring Imaging Cherenkov detectors), is a good candidate. As the blue lines shown in Fig.~\ref{fig:chp11_det_RICH}, with a 1~mrad Cherenkov angle resolution, an aerogel RICH of $n = 1.02$ could achieve a 3$\sigma$ separation of the $\pi/K$ mesons up to almost 15~GeV. A Cherenkov detector with a further small refractive index, such as $n = 1.0014$ with $C_4F_{10}$, covering high momentum range as shown with the green lines in Fig.~\ref{fig:chp11_det_RICH}, is unnecessary in EicC. For the other regions, where the hadron momenta are smaller than 6 GeV, a compact DIRC (Detection of Internally Reflected Cherenkov light) detector~\cite{ADAM2005281,Ali_2020} could be used. As the red lines shown in Fig.~\ref{fig:chp11_det_RICH}, with a 1~mrad (round filled dot) Cherenkov angle resolution, the $\pi$/K separation power at 3$\sigma$ for a DIRC detector of $n = 1.47$ could reach up to 6~GeV. 

For EM-Calorimeters, different types of calorimeters should be adopted at different rapidity coverage. A typical configuration could be sampling calorimeters for most of the regions, featuring relatively low-performance at a low price, and costly homogenous crystal calorimeters for small region at the $e$-Endcap side. A compromise between physics and budget needs to be found in the deployment of lower price and higher-quality calorimeters in different regions.

\begin{figure}[htbp]
\centering
\includegraphics[width=1.6\textwidth,angle =90]{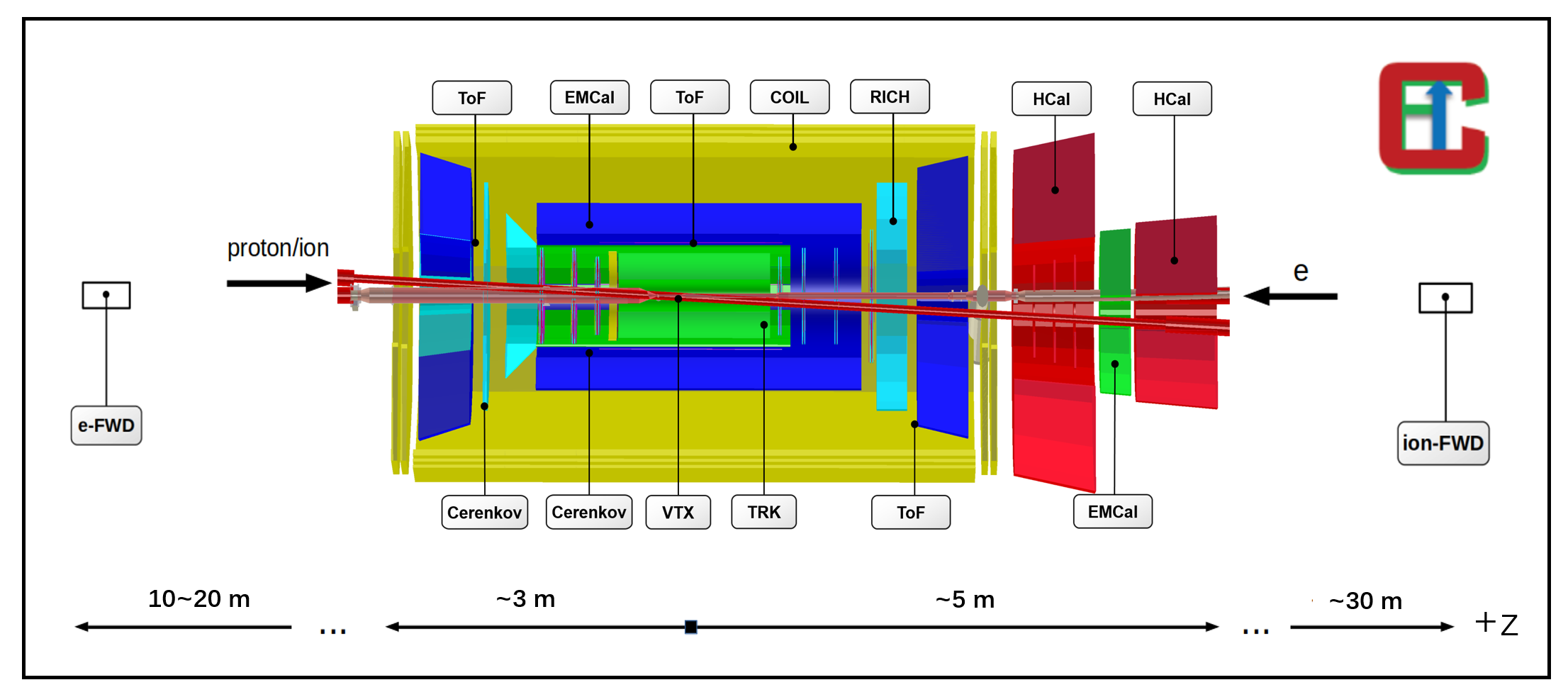}
\caption{\label{fig:detector_conceptual}
Conceptual design of the EicC detector.
}
\end{figure}

A conceptual design of the EicC detector is shown in Fig.~\ref{fig:detector_conceptual}. In the current conceptual design, forward detectors are also considered at small angles. These detectors are crucial to many important physics programs of EicC. As a doubly polarized high luminosity machine, the polarization and luminosity measurement are important and are designed with integration to the forward detectors. Starting from the conceptual design, the detector requirements for each sub-detector will be derived iteratively with detector and physics simulations.

As a general-purpose spectrometer, the EicC detector design is facing difficulties and challenges. Detector R\&D will be started at the early stage, including the R\&D of Cherenkov detector, tracking detector, calorimeters, super-conducting solenoid, the data acquisition system, etc.

\section*{Acknowledgement}
In 2018, we started the process of writing an EicC White Paper in order to, 
on one hand, unite the Chinese QCD and hadron physics community and, 
on the other hand, to establish an electron-ion collider whose somewhat 
lower center-of-mass energy serves to complement the US EIC physics 
program.  It would not have been possible for us to complete this document 
within two-and-a-half years without generous assistance from many 
colleagues all over the world.  We are grateful for support from the 
Institute of Modern Physics; and sincerely thank S. J. Brodsky, J.-P. Chen, 
A. Deshpande, H. Y. Gao, S. Goloskokov, T. Horn, X. D. Ji, S. Joosten, V. Kubarovsky, Z. E. Meziani, 
J. W. Qiu, E. Sichtermann, Z. H. Ye, F. Yuan, Y. S. Yuan, J. X. Zhang, X. B. Zhao, 
Z. W. Zhao and F. Zimmermann for helpful discussions and valuable advice. 
We would also like to thank the referees from Frontiers of Physics journal 
for their critical reading of the early version of the document and constructive suggestions.
During the studies necessary for the preparation of this document, we 
were granted access to an array of software packages written by others, including 
DJANGOH, DSSV14, eSTARlight, LAGER, LHAPDF, MILOU, NNPDF, PARTONS, and Pythia. 


\bibliographystyle{unsrt}
\bibliography{reference}


\end{document}